\documentclass[11pt,a4paper]{report}
\usepackage{axodraw}
\newcommand{\ps}{p\hspace{-0.42em}/}
\newcommand{\qs}{q\hspace{-0.42em}/}
\newcommand{\rs}{r\hspace{-0.42em}/}

\def\nl{\nonumber\\}

\def\nln{\nonumber\\*[-1ex]\phantom{\fbox{\rule{0em}{2ex}}}}

\def\beq{\begin{equation}}
\def\eeq{\end{equation}}
\def\beqar{\begin{eqnarray}}
\def\eeqar{\end{eqnarray}}
\def\barr#1{\begin{array}{#1}}
\def\earr{\end{array}}
\def\bfi{\begin{figure}}
\def\efi{\end{figure}}
\def\btab{\begin{table}}
\def\etab{\end{table}}
\def\bce{\begin{center}}
\def\ece{\end{center}}
\def\nn{\nonumber}

\def\text{\textstyle}
\def\arraystretch{1.4}

\def\de{\delta}

\def\la{\lambda}
\def\si{\sigma}
\def\Si{\Sigma}

\def\De{\Delta}

\def\refeq#1{\mbox{(\ref{#1})}}
\def\reffi#1{\mbox{Fig.~\ref{#1}}}
\def\reffis#1{\mbox{Figs.~\ref{#1}}}

\def\refse#1{\mbox{Sect.~\ref{#1}}}
\def\refses#1{\mbox{Sects.~\ref{#1}}}

\def\refapp#1{\mbox{Appendix~\ref{#1}}}
\def\citere#1{\mbox{Ref.~\cite{#1}}}
\def\citeres#1{\mbox{Refs.~\cite{#1}}}

\def\refch#1{\mbox{Ch.~\ref{#1}}}

\def\solid{\raise.9mm\hbox{\protect\rule{1.1cm}{.2mm}}}
\def\dash{\raise.9mm\hbox{\protect\rule{2mm}{.2mm}}\hspace*{1mm}}

\newcommand{\GeV}{\unskip\,\mathrm{GeV}}

\newcommand{\TeV}{\unskip\,\mathrm{TeV}}

\def\mathswitchr#1{\relax\ifmmode{\mathrm{#1}}\else$\mathrm{#1}$\fi}

\newcommand{\PW}{\mathswitchr W}
\newcommand{\PZ}{\mathswitchr Z}

\newcommand{\PA}{\mathswitchr A}

\newcommand{\PH}{\mathswitchr H}
\newcommand{\Pe}{\mathswitchr e}

\newcommand{\Pd}{\mathswitchr d}
\newcommand{\Pf}{\mathswitchr f}
\newcommand{\Pfbar}{\bar \mathswitchr f}

\newcommand{\Pu}{\mathswitchr u}
\newcommand{\Pp}{\mathswitchr p}
\newcommand{\Ps}{\mathswitchr s}
\newcommand{\Pb}{\mathswitchr b}
\newcommand{\Pc}{\mathswitchr c}
\newcommand{\Pt}{\mathswitchr t}

\newcommand{\Pep}{\mathswitchr {e^+}}
\newcommand{\Pem}{\mathswitchr {e^-}}

\newcommand{\PWp}{\mathswitchr {W^+}}
\newcommand{\PWm}{\mathswitchr {W^-}}
\newcommand{\PWpm}{\mathswitchr {W^\pm}}

\def\mathswitch#1{\relax\ifmmode#1\else$#1$\fi}

\newcommand{\MW}{\mathswitch {M_\PW}}

\newcommand{\MZ}{\mathswitch {M_\PZ}}
\newcommand{\MH}{\mathswitch {M_\PH}}

\newcommand{\Mb}{\mathswitch {m_\Pb}}
\newcommand{\Mt}{\mathswitch {m_\Pt}}

\newcommand{\thw}{\mathswitch {\theta_\mathrm{w}}}
\newcommand{\NCf}{\mathswitch {N_{\mathrm{C}}^f}}
\newcommand{\NCt}{\mathswitch {N_{\mathrm{C}}^t}}
\newcommand{\scrs}{\scriptscriptstyle}
\newcommand{\sw}{\mathswitch {s_{\scrs\PW}}}
\newcommand{\cw}{\mathswitch {c_{\scrs\PW}}}

\newcommand{\bew}{b^{\ew}}
\newcommand{\besw}{\tilde{b}^{\ew}}
\newcommand{\cew}{C^{\ew}}
\newcommand{\ctwo}{C^{\SUtwo}}
\newcommand{\csew}{\tilde{C}^{\ew}}
\newcommand{\dew}{D^{\ew}}

\newcommand{\ckm}{{\bf V}}

\newcommand{\vev}{{\bf v}}

\def\ie{i.e.\ }

\def\cf{cf.\ }

\renewcommand{\O}{{\cal O}}

\newcommand{\LA}{\stackrel{\mathrm{LA}}{=}}

\newcommand{\Tr}{\mathrm{Tr}}

\newcommand{\U}{\mathrm{U}}
\newcommand{\SUtwo}{\mathrm{SU(2)}}
\newcommand{\Uone}{\mathrm{U}(1)}
\newcommand{\ewgroup}{\SUtwo\times\Uone}

\newcommand{\rD}{\mathrm{D}}
\newcommand{\rR}{\mathrm{R}}
\newcommand{\rL}{\mathrm{L}}
\newcommand{\rT}{{\mathrm{T}}}

\newcommand{\ri}{\mathrm{i}}

\newcommand{\rd}{{\mathrm{d}}}

\newcommand{\coll}{{\mathrm{coll}}}

\newcommand{\elm}{\mathrm{em}}
\newcommand{\ew}{\mathrm{ew}}

\newcommand{\Htop}{\PH,\Pt}

\newcommand{\M}{{\cal {M}}}
\renewcommand{\L}{{\cal L}}
\newcommand{\F}{{\cal {F}}}
\newcommand{\D}{{\cal {D}}}

\newcommand{\DL}{{\mathrm{DL}}}

\newcommand{\SC}{{\mathrm{LSC}}}
\renewcommand{\SS}{{\mathrm{SSC}}}
\newcommand{\cc}{{\mathrm{C}}}

\newcommand{\pre}{{\mathrm{PR}}}
\newcommand{\Yuk}{{\mathrm{Yuk}}}

\newcommand{\ddl}{\frac{\mathrm{d}^Dl}{(2\pi)^D}}
\newcommand{\ddq}{\frac{\mathrm{d}^Dq}{(2\pi)^D}}
\newcommand{\brs}{\it{s}}

\def\Re{\mathop{\mathrm{Re}}\nolimits}

\newcommand{\Psibar}{\bar{\Psi}}

\newcommand{\ls}{l(s)}

\newcommand{\lsl}{l_{\cc}}
\newcommand{\lpr}{l_{\pre}}
\newcommand{\lYuk}{l_{\Yuk}}
\newcommand{\lZ}{l_{\PZ}}

\newcommand{\lemphi}{l^\elm(m_{\varphi}^2)}

\newcommand{\Ls}{L(s)}

\newcommand{\LrMi}{L(|r_{kl}|,M_i^2)}

\newcommand{\lrMi}{l(|r_{kl}|,M_i^2)}

\newcommand{\Lemphi}{L^\elm(s,\lambda^2,m_\varphi^2)}

\newcommand{\Lemftau}{L^\elm(s,\lambda^2,m_{f_\tau}^2)}
\newcommand{\Leme}{L^\elm(s,\lambda^2,m_e^2)}

\newcommand{\lrs}{\log{\left(\frac{|r_{kl}|}{s}\right)}}

\newcommand{\ltu}{\log{\frac{t}{u}}}
\newcommand{\lts}{\log{\frac{|t|}{s}}}
\newcommand{\lus}{\log{\frac{|u|}{s}}}

\newcommand{\NB}{N}
\newcommand{\sNB}{\tilde{N}}
\newcommand{\sV}{\tilde{V}}
\newcommand{\deone}{\de^{\Uone}}
\newcommand{\detwo}{\de^\SUtwo}
\newcommand{\sdeone}{\tilde{\de}^{\Uone}}
\newcommand{\sdetwo}{\tilde{\de}^\SUtwo}

\def\draftdate{\relax}
\def\mda{\relax}
\def\mua{\relax}
\def\mla{\relax}
\def\draft{
\def\thtystars{******************************}
\def\sixtystars{\thtystars\thtystars}
\typeout{}
\typeout{\sixtystars**}
\typeout{* Draft mode!
         For final version remove \protect\draft\space in source file *}
\typeout{\sixtystars**}
\typeout{}
\def\draftdate{\today}
\def\mua{\marginpar[\boldmath\hfil$\uparrow$]%
                   {\boldmath$\uparrow$\hfil}%
                    \typeout{marginpar: $\uparrow$}\ignorespaces}
\def\mda{\marginpar[\boldmath\hfil$\downarrow$]%
                   {\boldmath$\downarrow$\hfil}%
                    \typeout{marginpar: $\downarrow$}\ignorespaces}
\def\mla{\marginpar[\boldmath\hfil$\rightarrow$]%
                   {\boldmath$\leftarrow $\hfil}%
                    \typeout{marginpar: $\leftrightarrow$}\ignorespaces}
\def\Mua{\marginpar[\boldmath\hfil$\Uparrow$]%
                   {\boldmath$\Uparrow$\hfil}%
                    \typeout{marginpar: $\Uparrow$}\ignorespaces}
\def\Mda{\marginpar[\boldmath\hfil$\Downarrow$]%
                   {\boldmath$\Downarrow$\hfil}%
                    \typeout{marginpar: $\Downarrow$}\ignorespaces}
\def\Mla{\marginpar[\boldmath\hfil$\Rightarrow$]%
                   {\boldmath$\Leftarrow $\hfil}%
                    \typeout{marginpar: $\Leftrightarrow$}\ignorespaces}

\overfullrule 5pt
\oddsidemargin -15mm
\marginparwidth 29mm
}

\makeatletter

\def\eqnarray{\stepcounter{equation}\let\@currentlabel=\theequation
\global\@eqnswtrue
\global\@eqcnt\z@\tabskip\@centering\let\\=\@eqncr
$$\halign to \displaywidth\bgroup\hskip\@centering
  $\displaystyle\tabskip\z@{##}$\@eqnsel&\global\@eqcnt\@ne
  \hskip 2\arraycolsep \hfil${##}$\hfil
  &\global\@eqcnt\tw@ \hskip 2\arraycolsep $\displaystyle\tabskip\z@{##}$\hfil
   \tabskip\@centering&\llap{##}\tabskip\z@\cr}

\makeatother

\begin{document}
\pagestyle{empty}
\begin{titlepage}
\begin{center}
{\Huge{
\textsc{Electroweak Radiative \\Corrections at High Energies}
}\\[15mm]}
{\huge
{\bf  Dissertation} \\[2.5mm]
}{\LARGE{
zur}\\[1.5mm]
{
Erlangung der naturwissenschaftlichen Doktorw\"urde} 
\\[1.5mm]
{
(Dr.~sc.~nat.)}\\[7.5mm]
{
vorgelegt der}\\[1.5mm]
{
Mathematisch-naturwissenschaftlichen Fakult\"at}\\[1.5mm]
{
der}\\[1.5mm]
{
Universit\"at Z\"urich} \\[10mm]
 {
von}\\[1.5mm]
{{\bf Stefano POZZORINI}}\\[1.5mm]
 {
von}\\[1.5mm]
 {
Brissago TI}\\[15mm]
 {
Begutachtet von} \\[1.5mm]
 {
{\bf 
PD.~Dr.~Ansgar DENNER }\\[1.5mm]
 {\bf 
Prof.~Dr.~Daniel WYLER}
}
 \\[15mm]
{Z\"urich 2001}  }
\end{center}
\end{titlepage}

\thispagestyle{empty} 
{\small
\noindent
Die vorliegende Arbeit wurde von der Mathematisch-naturwissenschaftlichen 
Fakult\"at
der Universit\"at Z\"urich auf Antrag von Prof.~Dr.~Daniel Wyler
und von 
Prof.~Dr.~G\"unther Rasche als Dissertation angenommen.
}

\newpage
\hspace{4cm}
\begin{minipage}[t][5cm][r]{9cm}
{``Die wissenschaftliche Einstellung ist gewiss eine der gr\"ossten Errungenschaften der letzten 500 Jahre.
Sie bedeutet eine Haltung der Objektivit\"at.
Sie war eine mensch\-liche Einstellung, bei der es um Bescheidenheit ging und um die St\"arke, die Welt objektiv zu betrachten, das heisst, sie so zu sehen, wie sie ist, und nicht entstellt durch unsere eigenen W\"unsche, \"Angste und Vorstellungen.
Man musste den Mut haben, zu sehen und zu \"uberpr\"ufen, ob die gefundenen Daten unsere Vorstellung best\"atigten oder widerlegten,
und man musste den Mut haben, eine Theorie zu \"andern, wenn die Ergebnisse die Theorie nicht bewiesen.''} \\
Erich Fromm, {\em Die Pathologie der Normalit\"at}
\end{minipage}

\newpage
\pagestyle{plain}
\pagenumbering{roman}
\addcontentsline{toc}{chapter}{Abstract}
\centerline{\Large Abstract}
\vspace{3 cm}

The present work is concerned with the behaviour of one-loop electroweak 
corrections at high energies. 
By high energies we mean the energy range above the electroweak scale, $E \gg \MW$,
which will be explored by  the future particle colliders
such as the LHC or an $\Pep\Pem$ linear collider.
In this regime, radiative corrections are dominated by double or single logarithms 
of the ratio of the energy scale to the weak gauge-boson masses.
These contributions  increase with energy, and  at energies 
$E=0.5-1\TeV$ they typically  amount to 10 percent corrections to the lowest-order 
predictions.

In this PhD thesis  we investigate   
the {\em virtual} part of the electroweak one-loop corrections.
Infrared-finite  predictions can be obtained by including the well-known 
soft-photon Bremsstrahlung corrections.

We consider  electroweak processes involving arbitrary external particles
including chiral fermions, Higgs bosons, transverse and longitudinal gauge bosons.
However, we restrict ourselves to those processes
that are not mass-suppressed in the high-energy limit.
In this case the logarithmic electroweak corrections are {\em universal}, 
in the sense that they can be determined in a process-independent way.
The key feature is that they  originate from  restricted 
subsets of Feynman diagrams and from 
specific regions of loop momenta, so-called {\em leading regions}.

The main part of this work is dedicated to the {\em logarithmic mass singularities}.
These are restricted to Feynman diagrams involving 
virtual electroweak gauge bosons $\gamma,\PZ$ and $\PWpm$ coupled to external particles,
and originate from the region where the momenta of the virtual gauge bosons are  
{\em soft and/or collinear} to an external momentum. They are evaluated 
within the 't~Hooft--Feynman gauge.

We first determine the double-logarithmic mass singularities
originating from {\em soft and collinear} virtual
gauge bosons exchanged between pairs of external particles.
To this end we use the well-known {\em eikonal approximation}. 
The resulting double logarithms depend on the centre-of-mass 
energy as well as on the scattering angles.
 
Then, we determine the single logarithms that originate from 
{\em collinear or soft} virtual gauge bosons.
In particular, we proof the {\em factorization of 
collinear mass singularities} 
originating from loop diagrams involving collinear virtual gauge
bosons coupled to external particles.
As basic ingredients for this proof we derive  specific Ward
identities that we call {\em collinear Ward identities}.
These identities relate 
Green functions with arbitrary external particles
involving a gauge boson collinear to one of these. 
They are derived from the BRS invariance of the spontaneously broken
electroweak gauge theory.

We discuss in detail high-energy processes involving 
{\em longitudinally polarized  gauge bosons}.
These are treated using the
Goldstone-boson equivalence theorem, taking into account
the mixing between gauge bosons and would-be Goldstone bosons
that occurs in higher orders.
In this context,  we stress the role of the broken sector 
of the theory above the electroweak scale.

The remaining large logarithmic corrections 
result from the renormalization of the dimensionless parameters, \ie  
the gauge couplings, the top-quark Yukawa coupling, and the 
scalar self-coupling, at the scale $\MW$.

Finally, we apply our generic results 
in analytical and numerical form
to  following simple scattering processes:
$\Pep\Pem\to \mathrm{f}\mathrm{\bar{f}}$,
$\Pep\Pem\to \PWp\PWm$, $\Pep\Pem\to \PZ\PZ,\PZ\gamma,\gamma\gamma$ and
$\bar{\mathrm{d}}\mathrm{u}\to \PWp\PZ,\PWp\gamma$.

In all derivations we 
use the mass-eigenstate fields of the electroweak theory,
taking  care  of the mixing between the gauge-group eigenstates.
To this end, we transform the generic gauge-group generators
and other group-theoretical quantities in the basis of 
the mass-eigenstate fields. This is described in the appendices,
where  also the Feynman rules and the BRS transformations
are presented in generic  form.

\newpage


 \tableofcontents
\newpage

\chapter{Introduction}\label{ch:intro}

\pagenumbering{arabic}

\newcommand{\at}{{\bf \symbol{64}}\hspace{0.5cm}}
\newcommand{\nli}{\vspace{0.2cm} \newline {\bf \symbol{64}}\hspace{0.5cm}}
\section{Precision tests of the Electroweak Standard Model  at high energies}

%
The Glashow--Salam--Weinberg model \cite{EWSM}, known as the  Electroweak Standard Model 
(EWSM), describes the electromagnetic and the weak interactions through 
 a gauge theory with the symmetry group $\SUtwo_{\mathrm{w}}\times \Uone_Y$ that is spontaneously broken into $\Uone_\elm$. 
Since its birth in 1967 up to the present days, the model 
has been developed and tested  through an  intense dialogue between theory and experiment.
Here, we briefly mention the most significant measurements  
performed in the last  decade at the high-energy
colliders SLC, LEP, and Tevatron.
%
In these experiments the gauge structure of the
electroweak interactions predicted by the model 
has been directly tested at a  high level of accuracy.
\begin{itemize}
\item
The masses and decay widths of the weak gauge bosons \PZ~ and \PW,
as well as their couplings to leptons and hadrons have been measured 
with precision between the percent and the permille level or even better.
\item
The top quark has been discovered 
and the direct measurement of its mass
has confirmed the indirect determination obtained  from electroweak precision data.
%
\item
The non-abelian triple interactions of gauge bosons 
have been directly observed in gauge-boson pair-production events.
The corresponding couplings have been measured at the few-percent level of precision.

\item
The direct investigation of  quartic gauge-boson interactions 
through events with three gauge bosons in the final state 
is at the beginning.
\end{itemize}
At the present time,
global fits of all high-energy electroweak observables 
show no significant deviation from the predictions of the EWSM.
%

Despite this great experimental success, the model remains untested in one of its most interesting 
aspects:
the  generation of  the gauge-boson and the fermion masses through
the mechanism of  {\em spontaneous symmetry breaking}, also known as the Higgs--Kibble mechanism
\cite{Higgsmech}.
This theoretical construction is of vital importance for the predictive power of the EWSM 
since it represents the only known way to accommodate 
gauge-boson and fermion masses within the model without spoiling its renormalizability \cite{'tHooftLee}.
To this end one postulates the existence of a scalar Higgs boson that
acquires a  non-vanishing vacuum expectation value and
generates  the masses of the particles through gauge and Yukawa interactions.

Up to the present days, apart for a controversial excess of events 
measured at the end of LEP2 \cite{Barate:2000ts} which points to a Higgs mass $\MH\approx 115\GeV$, 
the Higgs boson has escaped direct observation. 
The lower bound for the Higgs mass has reached 
$113 \GeV$, and the upper bound obtained from electroweak precision data is at $212\GeV$,
 with $95\%$ confidence level \cite{Tournefier:2001qv}.

In the next decade, a new  generation of high-energy experiments will enable us to
observe or exclude the Standard Model Higgs boson and to clarify the nature of symmetry breaking.
These experiments will investigate quark and lepton collisions in a range of centre-of-mass energies 
up to the TeV scale, starting  with  run II of the Tevatron ($\mathrm{p\bar{p}}$ collisions at 2 $\TeV$), 
continuing at the  LHC \cite{Altarelli:2000ye} 
(\Pp\Pp~ collisions at 14 $\TeV$) 
and possibly at an  $\Pep\Pem$ linear collider (LC) \cite{Accomando:1997wt} (see for instance the  TESLA project \cite{Aguilar-Saavedra:2001rg} of a  500--800 $\GeV$ collider). 
The following direct observations and measurements are among the objectives 
of these future experiments: 
\begin{itemize}
\item the observation of the Higgs boson and the measurement of its properties: mass and  spin,
\item 
the investigation of  the Higgs interactions with gauge bosons and fermions,
\item 
the determination of the form of the Higgs potential and the  measurement of its parameters,
\item 
direct  measurements of the quartic  gauge-boson self-interactions,
\item
improved precision tests of the EWSM.
\end{itemize}
A key feature of the planned colliders  will be their high luminosity that 
guarantees  
an experimental precision at the percent (LHC) and up to the permille (LC) level. 
Experimental measurements of high precision have to be compared with theoretical predictions at the same level of accuracy.
In perturbative quantum field  theory this 
requires the evaluation of higher-order contributions, 
so-called {\em radiative} or {\em loop corrections},
up to the needed level of accuracy.

Radiative corrections represent a challenge that involves 
highly non-trivial computations and at the same time  an opportunity to access  and to investigate  
deep aspects of the theory.
In fact,  loop corrections represent quantum fluctuations 
and permit to test the theory at the quantum level.
Furthermore, they are sensitive to all sectors of the theory, 
including the Higgs sector or new physics. This permits to extract indirect 
informations about  particles that are not directly accessible.
We also stress that radiative corrections play a crucial role in the  
search of direct signals of new physics which can be disentangled  from the Standard Model 
background only if this latter is known with sufficient accuracy.


Analytical  techniques and  numerical implementations 
for the evaluation of electroweak corrections  have been developed 
and improved under the boost of the experimental progress.
At the one-loop level, 
a complete and well-defined scheme exists \cite{Passarino:1979jh,'tHooft:1979xw} for the algebraic reduction
and analytical evaluation of generic loop integrals.
Specific computer  codes have been developed for 
almost all 4-particle processes. 
For reactions involving more than 2 particles in the  final state a lot of work remains to be done.
Here, the major difficulty %
concerns the  large number of loop diagrams that are characterized by a complex 
algebraic and analytic structure  and by  
large gauge cancellations. 
At the two-loop level, no universal recipe is available for the evaluation of generic loop integrals.
Some two-loop contributions 
have been computed, for instance for the muon decay \cite{Freitas:2001zs},
but for the time being no complete two-loop calculation has been performed.

In view of the future high-energy experiments, 
increasing interest has been recently devoted to the study of 
universal  logarithmic  electroweak corrections in the energy range 
above the electroweak scale. 
An overview of the literature existing in this field is given in \refse{se:literature}.
The aim of the present work is to investigate these  universal electroweak corrections  
at the one-loop level  and for arbitrary processes. 
Our goals are
\begin{itemize}
\item
to provide an high-energy approximation of the one-loop corrections
that can reduce the theoretical error to the few-percent level,
\item
to have analytical results that are process-independent, 
easy to implement, and numerically stable; these results can also be used to  check  explicit one-loop calculations,
\item
to gain physical insight into  the leading part of the  electroweak corrections
above the electroweak scale, 
taking special care of the effects that are related to spontaneous symmetry breaking.
\end{itemize}
The approach that we adopt 
and the formalism that we develop 
should also provide a basis 
for the investigation of  universal two-loop corrections.

\section{Universal logarithms in electroweak radiative corrections}

In the LEP regime, at energies $\sqrt{s}\sim \MZ$, electroweak
radiative corrections 
are dominated by large electromagnetic effects from initial-state
radiation, by the contributions of the running electromagnetic
coupling, and by the corrections associated with the $\rho$ parameter.
These  corrections, relative to the lowest-order predictions, typically amount to $10\%$.
In the energy range above the electroweak scale, $\sqrt{s}\gg \MW$,
new leading contributions emerge: 
double-logarithmic
(DL) terms of the form $\alpha\log^2{(s/\MW^2)}$ (known as  Sudakov logarithms \cite{SUD}) and single-logarithmic (SL) terms of the form
$\alpha\log{(s/\MW^2)}$ involving the ratio of the energy to the
electroweak scale  (see the references in \refse{se:literature}).

For electroweak processes that are not mass-suppressed at high
energies, these  logarithmic corrections are universal, \ie
 in contrast to the non-universal non-logarithmic terms, they can be evaluated in a process-independent way.
On one hand, single logarithms originating from {\em short-distance scales} result
from the running 
of the dimensionless parameters, \ie the 
gauge, Yukawa, and scalar couplings, from the scale $\MW$ to the energy scale $\sqrt{s}$. 
On the other hand,
universal logarithms originating from the {\em long-distance scale}
$\MW\ll\sqrt{s}$ are expected to factorize, \ie they can be associated
with external lines or pairs of external lines in Feynman diagrams.  
They consist of DL and SL terms originating from {\em soft and collinear} and {\em collinear or soft} gauge bosons, respectively, coupled to external particles. 
These logarithmic corrections originating from long-distance effects are called soft and/or collinear singularities or {\em mass singularities}, since they are singular in the limit of massless gauge-bosons and massless external particles.

In gauge theories with massless particles, such as massless QED and QCD, the soft and/or collinear singularities  give rise to infinities in the virtual corrections. However, these infinities are  
cancelled by  the contribution of real soft and/or collinear gauge-boson radiation.
The real corrections need to be included in the definition of physical observables, since soft and/or collinear massless gauge bosons are degenerate with the massless external states and cannot be detected as separate 
particles.

In the EWSM 
owing to the finite masses of the weak gauge bosons, the virtual corrections originating from soft and/or collinear $\PZ$- and $\PW$-bosons give rise to large but not infinite logarithmic contributions.
Moreover,
the  $\PZ$- and $\PW$-boson masses  provide a physical cut-off for
real weak-boson emission, and for a sufficiently
good experimental resolution the massive gauge bosons can be detected
as separate 
particles.
Therefore,
soft and/or collinear weak-boson radiation
need not be included and, except for the electromagnetic infrared divergences,
the large logarithms originating from virtual electroweak corrections are of 
physical significance.

The logarithmic  electroweak corrections grow with energy. At energies $\sqrt{s}=0.5$--$1\TeV$,
the typical size of the DL and SL terms  is given by
\beq
\frac{\alpha}{4\pi \sw^2}\log^2{\frac{s}{\MW^2}}= 3.5 - 6.6 \%,\qquad \frac{\alpha}{4\pi \sw^2}\log{\frac{s}{\MW^2}}=1.0 - 1.3\%,
\eeq
 where $\sw^2\sim 0.22 $ is the squared sine of the weak mixing angle.
Furthermore there are DL terms of the  type
\beq
\frac{\alpha}{4\pi \sw^2}\log{\left(\frac{1\pm\cos{\theta_{kl}}}{2}\right)}\log{\frac{s}{\MW^2}}
\eeq
which depend on the angles $\theta_{kl}$ between the initial- and the final-state  momenta. 
These terms are comparable to the  SL terms if $75^\circ < \theta_{kl}< 105^\circ$ 
and twice as large at  $\theta_{kl}\approx 90^\circ\pm 45^\circ$.
If the experimental
precision is at the few-percent level like at the LHC, both DL and SL
contributions have to be included at the one-loop level. In view of
the precision objectives of a LC, between the percent and
the permille level, besides the complete one-loop
corrections also higher-order effects have to be taken into account.
The DL contributions represent a leading and negative correction,
whereas the SL ones often have opposite sign.
In the TeV range, the SL terms are numerically of the same size as the
DL terms and 
the compensation between DL and SL
corrections can be quite important. Depending on
the process and the energy, the SL contribution can be even larger
than the DL one.

\section{Existing literature 
}\label{se:literature}
The appearance of large logarithms in the  high-energy limit  of the electroweak  corrections is known  since many years (see, for instance, \citeres{Kuroda,eeWWhe}).
Their systematic investigation has started  in the last few years and  
in the following we give a short survey  of these  recent developments.
A  review on the recent literature  can  also be found in \citere{Denner:2001mn}.

\newcounter{listnumber00}
\begin{list}{\bf \arabic{listnumber00}.\hspace{1mm}}{\usecounter{listnumber00}
\setlength{\leftmargin}{0mm} \setlength{\labelsep}{0mm}
\setlength{\itemindent}{4mm}
}


\item {\em One-loop level} 
\nopagebreak
\newcommand{\etal}{{\it et al.~}}

Concerning the one-loop logarithmic corrections two different  strategies  have been adopted. 
On one hand, various  results have been extracted from existing and complete one-loop calculations 
in the high-energy limit.
On the other hand, the  origin and universal nature of the logarithmic contributions have been clarified 
and their generalization  to arbitrary electroweak processes has been  established.

\begin{itemize}
\renewcommand{\labelitemi}{\bf --\hspace{1mm}}
\item Beenakker \etal  have evaluated the high-energy limit of the complete electroweak corrections to 
the process $\Pep\Pem\to\PWp\PWm$
in \citere{eeWWhe}.

\item  The logarithmic corrections to $\Pep\Pem\to\Pf\Pfbar$ have been treated in the following papers using explicit  high-energy expansions for  vertex and box diagrams.  
Ciafaloni and Comelli  have  pointed out the role of the Sudakov DL corrections and discussed their origin in \citere{Ciafaloni:1999xg}.
Beccaria \etal have considered  the  complete logarithmic corrections and studied their impact on various observables: first for the case 
of light-fermions \cite{Beccaria:2000fk}  and then for bottom- \cite{Beccaria:2000xd} and top-quarks \cite{Beccaria:2001jz} in the final state;  in \citeres{Beccaria:2001jz,Beccaria:2001vb} they have also included SUSY logarithmic corrections, 
and in \citere{Beccaria:2001an} they have shown that,  at CLIC energies ($3\TeV$), the $\tan{\beta}$-dependence of SUSY logarithmic corrections  to $\Pep\Pem\to\Pt\bar{\Pt}$
can be exploited for a determination of $\tan{\beta}$; 
finally in  \citere{Beccaria:2001yf} they have pointed out the importance of the angular-dependent contributions.

\item 
Layssac and Renard  \cite{Layssac:2001ur} have evaluated the  complete logarithmic corrections for the process $\gamma\gamma\to f\bar{f}$ 
 including the case of heavy-quark production and considering also  SUSY contributions. 

\item

In this PhD thesis, we present a complete and
{\em process-independent} analysis of one-loop logarithmic electroweak corrections. 
We consider all sources of logarithmic corrections: 
the exchange of soft and/or collinear gauge bosons as well as the renormaliza\-tion-group  running of the gauge, scalar and Yukawa  couplings.
In particular, we prove the {\em factorization of collinear mass singularities}.
The results are summarized in simple analytic formulas  for 
the double, single and angular-dependent logarithmic 
corrections to {\em arbitrary} electoweak processes.
We also present  analytical and numerical applications 
for  the following processes:
$\Pep\Pem\to \mathrm{f}\mathrm{\bar{f}}$,
$\Pep\Pem\to \PWp\PWm 
,\PZ\PZ,\PZ\gamma,\gamma\gamma$ and
$\bar{\mathrm{d}}\mathrm{u}\to \PWp\PZ,\PWp\gamma$.
In this work, in addition to the results already published in \citeres{Denner:2001jv,Denner:2001gw}, 
we also consider the  logarithmic Higgs-mass and top-mass dependence in the large $\MH,\Mt$ limit including all contributions of the form $\alpha \log^n{(\MH/\MW)}$ and $\alpha\log^n{(\Mt/\MW)}$, with $n=1,2$.

\item The general method developed in \citeres{Denner:2001jv,Denner:2001gw}
has been applied in \citere{Accomando:2001fn} to study the electroweak logarithmic corrections to
$\PW\PZ$ and $\PW\gamma$ production at the LHC.
The corrections to the complete hadronic processes $\Pp\Pp\to \PWpm\gamma\to l\nu_l\gamma$ and $\Pp\Pp\to\PWpm\PZ \to l\nu_l \bar{l}'l'$ have been implemented in leading-pole approximation.
It has been checked that, in the region of large transverse momentum of the gauge bosons,
the  leading-pole approximation represents a sufficiently precise approach to evaluate the 
corrections for the LHC. In this region, the logarithmic electroweak corrections   
lower the theoretical predictions by 5--20\%.

\end{itemize}

\item {\em Resummation to all orders}\label{allorderresumm} 
\nopagebreak

The extension of the  one-loop logarithmic corrections to higher orders has been studied by means of resummation techniques 
that were derived within QCD. These techniques  have been applied to the electroweak theory in the symmetric phase.
{\em Leading} contributions of order $\alpha^n\log^{2n}{(s/\MW^2)}$,
{\em subleading} contributions of order  $\alpha^n\log^{2n-1}{(s/\MW^2)}$, and recently also  {\em sub-subleading} 
contributions of order  $\alpha^n\log^{2n-2}{(s/\MW^2)}$  have been studied.
\begin{itemize}
\renewcommand{\labelitemi}{\bf --\hspace{1mm}}

\item 
Fadin \etal  \cite{Fadin:2000bq}  have resummed  the leading contributions by means of the infrared evolution equation. 
Using the splitting function formalism,  Melles  has  extended this approach to the subleading level 
considering processes involving fermions,  transverse gauge bosons  \cite{Melles:2001gw}, would-be Goldstone bosons and Higgs bosons
\cite{Melles:2000ia};  the effect of the running gauge couplings at the subleading level has been discussed in \cite{Melles:2001mr}. 
A  review on  these  works can be found in \citere{Melles:2001ye}.  
Recently, a generalization including the  resummation of angular-dependent logarithmic corrections has been proposed \citere{Melles:2001dh}.

\item Ciafaloni and Comelli  have resummed  the leading contributions applying  the method of soft gauge-boson insertions. 
At first they  have considered the  decay $\PZ'\to f\bar{f}$ of an $\ewgroup$ singlet \cite{Ciafaloni:2000ub}.
Then they have investigated  fully inclusive scattering processes that are initiated by fermions \cite{Ciafaloni:2000BN1},
 transverse gauge bosons \cite{Ciafaloni:2001gm}
and  longitudinal gauge-bosons \cite{Ciafaloni:2001vt,Ciafaloni:2001vu}.
In this context they have pointed out the existence of 
violations of the Bloch--Nordsieck cancellations between virtual and real {\em weak} corrections.
These effects originate from the 
$\SUtwo$ non-singlet nature of the initial states as well as from mixing between weak-hypercharge-eigenstates \cite{Ciafaloni:2001vt}. 
Very recently, they have proposed an ansatz for the resummation of subleading logarithmic corrections to inclusive cross sections \cite{Ciafaloni:2001mu}.

\item  K\"uhn \etal  have  considered the process  $\Pep\Pem\to\Pf\Pfbar$ in the limit of massless fermions 
and have used  evolution equations to resum all logarithmic corrections including the angular-dependent ones:
first at the leading and subleading level \cite{Kuhn:2000},
and recently  up to the 
sub-subleading level \cite{Kuhn:2001hz}, including the  constant one-loop terms.
At $\TeV$ energies, large cancellations 
between the leading, subleading, and sub-subleading two-loop contributions have been observed. 
\end{itemize}

\item {\em Explicit two-loop calculations} 
\nopagebreak

Explicit electroweak calculations at the two-loop level are crucial in order to check the reliability 
of the resummation techniques. 
At the present time only  few results at the {\em leading} two-loop level are available.
\begin{itemize}
\renewcommand{\labelitemi}{\bf --\hspace{1mm}}
\item
The corrections to the decay $g\to\Pf_\kappa\Pfbar_\kappa$ of an $\ewgroup$ singlet into
massless chiral fermions have 
been evaluated in the abelian case $\kappa=\rR$ by Melles \cite{Melles:2000ed} 
and in the non-abelian  case $\kappa=\rL$ by Hori \etal \cite{Hori:2000tm}.

\item 
Beenakker and Werthenbach  have developed  a  Coulomb gauge-fixing  for massive gauge bosons \cite{Beenakker:2000na} that permits
to isolate leading higher-order logarithms  into  self-energy diagrams. This approach applies to  arbitrary  processes. 
Explicit two-loop   results for the processes $\Pep\Pem\to\Pf\Pfbar$ have been given in \citere{Beenakker:2000kb}.
Very recently, the calculation of leading two-loop logarithms for processes involving arbitrary external particles, \ie fermions, longitudinal gauge bosons, Higgs bosons and transverse gauge bosons
has been completed \cite{Beenakker:2001kf}.  
\end{itemize}
These  two-loop computations for fermionic processes are in agreement with the predictions 
obtained in \citere{Fadin:2000bq} 
by resumming the one-loop results.

\end{list}



\section{Spontaneous symmetry breaking and high-energy limit}
\label{intro:SSB}
In this section we want to explain our approach to the high-energy behaviour of the spontaneously broken electroweak theory.
In particular, we want to emphasize the role of  {\em gauge symmetry} in the high-energy limit of the spontaneously broken theory, and discuss the relation to an unbroken theory, where gauge symmetry is exact.

As can be seen in the detailed derivations in \refch{se:soft-coll} and \refch{factorization}, 
the universality of the long-distance logarithmic corrections arises from {\em approximate} charge-conservation 
relations and Ward identities for the electroweak matrix elements in the high-energy limit. 
With other words, universality is provided by approximate gauge symmetry in the high-energy limit.

It is important to note that these crucial symmetry relations are valid only in approximate form  and for matrix elements that are {\em not mass-suppressed},
\ie  matrix elements (with mass dimension $d$) that scale as $E^d$  in the high-energy limit $E\gg \MW$. 
The approximate symmetry relations  are violated by  mass-suppressed contributions of the order  
$\MW^nE^{d-n}$, $n>0$,
and these latter cannot be neglected if the matrix elements themselves are mass-suppressed.

To illustrate 
our approach to electroweak loop corrections in the high-energy limit, 
let us consider the electroweak Lagrangian (see \refapp{Feynrules})
\beq\label{symmbrokLag}
\mathcal{L_{\mathrm{ew}}}=\mathcal{L_{\mathrm{symm}}}+\mathcal{L}_{v}
\eeq
consisting of a  part $\mathcal{L}_{v}$, which contains couplings with mass-dimension proportional to the vacuum expectation value (vev) $v$ of the Higgs field\footnote{Also the negative mass term in the Higgs potential, 
with coefficient $\mu^2=\frac{\la_\PH}{4}v^2$ is part of $\mathcal{L}_{v}$.}, 
and a remaining manifestly symmetric part $\mathcal{L_{\mathrm{symm}}}$, which corresponds to a vanishing vev and depends only on dimensionless parameters. 
In the high-energy limit, it is natural  to  expect a correspondence between the 
spontaneously broken electroweak theory $\mathcal{L_{\mathrm{ew}}}$
and the unbroken $\ewgroup$ theory $\mathcal{L_{\mathrm{symm}}}$. 
This correspondence provides a useful intuitive picture.
We stress however, that we  {\em do not assume it a priori} as it has been done in the literature (see point \ref{allorderresumm} in \refse{se:literature}, and in particular \citere{Melles:2000ia})
since we want {\em to verify it}.
To this end we proceed as follows.
We first derive  the electroweak  logarithmic corrections 
within the complete spontaneously broken theory $\mathcal{L_{\mathrm{ew}}}$.
Then, in the high-energy limit, we isolate that part of the results that 
exhibits a universal and  $\ewgroup$-symmetric form.
This  symmetric part of the results  will naturally indicate the correspondence 
between the high-energy broken theory and the unbroken one, whereas
the remainig non-symmetric part  will indicate effects originating from $\mathcal{L}_{v}$.

This approach 
is motivated by 
the following  important features that distinguish the high-energy electroweak theory from an unbroken $\ewgroup$ theory.

\newcounter{listnumber0}
\begin{list}{\bf \arabic{listnumber0}.\hspace{1mm}}{\usecounter{listnumber0}
\setlength{\leftmargin}{0mm} \setlength{\labelsep}{0mm}
\setlength{\itemindent}{4mm}
}


\item {\em The mixing between the gauge-group eigenstates} 
\nopagebreak

Owing to symmetry breaking, the physical mass eigenstates originate from  mixing between the gauge-group eigenstates.
Obviously, we are interested in matrix elements involving the physical mass eigenstates and not the gauge-group eigenstates. At the same time we want to keep our formalism as symmetric as possible, 
so that invariants of the gauge group, such as the Casimir operator or the coefficients of the beta functions  can be easily recognized in the final results.
To this end, in the Feynman rules (see \refapp{Feynrules}) and in the gauge transformations (see \refapp{BRStra}), the generators of the gauge group are expressed in their generic matrix form.  Mixing is implemented  in a natural way by transforming the generators and all other group-theoretical quantities  
 in the basis corresponding to the mass-eigenstate fields (see \refapp{app:representations}).

\item {\em The definition of the asymptotic states 
and of the parameters at physical mass scales} 
\nopagebreak

Symmetry breaking introduces a fundamental mass scale in the theory, 
and  generates the physical masses of the particles.
These masses 
enter the definition of  the  on-shell asymptotic states
as physical  renormalization scale.
In particular, they enter the field (or wave-function) renormalization constants, 
which must be  fixed  such that the on-shell asymptotic fields 
have the correct normalization and do not mix.
For the definition of the fields and of the paramaters we adopt the on-shell scheme \cite{Denner:1993kt},
where the renormalized fields correspond to physical fields and 
all parameters 
are related to the  electromagnetic coupling constant
 and to the  masses of the particles.

\item {\em The gap between the gauge-boson masses} 
\nopagebreak

Another non-trivial feature emerging from symmetry breaking is the gap between the photon mass and the weak scale $\MW$. 
Owing to this gap,  electroweak corrections are manifestly non-symmetric 
at low energies, where they  are often  splitted 
into an electromagnetic part (photon loops) and a remaining weak part.
At high-energies, instead,  it is more convenient to adopt a different splitting of the corrections, which reflects the $\ewgroup$ symmetry. For this reason 
we  distinguish  a part originating from above the electroweak scale, which is called symmetric electroweak (ew) part,  from a remaining part originating from below the electroweak scale, which is called pure electromagnetic (em) part.
This splitting is done only in the final logarithmic results.
In practice, the logarithms resulting
from the electromagnetic and from the $\PZ$-boson loops are splitted into two
parts: the contributions corresponding to a fictitious heavy photon and a
$\PZ$-boson with mass $\MW$ are added to the $\PW$-boson loops
resulting in the ``symmetric electroweak'' (ew) contribution. 
The large logarithms originating in the photon loops owing to the gap
between the electromagnetic and the weak scale are denoted as
``pure electromagnetic'' (em) contribution.
The remaining logarithms originating from the difference between the
$\PZ$-boson mass and the mass of the $\PW$-boson are given separately.

\item {\em The presence of longitudinal gauge bosons as physical asymptotic states} 
\nopagebreak

Longitudinally polarized  gauge-boson states are not present in a symmetric gauge theory and 
represent a characteristic feature of spontaneous symmetry breaking.
They arise  together with gauge-boson masses in the Higgs mechanism \cite{Higgsmech}
owing to the  gauge interactions of the vacuum. 
In the high-energy limit, longitudinal gauge bosons are related to the corresponding  unphysical components of the scalar doublet,
the so-called would-be Goldstone bosons, via the Goldstone-boson equivalence theorem (GBET) \cite{et}. 
This represents a {\em non-trivial} correspondence  between  longitudinal gauge bosons in the broken theory and 
would-be Goldstone bosons in the  unbroken theory. 
Here we stress the following:
\begin{itemize}
\item\hspace{1mm}
the correspondence can be established only in one direction  starting from the broken theory, because
the GBET  follows from   Ward identities of the spontaneously broken theory and is  based on 
$\mathcal{L}_{v}$. 
\item\hspace{1mm}
The well-known lowest-order form of the GBET is modified by non-trivial  higher-order corrections \cite{etcorr}.
These corrections  are related to mixing-energies between longitudinal gauge bosons and would-be Goldstone bosons
that  originate from  $\mathcal{L}_{v}$. 
\end{itemize}
A  detailed  discussion is given in  \refse{se:gbetintro}. 
\end{list}


\section{Content and organization}
The universal one-loop results for high-energy electroweak DL
and SL
correcti\-ons and  application to simple processes
have been published 
in \citere{Denner:2001jv}. 
In \citere{Denner:2001gw} we have presented the proof of 
factorization of collinear mass singularities and the derivation
of the collinear Ward identities. In addition to the content of these
papers, in this thesis we include all universal logarithms 
$\log{(\Mt/\MW)}$, $\log{(\MH/\MW)}$ depending on the top and Higgs masses,
and we  apply our results  also to the processes
$\bar{\mathrm{d}}\mathrm{u} \rightarrow \mathrm{W}^+\mathrm{A}$ 
and $\bar{\mathrm{d}}\mathrm{u} \rightarrow \mathrm{W}^+\mathrm{Z}$.

The content is organized as follows. 
In \refch{ch:univllogs} we fix our basic notation and conventions, we define 
the logarithmic approximation, we discuss the Goldstone-boson equivalence 
theorem  and describe the form and origin of universal logarithmic corrections.
The DL mass singularities are treated in \refch{se:soft-coll},
where we show that they originate from soft-collinear gauge bosons coupling to 
external legs and we evaluate them using the eikonal approximation.

Chapters \ref{factorization} to
\ref{ch:CWI} concern  the SL mass singularities
that are related to the external particles.
In \refch{factorization} we treat the  contributions 
 originating from loop diagrams: we show that  
they are restricted 
to those diagrams involving collinear virtual gauge bosons 
coupling to the external legs, and we prove that they factorize.
The derivation of the 
collinear Ward identities used for this proof  
is  postponed to \refch{ch:CWI}.
In \refch{FRCSllogs} we give the SL mass singularities
originating from wave-function renormalization as well as the corrections to the GBET.
These contributions are  combined with the contributions presented in \refch{factorization} 
to form a gauge-invariant result.

The logarithmic corrections originating from parameter renormalization are presented in 
\refch{Ch:PRllogs}.
Finally, in \refch{ch:applicat}, 
 we discuss some
applications of our general results to simple specific processes and present 
numerical evaluations.

In the appendices we define all generic quantities appearing in our formulas.
In \refapp{app:GFs} we fix our conventions for Green functions and vertex functions.
All needed  group-theoretical quantities are given in  
\refapp{app:representations} in the basis of mass-eigenstate fields.
The Feynman rules and the BRS transformations of the fields are given 
in generic form 
in  \refapp{Feynrules} and \refapp{BRStra}, respectively.
In \refapp{app:transvRG} we show that in the case of transverse gauge-boson production
the large logarithms originating from parameter renormalization and the 
SL mass singularities 
associated with the final transverse gauge bosons 
compensate each other. Those 2-point functions 
that enter the field-renormalization constants are listed in  
\refapp{app:2pointlogapp} in logarithmic approximation.

The content of each chapter is summarized in the corresponding introduction.

\chapter{Universal logarithmic corrections at high energies}
\label{ch:univllogs}
In this chapter we introduce our basic notation, we define the high-energy (logarithmic) approximation, and specify our approach to logarithmic corrections at high-energies. In particular,  we introduce the Goldstone-boson equivalence theorem (GBET), which is used for the  description of longitudinal gauge bosons in the high-energy limit. Finally, we discuss the form and origin of the universal logarithmic corrections.

\section{Notation and conventions}
\label{se:not}
Our notation for fields and particles is as follows.
Chiral fermions and antifermions are denoted 
by $f^\kappa_{j,\si}$ and $\bar{f}^\kappa_{j,\si}$, respectively, where  $f=Q,L$ corresponds to  quarks and leptons.
For fermionic fields we use the symbol $\Psi$ instead of $f$, which is used for fermionic particles.
The chirality is specified by
$\kappa=\rR,\rL$.  The index $\si=\pm$ determines the weak isospin, and $j=1,2,3$ is the generation index. 
The gauge bosons are denoted by $V^a=\PA,\PZ,\PWpm$, and can be transversely ($\rT$) or longitudinally ($\rL$) polarized. 
The components of the scalar
doublet are denoted by  $\Phi_i=H,\chi,\phi^\pm$ and consist of the physical Higgs particle $\PH$ and the
unphysical would-be Goldstone bosons $\chi,\phi^\pm$, which are used to
describe the longitudinally polarized massive gauge bosons
$\PZ_\rL$ and $\PW^\pm_\rL$  with help of the GBET. 
The above fields (or particles) are collectively represented as components $\varphi_{i}$ of one  multiplet $\varphi$.

As a convention, we consider electroweak processes\footnote{Note that in the notation we use to denote scattering processes  like \refeq{process}, the symbol $\varphi_{i}(p)$ has to be understood as the particle associated to the field $\varphi_{i}$, and  $p$ specifies the corresponding momentum.
This is in contrast with our usual notation, where $\varphi_{i}(p)$
denote the Fourier component of a quantum or
 classical field.} 
\beq \label{process}
\varphi_{i_1}(p_1)\dots \varphi_{i_n}(p_n)\rightarrow 0,
\eeq
involving $n$ arbitrary incoming particles $\varphi_{i_1},\dots ,\varphi_{i_n}$
with incoming  momenta $p_1,\dots ,p_n$.
The predictions for general  processes, 
\beq \label{22proc}
\varphi_{i_1}(p^{\mathrm{in}}_1)\ldots\varphi_{i_m}(p^{\mathrm{in}}_m)
\rightarrow 
\varphi_{j_1}(p^{\mathrm{out}}_1)\dots \varphi_{j_{n-m}}(p^{\mathrm{out}}_{n-m}),
\eeq
can be obtained  by crossing symmetry from our predictions for the
$n\rightarrow 0$ process 
\beq\label{crossedproc}
\varphi_{i_1}(p^{\mathrm{in}}_1)\ldots\varphi_{i_m}(p^{\mathrm{in}}_m)
\bar{\varphi}_{j_1}(-p^{\mathrm{out}}_1)\dots\bar{\varphi}_{j_{n-m}}(-p^{\mathrm{out}}_{n-m})\rightarrow 0,
\eeq 
where $\bar{\varphi}_i$ represents the charge conjugate of
$\varphi_{i}$. Thus, outgoing particles (antiparticles) are
substituted by incoming antiparticles (particles) and the
corresponding momenta are reversed.  These substitutions can be
directly applied to our results. 

The matrix element for the process \refeq{process} is given by 
\begin{equation}\label{Bornampli}
\M^{\varphi_{i_1} \ldots \,\varphi_{i_n}}(p_1,\ldots, p_n)=
G^{\underline{\varphi}_{i_1}\ldots\,\underline{\varphi}_{i_n}}(p_1,\ldots, p_n)
\prod_{k=1}^n v_{\varphi_{i_k}}(p_{k}).
\end{equation}
On the right-hand side (rhs), we have the  Green function\footnote{
Note that the wave-function renormalization factors that usually appear on the rhs of \refeq{Bornampli} have been set equal to one, since  we  assume that the renormalized on-shell fields correspond to physical fields, \ie that they have been  normalized such that the  residues of the corresponding propagators equal 1.  
} $G$ 
contracted with the wave functions $v_{\varphi_{i_k}}(p_k)$ of the external particles. These latter equal 1 for scalars and are 
given by the Dirac-spinors for fermions and the polarization vectors for gauge bosons. 
The field arguments of the Green function $G$ are underlined. This 
indicates that the corresponding external legs are truncated. This and other 
conventions for Green functions are listed in  \refapp{app:GFs}.
The lowest-order (LO) matrix elements are denoted by $\M_0$.

The generators of the $\ewgroup$ gauge group are denoted by $I^{V^a}$.
In terms of the electric charge $Q$ and weak isospin $T^a$ they are given by
\beq\label{generatorsdef}
I^A=-Q,\qquad I^Z=\frac{T^3-\sw^2 Q}{\sw\cw},\qquad
I^\pm=
\frac{T^1\pm\ri T^2}{\sqrt{2}\sw}
\eeq
and depend on the sine $\sw=\sin{\thw}$ and cosine $\cw=\cos{\thw}$
of the weak mixing angle $\thw$, which are fixed 
by $\cw^2=1-\sw^2=\MW^2/\MZ^2$.  

Infinitesimal global transformations of the fields $\varphi_i$ are determined by the matrices $I^{V^a}_{\varphi_i\varphi_{i'}}$ in the corresponding representation and read
\beq \label{generators}
\delta  \varphi_{i}=
\ri e \sum_{V^a=A,Z,W^\pm}\sum_{\varphi_{i'}} I^{V^a}_{\varphi_i\varphi_{i'}}\de \theta^{V^a}\,\varphi_{i'},
\eeq
where $\de \theta^{V^a}$ are the infinitesimal gauge parameters.
The generators \refeq{generatorsdef} determine the gauge couplings. To be precise, the matrices $\ri eI^{V^a}_{\varphi_i\varphi_{i'}}$ are the couplings corresponding  to the gauge vertices  $V^a\bar{\varphi}_i\varphi_{i'}$,
where all fields are incoming.

The explicit form of the generators and other
  group-theoretical quantities is given in \refapp{app:representations}, and a
 detailed list of the Feynman rules can be found in \refapp{Feynrules}.


\section{High-energy limit and  logarithmic approximation}
\label{se:helimit}
For the process \refeq{process}, we restrict ourselves to following kinematic region.
The external momenta are considered to be on
shell, $p_k^2=M_{\varphi_k}^2$, whereas 
all other invariants are assumed to be much
larger than the gauge-boson masses, 
\beq \label{Sudaklim} 
\left(\sum_{l=1}^N p_{k_l}\right)^2\sim s \gg\MW^2, \qquad 1<N<n-1,\qquad
{k_l}\ne {k_{l'}} \ \mbox{ for }\ l\ne l'.
\eeq 
This corresponds to  high centre-of-mass energy $E=\sqrt{s}\gg \MW$ and not too small scattering angles.
In this region, matrix elements can be  expanded in the small parameters $M_i/E\ll 1$,  where $M_i$ represent the various  mass scales that enter the matrix element.

As already stressed in the introduction, we restrict ourselves  to processes  with Born matrix elements that are not mass-suppressed, \ie of order\footnote{Obviously, for  a precise Born-level prediction 
the exact matrix elements  have to be used, including all mass terms.} 
\beq\label{scaling}
\M_0^{\varphi_{i_1} \ldots \varphi_{i_n}}(p_1,\ldots, p_n)\sim E^d,
\eeq
where  $d$ is the  mass dimension of the matrix element. 
At one-loop level, we restrict ourselves to  double (DL) and single (SL) logarithmic mass-singular corrections, \ie corrections of the order
\beqar\label{LAscaling}
\de^{\mathrm{DL}}\M^{\varphi_{i_1} \ldots \varphi_{i_n}}(p_1,\ldots, p_n)&\sim& E^d L,\nl
\de^{\mathrm{SL}}\M^{\varphi_{i_1} \ldots \varphi_{i_n}}(p_1,\ldots, p_n)&\sim& E^d l,
\eeqar
where $L$ and $l$ represent double and single logarithms of the form
\beq\label{logdef}
\LrMi:=\frac{\alpha}{4\pi}\log^2{\frac{|r_{kl}|}{M_i^2}},\qquad
\lrMi:=\frac{\alpha}{4\pi}\log{\frac{|r_{kl}|}{M_i^2}},
\eeq 
depending on different masses $M_i$ and invariants
\beq \label{SudaklimB} 
r_{kl}=(p_k+p_l)^2 \approx 2p_kp_l \gg \MW^2.
\eeq
In the high-energy and fixed-angle limit, the contribution  of order \refeq{LAscaling} represent the leading part of the one-loop corrections. This part is universal, \ie it can be predicted in a process-independent way.
The remaining part, instead,  is in general non-universal and will be neglected.
In particular, we do not consider mass-suppressed logarithmic contributions of the order $\MW^nE^{d-n}L$ and $\MW^nE^{d-n}l$ with $n>0$. 
We also neglect all corrections of the order $\alpha E^d$, \ie corrections that are constant relative to the Born matrix element.

The logarithms \refeq{logdef} originating  from various  Feynman diagrams (and counterterms), depend on different energy scales  $r_{kl}$ and mass scales  $M_i$, as well as on the scale $\mu$ introduced by dimensional regularization\footnote{Note that $\mu$ does not correspond to the scale  of renormalization.}.
These scales  are characterized by following hierarchy
\beq
\mu^2=s\sim r_{kl}\gg\Mt^2,\MH^2 > \MW^2\sim\MZ^2 \gg m^2_{f\ne\Pt} \gg  \lambda^2,
\eeq
where the lightest scale $\la$ corresponds to the fictitious photon mass used to regularize infrared (IR) singularities, $m_{f\ne\Pt}$ denotes light-fermion masses, and we have set  $\mu^2=s$. The choice of the scale $\mu$ is free, since
the $S$ matrix in independent of $\mu$. We  choose $\mu^2=s$ so that 
we can restrict ourselves to the 
{\em mass-singular logarithms} $\log{(\mu^2/M_i^2)}$ or $\log{(s/M_i^2)}$, whereas
the logarithms $\log{(\mu^2/s)}$ originating from loop diagrams that are not mass-singular  
can be neglected.

In order to organize all one-loop results in the most symmetric way, we split all logarithms into a ``symmetric electroweak'' (ew) 
part given by logarithms of the ratio between the energy and the electroweak scale  and a remaining part. To be precise,
 all double and single  logarithms of the type \refeq{logdef}
are written in terms of
\beq \label{dslogs}
\Ls:=\frac{\alpha}{4\pi}\log^2{\frac{s}{\MW^2}}
,\qquad
\ls:=\frac{\alpha}{4\pi}\log{\frac{s}{\MW^2}}
,
\eeq
plus remaining  logarithms of mass ratios and ratios of invariants.  
\begin{itemize}
\item
All DL contributions proportional to
$\Ls$ and  $\ls\log(|r_{kl}|/s)$ as well as the SL contributions
proportional to $\ls$ are called ``symmetric-electroweak'' (ew) contributions.
They can be intuitively understood as the part of the corrections originating from above the electroweak scale.
\item
All DL and SL contributions containing  logarithms $\log{(\MW^2/\la^2)}$ and $\log(\MW^2/m_f^2)$, which  involve
 the photon mass $\la$ or  masses of light charged fermions are called  ``pure electromagnetic'' (em) part.
This part includes also energy-dependent double logarithms of the type  
$\ls\log{(\MW^2/\la^2)}$ and   
$\ls\log(\MW^2/m_f^2)$. 
\item
We also take into account the logarithmic dependence on 
$\MH$ and $\Mt$ originating from diagrams involving Higgs bosons and top quarks.
The corresponding SL terms are usually split as follows
\beq
\log{\frac{s}{M_i^2}}=\log{\frac{s}{\MW^2}}-\log{\frac{M_i^2}{\MW^2}}
,\qquad M_i=\MH,\Mt.
\eeq
For the Higgs mass we only  assume that 
$s\gg\MH>\MW$. 
In diagrams involving both top quarks and Higgs bosons the scale 
of the logarithms is determined by the largest mass
\beq\label{htopmassdef}
M_{\Htop}:=\max{(\MH,\Mt)},
\eeq  
which can be either $\Mt$ or $\MH$.
\item Since $\MW\sim\MZ$ we neglect  all single logarithms 
$\alpha\log(\MZ^2/\MW^2)$. The logarithmic dependence on $\MZ$ is only considered at the DL level by taking into account the contributions 
$\ls\log(\MZ^2/\MW^2)$, which grow with energy.
\end{itemize}
The remaining logarithmic contributions are  neglected. In particular,
we neglect  the pure angular-dependent contributions $\alpha\log{(|r_{kl}|/s)}$ and
$\alpha\log^2{(|r_{kl}|/s)}$ which are small in the limit \refeq{Sudaklim}.

\section{Longitudinal gauge bosons in the high-energy limit}
\label{se:gbetintro}
In this section we introduce and discuss the GBET, 
which is used to treat longitudinal gauge bosons in the  high-energy limit. 
Longitudinal gauge bosons 
are particularly interesting since these states 
originate from the  symmetry breaking mechanism and are not present  
in symmetric gauge theories.
Therefore, 
in processes involving longitudinal gauge bosons,
the broken sector of the theory plays a crucial role 
also at energies above the symmetry-breaking scale.

Let us consider  the process
\beq \label{LGBprocess}
V^{a_1}_\rL(q_1) \ldots V^{a_{m}}_\rL(q_m)
\varphi_{i_{1}}(p_{1})\dots \varphi_{i_n}(p_{n})
\rightarrow 0,
\eeq
involving $m$ longitudinal gauge bosons $V^{a_k}_\rL=Z_\rL,W^\pm_\rL$ 
and $n$ other arbitrary external states $\varphi_{i_k}$.
The corresponding matrix element is  given by 
\beqar \label{loGBmatel0} 
\lefteqn{\M^{V^{a_1}_\rL \ldots\, V^{a_{m}}_\rL \varphi_{i_{1}} \ldots\,\varphi_{i_n}
}(q_1,\dots,q_m,p_1,\ldots,p_{n})=}\quad&&\nl&=& 
\left[\prod_{k=1}^{m}\epsilon_\rL^{\mu_k}(q_k)\right]
G_{\mu_1\dots \mu_{m} }^{\underline{V}^{a_1} \ldots\, \underline{V}^{a_{m}}\,\underline{\varphi}_{i_{1}}\ldots\, \underline{\varphi}_{i_n}}(q_1,\dots,q_m,p_1,\ldots,p_{n}) \prod_{k=1}^n v_{\varphi_{i_k}}(p_{k}),\nln
\eeqar
where  the amputated Green function  
$G_{\mu_1\dots \mu_{m} }^{\underline{V}^{a_1} \ldots \underline{V}^{a_{m}}}$ is contracted with the longitudinal polarization vectors
\beq \label{longplovec}
\epsilon_\rL^{\mu_k}(q_k)=
\frac{q_k^{\mu_k}}{M_{V^{a_k}}}+\O\left(\frac{M_{V^{a_k}}}{q_k^0}\right),
\eeq 
with $q^0_k\sim E =\sqrt{s}$.
In the high-energy limit $M/E\ll 1$, each polarization vector 
%
yields a contribution  of order $E/M$. 
Therefore,  in order to determine the $\O(E^d)$ contribution to the  matrix element \refeq{loGBmatel0},
the Green function on the rhs of \refeq{loGBmatel0} needs to be determined  up to the order $M^{m}{E^{d-m}}$. 
In this Green function the mass terms in the propagators as well as the couplings with mass dimension
cannot be neglected. This indicates that 
the   contributions originating from the broken sector of the theory ($\mathcal{L}_{v}$) 
are crucial. 

Since the leading part of the longitudinal polarization vector \refeq{longplovec}
is proportional to the momentum of the corresponding gauge-boson, the matrix elements \refeq{loGBmatel0}
can be simplified using electroweak 
Ward identities 
that relate amputated Green functions involving weak  gauge bosons $V^a=Z,W^\pm$
to amputated Green functions involving the corresponding would-be Goldstone bosons
$\Phi_a=\chi,\phi^\pm$  (a detailed derivation is given in  \refse{se:loggaugebos}). 
These Ward identities  lead to the well-known  {\em Goldstone-boson equivalence theorem} 
\cite{et}, 
\beqar\label{GBET1}
\lefteqn{\M^{V^{a_1}_\rL \ldots V^{a_{m}}_\rL  \varphi_{i_{1}} \ldots\,\varphi_{i_n}}(q_1,\dots,q_m,p_1,\ldots,p_{n})=}
\quad &&\\&=&
\left[\prod_{k=1}^{m} \ri^{(1-Q_{V^{a_k}})} A^{V^{a_k}}\right]
\M^{\Phi_{a_1} \ldots \Phi_{a_{m}}\varphi_{i_{1}}\ldots\, \varphi_{i_n}}
(q_1,\dots,q_m,p_1,\ldots,p_{n})
+\O(M E^{d-1})\nn
\eeqar
that  will be the basis for our description of longitudinal gauge bosons in terms of would-be Goldstone bosons, 
in the high-energy limit.
The factors 
\beq\label{GBETcorrdef}
A^{V^a}=1+ \de A^{V^a},
\eeq
appearing in \refeq{GBET1}, consist of a  trivial  
lowest-order contribution and non-trivial  loop contributions $\de A^{V^a}$ [see \refeq{GBETder14}]. 
These latter are the so-called corrections to the GBET \cite{etcorr} and
have to be evaluated in the needed order of perturbation theory. 
In one-loop approximation, the corrections to the process \refeq{LGBprocess} are obtained from
\beqar\label{GBETandcorr}
\de\M^{V^{a_1}_\rL \ldots V^{a_{m}}_\rL  \varphi_{i_{1}} \ldots\,\varphi_{i_n}}&=&
\left[\sum_{k=1}^{m} \de A^{V^{a_k}}\right]\M_0^{V^{a_1}_\rL \ldots V^{a_{m}}_\rL  \varphi_{i_{1}} \ldots\,\varphi_{i_n}}
\nl&&{}+
\left[\prod_{k=1}^{m} \ri^{(1-Q_{V^{a_k}})}\right]\de\M^{\Phi_{a_1} \ldots \Phi_{a_{m}}\varphi_{i_{1}}\ldots\, \varphi_{i_n}},
\eeqar
combining the  corrections to the GBET with the corrections to the matrix element involving would-be Goldstone bosons.

The main advantage of the GBET is to remove the complications related to the  
mass terms. 
In fact,  the Green function on the rhs of \refeq{GBET1}
needs to be evaluated in leading order $E^d$ only. 
In this Green function all mass terms, \ie all contributions related to $\mathcal{L}_{v}$ 
can be neglected. 
This does not mean that the broken sector of the theory does not play any role in the high-energy limit.
In particular we stress the following:
\begin{itemize}
\item The electroweak Ward identities that lead to the GBET are based on the broken sector of the theory.

\item In higher-orders of perturbation theory 
the corrections  $\de A^{V^a}$ to the GBET involve loop  diagrams that originate from  the broken sector $\mathcal{L}_{v}$ (see \refse{se:loggaugebos}). 
In particular they involve 
mixing-energies  between gauge bosons and would-be Goldstone bosons.
\end{itemize}
In general, to handle matrix elements involving longitudinal gauge bosons, we have to rely on Ward identities of the full spontaneously broken theory, where the broken contributions play a crucial role.
This is the case also in \refch{ch:CWI}, where we derive {\em collinear Ward identities}, used to determine the  collinear loop corrections to the Green functions involving external would-be Goldstone bosons.

\section{Form and origin of universal logarithmic corrections}
\label{se:formorigin}
In this section, we briefly discuss the form of universal logarithmic corrections, we specify the strategy that we adopt for their calculation, and explain how they are classified according to their origin. 

As  shown in the following chapters, in leading-logarithmic  order 
\refeq{LAscaling}, the  {\em virtual} one-loop corrections assume the general form
\beqar\label{factform}
\lefteqn{
\de\M^{\varphi_{i_1} \ldots \,\varphi_{i_n}}
(
\{\la_i\},
p_1,\ldots, p_n)
=
\sum_{\la_i} \de \la_i
\frac{\partial\M_0^{\varphi_{i_1} \ldots \,\varphi_{i_n}}}{\partial \la_i}
(\{\la_i\}, p_1,\ldots, p_n)
}\quad&&\nl
&&+
\sum_{\varphi_{i'_1}\dots\varphi_{i'_n}}\M_0^{\varphi_{i'_1} \ldots \,\varphi_{i'_n}}
(\{\la_i\}, p_1,\ldots, p_n)\,
\de^{\varphi_{i_1} \ldots \,\varphi_{i_n}}_{\varphi_{i'_1} \ldots \,\varphi_{i'_n}}
(\{\la_i\},p_1,\ldots, p_n)
.
\eeqar
On one hand we have the well-known  contributions that are related to the renormalization  $\de \la_i$
of the dimensionless coupling constants  $\la_i$.
On the other hand we have contributions that {\em factorize} in momentum space into Born matrix elements times universal correction factors 
$\de^{\varphi_{i_1} \ldots \,\varphi_{i_n}}_{\varphi_{i'_1} \ldots \,\varphi_{i'_n}}$.
These latter  are tensors with  $\ewgroup$ indices that can be associated, as we will see, to single external states or pairs of external states.

The dimensionless correction factors 
$\de^{\varphi_{i_1} \ldots \,\varphi_{i_n}}_{\varphi_{i'_1} \ldots \,\varphi_{i'_n}}$
and the  running $\de \la_i$ of the couplings 
are evaluated in logarithmic  accuracy, \ie to order  $L$ and $l$, in a process-independent way.
Instead, the Born matrix elements  on the rhs  of \refeq{factform},
which involve in general $\ewgroup$-transformed fields $\varphi_{i'_1} \ldots \varphi_{i'_n}$,
require an explicit evaluation for each specific process. 
Note that  these Born matrix elements need to be evaluated in  leading order \refeq{scaling} only.

The universal corrections 
\refeq{factform} are derived in one-loop order using dimensional regularization. As explained in \refse{se:helimit}, we choose the scale $\mu^2=s$, so that we can restrict ourselves to mass singularities.
These are shared between the renormalization counterterms
and the truncated loop-diagrams, depending on the gauge fixing and the renormalization scheme.

\begin{itemize}
\item 
The one-loop contributions from renormalization are related to
Born diagrams in a natural way.
On one hand, the contribution of coupling-constant counterterms $\de \la_i$ is equivalent 
to a shift of the couplings  at Born level, which leads to the first term on the rhs of \refeq{factform}.
On the other hand, the 
field renormalization constants (FRC's) 
lead to
\beq\label{WFRCsubllogfact}
\de^{\mathrm{WF}} \M^{\varphi_{i_1} \ldots \varphi_{i_n}} =\sum_{k=1}^n \sum_{\varphi_{i'_k}}
\frac{1}{2}\delta
  Z_{\varphi_{i'_k}
    \varphi_{i_k}}
\M_0^{\varphi_{i_1} \ldots \varphi_{i'_k} \ldots \varphi_{i_n}},
\eeq
with the well-known $\de Z/2$ factors for each external leg.
For parameter renormalization we adopt the on-shell scheme \cite{Denner:1993kt}
for definiteness. This can be easily changed.
The field renormalization
constants (FRC's) are fixed such that 
the fields do not mix and the residua of renormalized propagators are equal to one, 
\ie renormalized fields correspond to physical fields 
and no extra wave-function renormalization
constants are required \cite{Denner:1993kt}.

\item 
The remaining mass-singular corrections originate from truncated loop diagrams 
and factorize in the form corresponding to the second term on the rhs of \refeq{factform}.
To prove the factorization of these contributions much more effort is needed. 
A possible strategy  is to avoid non-trivial mass-singular loop diagrams by an appropriate gauge-fixing, such that all mass singularities are isolated into the FRC's. 
This approach has been developed  in \citere{Beenakker:2000kb}, using the Coulomb gauge for massive gauge bosons. In this way, factorization is obtained in a natural way as in \refeq{WFRCsubllogfact}. However, the evaluation of 
the FRC's
becomes highly non-trivial, owing to the subtleties related to this particular gauge fixing, and for the time being, explicit results are available only at the DL level, in this approach.
\end{itemize}
In this thesis we adopt the 't~Hooft-Feynman gauge. Here, factorization of the DL and SL mass-singular truncated loop diagrams has to be proved explicitly.
This proof is based on the two following general ideas: 
\begin{itemize}
\item Mass singularities originate from a restricted subset of Feynman diagrams and from specific regions of loop-momenta, so-called {\em leading regions}.
With \citere{Kinoshita:1962ur} one can easily prove that mass singularities in the electroweak theory are restricted to Feynman diagrams with virtual gauge bosons coupling to the external legs, and originate in those regions of loop-momenta where the virtual gauge bosons are {\em soft  and/or collinear} to the external legs.
\item Factorization of the Feynman diagrams involving soft and/or collinear virtual gauge bosons is 
a consequence of {\em gauge symmetry}. In particular, the factorized form 
 of the second term on the rhs of \refeq{factform}
is obtained using appropriate  charge-conservation relations and Ward identities of the electroweak theory. 
\end{itemize}
These ideas are developed in the following chapters, where the leading-logarithmic corrections are split according to the leading regions they originate from. The various parts are denoted as follows
\beq\label{desplitting}
\delta\M=\de^{\SC}\M+\de^{\SS}\M+\de^{\cc}\M+\de^\pre\M.\
\eeq
The first two terms correspond to DL contributions, \ie $\de^{\SC}+\de^{\SS}=\de^{\mathrm{DL}}$. 
These originate from those one-loop diagrams
  where soft--colline\-ar gauge bosons are exchanged between pairs of
  external legs, and are evaluated with the eikonal
  approximation in \refch{se:soft-coll}.
\begin{itemize}
\item
The leading part $\de^\SC$ consists of angular-independent double logarithms of the type $\Ls$ and similar.
\item
The subleading part $\de^\SS$ is given by angular-dependent double logarithms of type $\ls\log(|r_{kl}|/s)$ and similar.
\end{itemize}
The last two terms in \refeq{desplitting} are the SL contributions, \ie $\de^{\cc}+\de^{\pre}=\de^{\mathrm{SL}}$.
\begin{itemize}
\item The collinear or soft  single logarithms are denoted by 
$\de^\cc$. One part of $\de^\cc$ originates from renormalization (FRC's and corrections to the GBET) and is treated in \refch{FRCSllogs}. The remaining part originates from truncated loop diagrams involving external-leg emission of collinear virtual gauge bosons. The factorization 
of these diagrams is proved in \refch{factorization} using  the Ward identities derived in \refch{ch:CWI}.
  
\item The SL contributions  of UV origin, which are related to the parameter renormalization, are denoted  by $\de^\pre$   and are presented in \refch{Ch:PRllogs}.
Since we set $\mu^2=s$, these corrections   
are absorbed into the parameter-renormalization counterterms. 
They include the contributions of the charge and
  weak-mixing-angle renormalization constants, as well as the
  renormalization of dimensionless mass ratios associated with the
  top-Yukawa coupling and the scalar self-coupling.
\end{itemize}

\chapter{Soft-collinear mass singularities}\label{se:soft-coll}
In this chapter we treat the double-logarithmic (DL) mass singularities.
First, we show that the DL mass-singular corrections
originate from exchange  of {\em soft-collinear virtual gauge bosons} between pairs of external particles 
and evaluate them  using the well-known {\em eikonal approximation}.
Then, we show  that the {\em angular-independent} part of the  DL corrections, which we call  
leading soft-collinear part $\de^{\mathrm{LSC}}\M$, factorizes as a single sum over external particles
\beq\label{LSClogfact}
\de^{\mathrm{LSC}} \M^{\varphi_{i_1} \ldots \, \varphi_{i_n}}(p_1,\ldots,p_n) =\sum_{k=1}^n \sum_{\varphi_{i'_k}}
\M_0^{\varphi_{i_1} \ldots \, \varphi_{i'_k} \ldots \, \varphi_{i_n}}(p_1,\ldots,p_n)\,
\delta^{\mathrm{LSC}}_{\varphi_{i'_k}\varphi_{i_k}}.
\eeq
The remaining  {\em angular-dependent} part of the DL corrections, which we call subleading soft-collinear part  $\de^{\mathrm{SSC}}\M$,
is given by a double sum over pairs of external legs
\beq
\de^{\mathrm{SSC}} \M^{\varphi_{i_1} \ldots \, \varphi_{i_n}}(p_1,\ldots,p_n) =\sum_{k=1}^n \sum_{l<k} \sum_{\varphi_{i'_k},\varphi_{i'_l}}
\M_0^{\varphi_{i_1} \ldots \, \varphi_{i'_k}  \ldots \,\varphi_{i'_l} \ldots \,\varphi_{i_n}}(p_1,\ldots,p_n)\,
\delta^{\mathrm{SSC}}_{\varphi_{i'_k}\varphi_{i_k}\varphi_{i'_l}\varphi_{i_l}}.
\eeq
We note that these DL contributions represent the simplest part of the electroweak logarithmic corrections. 
At this level, in fact, the  mechanism of  symmetry breaking is  irrelevant in its details and 
only the lowest-order mixing in the neutral gauge sector has to be taken into account.
This gives rise to  mixing between matrix elements involving photons and \PZ--bosons  in the leading  DL corrections. 

\section{Double-logarithmic mass singularities and eikonal approximation}\label{se:eikapp}
We first have to determine the class of Feynman diagrams that potentially give rise to DL mass singularities.
To this end  we can use  the analysis of  scalar loop integrals made by Kinoshita \cite{Kinoshita:1962ur}.
Since we choose the  't~Hooft--Feynman gauge, where all propagators have the same pole-structure as scalar  propagators,
we can directly apply the result of \citere{Kinoshita:1962ur} to conclude that 
DL  mass-singularities 
originate only from the subset of loop diagrams where 
a virtual particle $\varphi_s$ is exchanged between two on-shell external particles $\varphi_{i_k}$ and  $\varphi_{i_l}$
\cite{Kinoshita:1962ur}, \ie diagrams of the type\footnote{Here and in the following, all on-shell external legs that are not involved in our argumentation are omitted in the graphical representation.}
\beq\label{genericDLdiagramA}
\vcenter{\hbox{\begin{picture}(80,80)(5,-40)
\Line(65,0)(25,30)
\Line(65,0)(25,-30)
\CArc(65,0)(32.5,143.13,216.87)
\Vertex(39,19.5){2}
\Vertex(39,-19.5){2}
\GCirc(65,0){15}{1}
\Text(30,0)[r]{\scriptsize$\varphi_s$}
\Text(20,27)[rb]{\scriptsize$\varphi_{i_k}$}
\Text(20,-27)[rt]{\scriptsize$\varphi_{i_l}$}
\Text(46,19)[lb]{\scriptsize$\varphi_{i'_k}$}
\Text(46,-19)[lt]{\scriptsize$\varphi_{i'_l}$}
\end{picture}}}.
\eeq

In general, the fields $\varphi_s$, $ \varphi_{i_k}$, $ \varphi_{i'_k}$, $ \varphi_{i_l}$, $ \varphi_{i'_l}$ in \refeq{genericDLdiagramA} may be fermions, antifermions, Higgs bosons, would-be Goldstone bosons or gauge bosons. 
For the moment, we consider all possible diagrams of the type \refeq{genericDLdiagramA} that are allowed by the electroweak Feynman rules.
These lead to loop integrals of the type
\newcommand{\Isoft}{J}
\beqar \label{softmasssingloop}
\lefteqn{\Isoft=-\ri (4\pi)^2\mu^{4-D}\times}\quad&&\nl&&
\int\ddq
\frac{N(q)}{(q^2-M_{\varphi_s}^2+\ri \varepsilon)[(p_k-q)^2-M_{\varphi_{i'_k}}^2+\ri \varepsilon][(p_l+q)^2-M_{\varphi_{i'_l}}^2+\ri \varepsilon]}.  
\eeqar
The part denoted by $N(q)$ is kept implicit. 
It consists of the lowest-order (LO) contribution from the ``white blob'' in \refeq{genericDLdiagramA}, 
of the wave-function 
(spinor or polarization vector) corresponding to the external lines $\varphi_{i_k}$, $\varphi_{i_l}$, of the couplings of the virtual particle $\varphi_s$, and of the numerators of
the $\varphi_{s}$, $\varphi_{i'_k}$ and $\varphi_{i'_l}$~propagators. 

The DL mass singularities originate from the denominators of the $\varphi_{s}$, $\varphi_{i'_k}$ and $\varphi_{i'_l}$~propagators, in the regions of integration where the $\varphi_{s}$ momentum becomes collinear to one of the external 
momenta and soft, \ie  $q^\mu\to xp^\mu_k$ or $q^\mu\to xp^\mu_l$, and  $x\to 0$.
The DL mass singularities can be extracted from \refeq{softmasssingloop}  using the eikonal approximation (see for instance \citere{Collins:1989gx}) which 
consists in the following simple prescription for the numerator $N(q)$:
\beqar\label{eikappdef}
&(1)\,&\mbox{Substitute $N(q)\rightarrow N(0)$, \ie  set  $q^\mu \rightarrow 0$.}\hspace{3cm}
\nl&(2)\,&
\mbox{Neglect all mass terms in $N(0)$.}
\eeqar
We stress that this approximation is not directly applicable in the presence of  external longitudinal gauge bosons. In this case, it is  not  possible to set all mass terms in $N(0)$ to zero, since the longitudinal polarization vectors \refeq{longplovec} are inversely proportional to the gauge-boson masses.
For this reason,  matrix elements involving longitudinal gauge bosons  $V^a_\rL=\PZ_\rL,\PWpm$ are treated with the GBET \refeq{GBET1}, \ie they are expressed by matrix elements involving the corresponding would-be Goldstone bosons $\Phi_a =\chi, \phi^\pm$.
In the 't~Hooft--Feynman gauge, as we will see in \refse{se:loggaugebos}, the one-loop corrections 
to the GBET 
involve only SL contributions.  Therefore, at DL level \refeq{LAscaling} we can use the GBET in its LO form
\beqar\label{DLGBET2}
\lefteqn{\de^\DL \M^{V^{a_1}_\rL \ldots V^{a_{m}}_\rL  \varphi_{i_{1}} \ldots\,\varphi_{i_n}}(q_1,\dots,q_m,p_1,\ldots,p_{n})=}
\quad &&\\&=&
\left[\prod_{k=1}^{m} \ri^{(1-Q_{V^{a_k}})}\right]
\de^\DL \M^{\Phi_{a_1} \ldots \Phi_{a_{m}}\varphi_{i_{1}}\ldots\, \varphi_{i_n}}
(q_1,\dots,q_m,p_1,\ldots,p_{n}),\nn
\eeqar
and we can apply the eikonal approximation to the matrix elements 
involving the would-be Goldstone bosons, since here all mass terms can be neglected, in the high-energy limit.

In the eikonal approximation \refeq{eikappdef}, which we denote by the subscript eik., 
the integral \refeq{softmasssingloop} simplifies into a scalar three-point function, usually called $C_0$. In the 
high-energy limit \refeq{SudaklimB}, and in logarithmic approximation (LA) this yields \cite{Denn3}  
\beqar \label{genericeikonalapp}
\Isoft_{\mathrm{eik.}}&\LA&
\frac{N_{\mathrm{eik.}}(0)}{(p_k+p_l)^2}\left\{
\frac{1}{2}\ln^2\left(\frac{-(p_k+p_l)^2-\ri\varepsilon}{M^2_{\varphi_s}-\ri\varepsilon}\right)
+\sum_{m=k,l}I_C(p_m^2,M_{\varphi_s},M_{\varphi_{i'_m}})
\right\},
\eeqar
where $I_C$  represents the dilogarithmic integral
\beq\label{dilogs}
I_C(p^2,M_{0},M_{1}):=-\int_0^1\frac{\mathrm{d}x}{x}\ln\left(1+\frac{M_1^2-M_0^2-p^2}{M_0^2-\ri\varepsilon}x+ \frac{p^2}{M_0^2-\ri\varepsilon}x^2\right).
\eeq
Such integrals   lead to large logarithms only when the mass $M^2_0$ is very small compared to $p^2$ and/or $M^2_1-p^2$. 

In order to evaluate the Feynman diagrams of the type \refeq{genericDLdiagramA} in the eikonal approximation \refeq{eikappdef},  we first concentrate on the vertex  $\varphi_{i_k}(p_k)\to \varphi_s(q)\varphi_{i'_k}(p_k-q)$ and we consider the
insertion of the propagator of the soft particle $\varphi_s$ into the external leg $\varphi_{i_k}$
\beqar\label{eikvertex}
\left[\hspace{3mm}
\vcenter{\hbox{\begin{picture}(80,80)(5,-40)
\Line(65,0)(5,0)
\Line(30,0)(30,-25)
\Vertex(30,-25){2}
\Vertex(30,0){2}
\GCirc(65,0){15}{1}
\Text(27,-15)[r]{\scriptsize$\varphi_s$}
\Text(18,3)[rb]{\scriptsize$\varphi_{i_k}$}
\Text(40,3)[cb]{\scriptsize$\varphi_{i'_k}$}
\end{picture}}}
\right]_{\mathrm{eik.}}&=&
v_{\varphi_{i_k}}(p_k)G^{\bar{\varphi}_{s}\varphi_{s}}_{\mathrm{eik.}}(q)
\ri\Gamma_{\mathrm{eik.}}^{\bar{\varphi}_{s}\bar{\varphi}_{i'_k}\varphi_{i_k}}(-q,-p_k+q,p_k)
G^{\bar{\varphi}_{i'_k}\varphi_{i'_k}}_{\mathrm{eik.}}(p_k-q)
\times\nl&&
G_0^{\underline{\varphi}_{i_1}\dots\,\underline{\varphi}_{i'_k}\dots\,\underline{\varphi}_{i_n}}(p_1, \dots ,p_{k}, \dots ,p_n)\prod_{m\neq k}v_{\varphi_{i_m}}(p_m).
\eeqar
The notation used here is described in detail in \refeq{Bornampli} and in \refapp{app:GFs}.
The first line on the rhs of \refeq{eikvertex} corresponds to the left part  of the diagram on the left-hand side (lhs),
with the  propagators and the vertex in eikonal approximation.
The second line 
corresponds to the ``white blob'' in the diagram on the lhs. 

We are interested in diagrams \refeq{eikvertex} involving external scalars (Higgs fields or would-be Goldstone bosons), fermions, antifermions or transverse gauge bosons, \ie $\varphi_{i_k}=\Phi_i$, $ f^\kappa_{j,\si}$, $ \bar{f}^\kappa_{j,\si}$ or $V^a_\rT$. For the internal lines $\varphi_s$, $\varphi_{i'_k}$, in general, we have to consider all possible combinations that are allowed by the 
electroweak Feynman rules (see \refapp{Feynrules}). However, it turns out that most of them are suppressed in the eikonal approximation.   
\begin{itemize}
\item
Firstly, all diagrams with vertices $\varphi_s \bar{\varphi}_{i'_k}\varphi_{i_k}$ that have couplings with mass dimension can be neglected. 
Only the 3-particle vertices that have dimensionless couplings need to be considered, 
\ie the vertices of the type
$VVV$, $V\Phi\Phi$, $V f\bar{f}$, $\Phi f\bar{f}$, and the scalar--ghost--antighost vertices $Vu\bar{u}$. 
\item
Secondly, in the limit $q^\mu  \to 0$, 
only those diagrams where the soft particle is a gauge-boson, $\varphi_s=V^a$, are not suppressed, whereas  all other diagrams are mass-suppressed. This  can be easily verified using the Feynman rules given in \refapp{Feynrules}, 
the transversality of the gauge-boson polarization vectors, the   Dirac equation for fermionic spinors, and the identity $\ps^2=p^2$. 
\end{itemize}
For the diagrams that are not suppressed 
one obtains 
\beq\label{eikvertexb}
\left[\hspace{3mm}
\vcenter{\hbox{\begin{picture}(80,80)(5,-40)
\Line(65,0)(5,0)
\Photon(30,0)(30,-25){2}{3}
\Vertex(30,0){2}
\GCirc(65,0){15}{1}
\Text(27,-15)[r]{\scriptsize$\bar{V}^a_\mu$}
\Text(18,3)[rb]{\scriptsize$\varphi_{i_k}$}
\Text(40,3)[cb]{\scriptsize$\varphi_{i'_k}$}
\end{picture}}}
\right]_{\mathrm{eik.}}=\frac{-eI^{\bar{V}^a}_{\varphi_{i'_k} \varphi_{i_k}} }{D_1}\,  2p^\mu_k \,
\M_0^{\varphi_{i_1}\dots\, \varphi_{i'_k}\dots\,\varphi_{i_n}}(p_1,\dots ,p_{k},\dots ,p_n),
\eeq
where 
$D_1=[(p_k-q)^2-M_{\varphi_{i'_k}}^2+\ri \varepsilon]$ is the denominator of the internal propagator, and  the propagator 
of the soft gauge boson has been omitted.  
Note that this expression for the insertion of an {\em outgoing}  soft gauge boson $V^a_\mu$ into the external leg of an {\em incoming}  particle (or antiparticle) $\varphi_{i_k}=\Phi_i, f^\kappa_{j,\si}, \bar{f}^\kappa_{j,\si}, V^a_\rT$ is proportional to its  {\em incoming} momentum $p^\mu_k$ and to the gauge coupling $ I^{\bar{V}^a}_{\varphi_{i'_k} \varphi_{i_k}}$.
Here we  illustrate how \refeq{eikvertexb} is obtained in  the case of external transverse gauge bosons. For  $\varphi_{i_k}=V_\rT^b$, 
we have
\beqar\label{eikvertexGB}
\left[\hspace{3mm}
\vcenter{\hbox{\begin{picture}(80,80)(5,-40)
\Photon(65,0)(30,0){2}{4}
\Photon(5,0)(30,0){2}{3}
\Photon(30,0)(30,-25){2}{3}
\Vertex(30,0){2}
\GCirc(65,0){15}{1}
\Text(27,-15)[r]{\scriptsize$\bar{V}^a_\mu$}
\Text(18,3)[rb]{\scriptsize$V_\rT^b$}
\Text(40,3)[cb]{\scriptsize$V^{b'}$}
\end{picture}}}
\right]_{\mathrm{eik.}}&=&
\varepsilon_\rT^\nu(p_k)
\ri e I^{\bar{V}^a}_{V^{b'} V^{b}}
\left[g^{\mu}_{\nu'}p_{k\nu}-2 g_{\nu'\nu}p_{k}^{\mu}+g_{\nu}^{\mu}p_{k\nu'}\right]
\times \nl&&
 \frac{-\ri g^{\nu'\nu''}}{D_1}
G_{0,\nu''}^{\underline{\varphi}_{i_1}\dots\,\underline{V}^{b'}\dots\,\underline{\varphi}_{i_n}}(p_1, \dots ,p_{k}, \dots ,p_n)\prod_{n\neq k}v_{\varphi_{i_n}}(p_n)\nl
&=&
\frac{-e I^{\bar{V}^a}_{V^{b'} V^{b}} }{D_1}
\left[-\varepsilon_\rT^\nu(p_k) p_{k\nu}g^{\mu\nu'}
+2p_{k}^\mu \varepsilon_\rT^{\nu'}(p_k)
-\varepsilon_\rT^{\mu}(p_k)p_{k}^{\nu'}\right]
\times \nl&&
G_{0,\nu'}^{\underline{\varphi}_{i_1}\dots\,\underline{V}^{b'}\dots\,\underline{\varphi}_{i_n}}(p_1, \dots ,p_{k}, \dots ,p_n)\prod_{n\neq k}v_{\varphi_{i_n}}(p_n).
\eeqar
The first term within the square brackets cancels 
owing to the transversality of the polarization vector $\varepsilon_\rT^\nu(p_k)p_{k\nu}=0$. 
The third term, which is proportional to $p_k^{\nu'}$, 
is suppressed owing to the Ward identity 
\mda
\beqar
\lefteqn{p_{k}^{\nu'} G_{0,\nu'}^{\underline{\varphi}_{i_1}\dots\,\underline{V}^{b'}\dots\,\underline{\varphi}_{i_n}}(p_1, \dots ,p_{k}, \dots ,p_n)\prod_{n\neq k}v_{\varphi_{i_n}}(p_n)=}\quad&&\nl
&=&\ri^{(1-Q_{V^{b'}})}M_{V^{b'}}\M_{0}^{\varphi_{i_1}\dots\,\Phi_{b'}\dots\,\varphi_{i_n}}(p_1, \dots ,p_{k}, \dots ,p_n)
,
\eeqar
\mua
and the second term, which is proportional to $\varepsilon_\rT^{\nu'}(p_k)$, leads to \refeq{eikvertexb}.

We note that here,  but also in other derivations, owing to gauge-boson emission $\varphi_{i_k} \rightarrow V^a \varphi_{i'_k}$, 
we obtain expressions where the wave function $v_{\varphi_{i_k}}(p_{k})$ with mass $p_k^2=M_{\varphi_{i_k}}^2$ is contracted 
with a line $\varphi_{i'_k}$ carrying a mass  $M_{\varphi_{i'_k}}^2$ that can be different from $M_{\varphi_{i_k}}^2$ [in \refeq{eikvertexGB} we have $\varphi_{i_k}=V^{b}$ and $\varphi_{i'_k}=V^{b'}$].
However, in the limit $s\gg M_{\varphi_{i_k}}^2,M_{\varphi_{i'_k}}^2$, such expressions are 
identified with matrix elements for $\varphi_{i'_k}$, \ie
\beqar \label{shortBornampli3}
G^{\underline{\varphi}_{i_1}\ldots\,\underline{\varphi}_{i'_k}\ldots\,\underline{\varphi}_{i_n}}(p_1,\ldots, p_n)
\prod_{j=1}^n v_{\varphi_{i_j}}(p_{j})
&\sim&
\M_0^{\varphi_{i_1} \ldots \,\varphi_{i'_k}\ldots \,\varphi_{i_n}}(p_1,\ldots,p_k,\ldots, p_n)
\eeqar
since for the polarization vectors of fermions and transverse gauge bosons\footnote{
Note that  \refeq{shortBornampli3b} is not valid for polarization vectors of longitudinal gauge bosons. 
However, these are treated by means of the GBET.
}
\beqar \label{shortBornampli3b}
v_{\varphi_{i_k}}(p_{k})
&=&
v_{\varphi_{i'_k}}(p_{k})
+\O\left(\frac{M}{\sqrt{s}} \right).
\eeqar
To be precise,  the correction of order $\O\left(M/\sqrt{s}\right)$ in \refeq{shortBornampli3b}
occurs only for fermionic polarization vectors, where $M\sim m_{\varphi_{i_k}}- m_{\varphi_{i'_k}}$. In the case of transverse gauge bosons such correction does not occur
since the corresponding polarization vectors are independent of the gauge-boson mass.


We can now consider the complete set of electroweak  Feynman diagrams leading to DL mass singularities, \ie the diagrams
where virtual gauge
bosons $V^a=\PA,\PZ,\PWp,\PWm$ 
are exchanged between all possible pairs of external
legs
\beq\label{genericDLdiagram}
\delta^\DL \M^{\varphi_{i_1} \ldots \varphi_{i_n}}=
\sum_{k=1}^n\sum_{l<
  k}\sum_{V^a=A,Z,W^\pm}
\sum_{\varphi_{i'_k},\varphi_{i'_l}}
\left[
\vcenter{\hbox{\begin{picture}(80,80)(5,-40)
\Line(65,0)(25,30)
\Line(65,0)(25,-30)
\PhotonArc(65,0)(32.5,143.13,216.87){2}{4.5}
\Vertex(39,19.5){2}
\Vertex(39,-19.5){2}
\GCirc(65,0){15}{1}
\Text(30,0)[r]{\scriptsize$V^a$}
\Text(20,27)[rb]{\scriptsize$\varphi_{i_k}$}
\Text(20,-27)[rt]{\scriptsize$\varphi_{i_l}$}
\Text(46,19)[lb]{\scriptsize$\varphi_{i'_k}$}
\Text(46,-19)[lt]{\scriptsize$\varphi_{i'_l}$}
\end{picture}}}
\right]_{\mathrm{eik.}}.
\eeq

Combining the soft gauge-boson insertions \refeq{eikvertexb} for the external legs  $\varphi_{i_k}$ and $\varphi_{i_l}$, and using the high-energy approximation \refeq{genericeikonalapp} one easily finds
\beqar \label{eikonalappA}
\lefteqn{\delta^\DL \M^{\varphi_{i_1} \ldots \varphi_{i_n}}=}\quad&&\nl&=&
\sum_{k=1}^n\sum_{l<
  k}\sum_{V^a=A,Z,W^\pm}
\sum_{\varphi_{i'_k}, \varphi_{i'_l}}
\int \frac{\rd^4q}{(2\pi)^4} \frac{-4\ri e^2
  p_kp_lI^{V^a}_{\varphi_{i'_k} \varphi_{i_k}} I^{\bar{V}^a}_{\varphi_{i'_l} \varphi_{i_l}} \M_0^{\varphi_{i_1} \ldots \,\varphi_{i'_k}
    \ldots \,\varphi_{i'_l} \ldots \,\varphi_{i_n}}}
{(q^2-M_{V^a}^2)[(p_k-q)^2-M_{\varphi_{i'_k}}^2][(p_l+q)^2-M_{\varphi_{i'_l}}^2]}\nl
&\LA&
\sum_{k=1}^n\sum_{l<
  k}\sum_{V^a=A,Z,W^\pm}
\sum_{\varphi_{i'_k}, \varphi_{i'_l}}
I^{V^a}_{\varphi_{i'_k} \varphi_{i_k}} I^{\bar{V}^a}_{\varphi_{i'_l} \varphi_{i_l}} \M_0^{\varphi_{i_1} \ldots \,\varphi_{i'_k}
    \ldots \,\varphi_{i'_l} \ldots \,\varphi_{i_n}}
\nl&&{}\times
\frac{\alpha}{4\pi}\left\{
\ln^2\left(\frac{|r_{kl}|}{M^2_{V^a}}\right)
+2\sum_{m=k,l}I_C(M^2_{\varphi_{i_m}},M_{V^a},M_{\varphi_{i'_m}})
\right\},
\eeqar
where $r_{kl}=(p_k+p_l)^2\sim 2p_kp_l$.
The integrals $I_C(M^2_{\varphi_{i_m}},M_{V^a},M_{\varphi_{i'_m}})$, defined by \refeq{dilogs},  
lead to large logarithms only  if the mass of the soft gauge boson $V^a$ is much smaller than one of the other masses. 
This occurs  for all photon-exchange diagrams
as well as for those weak-boson exchange diagrams 
where 
one of the particles $\varphi_{i_m},\varphi_{i'_m}$ is a top quark or a Higgs boson. 
In the case of photonic diagrams we have 
\beq\label{photondilogs}
I_C(M^2_{\varphi_{i}},\la, M_{\varphi_{i}})\LA
-\frac{1}{4}\ln^2{\frac{M^2_{\varphi_{i}}}{\la^2}},
\eeq
in LA, where $\la\ll M_{\varphi_{i}}$ is the photon-mass regulator.
For the diagrams with weak bosons coupling to top quarks the relevant integrals yield
\beqar\label{topDLcontri}
I_C(\Mt^2,\MZ,\Mt)
&\LA&-\frac{1}{4}\ln^2{\left(\frac{\Mt^2}{\MZ^2}\right)}
,\nl
I_C(\Mt^2,\MW,\Mb)
&\LA& 
I_C(\Mb^2,\MW,\Mt)
\LA-\frac{1}{2}\ln^2{\left(\frac{\Mt^2}{\MW^2}\right)},
\eeqar
in the heavy-top limit $\Mt\gg\MZ,\MW$. For the diagrams where 
weak bosons couple to Higgs bosons we have the logarithmic contributions
\beqar\label{HiggsDLcontri}
I_C(M_{V^a}^2,M_{V^a},\MH)
\LA
I_C(\MH^2,M_{V^a},M_{V^a})
\LA-\frac{1}{2}\ln^2{\left(\frac{\MH^2}{M_{V^a}^2}\right)},
\eeqar
in the heavy-Higgs limit $\MH\gg M_{V^a}=\MW,\MZ$.
Recall that the logarithmic contributions to the imaginary part of the corrections 
are not relevant at one-loop level and are therefore neglected in the logarithmic approximation (LA).
Using  \refeq{photondilogs} for the photonic diagrams we can rewrite \refeq{eikonalappA} as
\beqar \label{eikonalapp}
\lefteqn{\delta^\DL \M^{\varphi_{i_1} \ldots \varphi_{i_n}}
=\frac{\alpha}{4\pi}\sum_{k=1}^n\sum_{l\neq k}\sum_{V^a=A,Z,W^\pm} \sum_{\varphi_{i'_k}, \varphi_{i'_l}}
I^{V^a}_{\varphi_{i'_k} \varphi_{i_k}} I^{\bar{V}^a}_{\varphi_{i'_l} \varphi_{i_l}} \M_0^{\varphi_{i_1} \ldots \,\varphi_{i'_k} \ldots \,\varphi_{i'_l} \ldots \,\varphi_{i_n}}}&&\qquad
\nl&&{}\times
\left[
\frac{1}{2}\log^2{\left(\frac{|r_{kl}|}{M^2_{V^a}}\right)}-\de_{V^aA}\frac{1}{2}
\log^2{\left(\frac{M^2_{\varphi_{i_k}}}{\la^2}\right)}
+(1-\de_{V^aA})2I_C(M^2_{\varphi_{i_k}},M_{V^a},M_{\varphi_{i'_k}})
\right],\nln
\eeqar
where  $\de_{V^aA}$ represents the Kronecker symbol.
This formula  applies to external chiral fermions, Higgs bosons, and transverse gauge bosons. 
The DL corrections to processes involving external longitudinal gauge bosons $\varphi_{i}=\PZ_\rL,\PWpm$ are obtained
from the DL corrections to the processes involving the corresponding would-be Goldstone bosons $\varphi_{i} =\chi, \phi^\pm$ using \refeq{DLGBET2}.
In practice one has to use the gauge couplings to would-be Goldstone bosons on the rhs of \refeq{eikonalapp}.
 
Note that the DL terms $\log^2{|r_{kl}|/M^2_{V^a}}$ in \refeq{eikonalapp} are angular-dependent since the invariant $r_{kl}$ depends on the angle between the momenta $p_k$ and $p_l$. Writing
\beq \label{angsplit}
\log^2{\left(\frac{|r_{kl}|}{M^2}\right)}=
\log^2{\left(\frac{s}{M^2}\right)}+
2\log{\left(\frac{s}{M^2}\right)}\log{\left(\frac{|r_{kl}|}{s}\right)}+
\log^2{\left(\frac{|r_{kl}|}{s}\right)},
\eeq
the angular-dependent part can be isolated in logarithms of $r_{kl}/s$ which
lead to a  subleading soft--collinear ($\SS$) contribution of order
$\alpha\log(s/M^2) \log(|r_{kl}|/s)$. The terms $\alpha\log^2(|r_{kl}|/s)$ can be neglected if all invariants are of the same order, as we have assumed in \refeq{Sudaklim}.
The terms of type $\alpha\log^2(s/M^2)$ together with the 
additional contributions from photon
loops in \refeq{eikonalapp} are angular independent. These terms are denoted as the leading soft--collinear ($\SC$) contribution.

\section{Leading soft--collinear contributions}
The angular-independent leading soft--collinear corrections can be further simplified using (approximate) charge-conservation relations that follow from 
global $\ewgroup$ symmetry. 
The variation of the electroweak Lagrangian with respect to 
global $\SUtwo\times\Uone$ transformations leads to  following relations 
between  lowest-order  $S$-matrix elements 
\beq\label{globinvar}
\ri e \sum_{k=1}^n\sum_{\varphi_{i'_k}} I^{V^a}_{\varphi_{i'_k}\varphi_{i_k}} 
\M_0^{\varphi_{i_1} \ldots \,\varphi_{i'_k} \ldots \,\varphi_{i_n}}
=
\O\left(v^m E^{d-m}\right),\qquad m>0,
\eeq
where $d$ is the  mass-dimension of the matrix elements on the lhs. In the special case  $V^a=A$,  the rhs vanishes owing to exact $U(1)_\elm$ symmetry, \ie the electric charge is exactly conserved. 
For $V^a=Z,W^\pm$, instead, the rhs is in  general non-vanishing and contains terms proportional to the vev $v$ owing to spontaneous symmetry breaking.
However,
since we restrict ourselves to matrix elements that are not mass-suppressed \refeq{scaling}, these contributions can be neglected in the 
high-energy limit.

Using \refeq{globinvar}, the angular-independent ($\SC$) logarithms
in \refeq{eikonalapp} can be written as a single sum over external legs,
\beq \label{SCsum}
\de^{\SC} \M^{\varphi_{i_1} \ldots \,\varphi_{i_n}} =\sum_{k=1}^n \sum_{\varphi_{i''_k}}\delta^\SC_{\varphi_{i''_k}\varphi_{i_k}}
\M_0^{\varphi_{i_1} \ldots \,\varphi_{i''_k}\ldots \,\varphi_{i_n}}.
\end{equation} 
After evaluating the sum over  $A$, $Z$,  $W^+$ and $W^-$, in \refeq{eikonalapp}, the
correction factors read
\beqar \label{deSC} 
\de^\SC_{\varphi_{i''_k}\varphi_{i_k}}&=&
-\frac{\alpha}{8\pi} C^{\ew}_{\varphi_{i''_k}\varphi_{i_k}}\log^2{\left(\frac{s}{\MW^2}\right)}
+\de_{\varphi_{i''_k}\varphi_{i_k}}\left\{\frac{\alpha}{4\pi}
(I^Z)_{\varphi_{i_k}}^2 \log{\left(\frac{s}{\MW^2}\right)}
\log{\left(\frac{\MZ^2}{\MW^2}\right)} \right.
\nl&&\left.{}-\frac{1}{2} Q_{\varphi_{i_k}}^2L^\elm(s,\la^2,M^2_{\varphi_{i_k}})+ \de^{\SC,\mathrm{h}}_{\varphi_{i_k}}\right\}.
\eeqar
The first term represents the DL symmetric-electroweak part and is
proportional to the electroweak Casimir operator $\cew$ defined in
\refeq{CasimirEW}. This is always diagonal in the indices $\varphi_{i''_k}\varphi_{i_k}$, except in the case of  external transverse
neutral gauge bosons, where the non-diagonal components $\cew_{AZ}$ and $\cew_{ZA}$
give rise to mixing between amplitudes involving photons and \PZ~bosons [see \refeq{physadjointcasimir},\refeq{adjKron}]. 
The term proportional to $(I^Z)^2$
originates from \PZ-boson loops, owing to the difference between $\MW$ and $\MZ$. The term
\beq
L^\elm(s,\la^2,M^2_\varphi)
:=\frac{\alpha}{4\pi}\left\{
 2\log{\left(\frac{s}{\MW^2}\right)}\log{\left(\frac{\MW^2}{\la^2} \right)}
+\log^2{\left(\frac{\MW^2}{\la^2}\right)}
-\log^2{\left(\frac{M_{\varphi}^2}{\la^2}\right)}\right\}
\eeq
contains all logarithms of pure electromagnetic origin. Finally, there are additional 
logarithms which  originate from the last term on the second line in \refeq{eikonalapp}
through \refeq{topDLcontri} and \refeq{HiggsDLcontri}. 
Also such contributions factorize in the simple form given by \refeq{SCsum} and \refeq{deSC}, with
\beqar\label{htopDLcontri}
\de^{\SC,\mathrm{h}}_{\varphi_{i_k}}&=&-\frac{\alpha}{2\pi}\sum_{V^a=Z,W^\pm} \sum_{\varphi_{i'_k}}
|I^{V^a}_{\varphi_{i'_k}\varphi_{i_k}}|^2
I_C(M^2_{\varphi_{i_k}},M_{V^a},M_{\varphi_{i'_k}}).
\eeqar
This is due to the fact that such large logarithms occur 
only if  one of the particles $\varphi_{i_k},\varphi_{i'_k}$ in \refeq{eikonalapp}
is a top quark or a Higgs boson\footnote{
This permits to write
\beqar
\sum_{\varphi_{i'_k}}
I^{\bar{V}^a}_{\varphi_{i''_k}\varphi_{i'_k}}
I^{V^a}_{\varphi_{i'_k}\varphi_{i_k}}
I_C(M^2_{\varphi_{i_k}},M_{V^a},M_{\varphi_{i'_k}})
&\LA&
\de_{\varphi_{i''_k}\varphi_{i_k}}
\sum_{\varphi_{i'_k}}
|I^{V^a}_{\varphi_{i'_k}\varphi_{i_k}}|^2
I_C(M^2_{\varphi_{i_k}},M_{V^a},M_{\varphi_{i'_k}}).\nonumber
\eeqar
}.
In practice, owing to the explicit form 
of the \PZ~ and \PW~ couplings to the Higgs boson and to the top quark, 
for each weak boson  $V^a=Z,W^\pm$  only one virtual particle $\varphi_{i'_k}$ contributes
to the sum in \refeq{htopDLcontri}.
In the case of external heavy quarks, \refeq{htopDLcontri} yields
\beqar
\de^{\SC,\mathrm{h}}_{\Pt^\rL}
&\LA& \frac{\alpha}{4\pi}\left[\frac{(3\cw^2-\sw^2)^2}{72\sw^2\cw^2}\ln^2{\left(\frac{\Mt^2}{\MZ^2}\right)}
+\frac{1}{2\sw^2}\ln^2{\left(\frac{\Mt^2}{\MW^2}\right)}
\right],
\nl
\de^{\SC,\mathrm{h}}_{\Pb^\rL}
&\LA& \frac{\alpha}{4\pi}\frac{1}{2\sw^2}\ln^2{\left(\frac{\Mt^2}{\MW^2}\right)},
\qquad
\de^{\SC,\mathrm{h}}_{\Pt^\rR}
\LA \frac{\alpha}{4\pi}\frac{2\sw^2}{9\cw^2}\ln^2{\left(\frac{\Mt^2}{\MZ^2}\right)}.
\eeqar
Numerically 
\beq
\de^{\SC,\mathrm{h}}_{\Pt^\rL}\approx 
\de^{\SC,\mathrm{h}}_{\Pb^\rL}\approx 0.3 \%
,\qquad
\de^{\SC,\mathrm{h}}_{\Pt^\rR}\approx 0.01 \%.
\eeq
For external Goldstone-bosons (longitudinal gauge bosons), and for Higgs bosons we have
\beqar
\de^{\SC,\mathrm{h}}_{\phi^\pm}
&\LA& 
\frac{\alpha}{4\pi}
\frac{1}{4\sw^2}\ln^2{\left(\frac{\MH^2}{\MW^2}\right)}
,\qquad
\de^{\SC,\mathrm{h}}_{\chi}
\LA 
\frac{\alpha}{4\pi}
\frac{1}{4\sw^2\cw^2}\ln^2{\left(\frac{\MH^2}{\MZ^2}\right)}
,\nl
\de^{\SC,\mathrm{h}}_{H}
&\LA&
\frac{\alpha}{4\pi}\left[
\frac{1}{4\cw^2\sw^2}
\ln^2{\left(\frac{\MH^2}{\MZ^2}\right)}
+\frac{1}{2\sw^2}\ln^2{\left(\frac{\MH^2}{\MW^2}\right)}
\right],
\eeqar
and in the case  of a heavy Higgs boson, these contributions can reach the percent-level per line.
For instance at $ \MH=200 (500) \GeV$ we have 
\beq
\de^{\SC,\mathrm{h}}_{\phi^\pm}\approx 0.2 (0.9) \% ,
\qquad
\de^{\SC,\mathrm{h}}_{\chi}\approx   0.2 (1.0) \% ,
\qquad
\de^{\SC,\mathrm{h}}_{H}\approx  0.6 (2.7) \%.
\eeq

Formula \refeq{deSC} is in agreement with \citeres{Fadin:2000bq,Ciafaloni:2000ub} apart for 
the electromagnetic logarithm $\alpha\log^2(M^2_{\varphi_{i_k}}/\la^2)$, which is missing in \citere{Ciafaloni:2000ub}, 
and for the top-, Higgs-, and \PZ-mass dependent logarithms, which are missing in both these references.

\section{Subleading soft--collinear contributions}
The contribution of the second term of \refeq{angsplit} to
\refeq{eikonalapp} remains a sum over pairs of external legs,
\beq \label{SScorr}
\de^\SS \M^{\varphi_{i_1} \ldots \varphi_{i_n}} =\sum_{k=1}^n
\sum_{l<k}\sum_{V^a=A,Z,W^\pm}\sum_{\varphi_{i'_k},\varphi_{i'_l}}\delta^{V^a,\SS}_{\varphi_{i'_k}\varphi_{i_k} \varphi_{i'_l}\varphi_{i_l}}
\M_0^{\varphi_{i_1}\ldots \,\varphi_{i'_k}\ldots \,\varphi_{i'_l}\ldots \,\varphi_{i_n}},
\eeq
with angular-dependent terms. The exchange of soft, neutral gauge
bosons contributes with
\beqar \label{subdl1} 
\de^{A,\SS}_{\varphi_{i'_k}\varphi_{i_k} \varphi_{i'_l}\varphi_{i_l}}&=&
\frac{\alpha}{2\pi} \left[\log{\left(\frac{s}{\MW^2}\right)}+\log{\left(\frac{\MW^2}{\la^2}\right)}\right]\lrs I_{\varphi_{i'_k}\varphi_{i_k}}^AI_{\varphi_{i'_l}\varphi_{i_l}}^A,\nl
\delta^{Z,\SS}_{\varphi_{i'_k}\varphi_{i_k} \varphi_{i'_l}\varphi_{i_l}}&=&
\frac{\alpha}{2\pi}\log{\left(\frac{s}{\MW^2}\right)}\lrs I_{\varphi_{i'_k}\varphi_{i_k}}^ZI_{\varphi_{i'_l}\varphi_{i_l}}^Z.
\eeqar
The photon couplings are always diagonal, $I^A_{\varphi_{i'}\varphi_{i}}=\de_{\varphi_{i'}\varphi_{i}} I^A_{\varphi_{i}}$, and the \PZ-couplings are diagonal everywhere except in the neutral scalar sector. Here, the non-diagonal components  $I^Z_{H\chi}$ and $I^Z_{\chi H}$ [see \refeq{ZHcoup}] give rise to mixing between matrix elements involving external Higgs bosons and longitudinal \PZ~bosons.
The exchange of soft charged gauge bosons yields
\beq \label{subdl2} 
\delta^{W^\pm,\SS}_{\varphi_{i'_k}\varphi_{i_k} \varphi_{i'_l}\varphi_{i_l}} =
\frac{\alpha}{2\pi}\log{\left(\frac{s}{\MW^2}\right)}\lrs I_{\varphi_{i'_k}\varphi_{i_k}}^\pm I_{\varphi_{i'_l}\varphi_{i_l}}^{\mp},
\eeq
and owing to the non-diagonal matrices $I^\pm$ [\cf
\refeq{ferpmcoup}, \refeq{scapmcoup} and \refeq{gaupmcoup}], 
contributions of $\SUtwo$-transformed Born matrix elements (with $\varphi_{i'_k}\neq \varphi_{i_k}, \varphi_{i'_l}\neq \varphi_{i_l}$ ) appear on
the lhs of \refeq{SScorr}. 
In general,
these transformed Born matrix elements are not related to the 
original Born matrix element and have to be evaluated explicitly.

The $\SS$ corrections for external longitudinal gauge bosons are obtained from \refeq{SScorr} with the GBET \refeq{DLGBET2},
\ie the couplings and the Born matrix elements for would-be Goldstone bosons
have to be used on the rhs of \refeq{SScorr}. 

The application of the above formulas is illustrated in \refch{ch:applicat} for the case of 4-particle processes 
\beq \label{40process}
\varphi_{i_1}(p_1) \varphi_{i_2}(p_2)\varphi_{i_3}(p_3) \varphi_{i_4}(p_4)\rightarrow 0,
\eeq
where according to  our convention \refeq{process}, all particles and momenta are incoming. Owing to total-momentum conservation we have $r_{12}=r_{34}$,
$r_{13}=r_{24}$ and  $r_{14}=r_{23}$, 
so that \refeq{SScorr} reduces to
\beqar  \label{4fsubdl} 
\lefteqn{\de^\SS \M^{\varphi_{i_1}\varphi_{i_2}\varphi_{i_3}\varphi_{i_4}} =
\frac{\alpha}{2\pi}\sum_{V^a=A,Z,W^\pm}\sum_{\varphi_{i'_k}, \varphi_{i'_l}}  \left[\log{\left(\frac{s}{\MW^2}\right)}+\log{\left(\frac{\MW^2}{M^2_{V^a}}\right)}\right]\times} \quad&&\nl&&
\left\{\log{\left(\frac{|r_{12}|}{s}\right)}\left[
I^{V^a}_{\varphi_{i'_1}\varphi_{i_1}}
I^{\bar{V}^a}_{\varphi_{i'_2}\varphi_{i_2}} 
\M_0^{\varphi_{i'_1}\varphi_{i'_2}\varphi_{i_3}\varphi_{i_4}}
+I^{V^a}_{\varphi_{i'_3}\varphi_{i_3}}I^{\bar{V}^a}_{\varphi_{i'_4}\varphi_{i_4}} 
\M_0^{\varphi_{i_1}\varphi_{i_2}\varphi_{i'_3}\varphi_{i'_4}}
\right]\right.\hspace{1.5cm}\nl
&&\left.{}+\log{\left(\frac{|r_{13}|}{s}\right)}\left[
I^{V^a}_{\varphi_{i'_1}\varphi_{i_1}}I^{\bar{V}^a}_{\varphi_{i'_3}\varphi_{i_3}} 
\M_0^{\varphi_{i'_1}\varphi_{i_2}\varphi_{i'_3}\varphi_{i_4}}
+I^{V^a}_{\varphi_{i'_2}\varphi_{i_2}}I^{\bar{V}^a}_{\varphi_{i'_4}\varphi_{i_4}} 
\M_0^{\varphi_{i_1}\varphi_{i'_2}\varphi_{i_3}\varphi_{i'_4}}
\right]\right.\nl
&&\left.{}+\log{\left(\frac{|r_{14}|}{s}\right)}\left[
I^{V^a}_{\varphi_{i'_1}\varphi_{i_1}}I^{\bar{V}^a}_{\varphi_{i'_4}\varphi_{i_4}} 
\M_0^{\varphi_{i'_1}\varphi_{i_2}\varphi_{i_3}\varphi_{i'_4}}
+I^{V^a}_{\varphi_{i'_2}\varphi_{i_2}}I^{\bar{V}^a}_{\varphi_{i'_3}\varphi_{i_3}} 
\M_0^{\varphi_{i_1}\varphi_{i'_2}\varphi_{i'_3}\varphi_{i_4}}
\right]\right\},
\eeqar
and the logarithm with $r_{kl}=s$ vanishes.
The corrections for $2\rightarrow 2$ processes  are obtained 
by crossing symmetry from the corrections \refeq{4fsubdl}
for the $4\to 0$ processes \refeq{40process}, \ie by substituting 
outgoing particles (antiparticles) by the corresponding incoming
antiparticles (particles), as illustrated in \refeq{crossedproc}.

\chapter{Collinear mass singularities from loop diagrams}
\label{factorization}
\newcommand{\Oper}{O}
In this chapter we treat the single-logarithmic (SL) mass singularities that originate from truncated loop diagrams.
First, we show that these mass-singular corrections, which we denote by $\de^\coll\M$,
originate from external-leg emission of {\em collinear virtual gauge bosons}  and we specify a collinear approximation to evaluate them.
Then, we prove the factorization identities 
\beq\label{collsubllogfact}
\de^{\coll} \M^{\varphi_{i_1} \ldots \, \varphi_{i_n}}(p_1,\ldots,p_n) =\sum_{k=1}^n \sum_{\varphi_{i'_k}}
\M_0^{\varphi_{i_1} \ldots \, \varphi_{i'_k} \ldots \, \varphi_{i_n}}(p_1,\ldots,p_n)\,
\delta^\coll_{\varphi_{i'_k}\varphi_{i_k}},
\eeq
for arbitrary (non mass-suppressed) matrix elements involving
external Higgs bosons, would-be Goldstone bosons, fermions, antifermions as well as transverse gauge bosons.
These identities permit to  factorize the collinear singularities in the same form as 
FRC's for the external fields.
The gauge-dependent collinear factors $\de^\coll$ are evaluated within the 't~Hooft--Feyn\-man gauge.
In order to obtain the complete and gauge-independent SL corrections 
that are associated to the external particles one has to include the contributions of
the corresponding  FRC's and, in the case of processes involving longitudinal gauge bosons,  the corrections to the GBET. 
These contributions, which  factorize in an obvious way, are derived in \refch{FRCSllogs}.
In order to  prove \refeq{collsubllogfact}
we use the {\em collinear Ward identities} that are derived in \refch{ch:CWI}.

\section{Mass singularities in truncated loop diagrams}
\newcommand{\light}{l}
\newcommand{\pext}{p_k}
\newcommand{\pextmu}{p_{k\mu}}
\newcommand{\pextvec}{\vec{p}_{k}}
As we already observed in \refch{se:soft-coll}, owing to the simple pole-structure of the propagators
in the 't~Hooft--Feynman gauge, the class of loop diagrams that potentially give rise to
mass singularities can be easily determined by applying the
analysis of  scalar loop integrals made by Kinoshita \cite{Kinoshita:1962ur}.
This tells us that  mass-singular logarithmic corrections arise only from the subset  of loop diagrams 
where an external on-shell line splits into two 
internal virtual lines\footnote{In the diagrammatic representation  \refeq{colldiagram}
all on-shell external legs $\varphi_{i_l}$ with $l\neq k$, which are not involved in our argumentation, are omitted. All external lines have to be understood as truncated (trunc.); the self-energy and mixing-energy insertions in external legs and the corresponding mass singularities  enter the FRC's  and the corrections to the GBET in \refeq{subllogfact2}.}
\beqar\label{colldiagram}
\vcenter{\hbox{\begin{picture}(85,50)(0,-25)
\Line(5,0)(25,0)
\Line(25,0)(60.9,14.1)
\Line(25,0)(60.9,-14.1)
\Vertex(25,0){2}
\GCirc(65,0){15}{1}
\Text(40,-15)[t]{\scriptsize $\varphi_c$}
\Text(5,5)[lb]{\scriptsize $\varphi_{i_k}$}
\Text(40,15)[b]{\scriptsize $\varphi_{i'_k}$}
\end{picture}}}.
\eeqar
The diagrams of this type 
involve SL as well as DL mass singularities. These latter originate from the subset of diagrams of type \refeq{genericDLdiagramA} in the soft-collinear region,  and have been already evaluated in \refch{se:soft-coll} using the eikonal approximation (eik.). Here we restrict ourselves to the remaining SL corrections that are obtained from\footnote{The appropriate sums over the internal lines in \refeq{colldiagram2} are not specified for the moment.}
\beqar\label{colldiagram2}
\left[
\vcenter{\hbox{\begin{picture}(85,50)(0,-25)
\Line(5,0)(25,0)
\Line(25,0)(60.9,14.1)
\Line(25,0)(60.9,-14.1)
\Vertex(25,0){2}
\GCirc(65,0){15}{1}
\Text(40,-15)[t]{\scriptsize $\varphi_c$}
\Text(5,5)[lb]{\scriptsize $\varphi_{i_k}$}
\Text(40,15)[b]{\scriptsize $\varphi_{i'_k}$}
\end{picture}}}
\right]_{\mathrm{trunc.}}
-\quad
\sum_{l\neq k}\left[\vcenter{\hbox{\begin{picture}(80,80)(5,-40)
\Line(65,0)(25,30)
\Line(65,0)(25,-30)
\CArc(65,0)(32.5,143.13,216.87)
\Vertex(39,19.5){2}
\Vertex(39,-19.5){2}
\GCirc(65,0){15}{1}
\Text(30,0)[r]{\scriptsize$\varphi_c$}
\Text(20,27)[rb]{\scriptsize$\varphi_{i_k}$}
\Text(20,-27)[rt]{\scriptsize$\varphi_{i_l}$}
\Text(46,19)[lb]{\scriptsize$\varphi_{i'_k}$}
\Text(46,-19)[lt]{\scriptsize$\varphi_{i'_l}$}
\end{picture}}}
\right]_{\mathrm{eik.}},
\eeqar
\ie by subtracting the  eikonal contributions from \refeq{colldiagram}.

For the moment we consider splittings $\varphi_{i_k}(p)\rightarrow
\varphi_c(q)\varphi_{i'_k}(p_k-q)$ involving  arbitrary combinations of
fields that are allowed by the electroweak Feynman rules. These lead to loop integrals of the type
\newcommand{\ddqt}{\frac{\mathrm{d}^{D-2}q_\rT}{(2\pi)^{D-2}}}
\newcommand{\ddqtz}{\frac{\mathrm{d}^{2-2\varepsilon}q_\rT}{(2\pi)^{2-2\varepsilon}}}
\beqar \label{masssingloop2}
I&=&
-\ri (4\pi)^2\mu^{4-D} \int\ddq
\frac{N(q)}{(q^2-M_{\varphi_c}^2+\ri \varepsilon)[(p_k-q)^2-M_{\varphi_{i'_k}}^2+\ri \varepsilon]}.  
\eeqar
Here, only the denominators of the  $\varphi_j$ and $\varphi_k$~propagators are explicit, whereas the remaining part of the diagrams, denoted by $N(q)$, is kept implicit. 
Since the soft contributions have been  subtracted in \refeq{colldiagram2}, 
we can assume that $N(q)$ is not singular in the 
soft limit $q^\mu\rightarrow 0$. 
%
The mass singularity in \refeq{masssingloop2} originates from the
denominators of the $\varphi_{c}$ and $\varphi_{i'_k}$~propagators in the
collinear region $q^\mu\rightarrow xp_k^\mu$,
 where the squares of the momenta $\pext$ and $\pext-q$ 
are small compared to the energy squared $\pext^2,(\pext-q)^2\ll s$. 
In order to show this, and to fix a precise prescription for 
extracting the part of the function $N(q)$ 
that enters the mass-singular part of \refeq{masssingloop2}, 
we introduce a Sudakov
parametrization  \cite{SUD} for the loop momentum  
\beq\label{Sudparam}
q^\mu=x\pext^\mu+y \light^\mu +q^\mu_\rT,
\eeq
where $\pext^\mu$ and the light-like four-vector $\light^\mu$,
\beq
\pext^\mu=(\pext^0,\pextvec),\qquad
 \light^\mu=(\pext^0,-\pext^0\pextvec/|\pextvec|),
\eeq
describe the component of $q^\mu$ which is collinear to the external momentum, whereas the
space-like vector $q^\mu_\rT$ with   
\beq
 q_\rT \pext= q_\rT\light =0,\qquad q_\rT^2=-|\vec{q}_\rT|^2
\eeq
represents the perpendicular component. In this parametrization we get 
\beq \label{masssingloop4}
I=
-4\ri(\pext\light) \mu^{4-D}
\int\mathrm{d}x\int\mathrm{d}y \int\ddqt \frac{N(q)}{(q^2-M_{\varphi_{c}}^2+\ri\varepsilon)[(\pext-q)^2-M_{\varphi_{i'_k}}^2+\ri\varepsilon]}.
\eeq
The denominators of the propagators read
\beqar
q^2-M_{\varphi_{c}}^2+\ri\varepsilon&=&x^2\pext^2+2xy(\pext\light)-|\vec{q}_\rT|^2-M_{\varphi_{c}}^2+\ri\varepsilon,
\nl
(\pext-q)^2-M_{\varphi_{i'_k}}^2+\ri\varepsilon&=&(1-x)^2\pext^2+2(x-1)y(\pext\light)-|\vec{q}_\rT|^2-M_{\varphi_{i'_k}}^2+\ri\varepsilon,
\eeqar
and are linear in  the variable $y$ owing to $\light^2=0$. 
For $x\not\in\{0,1\}$, the $y$~integral  has
single poles at
\beqar\label{poles}
y_0&=&\frac{|\vec{q}_\rT|^2-x^2\pext^2+M_{\varphi_{c}}^2-\ri\varepsilon}{2x(\pext\light)},\qquad x\neq 0,
\nl
y_1&=&\frac{|\vec{q}_\rT|^2-(1-x)^2\pext^2+M_{\varphi_{i'_k}}^2-\ri\varepsilon}{2(x-1)(\pext\light)},\qquad x\neq 1,
\eeqar
and is  non-zero only when the poles lie in opposite
complex half-planes, \ie for $0<x<1$. Therefore we have
\beq \label{masssingloop5}
I=
-\ri\frac{\mu^{4-D}}{(\pext\light)}
\int_0^1\frac{\mathrm{d}x}{x(x-1)} \int\ddqt \int\mathrm{d}y\, \frac{N(x,y,q_\rT)}{(y-y_0)(y-y_1)},
\eeq
and  closing the $y$~contour around one of the two poles we obtain
\beqar \label{masssingloop6}
I&=&-\frac{2\pi\mu^{4-D}}{(\pext\light)}\int_0^1\frac{\mathrm{d}x}{x(x-1)} \int\ddqt  \frac{ N(x,y_i,q_\rT)}{y_0-y_1}
\nl&=&
4\pi\mu^{4-D}\int_0^1\mathrm{d}x \int\ddqt \frac{N(x,y_i,q_\rT)}{|\vec{q}_\rT|^2+\Delta(x)},
\eeqar
where in the vicinity of $x=1,0$  the contour has to be closed around the pole at $y_i=y_0,y_1$, respectively. 
In the collinear region $|q_\rT|\rightarrow0$,
the transverse momentum integral in \refeq{masssingloop6}
exhibits a logarithmic singularity 
that is regulated by the mass terms in 
\beq\label{Delta}
\Delta(x)=(1-x)M_{\varphi_{c}}^2+xM_{\varphi_{i'_k}}^2-x(1-x)\pext^2.
\eeq
In leading approximation, we restrict ourselves to logarithmic
mass-singular contributions in \refeq{masssingloop6}.  Terms 
of order $|\vec{q}_\rT|^2$, $\pext^2$, $M_{\varphi_{c}}$ or $M_{\varphi_{i'_k}}$ 
are neglected in $N(q)$.
Since the relevant pole, $y_0$ or
$y_1$, is of order $|\vec{q}_\rT|^2/(\pext\light)$,
also contributions proportional to $y$ can be discarded. We therefore arrive
at the following simple 
recipe for $N(q)$ 
\beqar\label{collappdef}
&(1)\,&\mbox{Substitute $N(x,y,q_\rT)\rightarrow N(x,0,0)$, \ie  replace  $q^\mu \rightarrow x\pext^\mu$,}\hspace{3cm}
\nl&(2)\,&
\mbox{Neglect all mass 
terms in $N(x,0,0)$,}
\eeqar
which we denote as {\em collinear approximation}.
As already discussed for the 
eikonal approximation \refeq{eikappdef}, if the external state $\varphi_{i_k}$ is a longitudinal gauge boson, 
then the mass terms cannot be neglected. 
In this case  the GBET \refeq{GBET1} has to be used.

In collinear approximation,  the $q_\rT$ integration in \refeq{masssingloop6} 
can be easily performed, and expanding in $\varepsilon=4-D$ we obtain the leading
contribution 
\beqar \label{masssingloop7}
I&=&\Gamma(\varepsilon)\int_0^1\mathrm{d}x \left(\frac{4\pi\mu^2}{\De(x)}\right)^\varepsilon N(x,0,0)
\nl&=&
\frac{1}{\varepsilon}+\int_0^1\mathrm{d}x\, \log{\left(\frac{\mu^2}{\De(x)}\right)} N(x,0,0)-\gamma+\log{4\pi}+\O(\varepsilon).
\eeqar
Finally, omitting the ultraviolet singularity, which cancels in observables,
neglecting constant terms, and performing the integral, we obtain
\beq \label{masssingloop8}
I\LA \log{\left(\frac{\mu^2}{M^2}\right)}\int_0^1\mathrm{d}x\,  N(x,0,0),
\eeq
 in logarithmic approximation (LA).
The scale in the logarithm is of the order of 
the largest mass in
\refeq{Delta}, 
\beq
M^2\sim\max{(\pext^2,M_{\varphi_{c}}^2,M_{\varphi_{i'_k}}^2)}.
\eeq

\section{Factorization of collinear singularities}
\label{sec:Fact}
In this section,  we apply the collinear approximation \refeq{collappdef} to the complete set of electroweak  Feynman diagrams 
of the type  \refeq{colldiagram}, which  lead  to SL collinear  mass singularities that are associated to an external leg $\varphi_{i_k}$.
First, we concentrate on the splitting $\varphi_{i_k}(p_k)\rightarrow \varphi_c(q)\varphi_{i'_k}(p_k-q)$ of the external particle $\varphi_{i_k}$ into the propagators of the collinear particles $\varphi_c$ and $\varphi_{i'_k}$
\beqar\label{collvertex}
\left[\hspace{3mm}
\vcenter{\hbox{\begin{picture}(70,70)(5,-40)
\Line(60,0)(5,0)
\Line(30,0)(30,-25)
\Vertex(30,-25){2}
\Vertex(60,0){2}
\Vertex(30,0){2}
\Text(25,-15)[r]{\scriptsize$\varphi_c$}
\Text(18,3)[rb]{\scriptsize$\varphi_{i_k}$}
\Text(45,3)[cb]{\scriptsize$\varphi_{i'_k}$}
\end{picture}}}
\right]_{\mathrm{coll.}}&=&
v_{\varphi_{i_k}}(p_k)G^{\bar{\varphi}_{c}\varphi_{c}}_{\mathrm{coll.}}(q)
\ri\Gamma_{\mathrm{coll.}}^{\bar{\varphi}_{c}\bar{\varphi}_{i'_k}\varphi_{i_k}}(-q,-p_k+q,p_k)
G^{\bar{\varphi}_{i'_k}\varphi_{i'_k}}_{\mathrm{coll.}}(p_k-q).\nln
\eeqar
The notation used here is the same as for the eikonal vertex \refeq{eikvertex}.
The propagators and the vertex are evaluated in collinear approximation (coll.).
We are interested in diagrams \refeq{collvertex}
involving external scalar bosons ($\varphi_{i_k}=\Phi_i$), fermions and  antifermions ($\varphi_{i_k}= f^\kappa_{j,\si}$, $ \bar{f}^\kappa_{j,\si}$) as well as  
transverse gauge bosons  ($\varphi_{i_k}=V^a_\rT$). 
In general, for the internal lines $\varphi_c$, $\varphi_{i'_k}$, we have to consider all possible combinations 
generated by vertices $\varphi_c \bar{\varphi}_{i'_k}\varphi_{i_k}$ of the electroweak Feynman rules (see  \refapp{Feynrules}). 
However, if one applies the  collinear approximation \refeq{collappdef},
it turns out that only a restricted subset of  Feynman diagrams yields non-suppressed contributions.
\begin{itemize} 
\item First, we can restrict ourselves to the 3-particle vertices  that have dimensionless couplings, 
\ie the vertices of the type
$VVV$, $V\Phi\Phi$, $Vf\bar{f}$, $\Phi f\bar{f}$ and the scalar--ghost--antighost vertices $Vu\bar{u}$.  
\item
Secondly, in the collinear approximation it turns out that the splittings $V_\rT \to \Phi\Phi$, $V_\rT \to f\bar{f}$ and $V_\rT \to u\bar{u}$ are suppressed  owing to the transversality of the gauge-bosons polarization vectors, the splittings $\Phi\to f\bar{f}$ are suppressed owing to $\ps^2=p^2\ll s$, and also the splittings  $\bar{f}\to\Phi\bar{f}$ and $f\to\Phi f$ are suppressed owing to the Dirac equation for fermionic spinors.
\end{itemize}
Therefore, we need to consider only the splittings 
\beq\label{collGBsplittings}
\varphi_{i_k}(p_k) \rightarrow V_\mu^a(q) \varphi_{i'_k}(p_k-q),
\eeq
where  virtual gauge bosons $V^a=A,Z,W^+,W^-$ are emitted and $\varphi_{i_k}$ and $\varphi_{i'_k}$ are both fermions, gauge bosons, or scalars.

\subsection{Generic factorization identities}
In order to proceed  we introduce some shorthand notations for matrix elements and Green functions.
When we concentrate on a specific external leg $\varphi_{i_k}$, 
only the labels corresponding to this external leg are kept explicit.
For the  matrix element \refeq{Bornampli}
we use the shorthand 
\beq \label{shortBornampli}
\M^{\varphi_{i_k}}(p_{k})= v_{\varphi_{i_k}}(p_{k})G^{\underline{\varphi}_{i_k}\underline{\Oper}}(p_{k},r)=v_{\varphi_{i_k}}(p_{k})G^{\underline{\varphi}_{i_k}}(p_{k}),
\eeq
where the product of fields 
\beq\label{defoper}
\Oper(r)=\prod_{l\neq k}\varphi_{i_l}(p_l),\qquad r=\sum_{l\neq k}p_l,
\eeq
represents the remaining external legs  $\varphi_{i_l}$ with $l \neq k$. 
These are always assumed to be on-shell and contracted with the
corresponding wave functions, which are 
suppressed in the notation. Moreover, also 
$\Oper$ and the corresponding total  momentum $r$ are often not written.

In the following, we consider the diagrams of type \refeq{colldiagram2}
corresponding to the splittings \refeq{collGBsplittings} and
we derive the  factorization identities
\beqar\label{collfactorization}
\lefteqn{
\delta^\coll \M^{\varphi_{i_k}}(p_k)=}\qquad\nl
&=&\sum_{V^a=A,Z,W^\pm} \sum_{\varphi_{i'_k}}\left\{
\left[
\vcenter{\hbox{\begin{picture}(85,60)(0,-30)
\Line(5,0)(25,0)
\Line(25,0)(60.9,14.1)
\Photon(25,0)(60.9,-14.1){-2}{3}
\Vertex(25,0){2}
\GCirc(65,0){15}{1}
\Text(35,-14)[lt]{\scriptsize$V^a$}
\Text(35,+12)[lb]{\scriptsize$\varphi_{i'_k}$}
\Text(5,5)[lb]{\scriptsize $\varphi_{i_k}$}
\end{picture}}}
\right]_{\mathrm{trunc.}}
- \sum_{l\neq k} \sum_{\varphi_{i'_l}}\left[
\vcenter{\hbox{\begin{picture}(80,80)(5,-40)
\Line(65,0)(25,30)
\Line(65,0)(25,-30)
\PhotonArc(65,0)(32.5,143.13,216.87){2}{4.5}
\Vertex(39,19.5){2}
\Vertex(39,-19.5){2}
\GCirc(65,0){15}{1}
\Text(30,0)[r]{\scriptsize$V^a$}
\Text(20,27)[rb]{\scriptsize$\varphi_{i_k}$}
\Text(20,-27)[rt]{\scriptsize$\varphi_{i_l}$}
\Text(46,19)[lb]{\scriptsize$\varphi_{i'_k}$}
\Text(46,-19)[lt]{\scriptsize$\varphi_{i'_l}$}
\end{picture}}}
\right]_{\mathrm{eik.}}\right\}_{\mathrm{coll.}} 
\nl&=&
\sum_{\varphi_{i''_k}}
\vcenter{\hbox{\begin{picture}(80,60)(0,-30)
\Line(55,0)(10,0)
\GCirc(55,0){15}{1}
\Text(10,5)[lb]{\scriptsize$\varphi_{i''_k}$}
\end{picture}}}
\de^\coll_{\varphi_{i''_k}\varphi_{i_k}},
\eeqar
%
The detailed proof of these identities
depends on the spin of the external
particles. 
However, its basic structure can
be sketched in a universal way and consists of two main steps:
\begin{itemize}
\item After explicit evaluation of the  external vertices\footnote{Here the gauge-boson propagator has been omitted.} 
\beqar\label{gcollvertex}
\left[\hspace{3mm}
\vcenter{\hbox{\begin{picture}(70,70)(5,-40)
\Line(60,0)(5,0)
\Photon(30,0)(30,-25){2}{3}
\Vertex(60,0){2}
\Vertex(30,0){2}
\Text(25,-15)[r]{\scriptsize$\bar{V}_\mu^a$}
\Text(18,3)[rb]{\scriptsize$\varphi_{i_k}$}
\Text(45,3)[cb]{\scriptsize$\varphi_{i'_k}$}
\end{picture}}}
\right]_{\mathrm{coll.}}&=&
v_{\varphi_{i_k}}(p_k)
\ri\Gamma_{\mathrm{coll.}}^{\bar{V}_\mu^a\bar{\varphi}_{i'_k}\varphi_{i_k}}(-q,-p_k+q,p_k)
G^{\bar{\varphi}_{i'_k}\varphi_{i'_k}}_{\mathrm{coll.}}(p_k-q)\nln
\eeqar
in collinear approximation, and after explicit subtraction of the eikonal contributions obtained from \refeq{eikonalappA}, 
the lhs of
  \refeq{collfactorization} turns into\footnote{Here and in the
    following the $+\ri\varepsilon$ prescription of the propagators is
    suppressed in the notation.}
\beqar \label{UNIVcollamp}
\lefteqn{
\delta^\coll \M^{\varphi_{i_k}}(p_k)= \sum_{V^a=A,Z,W^\pm} \sum_{\varphi_{i'_k}} \mu^{4-D}\int\ddq
\frac{-\ri  e I^{\bar{V}^a}_{\varphi_{i'_k}\varphi_{i_k}}}{(q^2-M_{V^a}^2)[(p_k-q)^2-M_{\varphi_{i'_k}}^2]} \,K_{\varphi_{i_k}}} \quad
\nl&&\times \lim_{q^\mu\rightarrow xp_k^\mu} 
q^\mu
\left\{
\vcenter{\hbox{\begin{picture}(85,80)(12,20)
\Line(55,60)(15,60)
\Text(35,65)[b]{\scriptsize $\varphi_{i'_k}(p_k-q)$}
\Photon(30,30)(70,60){1}{6}
\Text(55,35)[t]{\scriptsize $V^a_\mu(q)$}
\GCirc(70,60){15}{1}
\end{picture}}}
- \sum_{\varphi_j}
\vcenter{\hbox{\begin{picture}(106,80)(-18,20)
\Line(55,60)(35,60)
\Line(35,60)(0,60)
\Text(30,65)[br]{\scriptsize $ \varphi_{i'_k}(p_k-q)$}
\Text(33,68)[bl]{\scriptsize $ \varphi_{j}(p_k)$}
\Photon(10,30)(35,60){1}{5}
\Vertex(35,60){2}
\Text(35,40)[t]{\scriptsize $V^a_\mu(q)$}
\GCirc(70,60){15}{1}
\end{picture}}}
\right\},
\hspace{2cm}
\eeqar     
with  $K_{\varphi_{i_k}}=1$ for scalar bosons and transverse gauge bosons 
and $K_{\varphi_{i_k}}=2$ for fermions. 
The first diagram appearing in
\refeq{UNIVcollamp} results from the first diagram of
\refeq{collfactorization} by omitting the external  vertex and
propagator \refeq{gcollvertex}.
The second diagram in \refeq{UNIVcollamp} originates from
the truncation of the self-energy and mixing-energy
($\varphi_i \varphi_j)$ insertions in the first diagram of
\refeq{collfactorization}.  Equation \refeq{UNIVcollamp} is derived in
\refses{se:fac_scal}--\ref{se:fac_fer}.

\item The contraction of the diagrams between the curly brackets on
  the rhs  of \refeq{UNIVcollamp} with the gauge-boson momentum
  $q^\mu$ can be simplified using the {\em collinear Ward identities}
\beqar\label{UNIVcollwi}
\lefteqn{\lim_{q^\mu\rightarrow xp_k^\mu} q^\mu
\left\{
\vcenter{\hbox{\begin{picture}(85,80)(12,20)
\Line(55,60)(15,60)
\Text(35,65)[b]{\scriptsize $\varphi_{i'_k}(p_k-q)$}
\Photon(30,30)(70,60){1}{6}
\Text(55,35)[t]{\scriptsize $V^a_\mu(q)$}
\GCirc(70,60){15}{1}
\end{picture}}}
- \sum_{\varphi_j}
\vcenter{\hbox{\begin{picture}(100,80)(-10,20)
\Line(55,60)(35,60)
\Line(35,60)(0,60)
\Text(30,65)[br]{\scriptsize $ \varphi_{i'_k}(p_k-q)$}
\Text(33,68)[bl]{\scriptsize $ \varphi_{j}(p_k)$}
\Photon(10,30)(35,60){1}{5}
\Vertex(35,60){2}
\Text(35,40)[t]{\scriptsize $V^a_\mu(q)$}
\GCirc(70,60){15}{1}
\end{picture}}}
\right\}}\quad&&\hspace{10cm}
\nl&&\quad
=\sum_{\varphi_{i_k''}}
\vcenter{\hbox{\begin{picture}(80,80)(0,-40)
\Line(55,0)(10,0)
\GCirc(55,0){15}{1}
\Text(10,5)[lb]{\scriptsize $\varphi_{i_k''}(p_k)$}
\end{picture}}}
 e I^{V^a}_{\varphi_{i_k''}\varphi_{i'_k}},
\eeqar
which are fulfilled in the collinear approximation and valid
up to mass-suppressed terms. These Ward identities
are derived in \refch{ch:CWI} using the BRS invariance of the
spontaneously broken $\SUtwo\times\Uone$ Lagrangian.

For the Green functions
corresponding to the diagrams within the curly brackets in
\refeq{UNIVcollamp} we introduce the shorthand
\beqar \label{subtractGF}
G^{[\underline{V}^a\underline{\varphi}_{i}] \underline{\Oper}}_\mu(q,p-q,r)&=&G^{\underline{V}^a\underline{\varphi}_i \underline{\Oper}}_\mu(q,p-q,r) 
- \sum_{\varphi_j}G^{\underline{V}^a \underline{\varphi}_i \varphi_j}_{\mu}(q,p-q,-p)
 G^{\underline{\varphi}_j \underline{\Oper}}(p,r)\nln
\eeqar
that will be used in the next sections.
\end{itemize}
Combining \refeq{UNIVcollwi} with \refeq{UNIVcollamp}, 
we obtain
\refeq{collfactorization}  with the collinear factor
\beqar\label{loopLA1}
\de^\coll_{\varphi_{i''}\varphi_{i}}
&=&
\sum_{V^a=A,Z,W^\pm}\sum_{\varphi_{i'}}
\mu^{4-D}\int\ddq
\frac{-\ri K_{\varphi_{i}} e^2I^{V^a}_{\varphi_{i''}\varphi_{i'}}I^{\bar{V}^a}_{\varphi_{i'}
\varphi_{i}}}{(q^2-M_{V^a}^2)[(p-q)^2-M_{\varphi_{i'}}^2]}
\nl&\LA&
\frac{\alpha}{4\pi}K_{\varphi_{i}}
\sum_{V^a=A,Z,W^\pm}\sum_{\varphi_{i'}}
I^{V^a}_{\varphi_{i''}\varphi_{i'}}I^{\bar{V}^a}_{\varphi_{i'}\varphi_{i}}
\log{\left(\frac{\mu^2}{\max{(M_{V^a}^2,M_{\varphi_{i}}^2,M_{\varphi_{i'}}^2)}}\right)},
\eeqar
where  we have used the logarithmic approximation \refeq{masssingloop8} for the loop integral \refeq{masssingloop2}.
The scale in the logarithms is determined by the largest mass in the loop.
For  photonic diagrams ($V^a=A$) it is given by the external mass $M_{\varphi_{i}}$, and for virtual massive gauge bosons $V^a=Z,W^+,W^-$ it is given  by  $\MW \sim \MZ$, $\MH$ or $\Mt$, depending on the diagrams.
Therefore we can write 
\beqar
\log{\left(\frac{\mu^2}{\max{(M_{V^a}^2,M_{\varphi_{i}}^2,M_{\varphi_{i'}}^2)}}\right)}
&=&\log{\frac{\mu^2}{\MW^2}}+\de_{V^a A}\log{\frac{\MW^2}{M_{\varphi_{i}}^2}}
\nl&&{}-
(1-\de_{V^a A})\log{\left(\frac{\max{(M_{V^a}^2,M_{\varphi_{i}}^2,M_{\varphi_{i'}}^2)}}{\MW^2}\right)},
\eeqar
so that we arrive at
\beqar\label{loopLA}
\de^\coll_{\varphi_{i''}\varphi_{i}}&\LA&
\frac{\alpha}{4\pi}K_{\varphi_{i}}\left\{
\cew_{\varphi_{i''}\varphi_{i}}\log{\frac{\mu^2}{\MW^2}}+ \de_{\varphi_{i''}\varphi_{i}}Q^2_{\varphi_{i}}\log{\frac{\MW^2}{M_{\varphi_i}^2}}
\right.\nl
&&\hspace{1cm}{}-\left.
\sum_{V^a=Z,W^\pm}\sum_{\varphi_{i'}}I^{V^a}_{\varphi_{i''}\varphi_{i'}}I^{\bar{V}^a}_{\varphi_{i'}\varphi_{i}}
\log{\left(\frac{\max{(\MW^2,M_{\varphi_{i}}^2,M_{\varphi_{i'}}^2)}}{\MW^2}\right)}\right\},
\eeqar
where the first term is a symmetric electroweak contribution  proportional to   the effective {\em electroweak Casimir operator} $\cew$ defined in \refeq{CasimirEW}. The second term is a purely electromagnetic contribution, and 
the remaining terms receive  contributions  only from diagrams with $\varphi_i$ and/or $\varphi_{i'}$ equal to a top quark or to a Higgs boson.


In the following sections we present the explicit derivations of \refeq{collfactorization} and \refeq{loopLA1} 
for external scalars, transverse gauge bosons and fermions.

\subsection{Factorization for scalars}
\label{se:fac_scal}
We first  consider the collinear enhancements  generated by the virtual splittings
\beq
\Phi_{i_k}(p_k) \rightarrow V_\mu^a(q) \Phi_{i'_k}(p_k-q),
\eeq
where an incoming on-shell Higgs boson or would-be  Goldstone boson
$\Phi_{i_k}=H,\chi,\phi^\pm$ emits a virtual collinear gauge boson
$V^a=A,Z,W^+,W^-$. 
In the collinear approximation $q^\mu\to x p_k^\mu$,  the vertices \refeq{gcollvertex} corresponding to these splittings give
\beq\label{scacollvertex}
\left[\hspace{3mm}
\vcenter{\hbox{\begin{picture}(70,70)(5,-40)
\DashLine(60,0)(5,0){4}
\Photon(30,0)(30,-25){2}{3}
\Vertex(60,0){2}
\Vertex(30,0){2}
\Text(25,-15)[r]{\scriptsize$\bar{V}^a_\mu$}
\Text(18,3)[rb]{\scriptsize$\Phi_{i_k}$}
\Text(45,3)[cb]{\scriptsize$\Phi_{i'_k}$}
\end{picture}}}
\right]_{\mathrm{coll.}}=
\lim_{q^\mu\rightarrow xp_k^\mu}
\frac{- e I^{\bar{V}^a}_{\Phi_{i'_k} \Phi_{i_k}}}{D_1}
(2p_k-q)_\mu 
=
\frac{- e  I^{\bar{V}^a}_{\Phi_{i'_k} \Phi_{i_k}}}{D_1} \left(\frac{2}{x}-1\right)q_\mu ,
\eeq
where $D_1=[(p_k-q)^2-M_{\Phi_{i'_k}}^2+\ri \varepsilon]$.
With this and with  \refeq{eikonalappA}
for the eikonal contributions  the amplitude 
\beqar\label{SCAcollfactorization}
\delta^\coll \M^{\Phi_{i_k}}(p_k)&=&\sum_{V^a} \sum_{\Phi_{i'_k}} \left\{
\left[
\vcenter{\hbox{\begin{picture}(85,60)(0,-30)
\DashLine(5,0)(25,0){4}
\DashLine(25,0)(60.9,14.1){4}
\Photon(25,0)(60.9,-14.1){-2}{3}
\Vertex(25,0){2}
\GCirc(65,0){15}{1}
\Text(35,-14)[lt]{\scriptsize$V^a$}
\Text(35,+12)[lb]{\scriptsize$\Phi_{i'_k}$}
\Text(5,5)[lb]{\scriptsize $\Phi_{i_k}$}
\end{picture}}}
\right]_{\mathrm{trunc.}}
- \sum_{l\neq k}\sum_{\varphi_{i'_l}}\left[
\vcenter{\hbox{\begin{picture}(80,80)(5,-40)
\DashLine(25,30)(39,19.5){4}\DashLine(39,19.5)(53,9){4}
\Line(65,0)(25,-30)
\PhotonArc(65,0)(32.5,143.13,216.87){2}{4.5}
\Vertex(39,19.5){2}
\Vertex(39,-19.5){2}
\GCirc(65,0){15}{1}
\Text(30,0)[r]{\scriptsize$V^a$}
\Text(20,27)[rb]{\scriptsize$\Phi_{i_k}$}
\Text(46,19)[lb]{\scriptsize$\Phi_{i'_k}$}
\Text(20,-27)[rt]{\scriptsize$\varphi_{i_l}$}
\Text(46,-19)[lt]{\scriptsize$\varphi_{i'_l}$}
\end{picture}}}
\right]_{\mathrm{eik.}}\right\}_{\mathrm{coll.}}
\nln
\eeqar
reads 
\beqar \label{SCAcollamp}
\delta^\coll \M^{\Phi_{i_k}}(p_k)&=& \sum_{V^a=A,Z,W^\pm} \sum_{\Phi_{i'_k}=H,\chi,\phi^\pm} \mu^{4-D}\int\ddq
\frac{\ri e I^{\bar{V}^a}_{\Phi_{i'_k}\Phi_{i_k}}}{(q^2-M_{V^a}^2)[(p_k-q)^2-M_{\Phi_{i'_k}}^2]}
\nl&&\times\lim_{q^\mu\rightarrow xp_k^\mu} \Biggl\{
\left(\frac{2}{x}-1\right)q^\mu
G_{\mu}^{[\underline{V}^a\underline{\Phi}_{i'_k}]}(q,p_k-q)
\nl&&\hspace{2cm}
{}+2p_k^\mu \sum_{l\neq k}\sum_{\varphi_{i'_l}}\frac{ 2e p_{l\mu} I^{V^a}_{\varphi_{i'_l}\varphi_{i_l}}  }{[(p_l+q)^2-M_{\varphi_{i'_l}}^2]}
\M_0^{\Phi_{i'_k}\varphi_{i'_l}}(p_k,p_l)
\Biggr\}.
\eeqar     
According to the definition \refeq{subtractGF}, we have
\beqar\label{SCAampdiag}
\lefteqn{G^{[\underline{V}^a \underline{\Phi}_i]}_{\mu}(q,p-q)=}\quad\nl&&
\vcenter{\hbox{\begin{picture}(90,60)(0,30)
\DashLine(55,60)(10,60){4}
\Text(35,65)[b]{\scriptsize $\Phi_i(p-q)$}
\Photon(30,30)(70,60){2}{6}
\Text(55,35)[t]{\scriptsize $V^a_\mu(q)$}
\GCirc(70,60){15}{1}
\end{picture}}}
-\sum_{\Phi_j}
\vcenter{\hbox{\begin{picture}(95,60)(0,30)
\DashLine(70,60)(35,60){4}
\DashLine(35,60)(5,60){4}
\Text(35,65)[br]{\scriptsize $\Phi_i(p-q)$}
\Text(40,65)[bl]{\scriptsize $\Phi_j(p)$}
\Photon(10,30)(35,60){2}{5}
\GCirc(35,60){2}{0}
\Text(35,40)[t]{\scriptsize $V^a_\mu(q)$}
\GCirc(75,60){15}{1}
\end{picture}}}
-\sum_{V^c}
\vcenter{\hbox{\begin{picture}(95,60)(0,30)
\Photon(70,60)(35,60){2}{4}
\DashLine(35,60)(5,60){4}
\Text(35,65)[br]{\scriptsize $\Phi_i(p-q)$}
\Text(40,65)[bl]{\scriptsize $V^c(p)$}
\Photon(10,30)(35,60){2}{5}
\GCirc(35,60){2}{0}
\Text(35,40)[t]{\scriptsize $V^a_\mu(q)$}
\GCirc(75,60){15}{1}
\end{picture}}}.
\eeqar
Note that the subtracted contributions, when inserted in
\refeq{SCAcollfactorization}, correspond to external scalar
self-energies ($\Phi\Phi$) and scalar--vector mixing-energies ($\Phi
V$).

The expression between the curly brackets in
\refeq{SCAcollamp} can be simplified using\footnote{Since the soft-collinear contributions are subtracted, we do not need a regularization of  $1/x$ for $x\rightarrow 0$.},
\beq\label{colleik}
\lim_{q^\mu\rightarrow xp_k^\mu}\frac{ 2p_k p_{l}}{[(p_l+q)^2-M_{\varphi_{i'_l}}^2]}=\frac{1}{x}+\O\left(\frac{M^2}{s}\right),
\eeq
and the  {\em collinear Ward identity} \refeq{SCACWIres} for scalar bosons ($\varphi_i=\Phi_{i}$). This yields
\beqar
\lim_{q^\mu\rightarrow xp_k^\mu} \biggl\{\ldots\biggr\} &=&\lim_{q^\mu\rightarrow x p_k^\mu}
\left\{
\left(\frac{2}{x}-1\right)q^\mu
G_{\mu}^{[\underline{V}^a\underline{\Phi}_{i'_k}]}(q,p_k-q)
\right.\nl&&\left.{}
+\frac{2e}{x}\sum_{l\neq k}\sum_{\varphi_{i'_l}} I^{V^a}_{\varphi_{i'_l}\varphi_{i_l}}
\M_0^{\Phi_{i'_k}\varphi_{i'_l}}(p_k,p_l)
\right\}=\nl
&=&-e\sum_{\Phi_{i''_k}}I^{V^a}_{\Phi_{i''_k}\Phi_{i'_k}}
\M_0^{\Phi_{i''_k}}(p_k)\nl
&&{}+
\frac{2e}{x}\left\{\sum_{\Phi_{i''_k}}
I^{V^a}_{\Phi_{i''_k}\Phi_{i'_k}}\M_0^{\Phi_{i''_k}}(p_k)
+\sum_{l\neq k} \sum_{\varphi_{i'_l}}I^{V^a}_{\varphi_{i'_l}\varphi_{i_l}}
\M_0^{\Phi_{i'_k}\varphi_{i'_l}}(p_k,p_l)
\right\}.
\eeqar
Owing to global $\ewgroup$ invariance \refeq{globinvar}  
%
the part proportional to $1/x$ is mass-suppressed as expected, since the
soft-collinear contributions have been subtracted.  Thus,
\refeq{SCAcollamp} turns into
\beqar
\delta^\coll \M^{\Phi_{i_k}}(p_k) &=&  \sum_{V^a,\Phi_{i'_k},\Phi_{i''_k}}\mu^{4-D}\int\ddq
\frac{-\ri e^2 I^{V^a}_{\Phi_{i''_k}\Phi_{i'_k}}I^{\bar{V}^a}_{\Phi_{i'_k}\Phi_{i_k}}}{(q^2-M_{V^a}^2)[(p_k-q)^2-M_{\Phi_{i'_k}}^2]}
\M_0^{\Phi_{i''_k}}(p_k),\nl
\eeqar   
and adopting the logarithmic approximation as in  \refeq{loopLA1}--\refeq{loopLA}
we obtain the collinear factor
\beqar \label{scalarcollfact}
\de^\coll_{\Phi_{i''}\Phi_{i}}&\LA&
\frac{\alpha}{4\pi}\de_{\Phi_{i''}\Phi_{i}}\left\{
\cew_{\Phi}\log{\frac{\mu^2}{\MW^2}}
\right.\nl&&\left.\hspace{1.4cm}
{}+\sum_{V^a=Z,W^\pm}\sum_{\Phi_{i'}}I^{V^a}_{\Phi_{i}\Phi_{i'}}I^{\bar{V}^a}_{\Phi_{i'}\Phi_{i}}
\log{\left(\frac{\MW^2}{\max{(\MW^2,M_{\Phi_{i}}^2,M_{\Phi_{i'}}^2)}}\right)}\right\}.
\eeqar
For Higgs bosons we simply have
\beq\label{scalarcollfactH}
\de^\coll_{HH}\LA
\frac{\alpha}{4\pi}\cew_{\Phi}\log{\frac{\mu^2}{\MH^2}},
\eeq
whereas  for the would-be Goldstone bosons $\Phi_b=\chi,\phi^\pm$ associated to the gauge bosons $V^b=Z,W^\pm$, using \refeq{Higgsgaugecoup} we have 
\beq\label{scalarcollfactwbgb}
\de^\coll_{\Phi_b\Phi_b}\LA
\frac{\alpha}{4\pi}\left\{\cew_{\Phi}\log{\frac{\mu^2}{\MW^2}}
-\frac{1}{4\sw^2}\frac{M_{V^b}^2}{\MW^2}\log{\frac{\MH^2}{\MW^2}}
\right\}.
\eeq

\subsection{Factorization for transverse gauge bosons}
\label{se:fac_vec}
Next, we consider the collinear enhancements generated by the virtual
splittings
\beq\label{GAUGEsplitting}
V^{b_k}_\nu(p_k) \rightarrow V_\mu^a(q) V^{b'_k}_{\nu'}(p_k-q),
\eeq
where an incoming on-shell transverse gauge boson $V_\rT^{b_k}=A_\rT,Z_\rT,W_\rT^\pm$ 
emits  a virtual collinear gauge boson $V^a=A,Z,W^+,W^-$.
In the collinear approximation $q^\mu\to x p_k^\mu$,  the vertices \refeq{gcollvertex} corresponding to these splittings give
\beqar\label{gaucollvertex}
\lefteqn{\left[\hspace{3mm}
\vcenter{\hbox{\begin{picture}(70,70)(5,-40)
\Photon(60,0)(5,0){2}{5}
\Photon(30,0)(30,-25){2}{3}
\Vertex(60,0){2}
\Vertex(30,0){2}
\Text(25,-15)[r]{\scriptsize$\bar{V}^a_\mu$}
\Text(18,3)[rb]{\scriptsize$V^{b}_\rT$}
\Text(45,3)[cb]{\scriptsize$V^{b'}_{\nu'}$}
\end{picture}}}
\right]_{\mathrm{coll.}}
=}\quad&&\nl&=&
\lim_{q^\mu\rightarrow xp_k^\mu}
\varepsilon^\nu_\rT(p_k)
\ri  e I^{\bar{V}^a}_{V^{b'}V^{b}}
\left[ g^{\mu\nu'}(p_k-2q)_\nu+g^{\nu'}_{\nu}(-2p_k+q)^\mu+g_{\nu}^{\mu}(p_k+q)^{\nu'}\right]
\frac{-\ri}{D_1}\nl
&=&
\lim_{q^\mu\rightarrow xp_k^\mu}
\frac{-eI^{\bar{V}^a}_{V^{b'}V^{b}}}{D_1}
\left[(2p_k-q)^\mu \varepsilon_{\rT}^{\nu'}(p_k)
-(p_k+q)^{\nu'}\varepsilon_{\rT}^{\mu}(p_k)\right]\nl
&=&
\frac{-eI^{\bar{V}^a}_{V^{b'}V^{b}}}{D_1}
\left[ \left(\frac{2}{x}-1\right)q^\mu\varepsilon_{\rT}^{\nu'}(p_k)
-\left(\frac{2}{1-x}-1\right)(p_k-q)^{\nu'}\varepsilon_{\rT}^{\mu}(p_k)\right],
\eeqar
where $D_1=[(p_k-q)^2-M_{V^{b'}}^2+\ri \varepsilon]$. Here we have used the transversality of the polarization vector which leads to $\varepsilon^\nu_\rT(p_k)\,(p_k-2q)_\nu=0$
in the limit $q^\mu \rightarrow x p_k^\mu$. 
In the fractions $2/x$ and $2/(1-x)$ we have isolated the terms
that correspond to the eikonal approximation 
for a soft-collinear $V^a$ gauge boson ($x\rightarrow 0$) and a soft-collinear $V^{b'}$ gauge boson ($x\rightarrow 1$), respectively.

The SL collinear corrections associated to external gauge bosons
are given by 
\beqar\label{GAUGEcollfactorization}
\delta^\coll \M^{V_\rT^{b_k}}(p_k)&=&\frac{1}{2}\sum_{V^a,V^{b'_k}} \left\{\left[
\vcenter{\hbox{\begin{picture}(85,60)(0,-30)
\Photon(5,0)(25,0){2}{2}
\Photon(25,0)(60.9,14.1){2}{3}
\Photon(25,0)(60.9,-14.1){-2}{3}
\Vertex(25,0){2}
\GCirc(65,0){15}{1}
\Text(35,-14)[lt]{\scriptsize$V^a$}
\Text(35,+12)[lb]{\scriptsize$V^{b'_k}$}
\Text(5,5)[lb]{\scriptsize $V_\rT^{b_k}$}
\end{picture}}}
\right]_{\mathrm{trunc.}}
\right.\nl&&\left.
{}- \sum_{l\neq k}\sum_{\varphi_{i'_l}}\left[
\vcenter{\hbox{\begin{picture}(80,80)(5,-40)
\Photon(25,30)(53,9){2}{4.5}
\Line(65,0)(25,-30)
\PhotonArc(65,0)(32.5,143.13,216.87){2}{4.5}
\Vertex(39,19.5){2}
\Text(46,19)[lb]{\scriptsize$V^{b'_k}$}
\Text(46,-19)[lt]{\scriptsize$\varphi_{i'_l}$}
\Vertex(39,-19.5){2}
\GCirc(65,0){15}{1}
\Text(30,0)[r]{\scriptsize$V^a$}
\Text(20,27)[rb]{\scriptsize$V_\rT^{b_k}$}
\Text(20,-27)[rt]{\scriptsize$\varphi_{i_l}$}
\end{picture}}}
+
\vcenter{\hbox{\begin{picture}(80,80)(5,-40)
\Photon(25,30)(53,9){2}{4.5}
\Line(65,0)(25,-30)
\PhotonArc(65,0)(32.5,143.13,216.87){2}{4.5}
\Vertex(39,19.5){2}
\Text(46,19)[lb]{\scriptsize$V^{a}$}
\Text(46,-19)[lt]{\scriptsize$\varphi_{i'_l}$}
\Vertex(39,-19.5){2}
\GCirc(65,0){15}{1}
\Text(30,0)[r]{\scriptsize$V^{b'_k}$}
\Text(20,27)[rb]{\scriptsize$V_\rT^{b_k}$}
\Text(20,-27)[rt]{\scriptsize$\varphi_{i_l}$}
\end{picture}}}
\right]_{\mathrm{eik.}}\right\}_{\mathrm{coll.}},  
\eeqar  
where the rhs  
has been written in a  manifestly
symmetric way with respect to an interchange of 
the gauge bosons $V^a$ and $V^{b'_k}$
resulting from the splitting \refeq{GAUGEsplitting}.  In particular,
the subtracted eikonal contributions have been decomposed into terms
originating from soft-collinear $V^a$ bosons ($q^\mu\rightarrow 0$) as well as
from soft-collinear  $V^{b'_k}$ bosons ($q^\mu\rightarrow p_k^\mu$).  The
symmetry factor $1/2$ compensates double counting in the sum over $V^a,
V^{b'_k}=A,Z,W^+,W^-$.  

Using  \refeq{gaucollvertex} and  \refeq{eikonalappA}
for the eikonal contributions we obtain
\newcommand{\GBvertex}{F}
\beqar \label{GAUGEcollamp}
\lefteqn{\delta^\coll \M^{V_\rT^{b_k}}(p_k)= \frac{1}{2}\sum_{V^a,V^{b'_k}} \mu^{4-D}\int\ddq
\frac{\ri e I^{\bar{V}^a}_{V^{b'_k}V^{b_k}}}{(q^2-M_{V^a}^2)[(p_k-q)^2-M_{V^{b'_k}}^2]}}\quad
\nl&&\times\lim_{q^\mu\rightarrow xp_k^\mu}
\Biggl\{
\left[ \left(\frac{2}{x}-1\right)q^\mu\varepsilon_{\rT}^{\nu'}(p_k)
-\left(\frac{2}{1-x}-1\right)(p_k-q)^{\nu'}\varepsilon_{\rT}^{\mu}(p_k)\right]
G_{\mu\nu'}^{[\underline{V}^a\underline{V}^{b'_k}]}(q,p_k-q)
\nl&&
\quad {}+\sum_{l\neq k}\sum_{\varphi_{i'_l}}
\Biggl[
\frac{ 4e p_{l}p_k I^{V^a}_{\varphi_{i'_l}\varphi_{i_l}}  }{[(p_l+q)^2-M_{\varphi_{i'_l}}^2]}
\M_0^{V_\rT^{b'_k}{\varphi}_{i'_l}}(p_k,p_l)
\nl&&
\quad\hspace{1cm} {}-
\frac{ 4e p_{l}p_k I^{V^{b'_k}}_{\varphi_{i'_l}\varphi_{i_l}}  }{[(p_l+p_k-q)^2-M_{\varphi_{i'_l}}^2]}
\M_0^{V_\rT^{a}\varphi_{i'_l}}(p_k,p_l)
\Biggr]
\Biggr\},
\eeqar     
where  
according to the definition
\refeq{subtractGF}, 
\beqar\label{GAUGEampdiag}
\lefteqn{G^{[\underline{V}^a \underline{V}^b]}_{\mu\nu}(q,p-q)=}\quad\nl&&
\vcenter{\hbox{\begin{picture}(90,60)(0,30)
\Photon(55,60)(10,60){2}{6}
\Text(30,65)[b]{\scriptsize $V^b_\nu(p-q)$}
\Photon(30,30)(70,60){2}{6}
\Text(55,35)[t]{\scriptsize $V^a_\mu(q)$}
\GCirc(70,60){15}{1}
\end{picture}}}
-\sum_{V^c}
\vcenter{\hbox{\begin{picture}(95,60)(0,30)
\Photon(70,60)(35,60){2}{4}
\Photon(35,60)(5,60){2}{4}
\Text(35,65)[br]{\scriptsize $V^b_\nu(p-q)$}
\Text(40,65)[bl]{\scriptsize $V^c(p)$}
\Photon(10,30)(35,60){2}{5}
\GCirc(35,60){2}{0}
\Text(35,40)[t]{\scriptsize $V^a_\mu(q)$}
\GCirc(75,60){15}{1}
\end{picture}}}
-\sum_{\Phi_j}
\vcenter{\hbox{\begin{picture}(95,60)(0,30)
\DashLine(70,60)(35,60){3}
\Photon(35,60)(5,60){2}{4}
\Text(35,65)[br]{\scriptsize $V^b_\nu(p-q)$}
\Text(40,65)[bl]{\scriptsize $\Phi_j(p)$}
\Photon(10,30)(35,60){2}{5}
\GCirc(35,60){2}{0}
\Text(35,40)[t]{\scriptsize $V^a_\mu(q)$}
\GCirc(75,60){15}{1}
\end{picture}}}.
\eeqar
The expression
between the curly brackets on the rhs  of \refeq{GAUGEcollamp}
can be simplified by means of \refeq{colleik},
\beq\label{colleikb}
\lim_{q^\mu\rightarrow xp_k^\mu}\frac{ 2p_k p_{l}}{[(p_l+p_k-q)^2-M_{\varphi_{i'_l}}^2]}=\frac{1}{1-x}+\O\left(\frac{M^2}{s}\right),
\eeq
and by using the {\em collinear Ward identity} \refeq{SCACWIres}
for gauge bosons ($\varphi_i=V^b_\nu$)
and the equivalent identity 
\beq\label{GAUGECWIres2} 
\lim_{q^\mu \rightarrow xp^\mu}(p-q)^{\nu}\varepsilon_\rT^\mu(p) 
 G^{[\underline{V}^a \underline{V}^{b}] \underline{\Oper}}_{\mu\nu}(q,p-q,r)
= e \sum_{V^{b'}} \M_0^{V_\rT^{b'} \Oper}(p,r) I^{V^{b}}_{V^{b'}V^a}.
\eeq
This leads to
\beqar \label{GAUGEfacteqb}
\lefteqn{\lim_{q^\mu\rightarrow xp_k^\mu} \biggl\{\ldots\biggr\} =}\quad
\nl&=&{}
\lim_{q^\mu\rightarrow xp_k^\mu}
\biggl\{
\left[\left(\frac{2}{x}-1\right)\varepsilon^{\nu'}_\rT(p_k)q^\mu
-\left(\frac{2}{1-x}-1\right)\varepsilon^{\mu}_\rT(p_k)(p-q)^{\nu'}
\right]
G_{\mu\nu'}^{[\underline{V}^a\underline{V}^{b'_k}]}(q,p_k-q)
\nl&&{}
+\sum_{l\neq k}\sum_{\varphi_{i'_l}}
\biggl[
\frac{2e}{x} I^{V^a}_{\varphi_{i'_l}\varphi_{i_l}}
\M_0^{V_\rT^{b'_k}\varphi_{i'_l}}(p_k,p_l)
-\frac{2e}{1-x}I^{V^{b'_k}}_{\varphi_{i'_l}\varphi_{i_l}} 
\M_0^{V_\rT^{a}\varphi_{i'_l}}(p_k,p_l)
\biggr]\biggr\}
\nl&=&
-e\sum_{V^{b''_k}} \left[I^{V^a}_{V^{b''_k}V^{b'_k}}-I^{V^{b'_k}}_{V^{b''_k}V^{a}}\right]\M_0^{V_\rT^{b''_k}}(p_k)
\nl&&{}
+
\frac{2e}{x}\biggl(\sum_{V^{b''_k}} I^{V^a}_{V^{b''_k}V^{b'_k}}\M_0^{V_\rT^{b''_k}}(p_k)
+ \sum_{l\neq k}\sum_{\varphi_{i'_l}}I^{V^a}_{\varphi_{i'_l}\varphi_{i_l}}
\M_0^{V_\rT^{b'_k}\varphi_{i'_l}}(p_k,p_l)\biggr)
\nl&&
{}-\frac{2e}{1-x}\biggl(\sum_{V^{b''_k}} I^{V^{b'_k}}_{V^{b''_k}V^{a}} \M_0^{V_\rT^{b''_k}}(p_k)
+\sum_{l\neq k}\sum_{\varphi_{i'_l}}
I^{{V}^{b'_k}}_{\varphi_{i'_l}\varphi_{i_l}} 
\M_0^{V_\rT^{a}\varphi_{i'_l}}(p_k,p_l)
\biggr).
\eeqar  
Again, the soft-collinear  terms proportional to $1/x$ and $1/(1-x)$ 
are mass-suppressed owing to global $\ewgroup$ invariance
\refeq{globinvar}, so that only the first term in \refeq{GAUGEfacteqb}
remains. Inserting this into \refeq{GAUGEcollamp} with
$I^{V^{b}}_{V^{c}V^{a}}=-I^{V^a}_{V^{c}V^{b}}$, we find
\beqar\label{gaugecollfactint}
\delta^\coll \M^{V_\rT^{b_k}}(p_k) &=&
\sum_{V^a,V^{b'_k},V^{b''_k}}\mu^{4-D}\int\ddq
\frac{-\ri e^2 I^{V^a}_{V^{b''_k}V^{b'_k}}I^{\bar{V}^a}_{V^{b'_k}V^{b_k}}}{(q^2-M_{V^a}^2)[(p_k-q)^2-M_{V^{b'_k}}^2]}
\M_0^{V_\rT^{b''_k}}(p_k).\nln
\eeqar   
Adopting the logarithmic approximation as in   \refeq{loopLA1}--\refeq{loopLA} we obtain the collinear factor
\beq \label{tragbcollfact}
\de^\coll_{V_\rT^{b''_k}V_\rT^{b_k}}\LA
\frac{\alpha}{4\pi}\cew_{V^{b''_k}V^{b_k}}\log{\frac{\mu^2}{\MW^2}},
\eeq
since in all diagrams contributing to \refeq{gaugecollfactint}
the largest mass scale is given by $\MW\sim\MZ$.


\subsection{Factorization for fermions}
\label{se:fac_fer}
We finally 
consider the collinear enhancements  generated by the virtual splittings 
\beq
f^\kappa_{j,\si}(p_k) \rightarrow V_\mu^a(q) f^\kappa_{j',\si'}(p_k-q),
\eeq
where  a virtual
collinear gauge boson $V^a=A,Z,W^+,W^-$ is emitted by an
incoming on-shell fermion $f^\kappa_{j,\si}$, \ie a quark or lepton
$f=Q,L$, with chirality $\kappa=\rL,\rR$, isospin index $\si=\pm$, and
generation index $j=1,2,3$.  
In the collinear approximation $q^\mu\to x p_k^\mu$,  the vertices \refeq{gcollvertex} corresponding to these splittings give
\beqar\label{fercollvertex}
\lefteqn{\left[\hspace{3mm}
\vcenter{\hbox{\begin{picture}(70,70)(5,-40)
\ArrowLine(30,0)(60,0)
\ArrowLine(5,0)(30,0)
\Photon(30,0)(30,-25){2}{3}
\Vertex(60,0){2}
\Vertex(30,0){2}
\Text(25,-15)[r]{\scriptsize$\bar{V}^a_\mu$}
\Text(18,3)[rb]{\scriptsize$f^\kappa_{j,\si}$}
\Text(45,3)[cb]{\scriptsize$f^\kappa_{j',\si'}$}
\end{picture}}}
\right]_{\mathrm{coll.}}=
\lim_{q^\mu\rightarrow xp_k^\mu}
\frac{-\ri(\ps_k-\qs)}{D_1}(-\ri e \gamma_\mu)
I^{\bar{V}^a}_{f^\kappa_{j',\si'} f^\kappa_{j,\si}} u_\kappa(p_k)
=}\quad\nl
&=&
\lim_{q^\mu\rightarrow xp_k^\mu}
\frac{- e I^{\bar{V}^a}_{f^\kappa_{j',\si'} f^\kappa_{j,\si}}}{ D_1} u_\kappa(p_k) (2p_k-2q)_\mu
=\frac{-e I^{\bar{V}^a}_{f^\kappa_{j',\si'} f^\kappa_{j,\si}}}{D_1} u_\kappa(p_k) \left(\frac{2}{x}-2\right)q_\mu,
\eeqar
where $D_1=[(p_k-q)^2-m_{f_{j',\si'}}^2+\ri \varepsilon]$, and the fermionic gauge couplings are defined in \refeq{fermgenerators}. Here we have used the Dirac equation which implies
\beqar
(\ps_k-\qs)\gamma^\mu u_\kappa(p_k)&\to& 
\left[\left\{(\ps_k-\qs),\gamma^\mu\right\} +\O(m_{j,\si})\right] u_\kappa(p_k)\nl
&=& 
2(p_k-q)^\mu u_\kappa(p_k)+ \O(m_{j,\si})u_\kappa (p_k)
\eeqar
for $q^\mu\to x p_k^\mu$.
With \refeq{fercollvertex} and with  \refeq{eikonalappA}
for the eikonal contributions  the amplitude 
\beqar\label{FERcollfactorization}
\delta^\coll \M^{f^\kappa_{j,\si}}(p_k)&=&
\sum_{V^a}\sum_{j'\si'}
\left\{
\left[\vcenter{\hbox{\begin{picture}(85,60)(0,-30)
\ArrowLine(5,0)(25,0)
\ArrowLine(25,0)(60.9,14.1)
\Photon(25,0)(60.9,-14.1){-2}{3}
\Vertex(25,0){2}
\GCirc(65,0){15}{1}
\Text(35,-14)[lt]{\scriptsize$V^a$}
\Text(5,5)[lb]{\scriptsize $f_{j,\si}$}
\Text(35,+12)[lb]{\scriptsize$f_{j',\si'}$}
\end{picture}}}
\right]_{\mathrm{trunc.}}
- \sum_{l\neq k}\sum_{\varphi_{i'_l}}\left[
\vcenter{\hbox{\begin{picture}(80,80)(5,-40)
\ArrowLine(25,30)(39,19.5)\ArrowLine(39,19.5)(53,9)
\Line(65,0)(25,-30)
\PhotonArc(65,0)(32.5,143.13,216.87){2}{4.5}
\Vertex(39,19.5){2}
\Vertex(39,-19.5){2}
\GCirc(65,0){15}{1}
\Text(30,0)[r]{\scriptsize$V^a$}
\Text(20,27)[rb]{\scriptsize$f_{j,\si}$}
\Text(20,-27)[rt]{\scriptsize$\varphi_{i_l}$}
\Text(46,19)[lb]{\scriptsize$f_{j',\si'}$}
\Text(46,-19)[lt]{\scriptsize$\varphi_{i'_l}$}
\end{picture}}}
\right]_{\mathrm{eik.}}\right\}_{\mathrm{coll.}}\nln
\eeqar
reads
\beqar \label{FERcollamp}
\delta^\coll \M^{f^\kappa_{j,\si}}(p_k)&=& \sum_{V^a=A,Z,W^\pm} \sum_{j',\si'} \mu^{4-D}\int\ddq
\frac{\ri e I^{\bar{V}^a}_{\si'\si}U^{\bar{V}^a}_{j'j}}{(q^2-M_{V^a}^2)[(p_k-q)^2-m_{f_{j',\si'}}^2]}
\nl&&\times\lim_{q^\mu\rightarrow xp_k^\mu} \Biggl\{
\left(\frac{2}{x}-2\right)q^\mu G_{\mu}^{[\underline{V}^a\underline{\Psi}_{j',\si'}^\kappa]}(q,p_k-q)
 u(p_k)
\nl&&
{}+\sum_{l\neq k}\sum_{\varphi_{i'_l}}\frac{ 4e p_{l}p_k I^{V^a}_{\varphi_{i'_l}\varphi_{i_l}}  }{[(p_l+q)^2-M_{\varphi_{i'_l}}^2]}
\M_0^{f_{j',\si'}^\kappa\varphi_{i'_l}}(p_k,p_l)
\Biggr\},\hspace{2cm}
\eeqar   
where 
$U^{V^a}$ is the unitary mixing matrix defined in \refeq{Umatrix}.  
According to the definition \refeq{subtractGF}
the Green function
$G^{[\underline{V}^a\underline{\Psi}^\kappa_{j,\si}]}_{\mu}$ 
is diagrammatically given by
\beqar\label{FERampdiag}
G^{[\underline{V}^a \underline{\Psi}_{j,\si}^\kappa]}_{\mu}(q,p-q)=
\vcenter{\hbox{\begin{picture}(90,60)(0,30)
\ArrowLine(10,60)(55,60)
\Text(35,65)[b]{\scriptsize $\Psi^\kappa_{j,\si}(p-q)$}
\Photon(30,30)(70,60){2}{6}
\Text(55,35)[t]{\scriptsize $V^a_\mu(q)$}
\GCirc(70,60){15}{1}
\end{picture}}}
-\sum_{\Psibar}\ 
\vcenter{\hbox{\begin{picture}(95,60)(0,30)
\ArrowLine(35,60)(70,60)
\ArrowLine(5,60)(35,60)
\Text(35,65)[br]{\scriptsize $\Psi^\kappa_{j,\si}(p-q)$}
\Text(40,65)[bl]{\scriptsize $\Psi(p)$}
\Photon(10,30)(35,60){2}{5}
\GCirc(35,60){2}{0}
\Text(35,40)[t]{\scriptsize $V^a_\mu(q)$}
\GCirc(75,60){15}{1}
\end{picture}}}.
\eeqar
With \refeq{colleik} and with the {\em collinear Ward identity} \refeq{FERCWIres} for fermions,  
the expression between the curly brackets in
\refeq{FERcollamp} simplifies into 
\beqar
\lim_{q^\mu\rightarrow xp_k^\mu} \biggl\{\ldots\biggr\} 
&=&\lim_{q^\mu\rightarrow x p_k^\mu}
\left(\frac{2}{x}-2\right)q^\mu G_{\mu}^{[\underline{V}^a\underline{\Psi}_{j',\si'}^\kappa]}(q,p_k-q)u(p_k)
\nl&&+{}
\frac{2e}{x}\sum_{l\neq k}\sum_{\varphi_{i'_l}} I^{V^a}_{\varphi_{i'_l}\varphi_{i_l}}
\M_0^{f_{j',\si'}^\kappa\varphi_{i'_l}}(p_k,p_l)
\nl&=&
-2 e\sum_{j'',\si''}I^{V^a}_{\si''\si'}U^{V^a}_{j''j'} \M_0^{f_{j'',\si''}^\kappa}(p_k)
\nl&&{}+
\frac{2e}{x}\left\{\sum_{j'',\si''}I^{V^a}_{\si''\si'}U^{V^a}_{j''j'} \M_0^{f_{j'',\si''}^\kappa}(p_k)
+\sum_{l\neq k} \sum_{\varphi_{i'_l}}I^{V^a}_{\varphi_{i'_l}\varphi_{i_l}}
\M_0^{f_{j',\si'}^\kappa\varphi_{i'_l}}(p_k,p_l)
\right\}.\quad\nn
\eeqar
Again, the soft-collinear contributions proportional to $1/x$ 
are mass-suppressed owing to global gauge invariance \refeq{globinvar}, and we obtain 
\beqar 
\delta^\coll\M^{f^\kappa_{j,\si}}(p_k) &=&
\sum_{V^a,j',j'',\si',\si''}\mu^{4-D}\int\ddq \frac{-2\ri e^2
  I^{V^a}_{\si''\si'}U^{V^a}_{j''j'}
  I^{\bar{V}^a}_{\si'\si}U^{\bar{V}^a}_{j'j}}
{(q^2-M_{V^a}^2)[(p_k-q)^2-m_{f_{j',\si'}}^2]}
\M_0^{f_{j'',\si''}^\kappa}(p_k).\nln 
\eeqar 
Performing  the logarithmic approximation as in   \refeq{loopLA1}--\refeq{loopLA} and using the unitarity of the mixing matrix,
$\sum_{j'}U^{V^a}_{j''j'}U^{\bar{V}^a}_{j'j}=\de_{j''j}$,
we obtain the collinear factor 
\beqar\label{fercollfactres0}
\de^\coll_{f^\kappa_{j'',\si''}f^\kappa_{j,\si}}&\LA& 
\de_{\si''\si}
\frac{\alpha}{2\pi} \left\{ 
\de_{j''j}\left[
\cew_{f^\kappa_{j,\si}}\log{\frac{\mu^2}{\MW^2}}
  +Q^2_{f_{j,\si}}\log{\frac{\MW^2}{m^2_{f_{j,\si}}}}\right]
\right.\nl &&\left.\hspace{0.5cm} {}-
\de_{fQ}\left\{\de_{j''j}\left[
\de_{\si +}\de_{j3} \left(\left(I^Z_{\Pt^\kappa}\right)^2+ \frac{\de_{\kappa\rL}}{2\sw^2}\right)\right]
+\de_{\si -}\frac{\de_{\kappa\rL}}{2\sw^2}\ckm^+_{j''3}\ckm_{3j}
\right\}\log{\frac{\Mt^2}{\MW^2}}
 \right\}. \nln
\eeqar
The first line contains the symmetric electroweak and the pure electromagnetic contributions. The  $\log{(\Mt/\MW)}$ terms in the second line originate only from diagrams involving real and/or virtual top quarks. The term proportional to
$\de_{\si +}\de_{j3}$ contributes only in the case of  external top quarks $Q^\kappa_{3,+}=\Pt^\kappa$, whereas the term proportional to $\de_{\kappa\rL}\de_{\si -}\ckm^+_{j''3}\ckm_{3j}$ contributes in the case of left-handed down quarks $Q^\rL_{j,-}$ of all three generations and gives also rise to mixing between the generations $j''\neq j$. However, owing to the phenomenological values
\beq\label{ckmvalues}
|\ckm_{33}|=0.999,\qquad
|\ckm_{j3}|<0.05,\qquad \mbox{for $j\neq 3$},
\eeq
of the quark-mixing matrix, and since 
\beq
\frac{\alpha}{4\pi} 
\frac{1}{\sw^2}\log{\frac{\Mt^2}{\MW^2}}\sim 0.004,
\eeq
the contributions to \refeq{fercollfactres0} with $j,j''\neq 3$ are smaller than $2 \times10^{-4}$ and we can use
\beq\label{ckmapprox}
\ckm^+_{j''3}\ckm_{3j}\sim \de_{j''j}\de_{j3},
\eeq
so that 
\beqar\label{fercollfactres}
\de^\coll_{f^\kappa_{j'',\si''}f^\kappa_{j,\si}}&\LA& 
\de_{\si''\si}\de_{j''j}
\frac{\alpha}{2\pi} \left\{ 
\cew_{f^\kappa_{j,\si}}\log{\frac{\mu^2}{\MW^2}}
  +Q^2_{f_{j,\si}}\log{\frac{\MW^2}{m^2_{f_{j,\si}}}}
\right.\nl &&\left.\hspace{1.5cm} {}+
\de_{fQ}\de_{j3}\left[
\de_{\si +} \left(I^Z_{\Pt^\kappa}\right)^2+ \frac{\de_{\kappa\rL}}{2\sw^2}\right]
\log{\frac{\MW^2}{\Mt^2}}
 \right\}.
\eeqar

\chapter{Mass singularities from wave-function renormalization}
\label{FRCSllogs}
\newcommand{\Dealpha}{\Delta \alpha (\MW^2)}
In this chapter we consider the mass singularities that originate from 
the  renormalization  of the asymptotic fields and from the corrections to the GBET.
These mass singularities appear as logarithms involving the ratio of the  scale $\mu$ of dimensional regularization to the renormalization scale $M$ for the on-shell fields, \ie  their  physical mass.
The corresponding corrections to $S$-matrix elements, which we generically denote by  $\de^{\mathrm{WF}}\M$, 
can be easily associated  to the external states in the form 
\beq\label{WFRCsubllogfact2}
\de^{\mathrm{WF}} \M^{\varphi_{i_1} \ldots\, \varphi_{i_n}}(p_1,\ldots,p_n)  =\sum_{k=1}^n \sum_{\varphi_{i'_k}}
\M_0^{\varphi_{i_1} \ldots\, \varphi_{i'_k} \ldots\, \varphi_{i_n}}(p_1,\ldots,p_n)\,
\de^{\mathrm{WF}}_{\varphi_{i'_k} \varphi_{i_k}}.
\eeq
In \refse{se:FRRCcoll}, we consider the contributions associated to external  Higgs bosons, transverse gauge-bosons and fermions or antifermions, which are simply given by the corresponding field-renormalization constants (FRC's) $\de Z$ as 
\beq\label{WFRCcontribution}
\de^{\mathrm{WF}}_{\varphi_{i'_k} \varphi_{i_k}}= \frac{1}{2}\delta  Z_{\varphi_{i'_k} \varphi_{i_k}}.
\eeq
In \refse{se:loggaugebos} we consider the contributions associated to external {\em longitudinal gauge bosons} $V_\rL^b= Z_\rL,W^\pm_\rL$. 
These have to be treated with  the GBET. Therefore, as can be seen in \refeq{GBETandcorr}, 
one has to take into account the FRC's for would-be Goldstone bosons $\Phi_b=\chi,\phi^\pm$ together with the  
the corrections $\de A^{V^b}$ to the GBET, resulting into the effective factors 
\beq\label{GBETCPRR}
 \delta^{\mathrm{WF}}_{V_\rL^{b'}V_\rL^{b}}=
\frac{1}{2}\delta  Z_{\Phi_{b'} \Phi_{b}}+\de A^{V^b}\de_{V^{b'}V^{b }},
\eeq
where $\de_{V^{b'}V^{b }}$ represents the Kronecker symbol.
Since the unphysical would-be Goldstone bosons can be kept unrenormalized, we will restrict ourselves to the   
corrections to the GBET, which involve  the FRC's  of 
the gauge bosons as well as  mixing-energies between gauge bosons and would-be Goldstone bosons.

Combining the contributions \refeq{WFRCsubllogfact2} with the collinear mass singularities  \refeq{collsubllogfact} that originate from  truncated loop diagrams we obtain the complete and gauge-invariant  SL mass-singular corrections 
that are associated with external states 
\beq\label{completeCCcorr}
\de^\cc \M^{\varphi_{i_1} \ldots\, \varphi_{i_n}}(p_1,\ldots,p_n)  =\sum_{k=1}^n \sum_{\varphi_{i'_k}}
\M_0^{\varphi_{i_1} \ldots\, \varphi_{i'_k} \ldots\, \varphi_{i_n}}(p_1,\ldots,p_n)\,
\de^\cc_{\varphi_{i'_k} \varphi_{i_k}},
\eeq
with 
\beq\label{subllogfact2}
\delta^\cc_{\varphi_{i'_k}\varphi_{i_k}}=\left.\left(\de^{\mathrm{WF}}_{\varphi_{i'_k}
    \varphi_{i_k}}+\delta^\coll_{\varphi_{i'_k}\varphi_{i_k}}\right)\right|_{\mu^2=s}.
\eeq
As explained in the introduction, we  exploit  the  $\mu$-independence of the  $S$ matrix and  
 choose the scale $\mu^2=s$ in order to suppress all logarithms of the type $\log{(r_{kl}/\mu^2)}$ which result from loop diagrams.

Since we apply the GBET, for external longitudinal gauge bosons 
  the collinear factors $\de^\coll$ for the corresponding would-be Goldstone bosons have to be used
in \refeq{subllogfact2}.
In \refse{se:compare}, we compare our  complete corrections for longitudinal gauge bosons with the 
corrections for physical would-be Goldstone bosons in the symmetric phase of the electroweak theory.

 
The contributions described above are derived in the following 
using  the 
't~Hooft--Feynman gauge.

\section{Wave-function renormalization and collinear mass singularities}\label{se:FRRCcoll}
The contributions \refeq{WFRCcontribution} from FRC's originate from the 
renormalization of the bare fields 
\beq
\varphi_{i_k,0}=\sum_{\varphi_{i'_k}}\left(\de_{\varphi_{i_k}\varphi_{i'_k}}+\frac{1}{2}\de Z_{\varphi_{i_k}\varphi_{i'_k}}\right)\varphi_{i'_k},
\eeq
which leads to the relation 
\beq\label{WFRCGF}
G_0^{\varphi_{i_1} \ldots\, \varphi_{i_n}}
=\sum_{\varphi_{i'_1},\dots,\varphi_{i'_n}}\prod_{k=1}^n \left[
\de_{\varphi_{i_k}\varphi_{i'_k}}+\frac{1}{2}\de Z_{\varphi_{i_k}\varphi_{i'_k}}\right]
G^{\varphi_{i_1} \ldots\, \varphi_{i'_k} \ldots\, \varphi_{i_n}},
\eeq
between bare and renormalized connected Green functions\footnote{Note that in the renormalized Green function in \refeq{WFRCGF}
the parameters are kept unrenormalized. The parameter-renormalization is discussed in \refch{Ch:PRllogs}.}.
The corresponding one-loop relation \refeq{WFRCsubllogfact2} between  matrix elements is simply obtained by truncation of \refeq{WFRCGF}.

In the \refses{se:WFRCHiggs}--\ref{se:WFRCfer} we evaluate the   FRC's $\de Z$ for Higgs bosons, gauge bosons, and fermions,
and  combine them with the corresponding collinear factors as in \refeq{subllogfact2}.
First, we specify the on-shell renormalization conditions \cite{Denner:1993kt}
for the fields and the resulting  relations between  FRC's and self-energies.
Then, we present the FRC's that have been obtained by 
an explicit evaluation of the electroweak self-energies, using  the generic Feynman rules listed in \refapp{Feynrules} and the logarithmic approximation (LA) for 2-point functions and their derivatives  that are summarized in \refapp{app:2pointlogapp}. 
These results  are in agreement with the exact electroweak FRC's \cite{Denner:1993kt} in the high-energy limit.

\subsection{Higgs bosons}\label{se:WFRCHiggs}
\newcommand{\tRe}{\widetilde{\Re}}
The bare Higgs field  $H_0$ and its bare mass $M_{\PH,0}$  are renormalized by the transformations 
\beq\label{higgsren}
H_0=\left(1+\frac{1}{2}\de Z_{HH}\right) H, \qquad
M^2_{\PH,0}=\MH^2+\de \MH^2,
\eeq
and the renormalized one-particle irreducible (1PI) 2-point function reads
\beq
\ri \Gamma^{HH}(p^2)=\ri(1+\de Z_{HH})\left(p^2-(M^2_{\PH}+\de M^2_{\PH})\right)+\ri\Sigma_0^{HH}(p^2).
\eeq
The FRC $\de Z_{HH}$ is fixed by the renormalization condition
\beq\label{Higgswfrcon}
\tRe \left.\frac{\partial\Gamma^{HH}(p^2)}{\partial p^2}\right|_{p^2=\MH^2}=1,
\eeq
so that the residuum of the Higgs propagator equals one. Here $\tRe$ removes only the absorptive part of loop  integrals and does not affect the imaginary part of the components of the quark-mixing matrix. 
The Higgs-mass  counterterm is determined by 
\beq\label{Hmassrencond}
\left. \tRe\, \Gamma^{HH}(p^2)\right|_{p^2=\MH^2}=0,
\eeq
so that the renormalized Higgs mass corresponds to the pole of the propagator, and it reads
\beq\label{higgsmassCT}
\de \MH = \tRe\, \Gamma^{HH}(\MH^2).
\eeq
Owing to \refeq{Higgswfrcon}, the FRC is given by the bare Higgs self-energy as
\beq\label{HWFRCdef}
\delta Z_{HH}=-\left.\frac{\partial}{\partial p^2}\tRe\, \Sigma_0^{HH}(p^2)\right|_{p^2=\MH^2}.
\eeq
The following diagrams contribute to the Higgs self-energy
\beqar\label{Hselfenergydiag}
\hspace{-1cm}
\Sigma_0^{HH}&=& \quad
\vcenter{\hbox{\begin{picture}(60,60)(0,-10)
\Text(3,23)[bc]{\scriptsize $H$} 
\Text(57,23)[bc]{\scriptsize $H$} 
\Text(32,0)[tc]{\scriptsize $\Phi_k$} 
\Text(32,40)[bc]{\scriptsize $V^a$} 
\DashLine(0,20)(15,20){2} 
\DashLine(45,20)(60,20){2} 
\Vertex(15,20){2}
\Vertex(45,20){2}
\PhotonArc(30,20)(15,0,180){1}{5} 
\DashCArc(30,20)(15,180,0){4} 
\end{picture}}}
\quad
+
\quad
\vcenter{\hbox{\begin{picture}(60,60)(0,-10)
\Text(3,23)[bc]{\scriptsize $H$} 
\Text(57,23)[bc]{\scriptsize $H$} 
\Text(32,0)[t]{\scriptsize $f_{j,\si}$} 
\Text(32,40)[b]{\scriptsize $\bar{f}_{j',\si'}$} 
\Vertex(15,20){2}
\Vertex(45,20){2}
\DashLine(0,20)(15,20){2} 
\DashLine(45,20)(60,20){2} 
\ArrowArc(30,20)(15,0,180) 
\ArrowArc(30,20)(15,180,0) 
\end{picture}}}
\quad +\quad
\vcenter{\hbox{\begin{picture}(60,60)(0,-10)
\Text(3,23)[bc]{\scriptsize $H$} 
\Text(57,23)[bc]{\scriptsize $H$} 
\Text(32,0)[t]{\scriptsize $V^a$} 
\Text(32,40)[b]{\scriptsize $V^b$} 
\Vertex(15,20){2}
\Vertex(45,20){2}
\DashLine(0,20)(15,20){2} 
\DashLine(45,20)(60,20){2} 
\PhotonArc(30,20)(15,0,180){1}{5} 
\PhotonArc(30,20)(15,180,0){1}{5} 
\end{picture}}}
\nl
&&\hspace{-2cm}
+\quad
\vcenter{\hbox{\begin{picture}(60,60)(0,-10)
\Text(3,23)[bc]{\scriptsize $H$} 
\Text(57,23)[bc]{\scriptsize $H$} 
\Text(32,0)[t]{\scriptsize $\Phi_k$} 
\Text(32,40)[b]{\scriptsize $\Phi_l$} 
\Vertex(15,20){2}
\Vertex(45,20){2}
\DashLine(0,20)(15,20){2} 
\DashLine(45,20)(60,20){2} 
\DashCArc(30,20)(15,0,180){4} 
\DashCArc(30,20)(15,180,0){4} 
\end{picture}}}
\quad +\quad
\vcenter{\hbox{\begin{picture}(60,60)(0,-10)
\Text(3,23)[bc]{\scriptsize $H$} 
\Text(57,23)[bc]{\scriptsize $H$} 
\Text(32,0)[t]{\scriptsize $u^a$} 
\Text(32,40)[b]{\scriptsize$\bar{u}^b$} 
\Vertex(15,20){2}
\Vertex(45,20){2}
\DashLine(0,20)(15,20){2} 
\DashLine(45,20)(60,20){2} 
\DashArrowArc(30,20)(15,0,180){1} 
\DashArrowArc(30,20)(15,180,0){1} 
\end{picture}}}
\quad +\quad
\vcenter{\hbox{\begin{picture}(60,60)(0,-10)
\Text(8,23)[bc]{\scriptsize $H$} 
\Text(52,23)[bc]{\scriptsize $H$} 
\Text(32,50)[bc]{\scriptsize $V^c$} 
\Vertex(30,20){2}
\DashLine(5,20)(30,20){2}
\DashLine(30,20)(55,20){2}
\PhotonArc(30,33)(13,0,360){2}{9} 
\end{picture}}}
\quad +\quad
\vcenter{\hbox{\begin{picture}(60,60)(0,-10)
\Text(8,23)[bc]{\scriptsize $H$} 
\Text(52,23)[bc]{\scriptsize $H$} 
\Text(32,50)[bc]{\scriptsize $\Phi_k$} 
\Vertex(30,20){2}
\DashLine(5,20)(30,20){2}
\DashLine(30,20)(55,20){2}
\DashCArc(30,33)(13,0,360){4} 
\end{picture}}}
\eeqar
Here and in the following we do not consider tadpole diagrams, 
since we assume that the 1PI Higgs 1-point function has been renormalized 
as in \refeq{tadrencond}, such that 
all tadpole diagrams are compensated by the tadpole counterterm $\de t$ and can be neglected.

The contribution of the diagrams \refeq{Hselfenergydiag} to \refeq{HWFRCdef} has been evaluated using the  2-point functions in LA given in \refapp{app:2pointlogapp}.
It turns out that only the first two diagrams contribute, whereas the diagrams  which contain couplings with mass dimension  do not give rise to large logarithmic contributions. In particular it has been checked that the terms $\frac{\MH^4}{\MW^2}\bar{B}'_0(\MH^2,\MW,\MW)$ that originate from the scalar diagram do not give large logarithms $\log{(\MH^2/\MW^2)}$ in the heavy Higgs limit $\MH\gg \MW$.
For the Higgs FRC we obtain 
\beqar\label{HWFRCres}
\delta Z_{HH}&\LA&\frac{\alpha}{4\pi}\left\{ 2\cew_{\Phi}\log{\frac{\mu^2}{\MH^2}}-\frac{\NCt}{2s_w^2}
\frac{\Mt^2}{\MW^2}\log{\frac{\mu^2}{M_{\Htop}^2}}\right\} 
\eeqar
in LA.  The  bosonic diagram gives the term proportional to $\cew_\Phi$, which  is the eigenvalue of the Casimir operator \refeq{casimirew} for the  scalar doublet.
The fermionic diagrams receive large contributions only from  virtual top quarks, with the colour factor $\NCt=3$.
The scale of the Yukawa logarithms is determined by the largest mass-scale in the loop, \ie by [see \refeq{htopmassdef}]
\beq
M_{\Htop}:=\max{(\MH,\Mt)}.
\eeq  

Combining \refeq{HWFRCres} with the collinear factor for Higgs bosons 
\refeq{scalarcollfact}  as in \refeq{subllogfact2} 
we obtain the complete  correction factor
\beq\label{Higgscc} 
\de^{\cc}_{HH}= \frac{\alpha}{4\pi}\left\{2\cew_\Phi\log{\frac{s}{\MH^2}}-\frac{3}{4\sw^2}\frac{\Mt^2}{\MW^2}\log{\frac{s}{M_{\Htop}^2}}\right\}.
\eeq

\subsection{Transverse gauge bosons}\label{se:WFRCtgb}
\newcommand{\antikro}{E}
\newcommand{\kroAA}{\de_{\NB A}\de_{\NB'A}}
\newcommand{\kroZZ}{\de_{\NB Z}\de_{\NB'Z}}

The bare physical gauge-boson fields and their masses 
are renormalized by the transformations
\beqar \label{phWFRCgb}
V^a_{\mu,0}&=&\sum_{V^{b}}\left( \de_{V^aV^b}+\frac{1}{2}\delta Z_{V^{a}V^{b}}\right)V_\mu^{b},\qquad
V^a,V^b=A,Z,W^+,W^-,
\\ \label{phmRCgb}
M^2_{V^a,0}&=& M^2_{V^a}+\de  M^2_{V^a},\qquad
V^a=Z,W^+,W^-,
\eeqar
with a non-diagonal  field-renormalization matrix $\de Z$.
Decomposing the renormalized 1PI 2-point function  into transverse and longitudinal parts, 
\beq
\Gamma^{V^a\bar{V}^b}_{\mu\nu}(p)= 
\left(g_{\mu\nu}-\frac{p_\mu p_\nu}{p^2}\right)\Gamma^{V^a\bar{V}^b}_{\rT}(p^2)
+\frac{p_\mu p_\nu}{p^2}\Gamma^{V^a\bar{V}^b}_{\rL}(p^2),
\eeq
the transverse part reads
\beqar\label{traGBprpagat}
\Gamma^{V^a\bar{V}^b}_{\rT}(p^2)&=&-p^2\left[\de_{V^aV^b}+\frac{1}{2}\de Z_{V^aV^b}+\frac{1}{2}\de Z_{V^bV^a}\right]
+\left(M^2_{V^a}+\de M^2_{V^a} \right)\de_{V^aV^b}
\nl&&{}
+\frac{M^2_{V^a}}{2}\de Z_{V^aV^b}+\frac{M^2_{V^b}}{2}\de Z_{V^bV^a}
-\Si^{V^a\bar{V}^b}_{\rT,0}(p^2),
\eeqar
and is constrained by the following 
renormalization conditions
\beqar\label{gbmassRC}
\left.
\tRe\, \Gamma^{V^a\bar{V}^b}_{\mu\nu}(p)
\varepsilon^\nu(p)\right|_{p^2=M^2_{V^a}}&=&0,\qquad \forall \quad V^a,V^b,
\\ \label{gbfieldRC}
\lim_{p^2\to M^2_{V^a}}\frac{1}{p^2 - M^2_{V^a}}
\tRe\, \Gamma^{V^a\bar{V}^a}_{\mu\nu}(p)
\varepsilon^\nu(p)&=&-\varepsilon_\mu(p).
\eeqar
The first condition \refeq{gbmassRC} fixes the mass counterterms 
\beq\label{gbmassCT}
\de M^2_{V^a}=\tRe\, \Gamma^{V^a\bar{V}^a}_{\rT,0}(M^2_{V^a}),\qquad
V^a=Z,W^+,W^-,
\eeq
such that the renormalized masses correspond to the poles 
of the propagators projected on the physical polarization vectors. 
Furthermore,  it guarantees that the on-shell fields do not mix by requiring for the 
non-diagonal components of the field-renormalization matrix
\beq\label{nondiagVVFRCS}
\de Z_{V^aV^b}= \frac{2 \tRe\, \Si^{V^b\bar{V}^a}_{\rT,0}(M^2_{V^b})}{M^2_{V^a}-M^2_{V^b}}
\qquad\mbox{for $V^a\neq V^b$}.
\eeq
The only non-vanishing components of \refeq{nondiagVVFRCS} are given by
the mixing-energies $\Si^{AZ}(p^2)=\Si^{ZA}(p^2)\neq 0$ and read 
\beq\label{nondiagsagbFRC}
\de Z_{AZ}= -2\tRe\, \frac{\Si^{AZ}_{\rT,0}(\MZ^2)}{\MZ^2},\qquad 
\de Z_{ZA}= 2\tRe\, \frac{\Si^{AZ}_{\rT,0}(0)}{\MZ^2}.
\eeq
The diagonal components of the field-renormalization matrix 
are determined by the renormalization condition \refeq{gbfieldRC}
and read
\beq\label{diagsagbFRC}
\de Z_{V^aV^a}=-\tRe \left.\frac{\partial
\Si^{V^a\bar{V}^a}_{\rT,0}(p^2)}{\partial p^2}\right|_{p^2=M^2_{V^a}}.
\eeq

The field renormalization constants 
are obtained by evaluating following  Feynman diagrams
\beqar\label{gaugeselfenergydiag}
-\ri \Si^{V^a\bar{V}^b}_{\mu\nu,0}&=&\hspace{3mm}
\vcenter{\hbox{\begin{picture}(60,60)(0,-10)
\Text(3,23)[bc]{\scriptsize $V_\mu^a$} 
\Text(57,23)[bc]{\scriptsize $\bar{V}_\nu^b$} 
\Text(32,0)[tc]{\scriptsize $V^c$} 
\Text(32,40)[bc]{\scriptsize $V^d$} 
\Vertex(15,20){2}
\Vertex(45,20){2}
\Photon(0,20)(15,20){2}{2}
\Photon(45,20)(60,20){2}{2}
\PhotonArc(30,20)(15,0,180){2}{5} 
\PhotonArc(30,20)(15,180,0){2}{5} 
\end{picture}}}
\hspace{3mm}
+
\hspace{3mm}
\vcenter{\hbox{\begin{picture}(60,60)(0,-10)
\Text(3,23)[bc]{\scriptsize $V_\mu^a$} 
\Text(57,23)[bc]{\scriptsize $\bar{V}_\nu^b$} 
\Text(32,0)[tc]{\scriptsize $u^c$} 
\Text(32,40)[bc]{\scriptsize $\bar{u}^d$} 
\Vertex(15,20){2}
\Vertex(45,20){2}
\Photon(0,20)(15,20){2}{2}
\Photon(45,20)(60,20){2}{2}
\DashArrowArc(30,20)(15,0,180){1} 
\DashArrowArc(30,20)(15,180,0){1} 
\end{picture}}}
\hspace{3mm}
+
\hspace{3mm}
\vcenter{\hbox{\begin{picture}(60,60)(0,-10)
\Text(3,23)[bc]{\scriptsize $V_\mu^a$} 
\Text(57,23)[bc]{\scriptsize $\bar{V}_\nu^b$} 
\Text(32,0)[tc]{\scriptsize $\Phi_k$} 
\Text(32,40)[bc]{\scriptsize $\Phi_j$} 
\Vertex(15,20){2}
\Vertex(45,20){2}
\Photon(0,20)(15,20){2}{2}
\Photon(45,20)(60,20){2}{2}
\DashCArc(30,20)(15,0,180){4} 
\DashCArc(30,20)(15,180,0){4}
\end{picture}}}
\hspace{3mm}
+
\hspace{3mm}
\vcenter{\hbox{\begin{picture}(60,60)(0,-10)
\Text(3,23)[bc]{\scriptsize $V_\mu^a$} 
\Text(57,23)[bc]{\scriptsize $\bar{V}_\nu^b$} 
\Text(32,0)[tc]{\scriptsize $f_{j,\si}$} 
\Text(32,40)[bc]{\scriptsize $\bar{f}_{j',\si'}$} 
\Vertex(15,20){2}
\Vertex(45,20){2}
\Photon(0,20)(15,20){2}{2}
\Photon(45,20)(60,20){2}{2}
\ArrowArc(30,20)(15,0,180) 
\ArrowArc(30,20)(15,180,0) 
\end{picture}}}
\nl&&\hspace{1.0cm}
{}
+
\hspace{3mm}
\vcenter{\hbox{\begin{picture}(60,60)(0,-10)
\Text(3,18)[bc]{\scriptsize $V_\mu^a$} 
\Text(57,18)[bc]{\scriptsize $\bar{V}_\nu^b$} 
\Text(32,50)[bc]{\scriptsize $V^c$} 
\Vertex(30,20){2}
\Photon(0,10)(30,20){2}{3}
\Photon(30,20)(60,10){2}{3}
\PhotonArc(30,33)(13,0,360){2}{9} 
\end{picture}}}
\hspace{3mm}
+
\hspace{3mm}
\vcenter{\hbox{\begin{picture}(60,60)(0,-10)
\Text(3,18)[bc]{\scriptsize $V_\mu^a$} 
\Text(57,18)[bc]{\scriptsize $\bar{V}_\nu^b$} 
\Text(32,50)[bc]{\scriptsize $\Phi_k$} 
\Vertex(30,20){2}
\Photon(0,10)(30,20){2}{3}
\Photon(30,20)(60,10){2}{3}
\DashCArc(30,33)(13,0,360){4} 
\end{picture}}}
\hspace{3mm}
+
\hspace{3mm}
\vcenter{\hbox{\begin{picture}(60,60)(0,-10)
\Text(3,23)[bc]{\scriptsize $V_\mu^a$} 
\Text(57,23)[bc]{\scriptsize $\bar{V}_\nu^b$} 
\Text(32,0)[tc]{\scriptsize $V^c$} 
\Text(32,40)[bc]{\scriptsize $\Phi_k$} 
\Vertex(15,20){2}
\Vertex(45,20){2}
\Photon(0,20)(15,20){2}{2}
\Photon(45,20)(60,20){2}{2}
\DashCArc(30,20)(15,0,180){4} 
\PhotonArc(30,20)(15,180,0){2}{5} 
\end{picture}}}
\hspace{3mm}
\eeqar
in LA. Also here tadpole diagrams are omitted since they are compensated by the tadpole renormalization \refeq{tadrencond}. 
The results for the diagonal components \refeq{diagsagbFRC} and the off-diagonal components \refeq{nondiagsagbFRC} 
can be  combined  as follows using the antisymmetric matrix $\antikro$  defined in \refeq{adjantiKron}, 
\beqar \label{gbdeltaZmat}
\delta{Z}_{V^aV^b}&\LA& \frac{\alpha}{4\pi}\left\{\left[\bew_{V^aV^b}-2\cew_{V^aV^b}
+\bew_{AZ}\antikro_{V^aV^b}\right]\log{\frac{\mu^2}{\MW^2}} 
+2\de_{V^aV^b}Q^2_{V^a} \log{\frac{\MW^2}{\la^2}}\right\} 
\nl&&\hspace{0cm}{}
-
\de_{V^a A}\de_{V^b A} \Dealpha + \delta{Z}^{\Htop}_{V^aV^b}.
\eeqar
The terms $\log{(\mu^2/\MW^2)}$ constitute  the symmetric electroweak part. 
They are  expressed in terms of the  coefficients of the $\beta$-function $\bew_{V^aV^b}$ 
defined in \refapp{app:betafunction} and the
electroweak Casimir operator $\cew_{V^aV^b}$ in the
adjoint representation \refeq{physadjointcasimir}.
We observe that this symmetric electroweak part
can be related to a diagonal renormalization matrix $\de \tilde{Z}$ for the gauge-group eigenstate fields 
\beq \label{symmWFRCgb}
\sV^a_{\mu,0}=\sum_{\sV^{b}}\left( \de_{\sV^a\sV^b}+\frac{1}{2}\delta \tilde{Z}_{\sV^{a}\sV^{b}} \right)\sV_\mu^{b},\qquad \sV^a,\sV^b=B,W^1,W^2,W^3.
\eeq
The relation between \refeq{phWFRCgb} and \refeq{symmWFRCgb} is obtained from the 
renormalization of the Weinberg rotation \refeq{defweinrot} and reads
\beq \label{0unbrokenWFREN} 
\delta Z= U(\thw)\delta \tilde{Z} U^{-1}(\thw)+2 \delta U(\thw) U^{-1}(\thw).
\eeq
The first term in \refeq{gbdeltaZmat} is symmetric and corresponds to the first term 
in \refeq{0unbrokenWFREN} that originates from  
\beq
\delta \tilde{Z}_{\sV^a\sV^b}=\frac{\alpha}{4\pi}\delta_{\sV^a\sV^b}\left[ \besw_{\sV^a}-2\csew_{\sV^a} \right]\log{\frac{\mu^2}{\MW^2}} 
\eeq
for the symmetric fields. This matrix $\de\tilde{Z}$ is diagonal, because the $\Uone$ and $\SUtwo$ 
components do not mix through self-energy loop diagrams. 
The antisymmetric part of \refeq{gbdeltaZmat}, \ie the term proportional to $\antikro_{V^aV^b}$,
is related  to the second term in \refeq{0unbrokenWFREN}, which corresponds to the  
renormalization of the weak mixing angle  in \refeq{defweinrot}. In fact, using the 
explicit expression \refeq{weinbergrenorm} for  the renormalization of the weak mixing angle in LA we have
\beq \label{1unbrokenWFREN} 
2 \left[\delta U(\thw) U^{-1}(\thw)\right]_{V^{a} V^{b}}= \frac{\cw}{\sw}\frac{\delta \cw^2}{\cw^2}\antikro_{V^{a} V^{b}}
\LA\frac{\alpha}{4\pi}\bew_{AZ}\antikro_{V^{a} V^{b}}\log{\frac{\mu^2}{\MW^2}},
\eeq
up to logarithmic terms of the type $\log(\Mt^2/\MW^2)$ and $\log(\MH^2/\MW^2)$.

The infrared-divergent logarithm $\log{(\MW^2/\la^2)}$ 
in the first line of \refeq{gbdeltaZmat}  originates from 
soft photons in the first and in the last diagram in \refeq{gaugeselfenergydiag}.
It contributes  only to the $\de Z_{W^\pm W^\pm}$ components.
The first term in the second line of  \refeq{gbdeltaZmat} 
is a contribution to the  $\de Z_{AA}$ component originating from light-fermion loops. It reads
\beq\label{runningal}
\Dealpha=  
\frac{\alpha}{3\pi}\sum_{f_{j,\si}\neq \Pt} \NCf Q^2_{f_{j,\si}}\log{\frac{\MW^2}{m^2_{f_{j,\si}}}},
\eeq
where the sum runs over the generations $j=1,2,3$ of leptons and
quarks $f=L,Q$ with isospin $\si=\pm$, omitting the top-quark
contribution and  $\NCf$ is the colour factor. This contribution corresponds to the 
running of the electromagnetic coupling constant from zero to the scale $\MW$.

Finally, the last term in \refeq{gbdeltaZmat} contains all logarithms 
$\log{(\MH^2/\MW^2)}$  and $\log{(\Mt^2/\MW^2)}$ 
from diagrams with virtual top quarks or Higgs bosons and reads
\newcommand{\topmat}{T}
\beqar
 \delta{Z}^{\Htop}_{V^aV^b}&=&\frac{\alpha}{4\pi}\left\{
 \de_{V^aV^b}\frac{M_{V^a}^2}{12\sw^2\MW^2}\log{\frac{\MH^2}{\MW^2}}
+\frac{2\NCt}{3} \topmat_{V^aV^b}\log{\frac{\Mt^2}{\MW^2}}
\right\},
\eeqar
where
\beqar\label{topWFRCgb1}
 \topmat_{V^aV^b}&=& \sum_{\kappa=\rR,\rL}
\left[\left(I^{V^a}I^{\bar{V}^b}+ Q^2_{V^a}I^{\bar{V}^b}I^{V^a}  \right)_{\Pt^\kappa\Pt^\kappa}+\left(I^{A}I^{Z}\right)_{\Pt^\kappa\Pt^\kappa}\antikro_{V^aV^b}\right]
,\nln
\eeqar
with components
\beqar\label{topWFRCgb2}
 \topmat_{W^\pm W^\pm}&=& \frac{1}{2\sw^2}
,\qquad
\topmat_{ZZ}= \frac{9-24\sw^2+32\sw^4}{36\sw^2\cw^2}
,\qquad
\topmat_{AZ}=\frac{16\sw^2-6}{9\sw\cw}
,\nl
\topmat_{ZA}&=&0
,\qquad
 \topmat_{AA}= \frac{8}{9}.
\eeqar

\newpage
Combining as in \refeq{subllogfact2} the FRC's \refeq{gbdeltaZmat}
with the collinear corrections \refeq{tragbcollfact} originating from loop diagrams
 we obtain the complete correction factor
\beqar \label{deccWT}
\delta^\cc_{V_\rT^aV_\rT^b}&=&\frac{\alpha}{4\pi}\left\{\frac{1}{2}
\left[\bew_{V^aV^b}+\bew_{AZ}\antikro_{V^aV^b}\right] \log{\frac{s}{\MW^2}} 
+\de_{V^aV^b}Q_{V^a}^2 \log{\frac{\MW^2}{\la^2}}
\right.\nl&&\left.\hspace{0.5cm}{}+
 \de_{V^aV^b}\frac{M_{V^a}^2}{24\sw^2\MW^2}\log{\frac{\MH^2}{\MW^2}}
+\topmat_{V^aV^b}\log{\frac{\Mt^2}{\MW^2}}
\right\}
-\frac{1}{2}\de_{V^a A}\de_{V^bA} \Dealpha.\nln
\eeqar
The symmetric part of the $\log{(s/\MW^2)}$ terms and the electromagnetic logarithms  
agree with \citere{Melles:2001gw}. In \citere{Melles:2001gw} however,
the antisymmetric part proportional to $\antikro_{V^aV^b}$ is not present. 
This part is a specific consequence of the on-shell renormalization condition \refeq{gbmassRC}, 
which ensures in particular that  on-shell photons and \PZ~bosons do not mix.   
The resulting off-diagonal components of  \refeq{deccWT} read
\beq\label{CCAZmixing}
\de^\cc_{AZ}=\frac{\alpha}{4\pi}\left\{\bew_{AZ}\log{\frac{s}{\MW^2}}
+\topmat_{AZ}\log{\frac{\Mt^2}{\MW^2}}\right\}
,\qquad
\de^\cc_{ZA} =0,
\eeq
and, owing to $\antikro_{AZ}=-\antikro_{ZA}=1$,
the correction factor for external photons does not involve mixing
with $\PZ$ bosons.

\subsection{Chiral fermions}\label{se:WFRCfer}
The bare fields of chiral fermions $\Psi^{\kappa}_{j,\si,0}$
and the bare fermionic masses $m_{f_{j,\si,0}}$
are renormalized by the transformations
\beqar\label{ferren}
f^{\kappa}_{j,\si,0}&=&\sum_{j'=1}^3\left(\de_{jj'}+
\frac{1}{2}\de Z_{f^{\kappa}_{j,\si}f^\kappa_{j',\si}}\right)f^\kappa_{j',\si},
\nl
m_{f_{j,\si},0}&=&m_{f_{j,\si}}+\de m_{f_{j,\si}},
\eeqar
where $f=Q,L$, $\kappa=\rR,\rL$, $j=1,2,3$, $\si=\pm$, and $\de Z$ is a matrix in the indices $jj'$.
The renormalized 1PI 2-point function can be decomposed as
\beq
\Gamma^{\Psi_{j,\si}\bar{\Psi}_{j',\si}}(p)=
\sum_{\kappa=\rL,\rR}\left[\ps \omega_\kappa 
\Gamma_{V,\kappa}^{\Psi_{j,\si}\bar{\Psi}_{j',\si}}(p^2)
+\omega_\kappa 
\Gamma_{S,\kappa}^{\Psi_{j,\si}\bar{\Psi}_{j',\si}}(p^2)\right],
\eeq
with  $\omega_{\rR,\rL}=\omega_{+,-}=(1\pm\gamma^5)/2$, and the components read
\beqar
\Gamma_{V,\kappa}^{\Psi_{j',\si}\bar{\Psi}_{j,\si}}(p^2)&=&
\de_{jj'}+ \frac{1}{2}\left(\de Z_{f^{\kappa}_{j,\si}f^\kappa_{j',\si}}+ \de Z^+_{f^{\kappa}_{j,\si}f^\kappa_{j',\si}}\right)
+\Si_{V,\kappa,0}^{\Psi_{j',\si}\bar{\Psi}_{j,\si}}(p^2)
\nl
\Gamma_{S,\kappa}^{\Psi_{j',\si}\bar{\Psi}_{j,\si}}(p^2)&=&
-\left(m_{f_{j,\si}}+\de m_{f_{j,\si}}\right)\de_{jj'}
- \frac{1}{2}\left(
m_{f_{j,\si}}\de Z_{f^{\kappa}_{j,\si}f^\kappa_{j',\si}}
+m_{f_{j',\si}} \de Z^+_{f^{\bar{\kappa}}_{j,\si}f^{\bar{\kappa}}_{j',\si}}
\right) \nl&&{}
+ \Si_{S,\kappa,0}^{\Psi_{j',\si}\bar{\Psi}_{j,\si}}(p^2),\nln
\eeqar
where $\bar{\kappa}=\rL,\rR$ for $\kappa=\rR,\rL$.
The counterterms are fixed by following renormalization conditions. Firstly, 
one requires
\beq\label{fermassRC}
\left.
\tRe\, \Gamma^{\Psi_{j',\si}\bar{\Psi}_{j,\si}}(p)
u_{j',\si}(p)\right|_{p^2= m^2_{f_{j',\si}}
}
=0,
\eeq
so that the on-shell fields do not mix and the poles of the propogators (projected on the Dirac spinors) coincide with the renormalized masses. 
Secondly, the normalization of the fermionic fields is fixed by
\beq
\lim_{p^2\to m^2_{f_{j,\si}}}\frac{\ps + m_{f_{j,\si}}}{p^2 - m^2_{f_{j,\si}}}
\tRe\, \Gamma^{\Psi_{j,\si}\bar{\Psi}_{j,\si}}(p)
u_{j,\si}(p)=u_{j,\si}(p).
\eeq
The matrices $\de Z$ resulting from these renormalization conditions have off-diagonal components
\beqar\label{nondiagferWFRC}
\de Z_{f^{\kappa}_{j,\si}f^\kappa_{j',\si}}
&=&
\frac{2}{m^2_{f_{j,\si}}-m^2_{f_{j',\si}}}\tRe\,
\left[m^2_{f_{j,\si}}
\Si_{V,\kappa,0}^{\Psi_{j',\si}\bar{\Psi}_{j,\si}}
(m^2_{f_{j,\si}})+
m_{f_{j,\si}}m_{f_{j',\si}}
\Si_{V,\bar{\kappa},0}^{\Psi_{j',\si}\bar{\Psi}_{j,\si}}
(m^2_{f_{j,\si}})
\right.\nl&&{}+\left.
m_{f_{j,\si}}
\Si_{S,\kappa,0}^{\Psi_{j',\si}\bar{\Psi}_{j,\si}}
(m^2_{f_{j,\si}})+
m_{f_{j',\si}}
\Si_{S,\bar{\kappa},0}^{\Psi_{j',\si}\bar{\Psi}_{j,\si}}
(m^2_{f_{j,\si}})
\right],\qquad \mbox{for $j\neq j'$,}
\eeqar
with  $\bar{\kappa}=\rL,\rR$ for $\kappa=\rR,\rL$, and diagonal components
\beqar\label{diagferWFRC}
\de Z_{f^{\kappa}_{j,\si}f^\kappa_{j,\si}}
&=&
-\tRe\, \Si_{V,\kappa,0}^{\Psi_{j,\si}\bar{\Psi}_{j,\si}}(m^2_{f_{j,\si}})
\nl&&
-m_{f_{j,\si}}\sum_{\kappa'=\rR,\rL}\frac{\partial}{\partial p^2}\tRe\,
\left[m_{f_{j,\si}} \Si_{V,\kappa',0}^{\Psi_{j,\si}\bar{\Psi}_{j,\si}}(p^2)
\left.
+\Si_{S,\kappa',0}^{\Psi_{j,\si}\bar{\Psi}_{j,\si}}(p^2)
\right]
\right|_{p^2= m^2_{f_{j,\si}}},
\eeqar
for $\kappa=\rR,\rL$. 
The mass counterterms are given by
\beqar\label{fermmassCT}
\de m_{f_{j,\si}}
&=& \frac{1}{2}\sum_{\kappa=\rR,\rL}\tRe\,
\left[
m_{f_{j,\si}}
\Si_{V,\kappa,0}^{\Psi_{j,\si}\bar{\Psi}_{j,\si}}(m^2_{f_{j,\si}})
+\Si_{S,\kappa,0}^{\Psi_{j,\si}\bar{\Psi}_{j,\si}}
(m^2_{f_{j,\si}})
\right].
\eeqar

The FRC's are determined by evaluating the contributions of the loop diagrams
\beq\label{ferselfenergydiag}
\ri \Sigma_0^{\Psi_{j',\si}\bar{\Psi}_{j,\si}}= \quad
\vcenter{\hbox{\begin{picture}(60,60)(0,-10)
\Text(3,23)[bc]{\scriptsize $f_{j',\si}$} 
\Text(57,23)[bc]{\scriptsize $\bar{f}_{j,\si}$} 
\Text(32,0)[tc]{\scriptsize $f_{j'',\si''}$} 
\Text(32,40)[bc]{\scriptsize $V^a$} 
\Vertex(15,20){2}
\Vertex(45,20){2}
\ArrowLine(0,20)(15,20)
\ArrowLine(45,20)(60,20)
\PhotonArc(30,20)(15,0,180){1}{5} 
\ArrowArc(30,20)(15,180,0) 
\end{picture}}}
\qquad
+
\qquad
\vcenter{\hbox{\begin{picture}(60,60)(0,-10)
\Text(3,23)[bc]{\scriptsize $f_{j',\si}$} 
\Text(57,23)[bc]{\scriptsize $\bar{f}_{j,\si}$} 
\Text(32,0)[tc]{\scriptsize $f_{j'',\si''}$} 
\Text(32,40)[bc]{\scriptsize $\Phi_k$} 
\Vertex(15,20){2}
\Vertex(45,20){2}
\ArrowLine(0,20)(15,20)
\ArrowLine(45,20)(60,20)
\DashCArc(30,20)(15,0,180){4} 
\ArrowArc(30,20)(15,180,0) 
\end{picture}}}
\eeq
in LA.
The non-diagonal components \refeq{nondiagferWFRC}
are given by the mixing-energies, 
and these can be generated  only by 
virtual charged gauge bosons and would-be Goldstone bosons
through the mixing matrix \refeq{Umatrix}.
In LA we have 
\beqar
\delta Z_{f^\kappa_{j,\si}f^\kappa_{j',\si}}&\sim&
\Si_{0}^{f_{j',\si}\bar{f}_{j,\si}}(m^2_{f^\kappa_{j,\si}})
\nl &\propto&
\sum_{j''}U^{f^\rL,W^\si}_{jj''}U^{f^\rL,W^{-\si}}_{j''j'}
\left[\log{\frac{\mu^2}{\MW^2}}+
\log{\left(\frac{\MW^2}{\max{(\MW^2,m_{f_{j,\si}}^2,m_{f_{j'',-\si}}^2)}}\right)}
\right]\nl
&=& \de_{fQ}\de_{\si -}\ckm^+_{j3}\ckm_{3j'}\log{\frac{\MW^2}{\Mt^2}},\qquad  
\mbox{for $j\neq j'$}. 
\eeqar
where, owing to the unitarity of the quark-mixing matrix,
only diagrams involving virtual top quarks ($f_{j'',-\si}=Q_{3,+}=\Pt$)
give non-vanishing contributions. The largest ones are those enhanced by the top Yukawa couplings and are of the order
\beq
\frac{\alpha}{4\pi} \frac{\Mt^2}{\sw^2\MW^2}|\ckm^+_{j3}\ckm_{3j'}|
\log{\frac{\Mt^2}{\MW^2}} < 10^{-3}, \qquad  
\mbox{for $j\neq j'$},
\eeq
owing to the small phenomenological values of the non-diagonal components of the quark-mixing matrix \refeq{ckmvalues}.
Therefore, with an accuracy at the percent level we can assume
\beq
\delta Z_{f^\kappa_{j,\si}f^\kappa_{j',\si}}\LA 0,\qquad
\mbox{for $j\neq j'$}. 
\eeq
For the diagonal components \refeq{diagferWFRC} we obtain
\beqar\label{fermWFRCres} 
\delta Z_{f^\kappa_{j,\si} f^\kappa_{j,\si}} &\LA&
\frac{\alpha}{4\pi}
\left[-\cew_{f^\kappa_{j,\si}}\log{\frac{\mu^2}{\MW^2}}+ 
Q_{f_{j,\si}}^2\left(2\log{\frac{\MW^2}{\la^2}}
-3\log{\frac{\MW^2}{m^2_{f_{j,\si}}}}\right)
\right]
+\delta Z^{\mathrm{top}}_{f^\kappa_{j,\si}} 
.
\eeqar
The first two terms originate from virtual gauge bosons.
The first one,  proportional to the eigenvalue of the Casimir operator \refeq{casimirew}
for fermions, is a symmetric electroweak contribution, and the second one
is a pure electromagnetic contribution.
The remaining part is the contribution from scalar and gauge-boson loops involving external and/or virtual top quarks. It reads
\beqar
\lefteqn{\delta Z^{\mathrm{top}}_{f^\kappa_{j,\si}}\LA}\hspace{2mm}&&\nl &=&
-\de_{fQ}\frac{\alpha}{4\pi}
\left\{ \frac{\Mt^2}{4\sw^2\MW^2}
\left[\left(
\de_{\si +}\de_{j3} 
(1+\delta_{\kappa\rR})
+\de_{\si -}\de_{\kappa\rL}\ckm^+_{j3}\ckm_{3j}
\right)\log{\frac{\mu^2}{\Mt^2}}   
+\frac{1}{2}\de_{\si +}\de_{j3} 
\log{\frac{\Mt^2}{M_{\Htop}^2}}
\right]
\right.\nl&&\hspace{1.5cm} {}+\left.
\left[
\de_{\si +}\de_{j3}\left( 
\left(I^Z_{\Pt^\kappa}\right)^2+ \frac{\de_{\kappa\rL}}{2\sw^2}\right)
+\de_{\si -}\frac{\de_{\kappa\rL}}{2\sw^2}\ckm^+_{j3}\ckm_{3j}
\right]\log{\frac{\MW^2}{\Mt^2}} 
\right\} 
\nl&\simeq&
-\de_{fQ}\de_{j3} \frac{\alpha}{4\pi}
\left\{ \frac{\Mt^2}{4\sw^2\MW^2}
\left[\left(\de_{\si +}(1+\delta_{\kappa\rR})
+\de_{\si -}\de_{\kappa\rL}\right)\log{\frac{\mu^2}{\Mt^2}}   
+\frac{1}{2}\de_{\si +} \log{\frac{\Mt^2}{M_{\Htop}^2}}\right]
\right.\nl&&\hspace{1.8cm} {}+\left.
\left[\de_{\si +} \left(I^Z_{\Pt^\kappa}\right)^2
+\frac{\de_{\kappa\rL}}{2\sw^2}
\right]\log{\frac{\MW^2}{\Mt^2}}\right\}.
\eeqar
Also here,  the non-diagonal elements  $\ckm_{3j}$ of the quark-mixing matrix \refeq{ckmvalues} lead to very  small contributions, and up to an error of the order $10^{-4}$ at $\mu=1\TeV$  we can set $\ckm_{3j}\simeq \de_{3j}$. 

Combining \refeq{fermWFRCres} with the collinear factor for fermions 
\refeq{fercollfactres} as in \refeq{subllogfact2}, we obtain the complete SL correction factor 
\beqar \label{deccfer}
\lefteqn{\de^{\cc}_{f^\kappa_{j'',\si''}
  f^\kappa_{j,\si}}=}\quad&&\\
&=&\de_{jj''}\de_{\si\si''}
\left\{
\frac{\alpha}{4\pi}\left[\frac{3}{2} \cew_{f^\kappa_{j,\si}} 
\log{\frac{s}{\MW^2}}
+Q_{f_{j,\si}}^2 
\left(\frac{1}{2}\log{\frac{\MW^2}{m^2_{f}}}
    +\log{\frac{\MW^2}{\la^2}}\right)
\right]
+
\de^{\cc,\mathrm{top}}_{f^\kappa_{j,\si}}
\right\},\nn
\eeqar 
with the top-quark contributions
\beqar
\de^{\cc,\mathrm{top}}_{f^\kappa_{j,\si}}
&=&\de_{fQ}\de_{j3} \frac{\alpha}{4\pi}
\left\{- \frac{\Mt^2}{8\sw^2\MW^2}
\left[\left(\de_{\si +}(1+\delta_{\kappa\rR})
+\de_{\si -}\de_{\kappa\rL}\right)\log{\frac{s}{\Mt^2}}   
+\frac{1}{2}\de_{\si +} \log{\frac{\Mt^2}{M_{\Htop}^2}}\right]
\right.\nl&&\hspace{1.8cm} {}+\frac{3}{2}\left.
\left[\de_{\si +} \left(I^Z_{\Pt^\kappa}\right)^2
+\frac{\de_{\kappa\rL}}{2\sw^2}
\right]\log{\frac{\MW^2}{\Mt^2}}\right\} .
\eeqar
These ladder are sizable only for  external heavy quarks $f^\kappa_{j,\si}= Q^\kappa_{3,+}=\Pt^\kappa$ and $f^\kappa_{j,\si}= Q^\rL_{3,-}=\Pb^\rL$, and give 
\beqar\label{dethfermWFRCres} 
\de^{\cc,\mathrm{top}}_{\Pt^\rL}&=&
\frac{\alpha}{4\pi}\left\{ -\frac{\Mt^2}{8\sw^2\MW^2}
\left[\log{\frac{s}{\Mt^2}}   
+\frac{1}{2}\log{\frac{\Mt^2}{M_{\Htop}^2}}\right]
+\frac{16\cw^4+10\cw^2+1}{24\sw^2\cw^2}
\log{\frac{\MW^2}{\Mt^2}} \right\} 
,\nl
\de^{\cc,\mathrm{top}}_{\Pt^\rR}&=&
\frac{\alpha}{4\pi}\left\{ -\frac{\Mt^2}{8\sw^2\MW^2}
\left[2\log{\frac{s}{\Mt^2}}   
+\frac{1}{2}\log{\frac{\Mt^2}{M_{\Htop}^2}}\right]
+\frac{2\sw^2}{3\cw^2}\log{\frac{\MW^2}{\Mt^2}} \right\} 
,\nl
\de^{\cc,\mathrm{top}}_{\Pb^\kappa}&=&
\frac{\alpha}{4\pi}\de_{\kappa\rL}
\left\{ -\frac{\Mt^2}{8\sw^2\MW^2}\log{\frac{s}{\Mt^2}}   
+\frac{3}{4\sw^2}\log{\frac{\MW^2}{\Mt^2}} 
\right\}.
\eeqar
In contrast to
the $\Mt^2$ corrections to the $\rho$ parameter, which are only
related to the (virtual) left-handed $(\Pt,\Pb)$ doublet, logarithmic
Yukawa contributions appear also for (external) right-handed top
quarks.

\section{Longitudinally polarized gauge bosons}
\label{se:loggaugebos}
As explained in \refse{se:gbetintro},
in the high-energy limit $\MW^2/s\to 0$, the matrix elements involving longitudinal gauge bosons 
$V_\rL^a=Z_\rL,W^\pm_\rL$
have to be expressed by matrix elements involving the corresponding would-be Goldstone bosons 
$\Phi_a=\chi,\phi^\pm$
via the GBET.
In this approach, the corrections to matrix elements involving longitudinal gauge bosons have a two-fold origin. 
On one hand there are corrections to the GBET ($\de A^{V^a}$), which involve the renormalization of the asymptotic gauge-boson fields 
($\de Z_{V^{a'}V^a}$)
as well as the effects of mixing  between gauge bosons and would-be Goldstone bosons.
On the other hand corrections to the matrix elements involving the would-be Goldstone bosons
originate from  the renormalization of the would-be-Goldstone-boson fields\footnote{Note that the unphysical would-be Goldstone bosons can be kept unrenormalized.}
($\de Z_{\Phi_{a'}\Phi_a}$)
and from truncated  loop diagrams ($\de^\coll_{\Phi_{a'}\Phi_a}$).


In \refse{subse:GBETcorr} we  derive the GBET  following  \citere{DennBohmJos} 
and we  relate the  corrections  to the GBET  to gauge-boson FRC's,  longitudinal gauge-boson self-energies and mixing-energies between gauge bosons and would-be Goldstone bosons.
In \refse{subse:LAGBETcorr} we give our results for the one-loop corrections 
to the GBET in logarithmic approximation and combine them with the corrections 
to matrix elements 
involving would-be Goldstone bosons as explained in \refeq{subllogfact2}.
Finally, in \refse{se:compare}, we compare our corrections associated with  longitudinal gauge bosons  with the corrections associated with  physical would-be Goldstone bosons in the symmetric phase of the electroweak theory.

\subsection{Corrections to the 
Goldstone-boson equivalence theorem}\label{subse:GBETcorr}
We consider a matrix element involving a longitudinal gauge boson $V^a_\rL$ with polarization vector \refeq{longplovec}, and $1-n$ other arbitrary physical fields,
\beqar \label{loGBmatel} 
\lefteqn{\M^{V^{a}_\rL \varphi_{i_2} \ldots \varphi_{i_n}}(p,p_2\dots,p_{n})= 
\epsilon_\rL^{\mu}(p)
\langle \underline{V}^{a}_\mu(p) \underline{\varphi}_{i_2}(p_2) \ldots \underline{\varphi}_{i_n}(p_n) \rangle
\prod_{k=2}^{n}v_{\varphi_{i_k}}(p_k)=}\quad&&\nl
&=&
\left[\frac{p^\mu}{M_{V^a}}+ \O\left(\frac{M_{V^a}}{p^0}\right)\right]
\langle \underline{V}^{a}_\mu(p) \underline{\varphi}_{i_2}(p_2) \ldots \underline{\varphi}_{i_n}(p_n) \rangle
\prod_{k=2}^{n}v_{\varphi_{i_k}}(p_k).
\eeqar
For the amputated Green functions we use the notation\footnote{More details concerning our notation for Green functions can be found in \refapp{app:GFs}.}
\beq
\langle \underline{V}^{a}_\mu(p) \underline{\varphi}_{i_2}(p_2) \ldots \underline{\varphi}_{i_n}(p_n) \rangle
=
G^{\underline{V}^{a}_\mu \underline{\varphi}_{i_2} \ldots \underline{\varphi}_{i_n}}(p,p_2,\ldots,p_n).
\eeq
In order to derive the GBET for the matrix element \refeq{loGBmatel},
we start from  following relation between bare Green functions that follows from the  BRS invariance  of the bare electroweak Lagrangian (\cf\refapp{BRStra}),
\beqar
0&=& \brs \left[\langle \bar{u}_{0}^{a}(x)\prod_{k=2}^n \varphi^+_{i_k,0}(x_k)\rangle\right]
\\&=&
-\frac{1}{\xi_a}\langle C_0^{\bar{V}^a}(x) \prod_{k=2}^n \varphi^+_{i_k,0}(x_k)\rangle
-\sum_{j=2}^n \langle\bar{u}_{0}^{a}(x) \varphi^+_{i_2,0}(x_2)\ldots \brs \varphi^+_{i_j,0}(x_j)\ldots
\varphi^+_{i_n,0}(x_n)\rangle,\nn
\eeqar
where the gauge-fixing term $C_0^{\bar{V}^a}$ \refeq{Gfix}
is generated by the BRS variation   $\brs \bar{u}^a_0$ \refeq{antighostbrstra} of the antighost field.
After Fourier transformation, amputation of the physical on-shell legs $\varphi_{i_k}$ and contraction with their polarization vectors,  
all terms generated by the BRS variation
of the physical on-shell fields  $\brs \varphi_{i_k}$ vanish and one obtains
\beqar\label{GBETder2}
0&=&p^\mu\langle\bar{V}^a_{\mu,0}(p)
\prod_{k=2}^n \underline{\varphi}^+_{i_k,0}(p_k)\rangle
\prod_{k=2}^{n}v_{\varphi_{i_k,0}}(p_k)
\nl&&{}+\ri^{(1-Q_{V^a})} \xi_{a,0} M_{V^a,0}\langle\Phi^+_{a,0} (p) 
\prod_{k=2}^n \underline{\varphi}^+_{i_k,0}(p_k)\rangle
\prod_{k=2}^{n}v_{\varphi_{i_k,0}}(p_k),
\eeqar
where $\Phi_a=\chi,\phi^\pm$ are the would-be Goldstone bosons associated to the  gauge fields $V^a=Z,W^\pm$.
In order to amputate these external lines $\Phi_a$ and $V^a$
we have to consider the full propagator for gauge bosons and would-be Goldstone bosons, including  mixing. This reads
\beqar\label{GBETder3}
G^{\varphi_i\varphi_j}_{(\mu\nu)}(p,-p)
&=&
\left(\begin{array}{c@{\qquad}c}
G^{V^aV^{a'}}_{\mu\nu}(p,-p) & G^{V^a\Phi_{b'}}_{\mu}(p,-p) \\ 
 G^{\Phi_bV^{a'}}_{\nu}(p,-p) & G^{\Phi_b\Phi_{b'}}(p,-p) 
\end{array}\right)
\nl&=&
\left(\begin{array}{c@{\qquad}c}
g^\rT_{\mu\nu} G^{V^aV^{a'}}_{\rT}(p^2) 
+g^\rL_{\mu\nu} G^{V^aV^{a'}}_{\rL}(p^2) 
& p^\mu G_{\rL}^{V^a\Phi_{b'}}(p^2) \\ 
 -p^\nu G_{\rL}^{\Phi_bV^{a'}}(p^2) & 
G^{\Phi_b\Phi_{b'}}(p^2) 
\end{array}\right),
\eeqar
with
\beq
g^\rT_{\mu\nu}=g_{\mu\nu}-\frac{p_\mu p_\nu}{p^2}
,\qquad
g^\rL_{\mu\nu}=\frac{p_\mu p_\nu}{p^2}.
\eeq 
In the four sectors of \refeq{GBETder3} we have  matrices with indices $V^a,V^{a'}=A,Z,W^\pm$ and $\Phi_b,\Phi_{b'}=\chi,\phi^\pm$, and these are  constrained  by the Ward identities \cite{DennBohmJos}
\beqar\label{GBETder4a}
\lefteqn{\hspace{-1.5cm}p^2G^{\bar{V}^aV^{b}}_{\rL,0}(p^2) 
-p^2\left[\ri^{(Q_{\bar{V}^a}+1)}\xi_{a,0}M_{V^a,0}
G_{\rL,0}^{\Phi^+_aV^{b}}(p^2) 
+\ri^{(Q_{V^b}+1)}\xi_{b,0}M_{V^b,0}
G_{\rL,0}^{\bar{V}^a\Phi_{b}}(p^2) 
\right]}
\quad&&\nl&&{}+
\ri^{(Q_{\bar{V}^a}+Q_{V^b})}\xi_{a,0} M_{V^a,0}\xi_{b,0} M_{V^b,0}
G_0^{\Phi^+_a\Phi_{b}}(p^2)
=-\ri\xi_{a,0} \de_{V^aV^b},
\eeqar
for $V^a,V^b=A,Z,W^\pm$.
Expressing \refeq{GBETder2} in terms of amputated Green functions times propagators and mixing propagators we have 
\beqar\label{GBETder4}
\lefteqn{\sum_{V^{b}}
\left\{G^{\bar{V}^aV^{b}}_{\rL,0}(p^2) 
-\ri^{(1-Q_{V^a})}\xi_{a,0} M_{V^a,0}G_{\rL,0}^{\Phi^+_a V^{b}}(p^2)\right\} 
p^\mu\langle\underline{V}^{b}_{\mu,0}(p)\underline{\Oper}\rangle
=}\quad&&\nl
&=&
-\sum_{\Phi_{b}}\left\{p^2 G_{\rL,0}^{\bar{V}^a\Phi_{b}}(p^2) 
+\ri^{(1-Q_{V^a})} \xi_{a,0} M_{V^a,0} G_0^{\Phi^+_{a}\Phi_{b}}(p^2) 
\right\}\langle\underline{\Phi}_{b,0}(p)\underline{\Oper}\rangle
,
\eeqar
where the fields $\prod_k \varphi_{i_k}$ have been denoted  by the shorthand $\Oper$, and their wave functions have been  omitted.
For $V^a=W^\pm, Z$, the only  mixing terms  with $V^b\neq V^a$ or $\Phi_b\neq\Phi_a$ in \refeq{GBETder4} that are allowed by charge conservation
are terms with $V^a=Z$ and $V^b=A$ in the first line. However, these terms cancel owing to the corresponding  Ward identity
\refeq{GBETder4a}, so that we arrive at
\beqar\label{GBETder5}
p^\mu\langle\underline{V}^{a}_{\mu,0}(p)\,\underline{\Oper}\rangle
&=&\ri^{(1-Q_{V^a})}M_{V^a,0}A^{V^a}_{0}
\langle\underline{\Phi}_{a,0}(p)\,\underline{\Oper}\rangle
,
\eeqar
for $V^a=Z,W^\pm$, with 
\beqar\label{GBETder6}
A^{V^a}_{0}&=&-\frac{\ri^{(Q_{V^a}-1)}p^2 G_{\rL,0}^{\bar{V}^a\Phi_{a}}(p^2) 
+\xi_{a,0} M_{V^a,0} G_0^{\Phi^+_{a}\Phi_{a}}(p^2)} 
{M_{V^a,0}\left[G^{\bar{V}^aV^{a}}_{\rL,0}(p^2) 
-\ri^{(1-Q_{V^a})}\xi_{a,0} M_{V^a,0}G_{\rL,0}^{\Phi^+_a V^{a}}(p^2)\right]}.
\eeqar
Note that owing to
\beqar
G_{\rL}^{ W^{-\si}W^{\si'}}&=& 
\de_{\si\si'}G_{\rL}^{ WW},\qquad
G^{ \phi^{-\si}\phi^{\si'}}= 
\de_{\si\si'}G^{ \phi\phi},
\nl
G_{\rL}^{ W^{-\si}\phi^{\si'}}&=& 
G_{\rL}^{\phi^{\si'} W^{-\si}}=
\si\de_{\si\si'}G_{\rL}^{ W\phi},
\eeqar
we have $A^{W^+}_{0}=A^{W^-}_{0}$.
The  constants \refeq{GBETder6} can be expressed through the full\footnote{Note that we here use the symbol $\Gamma^{\varphi_i\varphi_j}$ for the full 2-point function (including tadpole diagrams) which is normally only used for the 1PI part.} 2-point function $\Gamma^{\varphi_i\varphi_j}_{(\mu\nu),0}$ which  has the same structure as the  propagator \refeq{GBETder3} and is related to it by
\beq\label{GBETder8}
g^{(\nu\nu')}\sum_{\varphi_j} G^{\varphi_i\varphi_j}_{(\mu\nu),0}(p,-p)
\Gamma^{\varphi_j\varphi_k}_{(\nu'\rho),0}(p,-p)=
\ri \de^{\varphi_i\varphi_k} g_{(\mu\rho)},\qquad
\varphi_i,\varphi_j,\varphi_k= A,Z,W^\pm,\chi,\phi^\pm.
\eeq
We note that within a spontaneously broken theory this relation only applies to the full two-point function, 
where also the tadpole contributions have to be taken into account.
In the 't~Hooft gauge \refeq{Gfix}, the components of $\Gamma^{\varphi_j\varphi_k}_{0}$ are\footnote{In the notation of \citere{DennBohmJos} the mixing energies for charged gauge bosons read
\begin{displaymath}
\Gamma_{\rL}^{ W^{\si}\phi^{-\si'}}= 
\Gamma_{\rL}^{\phi^{-\si'} W^{\si}}=
\si\de_{\si\si'}\Gamma_{\rL}^{ W\phi}.\nonumber
\end{displaymath}
}
\beqar\label{GBETder9}
\Gamma^{\bar{V}^aV^b}_{\rT,0}(p^2)&=&-\de_{V^aV^b}(p^2-M^2_{V^a,0})
-\Si^{\bar{V}^aV^b}_{\rT,0}(p^2),\nl
\Gamma^{\bar{V}^aV^b}_{\rL,0}(p^2) &=&
-\frac{\de_{V^aV^b}}{\xi_{a,0}}(p^2-\xi_{a,0} M^2_{V^a,0})
-\Si^{\bar{V}^aV^b}_{\rL,0}(p^2)
, \nl
\Gamma^{\bar{V}^a\Phi_b}_{\rL,0}(p^2) &=&
\Gamma^{\Phi_b\bar{V}^a}_{\rL,0}(p^2) =\Si^{\bar{V}^a\Phi_b}_{\rL,0}(p^2), 
\nl
\Gamma^{\Phi^+_a\Phi_b}_{0}(p^2) &=&\de_{\Phi_a\Phi_b}(p^2-\xi_{a,0} M^2_{V^a,0})
+\Si^{\Phi^+_a\Phi_b}_{0}(p^2).
\eeqar
The  constants \refeq{GBETder6} can be expressed as \cite{DennBohmJos}
\beqar\label{GBETder10}
A^{V^a}_{0}&=&\frac{p^2/\xi_{a,0} + \Gamma_{\rL,0}^{V^a\bar{V}^a}(p^2)}
{M_{V^a,0}\left[M_{V^a,0}
-\ri^{(1+Q_{V^a})}
\Gamma^{V^{a}\Phi^+_a}_{\rL,0}(p^2)\right]}= 1+ \de A^{V^a}_{0},
\eeqar
and in  one-loop approximation
\beqar\label{GBETder11}
\de A^{V^a}_{0}&=&-\frac{\Si_{\rL,0}^{V^a\bar{V}^a}(p^2)}{M^2_{V^a,0}}
+\ri^{(1+Q_{V^a})}
\frac{\Si^{V^{a}\Phi^+_a}_{\rL,0}(p^2)}{M_{V^a,0}}.
\eeqar

We can now renormalize the relation \refeq{GBETder5} by including  
the counterterms originating from the renormalization  
of the gauge-boson fields \refeq{phWFRCgb} and of their masses \refeq{phmRCgb}.
The unphysical would-be Goldstone  fields are kept unrenormalized and 
the renormalization of the gauge parameters does not contribute 
since \refeq{GBETder5} and \refeq{GBETder11} are explicitly $\xi_a$--independent. 
In one-loop approximation, we obtain
\beqar\label{GBETder12}
\sum_{V^b} p^\mu\langle\underline{V}^{b}_{\mu}(p)\,\underline{\Oper}\rangle
\left( \de_{V^bV^a}-\frac{1}{2}\delta Z_{V^{b}V^{a}}\right)
&=&\ri^{(1-Q_{V^a})}(M_{V^a}+\de M_{V^a} )A^{V^a}_{0}
\langle\underline{\Phi}_{a}(p^2)\,\underline{\Oper}\rangle
,\nln
\eeqar
and since the only possible mixing terms between \PZ~bosons and photons, induced by $\de Z_{AZ}$,  cancel owing to the QED Ward identity 
$p^\mu \langle\underline{A}_{\mu}(p)\,\underline{\Oper}\rangle$, we arrive to the  renormalized relation
\beqar\label{GBETder13}
p^\mu\langle\underline{V}^{a}_{\mu}(p)\,\underline{\Oper}\rangle
&=&\ri^{(1-Q_{V^a})}M_{V^a} (1+\de A^{V^a})
\langle\underline{\Phi}_{a}(p^2)\,\underline{\Oper}\rangle
,\nln
\eeqar
with 
\beqar\label{GBETder14}
\de A^{V^a}&=&\de A^{V^a}_{0}+\frac{1}{2}\frac{\de M^2_{V^a}}{M^2_{V^a}} +
\frac{1}{2}\delta Z_{V^{a}V^{a}}\nl
&=&
-\frac{\Si_{\rL,0}^{V^a\bar{V}^a}(p^2)}{M^2_{V^a,0}}
+\ri^{(1+Q_{V^a})}
\frac{\Si^{V^{a}\Phi^+_a}_{\rL,0}(p^2)}{M_{V^a,0}}
+\frac{1}{2}\frac{\de M^2_{V^a}}{M^2_{V^a}} +
\frac{1}{2}\delta Z_{V^{a}V^{a}}.
\eeqar
For the matrix element \refeq{loGBmatel}, the relation \refeq{GBETder13} yields
\beq
\M^{V^{a}_\rL \varphi_{i_2} \ldots \varphi_{i_n}}(p,p_2\dots,p_{n})= 
\ri^{(1-Q_{V^a})}(1+\de A^{V^a})\M^{\Phi_{a} \varphi_{i_2} \ldots \varphi_{i_n}}(p,p_2\dots,p_{n}),
\eeq
up to terms suppressed by factors of order $M_{V^a}/E$.

\subsection{Logarithmic corrections to longitudinal gauge bosons}\label{subse:LAGBETcorr}
In the following, we present our results for the corrections  \refeq{GBETder14}
in LA. 
We observe that the tadpole diagrams give non-vanishing contributions to single terms of \refeq{GBETder14}. However, these contributions compensate each other in  \refeq{GBETder14} and in the following they are omitted everywhere.
The diagrams    \refeq{gaugeselfenergydiag} yield
\beqar\label{GBETder15a}
\frac{\Si_{\rL,0}^{V^a\bar{V}^a}(M^2_{V^a})}{M^2_{V^a}}&\LA&
-\frac{\alpha}{4\pi}\left\{
\left[2\cew_{V^aV^a}-4\cew_\Phi\right]\log{\frac{\mu^2}{\MW^2}}
+\frac{\NCt}{2\sw^2}\frac{\Mt^2}{\MW^2}\log{\frac{\mu^2}{\Mt^2}}
\right.\nl&&\left.\hspace{1cm}{}
+\frac{3M_{V^a}^2}{4\sw^2\MW^2}\log{\frac{\MH^2}{\MW^2}}
\right\},
\eeqar
for the longitudinal gauge-boson propagator, where 
$\cew_{V^aV^a}$ and $\cew_\Phi$ represent the electroweak Casimir operator
in the adjoint representation \refeq{physadjointcasimir}
and in the scalar representation \refeq{casimirew}, respectively.
For the mass counterterms \refeq{gbmassCT} we obtain
\beqar \label{massCT}
\frac{\de M^2_{V^a}}{M^2_{V^a}}&\LA&
-\frac{\alpha}{4\pi}\left\{
\left[\bew_{V^aV^a}-4\cew_\Phi\right]\log{\frac{\mu^2}{\MW^2}}
+\frac{\NCt}{2\sw^2}\frac{\Mt^2}{\MW^2}\log{\frac{\mu^2}{\Mt^2}}
\right.\nl&&\left.\hspace{1cm}{}
+\frac{10M_{V^a}^2}{12\sw^2\MW^2}\log{\frac{\MH^2}{\MW^2}}
+\frac{2\NCt}{3} \topmat_{V^aV^b}\log{\frac{\Mt^2}{\MW^2}}\right\}
, 
\eeqar
where $\bew_{V^aV^a}$
are the coefficients of the beta function defined in \refapp{app:betafunction},
and the matrix  $\topmat_{V^aV^b}$ is given in 
\refeq{topWFRCgb1} and  \refeq{topWFRCgb2}.
The FRC's $\de Z_{V^aV^a}$ are given in \refeq{gbdeltaZmat}. The mixing-energies are obtained from the diagrams
\beqar\label{mixingselfenergydiag}
\ri \Si^{V^a\Phi^+_a}_{\mu,0}(p)&=&\hspace{3mm}
\vcenter{\hbox{\begin{picture}(60,60)(0,-10)
\Text(3,23)[bc]{\scriptsize $V_\mu^a$} 
\Text(57,23)[bc]{\scriptsize $\Phi_a^+$} 
\Text(32,0)[tc]{\scriptsize $V^c$} 
\Text(32,40)[bc]{\scriptsize $V^d$} 
\Vertex(15,20){2}
\Vertex(45,20){2}
\Photon(0,20)(15,20){2}{2}
\DashLine(45,20)(60,20){2} 
\PhotonArc(30,20)(15,0,180){2}{5} 
\PhotonArc(30,20)(15,180,0){2}{5} 
\end{picture}}}
\hspace{3mm}
+
\hspace{3mm}
\vcenter{\hbox{\begin{picture}(60,60)(0,-10)
\Text(3,23)[bc]{\scriptsize $V_\mu^a$} 
\Text(57,23)[bc]{\scriptsize $\Phi_a^+$} 
\Text(32,0)[tc]{\scriptsize $u^c$} 
\Text(32,40)[bc]{\scriptsize $\bar{u}^d$} 
\Vertex(15,20){2}
\Vertex(45,20){2}
\Photon(0,20)(15,20){2}{2}
\DashLine(45,20)(60,20){2} 
\DashArrowArc(30,20)(15,0,180){1} 
\DashArrowArc(30,20)(15,180,0){1} 
\end{picture}}}
\hspace{3mm}
+
\hspace{3mm}
\vcenter{\hbox{\begin{picture}(60,60)(0,-10)
\Text(3,23)[bc]{\scriptsize $V_\mu^a$} 
\Text(57,23)[bc]{\scriptsize $\Phi_a^+$} 
\Text(32,0)[tc]{\scriptsize $V^c$} 
\Text(32,40)[bc]{\scriptsize $\Phi_k$} 
\Vertex(15,20){2}
\Vertex(45,20){2}
\Photon(0,20)(15,20){2}{2}
\DashLine(45,20)(60,20){2} 
\DashCArc(30,20)(15,0,180){4} 
\PhotonArc(30,20)(15,180,0){2}{5} 
\end{picture}}}
\nl&&\hspace{1.0cm}
{}
+
\hspace{3mm}
\vcenter{\hbox{\begin{picture}(60,60)(0,-10)
\Text(3,23)[bc]{\scriptsize $V_\mu^a$} 
\Text(57,23)[bc]{\scriptsize $\Phi_a^+$} 
\Text(32,0)[tc]{\scriptsize $\Phi_k$} 
\Text(32,40)[bc]{\scriptsize $\Phi_j$} 
\Vertex(15,20){2}
\Vertex(45,20){2}
\Photon(0,20)(15,20){2}{2}
\DashLine(45,20)(60,20){2} 
\DashCArc(30,20)(15,0,180){4} 
\DashCArc(30,20)(15,180,0){4}
\end{picture}}}
\hspace{3mm}
+
\hspace{3mm}
\vcenter{\hbox{\begin{picture}(60,60)(0,-10)
\Text(3,23)[bc]{\scriptsize $V_\mu^a$} 
\Text(57,23)[bc]{\scriptsize $\Phi_a^+$} 
\Text(32,0)[tc]{\scriptsize $f_{j,\si}$} 
\Text(32,40)[bc]{\scriptsize $\bar{f}_{j',\si'}$} 
\Vertex(15,20){2}
\Vertex(45,20){2}
\Photon(0,20)(15,20){2}{2}
\DashLine(45,20)(60,20){2} 
\ArrowArc(30,20)(15,0,180) 
\ArrowArc(30,20)(15,180,0) 
\end{picture}}}
\qquad,
\eeqar
and in LA we obtain
\beqar\label{GBETder15b}
\frac{\Si^{V^{a}\Phi^+_a}_{\rL,0}(M^2_{V^a})}{M^2_{V^a}}&\LA&
\ri^{(1-Q_{V^a})}\frac{\alpha}{4\pi}\left\{
\left[\cew_{V^aV^a}-3\cew_\Phi\right]\log{\frac{\mu^2}{\MW^2}}
+\frac{\NCt}{2\sw^2}\frac{\Mt^2}{\MW^2}\log{\frac{\mu^2}{\Mt^2}}
\right.\nl&&\left.\hspace{1.5cm}{}
+\frac{3 M_{V^a}^2}{4\sw^2\MW^2}\log{\frac{\MH^2}{\MW^2}}
\right\}.
\eeqar
The corrections \refeq{GBETder14} to the GBET are obtained by combining the above results and read
\beqar \label{eq:renetcorr} 
\de A^{V^a}&=&\frac{\alpha}{4\pi}\left\{
\cew_\Phi\log{\frac{\mu^2}{\MW^2}}-\frac{\NCt}{4\sw^2}\frac{\Mt^2}{\MW^2}\log{\frac{\mu^2}{\Mt^2}}
+Q_{V^a}^2\log{\frac{\MW^2}{\la^2}}
\right.\nl&&\left.\hspace{1cm}{}
+\frac{3 M_{V^a}^2}{8\sw^2\MW^2}\log{\frac{\MH^2}{\MW^2}}
\right\},
\eeqar
in LA.
Note that the $\mu$--dependent part is determined by the 
scalar eigenvalue $\cew_\Phi$ of the electroweak Casimir operator  
and by large $\Mt$-dependent contributions
originating from the mass counterterms \refeq{massCT}, which are 
proportional to the colour factor $\NCt=3$.

Finally,  the complete  SL mass-singular  corrections 
associated to external  longitudinal gauge bosons
are obtained by 
\beq
\delta^\cc_{V_\rL^{a'}V_\rL^{a}}=\delta_{V^{a'}V^{a}}\left[\de A^{V^a}+ \de^\coll_{\Phi_a\Phi_a}\right],
\eeq
\ie by including, as indicated in \refeq{subllogfact2},
the collinear corrections 
\refeq{scalarcollfactwbgb}   originating from loop-diagrams with external would-be Goldstone bosons.
The final result for this correction factor reads 
\beqar \label{longeq:coll} 
\delta^\cc_{V_\rL^{a'}V_\rL^{a}}&=&
\delta_{V^{a'}V^{a}}
\frac{\alpha}{4\pi}\left\{2\cew_\Phi\log{\frac{s}{\MW^2}}-\frac{3}{4\sw^2}\frac{\Mt^2}{\MW^2}\log{\frac{s}{\Mt^2}}   
+Q_{V^a}^2\log{\frac{\MW^2}{\la^2}}
\right.\nl&&\left.\hspace{1.5cm}{}
+\frac{M_{V^a}^2}{8\sw^2\MW^2}\log{\frac{\MH^2}{\MW^2}}
\right\}.
\eeqar
\subsection{Comparison between spontaneously broken and symmetric phase}\label{se:compare}
We stress that the  result \refeq{longeq:coll}, and in particular all mixing-energy  diagrams \refeq{mixingselfenergydiag}, receive contributions from the broken part of the electroweak Lagrangian. However, we observe the following.
If we  restrict ourselves to the 
symmetric electroweak part of \refeq{longeq:coll}, \ie to the $\log(s)$ terms, and  compare it with \refeq{Higgscc},
we observe that Higgs bosons and
longitudinal gauge bosons receive the same collinear SL corrections.
This suggests that, up to electromagnetic terms and $\log{(\MH/\MW)}$ terms,
the collinear corrections to longitudinal gauge bosons 
can be expressed as collinear  corrections \refeq{subllogfact2} 
to on-shell 
renormalized would-be Goldstone bosons with\footnote{Recall that the would-be Goldstone bosons have been kept unrenormalized in the above derivation.}
\beq\label{WBGBren}
\tRe \left. \frac{\partial \Gamma^{\Phi_{a}\Phi^+_{a}}(p^2)}{\partial p^2}\right|_{p^2=M^2_{V^a}}=1,
\eeq
\ie with FRC's
\beqar \label{wbGBWFRCa} 
\delta Z_{\Phi_{a}\Phi_{a}}&=&-\left. \frac{\partial}{\partial p^2} \tRe\,\Sigma_0^{\Phi_{a}\Phi^+_{a}}(p^2)\right|_{p^2=M^2_{V^a}}.
\eeqar
In fact, evaluating the diagrams
\beqar\label{wbgbselfenergydiag}
\hspace{-1cm}
\ri \Sigma_0^{\Phi_a\Phi^+_a}&=& \quad
\vcenter{\hbox{\begin{picture}(60,60)(0,-10)
\Text(3,23)[bc]{\scriptsize $\Phi_a$} 
\Text(57,23)[bc]{\scriptsize $\Phi^+_a$} 
\Text(32,0)[tc]{\scriptsize $\Phi_k$} 
\Text(32,40)[bc]{\scriptsize $V^b$} 
\DashLine(0,20)(15,20){2} 
\DashLine(45,20)(60,20){2} 
\Vertex(15,20){2}
\Vertex(45,20){2}
\PhotonArc(30,20)(15,0,180){1}{5} 
\DashCArc(30,20)(15,180,0){4} 
\end{picture}}}
\quad
+
\quad
\vcenter{\hbox{\begin{picture}(60,60)(0,-10)
\Text(3,23)[bc]{\scriptsize $\Phi_a$} 
\Text(57,23)[bc]{\scriptsize $\Phi^+_a$} 
\Text(32,0)[t]{\scriptsize $f_{j,\si}$} 
\Text(32,40)[b]{\scriptsize $\bar{f}_{j',\si'}$} 
\Vertex(15,20){2}
\Vertex(45,20){2}
\DashLine(0,20)(15,20){2} 
\DashLine(45,20)(60,20){2} 
\ArrowArc(30,20)(15,0,180) 
\ArrowArc(30,20)(15,180,0) 
\end{picture}}}
\quad
+
\quad
\vcenter{\hbox{\begin{picture}(60,60)(0,-10)
\Text(3,23)[bc]{\scriptsize $\Phi_a$} 
\Text(57,23)[bc]{\scriptsize $\Phi^+_a$} 
\Text(32,0)[t]{\scriptsize $V^c$} 
\Text(32,40)[b]{\scriptsize $V^b$} 
\Vertex(15,20){2}
\Vertex(45,20){2}
\DashLine(0,20)(15,20){2} 
\DashLine(45,20)(60,20){2} 
\PhotonArc(30,20)(15,0,180){1}{5} 
\PhotonArc(30,20)(15,180,0){1}{5} 
\end{picture}}}
\nl &&\hspace{-2cm}
+\quad
\vcenter{\hbox{\begin{picture}(60,60)(0,-10)
\Text(3,23)[bc]{\scriptsize $\Phi_a$} 
\Text(57,23)[bc]{\scriptsize $\Phi^+_a$} 
\Text(32,0)[t]{\scriptsize $\Phi_k$} 
\Text(32,40)[b]{\scriptsize $\Phi_l$} 
\Vertex(15,20){2}
\Vertex(45,20){2}
\DashLine(0,20)(15,20){2} 
\DashLine(45,20)(60,20){2} 
\DashCArc(30,20)(15,0,180){4} 
\DashCArc(30,20)(15,180,0){4} 
\end{picture}}}
\quad +\quad
\vcenter{\hbox{\begin{picture}(60,60)(0,-10)
\Text(3,23)[bc]{\scriptsize $\Phi_a$} 
\Text(57,23)[bc]{\scriptsize $\Phi^+_a$} 
\Text(32,0)[t]{\scriptsize $u^c$} 
\Text(32,40)[b]{\scriptsize$\bar{u}^b$} 
\Vertex(15,20){2}
\Vertex(45,20){2}
\DashLine(0,20)(15,20){2} 
\DashLine(45,20)(60,20){2} 
\DashArrowArc(30,20)(15,0,180){1} 
\DashArrowArc(30,20)(15,180,0){1} 
\end{picture}}}
\quad +\quad
\vcenter{\hbox{\begin{picture}(60,60)(0,-10)
\Text(8,23)[bc]{\scriptsize $\Phi_a$} 
\Text(52,23)[bc]{\scriptsize $\Phi^+_a$} 
\Text(32,50)[bc]{\scriptsize $V^c$} 
\Vertex(30,20){2}
\DashLine(5,20)(30,20){2}
\DashLine(30,20)(55,20){2}
\PhotonArc(30,33)(13,0,360){2}{9} 
\end{picture}}}
\quad +\quad
\vcenter{\hbox{\begin{picture}(60,60)(0,-10)
\Text(8,23)[bc]{\scriptsize $\Phi_a$} 
\Text(52,23)[bc]{\scriptsize $\Phi^+_a$} 
\Text(32,50)[bc]{\scriptsize $\Phi_k$} 
\Vertex(30,20){2}
\DashLine(5,20)(30,20){2}
\DashLine(30,20)(55,20){2}
\DashCArc(30,33)(13,0,360){4} 
\end{picture}}}
,
\eeqar
in LA, we obtain
\beqar \label{wbGBWFRCb} 
\frac{1}{2}\delta Z_{\Phi_{a}\Phi_{a}}&\LA&\frac{\alpha}{4\pi}\left\{
\cew_\Phi\log{\frac{\mu^2}{\MW^2}}-\frac{\NCt}{4\sw^2}\frac{\Mt^2}{\MW^2}\log{\frac{\mu^2}{\Mt^2}}
+\frac{ M_{V^a}^2}{4\sw^2\MW^2}\log{\frac{\MH^2}{\MW^2}}
\right\}.
\eeqar
The symmetric electroweak  part $\de Z^\ew$ of \refeq{wbGBWFRCb} 
corresponds to the corrections to the GBET \refeq{eq:renetcorr}, so that 
\beq \label{longeq:coll2} 
\delta^{\cc,\ew}_{V_\rL^{a'}V_\rL^{a}}\LA \left.\left(\frac{1}{2}\delta Z^\ew_{\Phi_{a'}\Phi^+_{a}}
+ \delta^{\coll,\ew}_{\Phi_{a'}\Phi_{a}}\right)\right|_{\mu^2=s}.
\eeq
Furthermore, also the electromagnetic $\log{(\MW^2/\la^2)}$ term can be correctly reproduced if one neglects the last diagram in the first line of \refeq{wbgbselfenergydiag}, which originates from spontaneous symmetry breaking. 
The remaining diagrams that 
originate from spontaneous symmetry breaking  do not affect \refeq{wbGBWFRCb}, as we already observed 
in the case of the Higgs FRC. 
Therefore, up to the $\log{(\MH/\MW)}$ terms,
{\em the collinear corections \refeq{longeq:coll} 
associated with longitudinal gauge bosons
correspond to the collinear corrections \refeq{longeq:coll2}
associated with physical scalar bosons belonging to a Higgs
doublet with vanishing vev.}  
This justifies, at the one-loop level, the symmetric approach
adopted in \citere{Melles:2000ia}, 
where high-energy longitudinal gauge bosons are described by
would-be Goldstone bosons as physical scalar bosons.

\chapter{Collinear Ward identitites}
\label{ch:CWI}
\newcommand{\linbrs}{X}
In this chapter we discuss and derive the {\em collinear Ward identities}
used in \refch{factorization} to prove the 
factorization of  collinear 
mass singularities. 
In \refse{se:genericWI} we present these identities
in generic form and make some important remarks concerning their applicability and their derivation.
In particular we stress that the mechanism of spontaneous symmetry breaking plays a 
non-trivial role in ensuring their validity.
The detailed derivations are performed  in  \refses{se:SCACWI} and \ref{se:FERCWI} using the BRS invariance of the spontaneously broken electroweak theory.


\section{Ward identities for generic fields}\label{se:genericWI}
Recall that we are interested in the mass-singular corrections to  generic processes  \refeq{process}.
As discussed in \refch{factorization}, these mass singularities originate from loop diagrams where one of the external particles splits into two collinear virtual particles, $\varphi_{i_k}(p_k) \rightarrow V_\mu^a(q) \varphi_{i'_k}(p_k-q)$, one of these being a gauge boson.
Here we treat the subprocesses that result from these splittings, \ie processes of the type
\beq\label{transition}
\varphi_{i_1}(p_1)\, \cdots \,\varphi_{i_{k-1}}(p_{k-1})\, V^a_\mu(q)\, \varphi_{i'_k}(p_k-q)\,\varphi_{i_{k+1}}(p_{k+1})\,\cdots \,\varphi_{i_n}(p_n)\rightarrow 0,
\eeq
where the momenta $p_k$, with $k=1,\dots, n$, are  relativistic and on-shell, whereas 
the momenta  $q$ and $p_k-q$ of the collinear particles  are quasi on-shell.
For the subprocess \refeq{transition} we consider the following lowest-order matrix element\footnote{The diagrammatic notation used in this chapter  as well as the shorthands for Green functions and matrix elements are defined in \refch{factorization}.}
\beqar \label{shortBornampli4}
\vcenter{\hbox{\begin{picture}(85,80)(12,20)
\Line(55,60)(15,60)
\Text(35,65)[b]{\scriptsize $\varphi_{i'_k}(p_k-q)$}
\Photon(30,30)(70,60){1}{6}
\Text(55,35)[t]{\scriptsize $V^a_\mu(q)$}
\GCirc(70,60){15}{1}
\end{picture}}}
&=&G_{\mu,0}^{\underline{\varphi}_{i_1}\ldots\,\underline{\varphi}_{i_{k-1}}\underline{V}^a\underline{\varphi}_{i'_k}\underline{\varphi}_{i_{k+1}}\,\ldots\, \underline{\varphi}_{i_n}}(p_1,\ldots,p_{k-1},q,p_k-q,p_{k+1},\ldots ,p_n)
\nl&&{}\times
\prod_{j=1}^n v_{\varphi_{i_j}}(p_{j})
,
\hspace{0cm}
\eeqar
where all external lines, except for the gauge boson,  are contracted with their 
wave functions.
The {\em collinear Ward identities} we want to prove read
\beqar\label{UNIVcollwi2}
\lim_{q^\mu\rightarrow xp_k^\mu} q^\mu
\left\{
\vcenter{\hbox{\begin{picture}(85,80)(12,20)
\Line(55,60)(15,60)
\Text(35,65)[b]{\scriptsize $\varphi_{i}(p-q)$}
\Photon(30,30)(70,60){1}{6}
\Text(55,35)[t]{\scriptsize $V^a_\mu(q)$}
\GCirc(70,60){15}{1}
\end{picture}}}
\hspace{-2mm}- \sum_{\varphi_j}
\vcenter{\hbox{\begin{picture}(100,80)(-10,20)
\Line(55,60)(35,60)
\Line(35,60)(0,60)
\Text(30,66)[br]{\scriptsize $ \varphi_{i}(p-q)$}
\Text(33,66)[bl]{\scriptsize $ \varphi_{j}(p)$}
\Photon(10,30)(35,60){1}{5}
\Vertex(35,60){2}
\Text(35,40)[t]{\scriptsize $V^a_\mu(q)$}
\GCirc(70,60){15}{1}
\end{picture}}}
\right\}
=\sum_{\varphi_{i'}}\hspace{-2mm}
\vcenter{\hbox{\begin{picture}(80,80)(0,-40)
\Line(55,0)(10,0)
\GCirc(55,0){15}{1}
\Text(10,5)[lb]{\scriptsize $\varphi_{i'}(p)$}
\end{picture}}}
 e I^{V^a}_{\varphi_{i'}\varphi_{i}},\nln
\eeqar
in diagrammatic notation. Equivalently we can write them as
\beq \label{CWIs}
\lim_{q^\mu \rightarrow xp^\mu}
q^\mu
v_{\varphi_i}(p)G^{[\underline{V}^a\underline{\varphi}_i]
  \underline{\Oper}}_{\mu,0}(q,p-q,r)= e
\sum_{\varphi_{i'}}\M_0^{\varphi_{i'}
\Oper}(p,r) I^{V^a}_{\varphi_{i'}\varphi_i},
\eeq
using the  shorthand notations 
introduced in \refeq{shortBornampli}, \refeq{defoper}, and \refeq{subtractGF}. 

A detailed derivation of \refeq{CWIs} 
is presented in \refse{se:SCACWI} 
for external scalar fields
($\varphi_i=\Phi_i$) and gauge bosons ($\varphi_i=V^a$) and in
\refse{se:FERCWI} for fermions ($\varphi_i=\Psi^\kappa_{j,\si}$). 
Here we anticipate 
the most important features and restrictions concerning these
collinear Ward identities:
\begin{itemize}
\item They are restricted to lowest-order (LO) matrix elements. The subscript $0$ indicating LO quantities in often omitted in the following. However,
we stress that all equations used in this chapter are only valid in LO.

\item They are realized in the high-energy limit \refeq{Sudaklim},
for  (quasi) on-shell external momenta $p$, 
and in the limit of collinear gauge boson momenta $q$,
\ie in the limit where
  $0< p^2,(p-q)^2\ll s$. All these limits have to be taken
  simultaneously.
  The wave function $v_{\varphi_i}(p)$ corresponds to a particle with
  mass $\sqrt{p^2}$.
\item They are valid only up to mass-suppressed terms, 
 to be precise terms of
  the order $M/\sqrt{s}$ (for fermions) or $M^2/s$ (for bosons)
  with respect to the leading terms appearing in \refeq{CWIs}, where 
$M^2\sim\max(p^2,M_{\varphi_i}^2,M^2_{V^a})$.
  Furthermore, they apply only to matrix elements that are not
  mass-suppressed. In other words, they apply to those matrix elements
  that arise from $\mathcal{L_{\mathrm{symm}}}$ in LO.
\item Their derivation is based on the BRS invariance of a spontaneously
  broken gauge theory (see \refapp{BRStra}). In particular, we used
  only the generic form of the BRS transformations of the fields, the
  form of the gauge-fixing term in an arbitrary `t~Hooft gauge,
\refeq{Gfix}, and the corresponding form of the tree-level
propagators.  Therefore, the result is valid for a general
spontaneously broken
gauge theory, in an  arbitrary `t~Hooft gauge.

\end{itemize}
It is important to observe that the identities \refeq{CWIs} do not
reflect the presence of the non-vanishing vev of the Higgs doublet.
In fact, they are identical to the identities obtained within a
symmetric gauge theory with massless gauge bosons. However,
spontaneous symmetry breaking plays a non-trivial role in ensuring the
validity of \refeq{CWIs}. 
It guarantees the cancellation
of mixing terms between gauge bosons and would-be Goldstone bosons.
In particular, we stress the following:
{\em extra contributions originating from $\L_{v}$ cannot be excluded
  a priori in \refeq{CWIs}}.  In fact, the corresponding
mass-suppressed couplings can in principle give extra leading
contributions if they are enhanced by propagators with small
invariants. We show that no such extra terms are left 
in the final result.
Such terms appear, however, in the derivation of the 
Ward identity for external
would-be 
Goldstone bosons ($\varphi_i=\Phi_i$) as ``extra contributions''
involving gauge bosons
($\varphi_{j}= V^a$), and in the derivation of the 
Ward identity for external gauge
bosons ($\varphi_i=V^a$) as ``extra contributions'' involving
would-be Goldstone bosons ($\varphi_{j}=\Phi_i$) [see \refeq{SCAMterm}].
Their cancellation is 
guaranteed by Ward identities \refeq{brokenWIs}
relating the electroweak vertex functions that involve explicit
factors with mass dimension.
In other words, {\em the validity of \refeq{CWIs} within a spontaneously
broken gauge theory is a non-trivial consequence of the symmetry of
the full theory.}


\section{Transverse gauge bosons, would-be Goldstone bosons and Higgs bosons}
\label{se:SCACWI}
The Ward identities for external scalar bosons
$\varphi_i=\Phi_i=H,\chi,\phi^\pm$ and transverse gauge bosons $\varphi_i=V^b=A,Z,W^\pm$
are of the same form.  Here we derive a generic Ward identity for
external bosonic fields $\varphi_i$ valid for $\varphi_i=\Phi_i$ as
well as $\varphi_i=V^b$.  In both cases mixing between would-be
Goldstone bosons and gauge bosons has to be taken into
account\footnote{For external Higgs bosons or photons all mixing
  terms vanish.}.  Therefore, we use the symbol $\tilde{\varphi}$ to
denote the mixing partner of $\varphi$, \ie we have
$(\varphi,\tilde{\varphi})=(\Phi,V)$ or
$(\varphi,\tilde{\varphi})=(V,\Phi)$.  The resulting Ward identities
read
\beqar\label{SCACWIdiag}
\lefteqn{\hspace{-3mm}
\lim_{q^\mu\rightarrow xp^\mu}
q^\mu \times
\left\{
\vcenter{\hbox{\begin{picture}(85,80)(5,20)
\DashLine(55,60)(10,60){4}
\Text(30,65)[b]{\scriptsize $\varphi_i(p-q)$}
\Photon(30,30)(70,60){1}{6}
\Text(55,35)[t]{\scriptsize $V^a_\mu(q)$}
\GCirc(70,60){15}{1}
\end{picture}}}
-\sum_{\varphi_{i'}}
\vcenter{\hbox{\begin{picture}(90,80)(5,20)
\DashLine(60,60)(35,60){4}
\DashLine(35,60)(10,60){4}
\Text(20,65)[b]{\scriptsize $ \varphi_i(p-q)$}
\Text(40,65)[bl]{\scriptsize $ \varphi_{i'}(p)$}
\Photon(10,30)(35,60){1}{5}
\Vertex(35,60){2}
\Text(35,40)[t]{\scriptsize $V^a_\mu(q)$}
\GCirc(75,60){15}{1}
\end{picture}}}
-\sum_{\tilde{\varphi}_j}
\vcenter{\hbox{\begin{picture}(90,80)(5,20)
\Photon(60,60)(35,60){2}{3.5}
\DashLine(35,60)(10,60){4}
\Text(20,65)[b]{\scriptsize $ \varphi_i(p-q)$}
\Text(40,65)[bl]{\scriptsize $\tilde{\varphi}_j(p)$}
\Photon(10,30)(35,60){1}{5}
\Vertex(35,60){2}
\Text(35,40)[t]{\scriptsize $V^a_\mu(q)$}
\GCirc(75,60){15}{1}
\end{picture}}}
\right\}=}\quad&&\hspace{13cm}
\nl&&= \sum_{\varphi_{i'}}\hspace{-2mm} 
\vcenter{\hbox{\begin{picture}(80,80)(10,20)
\DashLine(55,60)(25,60){4}
\Text(40,65)[b]{\scriptsize $\varphi_{i'}(p)$}
\GCirc(70,60){15}{1}
\end{picture}}}\hspace{2mm}
e I^{V^a}_{\varphi_{i'}\varphi_i}\quad
+ \O\left(M^2E^{d-2}\right),
\eeqar
where the diagrammatic representation corresponds to external scalars 
($\varphi=\Phi$),  $d$ is the
mass dimension of the matrix element and
 $M^2\sim\max(p^2,M_{\varphi_i}^2,M^2_{V^a})$.
The  proof of \refeq{SCACWIdiag}
is organized in six steps:

\newcounter{listnumber}
\begin{list}{\bf \arabic{listnumber}.\hspace{1mm}}{\usecounter{listnumber}
\setlength{\leftmargin}{0mm} \setlength{\labelsep}{0mm}
\setlength{\itemindent}{4mm}
}
\item {\bf Ward identity for connected Green functions}\nopagebreak

We start using the BRS invariance (\cf\refapp{BRStra}) of the electroweak Lagrangian. This implies the invariance of connected Green functions with respect 
to BRS transformations of their field arguments. For the 
Green function $\langle\bar{u}^{a}(x)\varphi^+_{i}(y)\Oper(z)\rangle$\footnote{Analogously to \refeq{defoper},
$\Oper(z)=\prod_{l\neq k}\varphi_{i_l}(z_l)$
represents a combination of arbitrary physical fields.}
we have the relation
\beqar\label{SCABRSinv}
\langle[\brs \bar{u}^{a}(x)]\varphi^+_{i}(y)\Oper(z)\rangle
-\langle\bar{u}^{a}(x)[\brs \varphi^+_{i}(y)]\Oper(z)\rangle
=\langle\bar{u}^{a}(x)\varphi^+_{i}(y)[\brs \Oper(z)]\rangle.
\eeqar
Since the BRS variation of the antighost field  \refeq{antighostbrstra} 
corresponds to the gauge-fixing term \refeq{Gfix} and 
using the  BRS variation of the physical fields \refeq{ccphysbrstra} 
we obtain the Ward identity
\beqar
\lefteqn{\frac{1}{\xi_a}\partial^\mu_{x}\langle\bar{V}^a_\mu(x)\varphi^+_{i}(y)\Oper(z)\rangle
-\ri e v\sum_{\Phi_j=H,\chi,\phi^\pm} I^{V^a}_{H\Phi_j}\langle\Phi_j(x) \varphi^+_{i}(y)\Oper(z)\rangle}\quad
\nl&&{}
+\sum_{V^b=A,Z,W^\pm}\biggl[\linbrs^{V^b}_{\varphi^+_i}
\langle\bar{u}^{a}(x)u^{b}(y)\Oper(z)\rangle
-\ri e\sum_{\varphi_{i'}}
\langle\bar{u}^{a}(x)u^{b}(y)\varphi^+_{i'}(y)\Oper(z)\rangle
I^{V^b}_{\varphi_{i'}\varphi_i}\biggr]
\nl&=&
-\langle\bar{u}^{a}(x)\varphi^+_{i}(y)[\brs \Oper(z)]\rangle
.
\eeqar
Fourier transformation of the variables $(x,y,z)$ to the incoming
momenta $(q,p-q,r)$ ($\partial_{x}^\mu \rightarrow \ri q^\mu$) gives
\beqar \label{PSWISCABOS}
\lefteqn{\frac{\ri}{\xi_a} q^\mu\langle\bar{V}^a_\mu(q)\varphi^+_{i}(p-q)\Oper(r)\rangle
{}-\ri e v\sum_{\Phi_j} I^{V^a}_{H\Phi_j}\langle\Phi_j(q) \varphi^+_{i}(p-q)\Oper(r)\rangle}\quad
\nl&&
{}+\sum_{V^b}\linbrs^{V^b}_{\varphi^+_i}
\langle\bar{u}^{a}(q)u^{b}(p-q)\Oper(r)\rangle
\nl&&{}
-\ri e\sum_{V^b,\varphi_{i'}}
\int \ddl
\langle\bar{u}^{a}(q)u^{b}(l)\varphi^+_{i'}(p-q-l)\Oper(r)\rangle
I^{V^b}_{\varphi_{i'}\varphi_i}
\nl&=&
-\langle\bar{u}^{a}(q)\varphi^+_{i}(p-q)[\brs \Oper(r)]\rangle.
\eeqar 
{}From now on, the rhs is omitted, 
since the BRS variation
of on-shell physical fields does not contribute to physical 
matrix elements.
This can be verified by truncation of the physical external legs
$\Oper(r)$ and contraction with the 
corresponding wave functions.  

\item{\bf Restriction to lowest order}\nopagebreak

A further simplification concerns the last term on
the lhs of \refeq{PSWISCABOS}. This originates from the BRS
variation $\brs \varphi^+_{i}(y)$ of the external scalar or vector 
field and contains an external ``BRS vertex'' connecting the fields
$u^{b}(y)\varphi^+_{i'}(y)$, which we represent by a small box in
\refeq{O1O2picture}.  When we restrict the relation \refeq{PSWISCABOS}
to LO connected
Green functions, this term simplifies into those tree diagrams where
the external ghost line is not connected to the scalar leg of the BRS
vertex by internal vertices,
\beqar\label{O1O2picture}
\vcenter{\hbox{\begin{picture}(90,65)(0,28)
\DashLine(55,62)(10,62){4}
\DashArrowLine(55,58)(10,58){1}
\EBox(8,58)(12,62)
\Text(5,60)[r]{\scriptsize $\brs \varphi_i^+(p-q)$}
\DashArrowLine(30,30)(70,60){1}
\Vertex(30,30){2}
\Text(55,35)[t]{\scriptsize $\bar{u}^{a}(q)$}
\GCirc(70,60){15}{1}
\end{picture}}}
=
\vcenter{\hbox{\begin{picture}(90,65)(0,28)
\DashLine(55,62)(25,62){4}
\EBox(23,58)(27,62)
\Text(40,65)[b]{\scriptsize $\varphi^+_{i'}(p)$}
\DashArrowLine(0,30)(23,58){1}
\Vertex(0,30){2}
\Text(25,35)[t]{\scriptsize $\bar{u}^a(q)$}
\GCirc(70,60){15}{1}
\Text(70,60)[c]{\small$O$}
\end{picture}}}
+\sum_{\Oper_1\neq \Oper}
\vcenter{\hbox{\begin{picture}(85,65)(10,28)
\DashLine(70,75)(25,62){4}
\EBox(23,58)(27,62)
\Text(35,75)[b]{\scriptsize $\varphi^+_{i'}(p+r_2)$}
\Vertex(25,30){2}
\DashArrowLine(25,30)(70,45){1}
\DashArrowLine(70,45)(27,58){1}
\Text(45,30)[t]{\scriptsize $\bar{u}^{a}(q)$}
\GCirc(70,75){10}{1}
\Text(70,75)[c]{\small $\Oper_1$}
\GCirc(70,45){10}{1}
\Text(70,45)[c]{\small $\Oper_2$}
\end{picture}}}.
\eeqar
We will see in the following that the relevant contributions result
only from the first diagram on the rhs of \refeq{O1O2picture},
where the ghosts are joined by a propagator and all on-shell legs
$\Oper(r)$ are connected to the leg $\varphi^+_{i'}$, which receives
momentum $p=-r$.  In the remaining diagrams, the on-shell legs are
distributed into two 
subsets $\Oper(r)=\Oper_1(r_1)\Oper_2(r_2)$ with momenta $r_1+r_2=r$.
One subset $\Oper_1$ interacts with the leg $\varphi^+_{i'}$, which
receives momentum $p+r_2=-r_1$. The other subset $\Oper_2$ interacts
with the ghost line. Therefore, in LO the last term on the lhs of
\refeq{PSWISCABOS} yields
\beqar\label{lhstermLO} 
\lefteqn{-\ri e\sum_{V^b,\varphi_{i'}}
\int \ddl
\langle\bar{u}^{a}(q)u^{b}(l)\varphi^+_{i'}(p-q-l)\Oper(r)\rangle
I^{V^b}_{\varphi_{i'}\varphi_i}
}\quad\nl
&=&
-\ri e\sum_{\varphi_{i'}}
\langle\bar{u}^{a}(q)u^{a}(-q)\rangle
\langle\varphi^+_{i'}(p)\Oper(r)\rangle
I^{V^a}_{\varphi_{i'}\varphi_i}
\nl 
&&{}-\ri e\sum_{V^b,\varphi_{i'}}
\sum_{\Oper_1\neq\Oper}
\langle\bar{u}^{a}(q)u^{b}(-q-r_2)\Oper_2(r_2)\rangle
\langle\varphi^+_{i'}(p+r_2)\Oper_1(r_1)\rangle
I^{V^b}_{\varphi_{i'}\varphi_i}, \eeqar and if we split off the
momentum-conservation $\de$-functions, \refeq{PSWISCABOS} becomes
\beqar\label{SCABRSWI}
\lefteqn{\frac{\ri}{\xi_a} q^\mu G^{\bar{V}^a {\varphi^+_i} \underline{\Oper}}_{\mu}(q,p-q,r)
-\ri e v \sum_{\Phi_j}I^{V^a}_{H\Phi_j}
G^{{\Phi}_j {\varphi^+_i} \underline{\Oper}}(q,p-q,r)}\quad\nl
\nl&&{}
+\sum_{V^b} \linbrs^{V^b}_{\varphi^+_i}
G^{\bar{u}^a {u}^b \underline{\Oper}}(q,p-q,r)
-\ri e \sum_{\varphi_{i'}}
G^{\bar{u}^a u^a}(q)G^{{\varphi^+_{i'}} \underline{\Oper}}(p,r)
I^{V^a}_{\varphi_{i'}\varphi_i} 
\nl&=&
\ri e\sum_{V^b,\varphi_{i'}} \sum_{\Oper_1\neq \Oper}
G^{{\varphi^+_{i'}} \underline{\Oper}_1}(p+r_2,r_1)
I^{V^b}_{\varphi_{i'}\varphi_i}
G^{\bar{u}^a {u}^b\underline{\Oper}_2}(q,-q-r_2,r_2)
. 
\eeqar

\item{\bf Enhanced internal propagators}\nopagebreak

Recall that we are interested in the on-shell and ``massless'' limit
$p^2\ll s$ of the above equation. Therefore, we have to take special
care of all terms that are enhanced in this limit, like internal
propagators carrying momentum $p$.  Since internal lines with small
invariants do not occur on the rhs of \refeq{SCABRSWI}, we now
concentrate on the lhs.  Using \refeq{subtractGF}, the first term
can be written as
\beqar
G_\mu^{\bar{V}^a{\varphi^+_i} \underline{\Oper}}(q,p-q,r) 
&=&G_\mu^{[\bar{V}^a{\varphi^+_{i}}] \underline{\Oper}}(q,p-q,r) 
\nl&&{}
+\sum_{\varphi_{i'}}G_\mu^{\bar{V}^a {\varphi^+_i} \varphi_{i'}}(q,p-q,-p)G^{\underline{\varphi}_{i'} \underline{\Oper}}(p,r)
\nl&&{}
+ \sum_{\tilde{\varphi}_j}G^{\bar{V}^a {\varphi^+_i} \tilde{\varphi}_j}_{\mu}(q,p-q,-p)G^{ \underline{\tilde{\varphi}}_j \underline{\Oper}}(p,r)
,
\eeqar
where for scalar $\varphi^+_i$ the sums run over scalar $\varphi_{i'}$
and vector $\tilde{\varphi}_j$ and vice versa if $\varphi^+_i$ is a
vector.
In this way the enhanced internal propagators with momentum $p$ are
isolated in the terms $G^{\bar{V}^a \varphi^+_i
  \varphi_{i'}}_{\mu}(q,p-q,-p)$ and $G^{\bar{V}^a \varphi^+_i
  \tilde{\varphi}_j}_{\mu}(q,p-q,-p)$, whereas the subtracted Green
functions $G_\mu^{[\bar{V}^a{\varphi}_{i}^+] \underline{\Oper}}$ contain
no enhancement by definition. A similar decomposition is used for the
second and third term on the lhs of \refeq{SCABRSWI}, whereas the
enhanced propagator contained in the last term is isolated by writing
\beq
G^{{\varphi^+_{i'}} \underline{\Oper}}(p,r)=
G^{{\varphi^+_{i'}}{\varphi}_{i'}}(p)
G^{\underline{\varphi}_{i'} \underline{\Oper}}(p,r).
\eeq
In this way, the lhs of \refeq{SCABRSWI} can be written as
\beqar\label{SCABRSWIb}
\lefteqn{
\frac{\ri}{\xi_a} q^\mu G^{[\bar{V}^a {\varphi^+_i}] \underline{\Oper}}_{\mu}(q,p-q,r)
-\ri e v \sum_{\Phi_j}I^{V^a}_{H\Phi_j}
G^{[{\Phi}_j {\varphi^+_i}] \underline{\Oper}}(q,p-q,r) +{}
}\quad
\nl&&
{}+\sum_{V^b} \linbrs^{V^b}_{\varphi^+_i}
G^{[\bar{u}^a {u}^b] \underline{\Oper}}(q,p-q,r)
+\sum_{\varphi_{i'}}S^{\bar{V}^a}_{\varphi^+_{i}\varphi_{i'}}
G^{\underline{\varphi}_{i'} \underline{\Oper}}(p,r)
+\sum_{\tilde{\varphi}_j}M^{\bar{V}^a}_{\varphi^+_{i}\tilde{\varphi}_j}
G^{\underline{\tilde{\varphi}}_j \underline{\Oper}}(p,r),\nln
\eeqar
where all enhanced terms are in the self-energy-like ($\varphi\varphi$) 
contributions
\beqar\label{SCASterm}
S^{\bar{V}^a}_{\varphi^+_{i}\varphi_{i'}}&=&
\frac{\ri}{\xi_a} q^\mu G^{\bar{V}^a {\varphi^+_i} \varphi_{i'}}_{\mu}(q,p-q,-p)
-\ri e v \sum_{\Phi_k}I^{V^a}_{H\Phi_k}
G^{{\Phi}_k {\varphi}^+_i \varphi_{i'}}(q,p-q,-p)
\nl&&
{}+ \sum_{V^b} \linbrs^{V^b}_{\varphi^+_i}
G^{\bar{u}^a {u}^b \varphi_{i'}}(q,p-q,-p)
-\ri e G^{\bar{u}^a u^a}(q)
G^{{\varphi^+_{i'}}{\varphi}_{i'}}(p)
I^{V^a}_{\varphi_{i'}\varphi_i}
\eeqar
and in the mixing-energy-like ($\varphi\tilde\varphi$)
 contributions
\beqar\label{SCAMterm}
M^{\bar{V}^a}_{\varphi^+_{i}\tilde{\varphi}_j}
&=&\frac{\ri}{\xi_a} q^\mu G^{\bar{V}^a {\varphi^+_i}\tilde{\varphi}_j}_{\mu}(q,p-q,-p)
-\ri e v \sum_{\Phi_k}I^{V^a}_{H\Phi_k} G^{{\Phi}_k {\varphi^+_i} \tilde{\varphi}_j}(q,p-q,-p)
\nl&&
{}+\sum_{V^b} \linbrs^{V^b}_{\varphi^+_i}G^{\bar{u}^a {u}^b \tilde{\varphi}_j}(q,p-q,-p).\hspace{4cm}
\eeqar
Note that here the terms originating from $\L_v$, \ie terms
proportional to the vev, are enhanced by the internal
$\tilde\varphi_j$ propagators and represent leading contributions to
\refeq{SCABRSWIb}.  

\item{\bf Two further Ward identities}\nopagebreak

In order to simplify \refeq{SCASterm} and
\refeq{SCAMterm}, and to check whether contributions proportional to
the vev survive, we have to derive two further Ward identities.

\begin{itemize}
\renewcommand{\labelitemi}{{\bf --\hspace{1mm}}}
\item For the self-energy-like
 contributions \refeq{SCASterm}  we exploit the BRS invariance of the
 Green function $\langle\bar{u}^{a}(x)\varphi^+_{i}(y)\varphi_{i'}(z)\rangle$:
\beq
\langle[\brs \bar{u}^{a}(x)]\varphi^+_{i}(y)\varphi_{i'}(z)\rangle
-\langle\bar{u}^{a}(x)[\brs \varphi^+_{i}(y)]\varphi_{i'}(z)\rangle
=\langle\bar{u}^{a}(x)\varphi^+_{i}(y)[\brs \varphi_{i'}(z)]\rangle.
\eeq 
Using the BRS variations \refeq{antighostbrstra},
\refeq{ccphysbrstra}, and \refeq{physbrstra}, we have
\beqar
\lefteqn{\frac{1}{\xi_a}\partial^\mu_{x}\langle\bar{V}^a_\mu(x)\varphi^+_{i}(y)\varphi_{i'}(z)\rangle
-\ri e v \sum_{\Phi_j}I^{V^a}_{H\Phi_j}\langle\Phi_j(x) \varphi^+_{i}(y)\varphi_{i'}(z)\rangle}\quad
\nl&&{}
+\sum_{V^b}\linbrs^{V^b}_{\varphi^+_i}
\langle\bar{u}^{a}(x)u^{b}(y)\varphi_{i'}(z)\rangle
-\ri e\sum_{V^b,\varphi_k}
\langle\bar{u}^{a}(x)u^{b}(y)\varphi^+_{k}(y)\varphi_{i'}(z)\rangle I^{V^b}_{\varphi_{k}\varphi_i}
\nl
&=&
-\sum_{V^b}\linbrs^{V^b}_{\varphi_{i'}}
\langle\bar{u}^{a}(x)\varphi^+_{i}(y)u^{b}(z)\rangle
-\ri e\sum_{V^b,\varphi_k}
I^{V^b}_{\varphi_{i'}\varphi_{k}}\langle\bar{u}^{a}(x)\varphi^+_{i}(y)u^{b}(z)\varphi_{k}(z)\rangle.\nln
\eeqar
In LO, the terms involving four fields reduce to  products of pairs of
propagators. After Fourier transformation we obtain  
\beqar\label{vertexWI}
\lefteqn{\frac{\ri}{\xi_a} q^\mu\langle\bar{V}^a_\mu(q)\varphi^+_{i}(p-q)\varphi_{i'}(-p)\rangle
-\ri e v \sum_{\Phi_j}I^{V^a}_{H\Phi_j}\langle\Phi_j(q) \varphi^+_{i}(p-q)\varphi_{i'}(-p)\rangle}\quad
\nl&&{}
+\sum_{V^b}\linbrs^{V^b}_{\varphi^+_i}
\langle\bar{u}^{a}(q)u^{b}(p-q)\varphi_{i'}(-p)\rangle
-\ri e  \langle\bar{u}^{a}(q)u^{a}(-q)\rangle
\langle\varphi^+_{i'}(p)\varphi_{i'}(-p)\rangle I^{V^a}_{\varphi_{i'}\varphi_i}
\nl&=&
-\sum_{V^b}\linbrs^{V^b}_{\varphi_{i'}}
\langle\bar{u}^{a}(q)\varphi^+_{i}(p-q)u^{b}(-p)\rangle
\nl&&{}
-\ri e \langle\bar{u}^{a}(q)u^{a}(-q)\rangle \langle\varphi^+_{i}(p-q) \varphi_{i}(-p+q)\rangle I^{V^a}_{\varphi_{i'}\varphi_i}
,
\eeqar
and we easily see that 
\beq  \label{CWISCABOSamp}
S^{\bar{V}^a}_{\varphi^+_{i}\varphi_{i'}}
=
-\sum_{V^b} \linbrs^{V^b}_{\varphi_{i'}}G^{\bar{u}^a {\varphi}^+_{i} u^b}(q,p-q,-p)
-\ri e 
G^{\bar{u}^a u^a}(q)G^{\varphi^+_i\varphi_i}(p-q)
I^{V^a}_{\varphi_{i'}\varphi_i}.
\eeq

\item For the mixing-energy-like contributions \refeq{SCAMterm} we use
  the BRS invariance of the Green function
  $\langle\bar{u}^{a}(x)\varphi^+_{i}(y)\tilde{\varphi}_j (z)\rangle$.
  The resulting WI is obtained from \refeq{vertexWI} by substituting
  $\varphi_{i'}\rightarrow \tilde{\varphi}_j$ and by neglecting the
  mixing propagators $\langle\varphi^+_{i}(p)\tilde{\varphi}_j
  (-p)\rangle$ which vanish in LO and reads
\beqar\label{brokenWIs}
\lefteqn{\frac{\ri}{\xi_a} q^\mu\langle\bar{V}^a_\mu(q)\varphi^+_{i}(p-q)\tilde{\varphi}_{j}(-p)\rangle
-\ri e v \sum_{\Phi_k}I^{V^a}_{H\Phi_k}\langle\Phi_k(q) \varphi^+_{i}(p-q)\tilde{\varphi}_{j}(-p)\rangle}\quad
\\&&{}
+\sum_{V^b}\linbrs^{V^b}_{\varphi^+_i}
\langle\bar{u}^{a}(q)u^{b}(p-q)\tilde{\varphi}_{j}(-p)\rangle
= -\sum_{V^b}\linbrs^{V^b}_{\tilde{\varphi}_{j}}
\langle\bar{u}^{a}(q)\varphi^+_{i}(p-q)u^{b}(-p)\rangle.\nonumber
\eeqar
This relation involves the $VV\Phi$ couplings as well as  other terms originating from  $\L_{v}$, and leads to
\beq  \label{CWISCABOS2}
M^{\bar{V}^a}_{\varphi^+_i\tilde{\varphi}_j}
=-\sum_{V^b} \linbrs^{V^b}_{\tilde{\varphi}_j} G^{\bar{u}^a {\varphi}^+_{i} u^b}(q,p-q,-p).
\eeq
\end{itemize}
Both \refeq{CWISCABOSamp} and \refeq{CWISCABOS2} contain the ghost
vertex function $G^{\bar{u}^a {\varphi}^+_{i} u^b}$, but when we
combine them in \refeq{SCABRSWIb} these ghost contributions cancel
owing to the LO identity that 
relates external would-be Goldstone bosons and gauge bosons,
\beqar 
\lefteqn{\left[\sum_{V^d} 
G_\mu^{\underline{V}^d \underline{\Oper}}(p,r)\linbrs^{V^b}_{V^d_\mu}
+ \sum_{\Phi_j}G^{\underline{\Phi}_{j} \underline{\Oper}}(p,r)\linbrs^{V^b}_{\Phi_j} \right]G^{\bar{u}^a {\varphi}^+_{i} u^b}(q,p-q,-p)}\quad
\nl&=& 
\left[-\ri p^\mu G_\mu^{\underline{V}^{b} \underline{\Oper}}(p,r)
+\ri e v \sum_{\Phi_{j}} I^{V^b}_{\Phi_{j}H} G^{\underline{\Phi}_{j}
  \underline{\Oper}}(p,r)\right]G^{\bar{u}^a {\varphi}^+_{i} u^b}(q,p-q,-p)
\nl&=&0.  
\eeqar
Thus, all terms originating from internal propagators with momentum $p$ cancel and the complete identity
\refeq{SCABRSWI} becomes
\beqar \label{CWISCABOSa0}
\lefteqn{
\frac{\ri}{\xi_a} q^\mu G^{[\bar{V}^a {\varphi^+_i}]
  \underline{\Oper}}_{\mu}(q,p-q,r) -\ri e v
\sum_{\Phi_j}I^{V^a}_{H\Phi_j} G^{[{\Phi}_j {\varphi^+_i}]
  \underline{\Oper}}(q,p-q,r)}\quad
 \nl&&{} 
+\sum_{V^b}
\linbrs^{V^b}_{\varphi^+_i} G^{[\bar{u}^a {u}^b]
  \underline{\Oper}}(q,p-q,r) - \ri e G^{\bar{u}^a
  u^a}(q)G^{\varphi^+_i\varphi_i}(p-q)
\sum_{\varphi_{i'}}G^{\underline{\varphi}_{i'} \underline{\Oper}}(p,r)
I^{V^a}_{\varphi_{i'}\varphi_i} \nl
&=& \ri e\sum_{V^b,\varphi_{i'}}
\sum_{\Oper_1\neq \Oper} G^{{\varphi}^+_{i'}
  \underline{\Oper}_1}(p+r_2,r_1) I^{V^b}_{\varphi_{i'}\varphi_i}
G^{\bar{u}^a {u}^b\underline{\Oper}_2}(q,-q-r_2,r_2) .  
\eeqar 

\item{\bf External-leg truncation}\nopagebreak

Now we can truncate the two remaining external legs.  To this end we
observe that [see \refeq{gaugeprop2}, \refeq{scaprop}]
the longitudinal part of the LO
gauge-boson propagator $G_\rL^{V^a\bar{V}^a}(q)$, the LO ghost
propagator, and the LO propagator of the associated would-be Goldstone
boson $\Phi_j$ are related by
\beq\label{lobgproprelat}
\frac{1}{\xi_a}
G_\rL^{V^a\bar{V}^a}(q)=G^{\bar{u}^au^a}(q)=-G^{\Phi^+_j\Phi_j}(q).
\eeq   
Using this relation, the leg with momentum $q$ is easily truncated by
multiplying the above identity with the longitudinal part of the
inverse gauge-boson propagator $-\ri\xi_a
\Gamma^{V^a\bar{V}^a}_\rL(q)$.  The leg with momentum $p-q$ is
truncated by multiplying \refeq{CWISCABOSa0} by the inverse
(scalar-boson or gauge-boson) propagator $-\ri
\Gamma^{\varphi_i\varphi_i^+}(p-q)$, and by using
\beq
\linbrs^{V^b}_{\varphi^+_i}G^{u^b\bar{u}^b}(p-q)=
c_{\varphi_i}G^{\varphi^+_i\varphi_i}(p-q)\linbrs^{V^b}_{\varphi^+_i}.
\eeq
with $c_{\Phi_i}=1$ and  $c_{V^b}=-1/\xi_b$.
The truncated identity reads
\beqar \label{CWISCABOSa}
\lefteqn{
\ri q^\mu G^{[\underline{V}^a \underline{\varphi}_i] \underline{\Oper}}_{\mu}(q,p-q,r)
+\ri e v \sum_{\Phi_j}I^{V^a}_{H\Phi_j}
G^{[\underline{\Phi}_j^+ \underline{\varphi}_i] \underline{\Oper}}(q,p-q,r)
}\quad
\nl&&{}
+\sum_{V^b} c_{\varphi_i} \linbrs^{V^b}_{\varphi^+_i}
G^{[\underline{u}^a \underline{\bar{u}}^b] \underline{\Oper}}(q,p-q,r)
-
\ri e \sum_{\varphi_{i'}}
G^{\underline{\varphi}_{i'} \underline{\Oper}}(p,r)
I^{V^a}_{\varphi_{i'}\varphi_i} 
\nl&=&
\ri e\sum_{V^b,\varphi_{i'}} \sum_{\Oper_1\neq \Oper}
\left[-\ri \Gamma^{\varphi_i\varphi_i^+}(p-q)
G^{{\varphi}^+_{i'} \underline{\Oper}_1}(-r_1,r_1) \right]
I^{V^b}_{\varphi_{i'}\varphi_i}
G^{\underline{u}^a{u}^b\underline{\Oper}_2}(q,-q-r_2,r_2).
\nln
\eeqar

\item{\bf Collinear limit}\nopagebreak

Finally,  we take the collinear limit
$q^\mu \rightarrow xp^\mu$ and assume 
\beq
M^2\sim\max(p^2,M_{\varphi_i}^2,M_{V^a}^2)\ll s.
\eeq
Furthermore,  we contract \refeq{CWISCABOSa} with the wave function $v_{\varphi_i}(p)$ of an on-shell external state with mass $\sqrt{p^2}$.
For scalar bosons the wave function is trivial ($v_\Phi(p)=1$),
whereas for external gauge bosons we consider transverse polarizations
$v_{V_\nu}(p)=\varepsilon_\rT^\nu(p)$.  
Then various terms in \refeq{CWISCABOSa} are
mass-suppressed.  The rhs is mass-suppressed owing to
\beq\label{rhsmsupp}
\lim_{q^\mu \rightarrow xp^\mu}
v_{\varphi_i}(p)\Gamma^{\varphi_i\varphi^+_i}(p-q)
G^{\varphi^+_{i'}\varphi_{i'}}(-r_1)
\sim\lim_{q^\mu \rightarrow xp^\mu}
\frac{(p-q)^2-M_{\varphi_i}^2}{r_1^2}
=\O\left(\frac{M^2}{s}\right),
\eeq
since  $(p-q)^2-M_{\varphi_i}^2\sim M^2$ in the
collinear limit, whereas  $r_1$ is a non-trivial combination of the external momenta, 
and like for all invariants \refeq{Sudaklim} we assume that $r_1^2\sim s$.
The second term on the the lhs 
of \refeq{CWISCABOSa} is proportional to the vev and therefore
mass-suppressed, and for the third term we have
\beq
\lim_{q^\mu \rightarrow xp^\mu} v_{\varphi_i}(p) \linbrs^{V^b}_{\varphi^+_i}=\O(M).
\eeq
For gauge bosons this is due to the transversality of the polarization vector
\beq
\lim_{q^\mu \rightarrow xp^\mu}\, (p-q)_\nu\varepsilon_\rT^\nu(p)=
0,
\eeq
whereas for scalar bosons $\linbrs^{V^b}_{\Phi^+_i}$ is explicitly
proportional to the vev.  The remaining leading terms give the result
\beqar \label{SCACWIres}
\lefteqn{\lim_{q^\mu \rightarrow xp^\mu}
q^\mu v_{\varphi_i}(p)
 G^{[\underline{V}^a \underline{\varphi}_i]
   \underline{\Oper}}_{\mu}(q,p-q,r)}\quad
\nl&=&
e \sum_{\varphi_{i'}}
v_{\varphi_i}(p)G^{\underline{\varphi}_{i'} \underline{\Oper}}(p,r)
I^{V^a}_{\varphi_{i'}\varphi_i} 
+\O\left(\frac{M^2}{s} \M^{\varphi_{i}\Oper}\right),
\nl&=&
e \sum_{\varphi_{i'}}
\M^{\varphi_{i'}\Oper}(p,r)
I^{V^a}_{\varphi_{i'}\varphi_i} 
+\O\left(\frac{M^2}{s} \M^{\varphi_{i}\Oper}\right),
\eeqar
which is the identity represented in \refeq{SCACWIdiag} in diagrammatic form.
Recall that, as noted in \refeq{shortBornampli3b},
the wave function 
for transverse gauge bosons is independent of their mass, \ie
$v_{\varphi_i}(p)=v_{\varphi_{i'}}(p)$.

\end{list}
\section{Chiral fermions and antifermions}\label{se:FERCWI}
The collinear Ward identities for chiral  fermions $\varphi_i=\Psi^\kappa_{j,\si}$
and antifermions $\varphi_i=\bar{\Psi}^\kappa_{j,\si}$ are derived in the same way 
as the identities \refeq{SCACWIdiag} for gauge bosons and scalar bosons.
Actually the derivation is  much simpler since no mixing contributions
have to be considered. 
Here we restrict ourselves to the case of antifermions, and we prove the identity 
\beqar\label{FERCWIdiag}
\lefteqn{
\lim_{q^\mu\rightarrow xp^\mu} q^\mu \times
\left\{
\vcenter{\hbox{\begin{picture}(85,80)(5,20)
\ArrowLine(55,60)(10,60)
\Text(35,65)[b]{\scriptsize $\bar{\Psi}^\kappa_{j,\sigma}(p-q)$}
\Photon(30,30)(70,60){1}{6}
\Text(55,35)[t]{\scriptsize $V^a_\mu(q)$}
\GCirc(70,60){15}{1}
\end{picture}}}
-
\vcenter{\hbox{\begin{picture}(85,80)(5,20)
\ArrowLine(55,60)(35,60)
\ArrowLine(35,60)(10,60)
\Text(30,65)[b]{\scriptsize $ \bar{\Psi}^\kappa_{j,\si}(p-q)$}
\Photon(10,30)(35,60){1}{5}
\GCirc(35,60){1}{0}
\Text(35,40)[t]{\scriptsize $V^a_\mu(q)$}
\GCirc(70,60){15}{1}
\end{picture}}}
\right\}=}\quad&&\nl
&&=  \sum_{j',\si'}
\vcenter{\hbox{\begin{picture}(80,80)(15,20)
\ArrowLine(55,60)(25,60)
\Text(40,65)[b]{\scriptsize $\bar{\Psi}^\kappa_{j',\si'}(p)$}
\GCirc(70,60){15}{1}
\end{picture}}}
eI^{V^a}_{\bar{f}^\kappa_{j',\si'} \bar{f}^\kappa_{j,\si}}
\quad
+\O\left(M E^{d-1}\right),
\eeqar
where  $d$ is the mass dimension of the matrix element and
 $M^2\sim\max(p^2,m_{f_{j,\si}}^2,M^2_{V^a})$.
The generalized gauge couplings to  fermions are defined in \refeq{fermgenerators} 
and they  include the quark mixing matrix. Owing to \refeq{transprel}, for antifermions we have 
\beq\label{fermgenerators2}
I^{V^a}_{\bar{f}^\kappa_{j',\si'} \bar{f}^\kappa_{j,\si}}=
-I^{V^a}_{f^\kappa_{j,\si} f^\kappa_{j',\si'}}=
- I^{V^a}_{f^\kappa_{\si} f^\kappa_{\si'}}
U^{f^\kappa,V^a}_{jj'}.
\eeq
In the following we will often omit the chirality $\kappa$ and  use the shorthand 
\beq
I^{V^a}_{\si\si'} U^{V^a}_{jj'}=I^{V^a}_{f^\kappa_{j,\si} f^\kappa_{j',\si'}}. 
\eeq
For the   proof of \refeq{FERCWIdiag} we follow the same steps as in
\refse{se:SCACWI}.

\newcounter{listnumber2}
\begin{list}{\bf \arabic{listnumber2}.\hspace{1mm}}{\usecounter{listnumber2}
\setlength{\leftmargin}{0mm} \setlength{\labelsep}{0mm}
\setlength{\itemindent}{4mm}
}
\item {\bf Ward identity for connected Green functions}\nopagebreak

We start from the BRS invariance of the connected GF 
$\langle\bar{u}^{a}(x)\Psi^\kappa_{j,\si}(y)\Oper(z)\rangle$\footnote{Note that connected Green functions involving the fermionic fields $\Psi$ are associated 
to incoming antifermions.} 
\beq
\langle [\brs \bar{u}^{a}(x)] \Psi^\kappa_{j,\sigma}(y)\Oper(z) \rangle
-\langle\bar{u}^{a}(x)[\brs \Psi^\kappa_{j,\sigma}(y)]\Oper(z)\rangle 
= -\langle\bar{u}^{a}(x)\Psi^\kappa_{j,\sigma}(y)[\brs \Oper(z)]\rangle,
\eeq
and with the  BRS variations \refeq{antighostbrstra} and \refeq{physbrstra} 
we have the Ward identity
\beqar
&&\frac{1}{\xi_a}\partial_{x}^\mu\langle\bar{V}^a_\mu(x)\Psi^\kappa_{j,\si}(y)\Oper(z)\rangle
-\ri e v \sum_{\Phi_i=H,\chi,\phi^\pm}I^{V^a}_{H\Phi_i}\langle\Phi_i(x) \Psi^\kappa_{j,\sigma}(y)\Oper(z)\rangle
\nl&&{}+
\ri e\sum_{V^b=A,Z,W^\pm}\sum_{j',\si'}
I^{V^b}_{f^\kappa_{j,\si} f^\kappa_{j',\si'}}
\langle\bar{u}^{a}(x)u^{b}(y)\Psi^\kappa_{j',\si'}(y)\Oper(z)\rangle
=\langle\bar{u}^{a}(x)\Psi^\kappa_{j,\si}(y)[\brs \Oper(z)]\rangle .
\nln 
\eeqar
Fourier transformation of the variables $(x,y,z)$ to the incoming
momenta $(q,p-q,r)$ ($\partial_{x}^\mu \rightarrow \ri q^\mu$) gives
\beqar\label{PSWIFER}
\lefteqn{
\frac{\ri}{\xi_a} q^\mu\langle\bar{V}^a_\mu(q)\Psi^\kappa_{j,\si}(p-q)\Oper(r)\rangle
-\ri e v \sum_{\Phi_i}I^{V^a}_{H\Phi_i}\langle\Phi_i(q) \Psi^\kappa_{j,\sigma}(p-q)\Oper(r)\rangle
}\quad&&\nl&&{}
+\ri e
\sum_{V^b,j',\si'}
I^{V^b}_{\si\si'} U^{V^b}_{jj'}
\int\frac{\rd^Dl}{(2\pi)^D} \langle\bar{u}^{a}(q)u^{b}(l)\Psi^\kappa_{j',\si'}(p-q-l)\Oper(r)\rangle
\nl&=&
 \langle\bar{u}^{a}(q)\Psi^\kappa_{j,\si}(p-q)[\brs \Oper(r)]\rangle.
\eeqar
Again we assume that all physical external legs denoted by $\Oper$
have been amputated and contracted with their wave functions, so that 
the rhs  of \refeq{PSWIFER}, which involves BRS variations of these 
on-shell physical fields, vanishes.

\item{\bf Restriction to lowest order}\nopagebreak

When  we restrict the relation  \refeq{PSWIFER} to lowest-order, 
the  last term on the lhs simplifies as  in \refeq{O1O2picture}
into
\beqar\label{O1O2ferpicture}
\vcenter{\hbox{\begin{picture}(90,65)(0,28)
\ArrowLine(55,62)(10,62)
\DashArrowLine(55,58)(10,58){1}
\EBox(8,58)(12,62)
\Text(35,65)[b]{\scriptsize $\brs \Psi^\kappa_{j,\si}(p-q)$}
\DashArrowLine(30,30)(70,60){1}
\Vertex(30,30){2}
\Text(55,35)[t]{\scriptsize $\bar{u}^{a}(q)$}
\GCirc(70,60){15}{1}
\end{picture}}}
=
\vcenter{\hbox{\begin{picture}(90,65)(0,28)
\ArrowLine(55,62)(25,62)
\EBox(23,58)(27,62)
\Text(40,65)[b]{\scriptsize $\Psi^\kappa_{j',\si'}(p)$}
\DashArrowLine(0,30)(23,58){1}
\Vertex(0,30){2}
\Text(25,35)[t]{\scriptsize $\bar{u}^a(q)$}
\GCirc(70,60){15}{1}
\Text(70,60)[c]{\small$O$}
\end{picture}}}
+\sum_{\Oper_1\neq \Oper}
\vcenter{\hbox{\begin{picture}(85,65)(10,28)
\ArrowLine(70,75)(25,62)
\EBox(23,58)(27,62)
\Text(35,75)[b]{\scriptsize $\Psi^\kappa_{j',\si'} (p+r_2)$}
\Vertex(25,30){2}
\DashArrowLine(25,30)(70,45){1}
\DashArrowLine(70,45)(27,58){1}
\Text(45,30)[t]{\scriptsize $\bar{u}^{a}(q)$}
\GCirc(70,75){10}{1}
\Text(70,75)[c]{\small $\Oper_1$}
\GCirc(70,45){10}{1}
\Text(70,45)[c]{\small $\Oper_2$}
\end{picture}}},
\eeqar
\ie
\beqar
\lefteqn{\ri e
\sum_{V^b,j',\si'}
I^{V^b}_{\si\si'} U^{V^b}_{jj'}
\int\frac{\rd^Dl}{(2\pi)^D} \langle\bar{u}^{a}(q)u^{b}(l)\Psi^\kappa_{j',\si'}(p-q-l)\Oper(r)\rangle=}\quad&&
\nl&=&\ri e \sum_{j',\si'}I^{V^a}_{\si\si'} U^{V^a}_{jj'}
\langle\bar{u}^{a}(q)u^{a}(-q)\rangle
\langle\Psi^\kappa_{j',\si'}(p)\Oper(r)\rangle\nl
&&{}+\ri e\sum_{V^b,j',\si'}
I^{V^b}_{\si\si'} U^{V^b}_{jj'} 
\sum_{\Oper_1\neq\Oper} 
\langle\bar{u}^{a}(q)u^{b}(-q-r_2)\Oper_2(r_2)\rangle  
\langle\Psi^\kappa_{j',\si'}(p+r_2)\Oper_1(r_1)\rangle,\nl
\eeqar
where the  operators $\Oper(r)$, $\Oper_1(r_1)$ and  $\Oper_2(r_2)$ are defined as in
\refeq{lhstermLO}.
Splitting off the
momentum-conservation $\de$-functions, \refeq{PSWIFER} becomes
\beqar\label{FERBRSWI}
\lefteqn{\hspace{-5mm}
\frac{\ri}{\xi_a} q^\mu 
G^{\bar{V}^a \Psi^\kappa_{j,\si} \underline{\Oper}}_{\mu}(q,p-q,r)
-\ri e v\sum_{\Phi_i} I^{V^a}_{H\Phi_i}
G^{\Phi_i \Psi^\kappa_{j,\si} \underline{\Oper}}(q,p-q,r)
}\quad&&\nl&&{}
+\ri e \sum_{j',\si'}I^{V^a}_{\si\si'} U^{V^a}_{jj'} 
G^{\bar{u}^a u^a}(q)
G^{\Psi^\kappa_{j',\si'}\underline{\Oper}}(p,r)
\nl &=&
-\ri e\sum_{V^b,j',\si'} I^{V^b}_{\si\si'} U^{V^b}_{jj'} \sum_{\Oper_1\neq \Oper}
G^{\Psi^\kappa_{j',\si'} \underline{\Oper}_1}(p+r_2,r_1)
G^{\bar{u}^a u^b\underline{\Oper}_2}(q,-q-r_2,r_2)
. 
\eeqar

\item{\bf Enhanced internal propagators}\nopagebreak

In order to identify all terms containing 
enhanced internal propagators with momentum $p$, we 
 rewrite the first term  on the lhs of \refeq{FERBRSWI}
using  \refeq{subtractGF} as
\beqar
\lefteqn{G_\mu^{\bar{V}^a  \Psi^\kappa_{j,\si} \underline{\Oper}}(q,p-q,r)=}\quad&&\nl 
&=&G_\mu^{[\bar{V}^a \Psi^\kappa_{j,\si}] \underline{\Oper}}(q,p-q,r) 
+\sum_{j',\si'}G_\mu^{\bar{V}^a  \Psi^\kappa_{j,\si} \bar{\Psi}^\kappa_{j',\si'}}(q,p-q,-p)
G^{ \underline{\bar{\Psi}}^\kappa_{j',\si'}  \underline{\Oper}}(p,r),
\eeqar
and the second term in the same form. For the third term we have
\beq
G^{  \Psi^\kappa_{j',\si'}  \underline{\Oper}}(p,r)=
G^{  \Psi^\kappa_{j',\si'} \bar{\Psi}^\kappa_{j',\si'} }(p)
G^{\underline{\bar{\Psi}}^\kappa_{j',\si'}\underline{\Oper}}(p,r).
\eeq
In this way, the lhs of \refeq{FERBRSWI} can be written as
\beqar\label{FERBRSWIb}
\lefteqn{\hspace{-5mm}
\frac{\ri}{\xi_a} q^\mu 
G^{[\bar{V}^a \Psi^\kappa_{j,\si}] \underline{\Oper}}_{\mu}(q,p-q,r)
-\ri e v\sum_{\Phi_i} I^{V^a}_{H\Phi_i}
G^{[\Phi_i \Psi^\kappa_{j,\si}] \underline{\Oper}}(q,p-q,r)
}\quad&&\nl&&{}
+ \sum_{j',\si'}
S^{\bar{V}^a}_{ \Psi^\kappa_{j,\si} \bar{\Psi}^\kappa_{j',\si'}}
G^{\underline{\bar{\Psi}}^\kappa_{j',\si'}\underline{\Oper}}(p,r),
\hspace{4cm}
\eeqar
where all enhanced propagators  are isolated in 
\beqar\label{FERSterm}
S^{\bar{V}^a}_{ \Psi^\kappa_{j,\si} \bar{\Psi}^\kappa_{j',\si'}}&=&
\frac{\ri}{\xi_a} q^\mu 
G_\mu^{\bar{V}^a \Psi^\kappa_{j,\si} \bar{\Psi}^\kappa_{j',\si'} }(q,p-q,-p)
-\ri e v \sum_{\Phi_i}I^{V^a}_{H\Phi_i}
G^{\Phi_i  \Psi^\kappa_{j,\si} \bar{\Psi}^\kappa_{j',\si'}}_{\mu}(q,p-q,-p)
\nl&&{}
+\ri e I^{V^a}_{\si\si'} U^{V^a}_{jj'} 
G^{\bar{u}^a u^a}(q)
G^{  \Psi^\kappa_{j',\si'} \bar{\Psi}^\kappa_{j',\si'} }(p).
\eeqar

\item{\bf A further Ward identity}\nopagebreak

In order to simplify \refeq{FERSterm} 
we derive a Ward identity for the $V^a\bar{\Psi}{\Psi}$ vertex. 
This is obtained from  the BRS invariance of the Green function 
$\langle\bar{u}^{a}(x)\Psi^\kappa_{j,\si}(y)\bar{\Psi}^\kappa_{j',\si'}(z)\rangle$, 
\beq
\langle[\brs \bar{u}^{a}(x)]\Psi^\kappa_{j,\si}(y)\bar{\Psi}^\kappa_{j',\si'}(z)\rangle
-\langle\bar{u}^{a}(x)[\brs \Psi^\kappa_{j,\si}(y)]\bar{\Psi}^\kappa_{j',\si'}(z)\rangle
=-\langle\bar{u}^{a}(x)\Psi^\kappa_{j,\si}(y)[\brs \bar{\Psi}^\kappa_{j',\si'}(z)]\rangle.
\eeq 
With the BRS variations \refeq{antighostbrstra},
\refeq{physbrstra}, and \refeq{ccphysbrstra}, we have
\beqar
\lefteqn{\hspace{-1cm}\frac{1}{\xi_a}\partial_{x}^\mu\langle\bar{V}^a_\mu(x)\Psi^\kappa_{j,\si}(y)\bar{\Psi}^\kappa_{j',\si'}(z)\rangle
-\ri e v \sum_{\Phi_i}I^{V^a}_{H\Phi_i}\langle\Phi_i(x) \Psi^\kappa_{j,\si}(y)\bar{\Psi}^\kappa_{j',\si'}(z)\rangle}\quad&&\nl
&&{}+\ri e\sum_{V^b,k,\rho}I^{V^b}_{\si\rho}U^{V^b}_{jk}
\langle\bar{u}^{a}(x)u^{b}(y)\Psi^\kappa_{k,\rho}(y)\bar{\Psi}^\kappa_{j',\si'}(z)\rangle\nl
&=&-\ri e\sum_{V^b,k,\rho}\langle\bar{u}^{a}(x)\Psi^\kappa_{j,\si}(y)u^{b}(z)\bar{\Psi}^\kappa_{k,\rho}(z)\rangle
I^{V^b}_{\rho\si'}U^{V^b}_{kj'},
\eeqar
and in LO the terms involving four fields reduce to  products of pairs of
propagators. After Fourier transformation 
\beqar
\lefteqn{\hspace{-1cm}\frac{\ri}{\xi_a} q^\mu\langle\bar{V}^a_\mu(q)\Psi^\kappa_{j,\si}(p-q)\bar{\Psi}^\kappa_{j',\si'}(k)\rangle
-\ri e v \sum_{\Phi_i}I^{V^a}_{H\Phi_i}\langle\Phi_i(q) \Psi^\kappa_{j,\si}(p-q)\bar{\Psi}^\kappa_{j',\si'}(k)\rangle}\quad&&\nl
&&{}
+\ri e I^{V^a}_{\si\si'} U^{V^a}_{jj'}\langle\bar{u}^{a}(q)u^{a}(-q)\rangle
\langle\Psi^\kappa_{j',\si'}(p)\bar{\Psi}^\kappa_{j',\si'}(k)\rangle
\nl
&=&+\ri e I^{V^a}_{\si\si'} U^{V^a}_{jj'}\langle\bar{u}^{a}(q)u^{a}(-q)\rangle
\langle\Psi^\kappa_{j,\si}(p-q) \bar{\Psi}^\kappa_{j,\si}(k+q)\rangle,
\eeqar
where the sign on the rhs has changed owing to exchange of the anti-commuting 
ghost and fermionic fields.
We see that 
\beq  \label{CWIFERBOSamp}
S^{\bar{V}^a}_{ \Psi^\kappa_{j,\si} \bar{\Psi}^\kappa_{j',\si'}}=
\ri e I^{V^a}_{\si\si'} U^{V^a}_{jj'} 
G^{\bar{u}^a u^a}(q)
G^{  \Psi^\kappa_{j,\si} \bar{\Psi}^\kappa_{j,\si}}(p-q),
\eeq
and the complete identity
\refeq{FERBRSWI} becomes
\beqar \label{CWIFERBOSa0}
\lefteqn{\hspace{-12mm}
\frac{\ri}{\xi_a} q^\mu 
G^{[\bar{V}^a \Psi^\kappa_{j,\si}] \underline{\Oper}}_{\mu}(q,p-q,r)
-\ri e v\sum_{\Phi_i} I^{V^a}_{H\Phi_i}
G^{[\Phi_i \Psi^\kappa_{j,\si}] \underline{\Oper}}(q,p-q,r)
}\quad&&\nl&&{}
+\ri e \sum_{j',\si'}I^{V^a}_{\si\si'} U^{V^a}_{jj'} 
G^{\bar{u}^a u^a}(q)
G^{  \Psi^\kappa_{j,\si} \bar{\Psi}^\kappa_{j,\si}}(p-q)
G^{\underline{\bar{\Psi}}^\kappa_{j',\si'}\underline{\Oper}}(p,r)
\nl &=&
-\ri e\sum_{V^b,j',\si'} I^{V^b}_{\si\si'} U^{V^b}_{jj'} \sum_{\Oper_1\neq \Oper}
G^{\Psi^\kappa_{j',\si'} \underline{\Oper}_1}(p+r_2,r_1)
G^{\bar{u}^a u^b\underline{\Oper}_2}(q,-q-r_2,r_2)
. 
\eeqar 

\item{\bf External-leg truncation}\nopagebreak

The two remaining external legs are truncated by 
multiplying \refeq{CWIFERBOSa0} with the 
longitudinal part of the inverse gauge-boson propagator $-\ri\xi_a
\Gamma^{V^a\bar{V}^a}_\rL(q)$, 
using \refeq{lobgproprelat}, and with the inverse fermionic propagator
$\ri \Gamma^{\bar{\Psi}_{j,\si} \Psi_{j,\si}}(p-q)$. 
The resulting truncated identity reads
\beqar \label{CWIFERa}
\lefteqn{\hspace{1cm}\ri q^\mu
G^{[\underline{V}^a \underline{\bar{\Psi}}_{j,\si}^\kappa] \underline{\Oper}}_{\mu}(q,p-q,r) 
+\ri e v \sum_{\Phi_i}I^{V^a}_{H\Phi_i}
G^{[\underline{\Phi}_i^+ \underline{\bar{\Psi}}^\kappa_{j,\si}]\underline{\Oper}}(q,p-q,r)
}\quad&&\nl
&&{}+\ri e \sum_{j',\si'}I^{V^a}_{\si\si'} U^{V^a}_{jj'}
G^{\underline{\bar{\Psi}}^\kappa_{j',\si'} \underline{\Oper}}(p,r)
\nl&=&
e\sum_{V^b,j',\si'} I^{V^b}_{\si\si'} U^{V^b}_{jj'}
 \sum_{\Oper_1\neq \Oper}
\Gamma^{\bar{\Psi}_{j,\si}\Psi_{j,\si}}(p-q) 
 G^{\Psi^\kappa_{j',\si'}  \underline{\Oper}_1}(-r_1,r_1)
G^{\underline{u}^a u^b\underline{\Oper}_2}(q,-q-r_2,r_2). 
\nln
\eeqar

\item{\bf Collinear limit}\nopagebreak

Finally,  we take the collinear limit
$q^\mu \rightarrow xp^\mu$, we assume 
\beq
M^2\sim\max{(p^2, m_{f_{j,\si}}^2, M_{V^a}^2)}\ll s,
\eeq
and  we contract \refeq{CWIFERa} with the spinor $\bar{v}(p)$  of an on-shell antifermion with mass $\sqrt{p^2}$.
Then the rhs  of \refeq{CWIFERa} is mass-suppressed owing to
\beq
\lim_{q^\mu \rightarrow xp^\mu} \bar{v}(p)
\Gamma^{\bar{\Psi}_{j,\si} \Psi_{j,\si}} (p-q) G^{\Psi_{j',\si'} \bar{\Psi}_{j',\si'} }(-r_1)
\propto \frac{M\bar{v}(p)\rs_1}{r_1^2} =\O\left(\frac{M}{\sqrt{s}}\right),
\eeq
since  $\bar{v}(p)[\ps-\qs -m_{f_{j,\si}}] \sim M$ in the
collinear limit, whereas  $r_1$ is a non-trivial combination of the external momenta, 
and like for all invariants \refeq{Sudaklim} we assume that $r_1^2\sim s$.
Also the second term on the the lhs is mass-suppressed since it  is 
proportional to the vev. The remaining leading terms give 
\beqar\label{ANTIFERCWIres}
\lefteqn{\lim_{q^\mu \rightarrow xp^\mu}
q^\mu \bar{v}(p)
G^{[\underline{V}^a\underline{\bar{\Psi}}_{j,\si}^\kappa] 
\underline{\Oper}}_{\mu}(q,p-q,r)=}\quad&&\nl
&=& -e \sum_{j',\si'}I^{V^a}_{\si\si'}U^{V^a}_{jj'}
\bar{v}(p)G^{\underline{\bar{\Psi}}^\kappa_{j',\si'}  \underline{\Oper}}(p,r)
+\O\left(\frac{M}{\sqrt{s}} \M^{\bar{f}_{j,\si}^\kappa\Oper}\right),\nl
&=& -e \sum_{j',\si'}I^{V^a}_{f^\kappa_\si f^\kappa_{\si'}}U^{f^\kappa,V^a}_{jj'}
\M^{\bar{f}_{j',\si'}^\kappa \Oper}(p,r)
+\O\left(\frac{M}{\sqrt{s}} \M^{\bar{f}_{j,\si}^\kappa\Oper}\right),
\eeqar
which is the identity represented in \refeq{FERCWIdiag}
with the gauge couplings \refeq{fermgenerators2} for antifermions.
Note that $\Psi_{j,\si}\ne\Psi_{j',\si'}$   for $V^a=W^\pm$   and, 
as noted in \refeq{shortBornampli3}  and \refeq{shortBornampli3b},
the mass of the wave function $\bar{v}(p)$ need not be equal to the masses of the fields $\Psi_{j,\si}$ or $\Psi_{j',\si'}$.

The analogous identity  for fermions reads
\beqar\label{FERCWIres}
\lim_{q^\mu \rightarrow xp^\mu} 
q^\mu G^{[\underline{V}^a\underline{\Psi}_{j,\si}^\kappa] 
\underline{\Oper}}_{\mu}(q,p-q,r)u(p)&=&e \sum_{j',\si'}
\M^{f_{j',\si'}^\kappa \Oper}(p,r)
I^{V^a}_{f^\kappa_{\si'} f^\kappa_{\si}}U^{f^\kappa,V^a}_{j'j}
+\O\left(\frac{M}{\sqrt{s}} \M^{f_{j,\si}^\kappa\Oper}\right),\nln
\eeqar
and can be derived exactly in the same way.

\end{list}

\chapter{Leading logarithms from parameter renormalization}
\label{Ch:PRllogs}

\newcommand{\eff}{\mathrm{eff}}
\newcommand{\gt}{\la_{\Pt}}
\newcommand{\gH}{\la_{\PH}}
\newcommand{\gHR}{\la_{\PH,0}}
\newcommand{\rt}{h_{\Pt}}
\newcommand{\rH}{h_{\PH}}

In this chapter we present  the one-loop logarithmic corrections that are related to the renormalization of the parameters. 
As we will see, at high-energies and for matrix elements that are not mass-suppressed,  one can restrict oneself to the 
running of the dimensionless coupling constants that appear 
in the symmetric phase of the electroweak theory.
However, in order to relate  these parameters to  physical 
quantities that are defined at the electroweak scale or below,
additional effects that are related to the renormalization of the vacuum expectation value 
(tadpole contributions) have to be taken into account.

\section{Logarithms connected to parameter renormalization}
\label{se:ren}
The running of the parameters from the renormalization scale $\mu_\rR$ to the energy scale $\sqrt{s}$ gives rise to  one-loop logarithmic corrections  of the type $\alpha\log{(s/\mu^2_\rR)}$.
These logarithms  are related to ultraviolet divergences and to the scale $\mu$ of dimensional regularization.
In the corrections to $S$-matrix elements they always appear 
in the combination  
\beq\label{runninglogs}
\log{\frac{s}{\mu^2}}+\log{\frac{\mu^2}{\mu^2_\rR}}=\log{\frac{s}{\mu^2_\rR}},
\eeq
with the contributions $\log{(s/\mu^2)}$ from loop diagrams,
and the  contributions $\log{(\mu^2/\mu^2_\rR)}$ from the counterterms. 
As a consequence of renormalizability, this combination is always $\mu$--independent.
In our approach, all large logarithms of the type \refeq{runninglogs}
that are related to the running of the parameters 
are absorbed into the counterterms by setting $\mu^2=s$.
The corresponding one-loop corrections to a given matrix element 
are obtained as 
\beq\label{PRbornmel}
\de^\pre \M^{\varphi_{i_1}\ldots\, \varphi_{i_n}} = \left. \sum_{\la_i}\frac{\partial
\M_0^{\varphi_{i_1}\ldots\, \varphi_{i_n}} }{\partial \la_i}
\,\de \la_i 
\, \right|_{\mu^2=s},
\eeq
where  $\de \la_i$ are  the counterterms that  renormalize the bare parameters
\beq
\la_{i,0}=\la_i + \de \la_i.
\eeq
These counterterms depend on the specific choice of the renormalization conditions.

In the high-energy limit, for processes that are not mass-suppressed,
and provided that matrix elements involving longitudinal gauge bosons are expressed using the GBET, 
the running of the  masses in the propagators or in couplings with mass dimension yields only mass-suppressed corrections. 
Then, one can restrict oneself to the renormalization of the {\em dimensionless parameters}
\beq\label{symmparamset}
\la_i= g_1,g_2,\gH, \gt,
\eeq
\ie the gauge couplings $g_1$,$g_2$, the Higgs self-coupling $\gH$,
and the top-quark Yukawa coupling\footnote{The Yukawa couplings $\la_f\propto m_f/\MW$ to light fermions $f\neq\Pt$ are not considered since they are very small.} $\gt$.
These are the  dimensionless coupling constants  that appear in the
symmetric phase of the electroweak theory. They represent a convenient set of
independent parameters in order to describe electroweak processes at high energies. However, they are not directly related to observable physical quantities.

Since we want to express the $S$-matrix elements in terms of physical 
parameters (coupling constants and masses) that are measured  at the electroweak scale or below,
we adopt the following set of independent parameters
\beq\label{onshellparamset}
\la_i= e, \cw , \rH, \rt, 
\eeq
involving the electric charge $e$, 
the cosine of the weak mixing angle  $\cw=\cos{\thw}$ defined as 
\beq \label{mixingangle}
\cw^2=1-\sw^2=\frac{\MW^2}{\MZ^2},
\eeq
and the mass ratios
\beqar \label{Htopratio}
\rH&=&\frac{\MH^2}{\MW^2}
,\qquad
\rt=\frac{\Mt}{\MW}.
\eeqar
For the renormalization of these parameters we adopt the on-shell scheme \cite{Denner:1993kt}, where the 
electric charge is defined in the Thomson limit of Compton scattering and the masses $\MW,\MZ,\Mt$, and $\MH$ are defined 
as the poles of  the  propagators of the physical fields  $\PW,\PZ,\Pt,$ and $\PH$, respectively.

Once the parameters \refeq{onshellparamset} and their counterterms have been determined 
in the on-shell scheme they can be translated into the parameters \refeq{symmparamset} 
and the corresponding counterterms. 
The parameters \refeq{symmparamset} are obtained by the following simple relations:
\beqar\label{gaugecouplings}
g_1&=&\frac{e}{\cw}, 
\qquad 
g_2=\frac{e}{\sw},\\
\gH&=& \frac{e^2}{{2}\sw^2}\rH
,\qquad
\gt=\frac{e}{\sqrt{2}\sw}\rt
.
\label{tophiggscouplings}
\eeqar
In one-loop approximation, 
the corresponding relations for the counterterms 
read
\beqar \label{gaugecouplings1CT}
\frac{\de g_1}{g_1}&=&\frac{\de e}{e} - \frac{1}{2}\frac{\de \cw^2}{\cw^2}
,\\
\label{gaugecouplings2CT}
\frac{\de g_2}{g_2}&=&\frac{\de e}{e} - \frac{1}{2}\frac{\de \sw^2}{\sw^2}
=\frac{\de e}{e} + \frac{\cw^2}{2\sw^2}\frac{\de \cw^2}{\cw^2}
,\\
\label{higgscouplingsCT}
\frac{\de \gH}{\gH}&=& 2\frac{\de g_2}{g_2} +\frac{\de\rH^\eff}{\rH}
,\\
\label{topyukcouplingsCT}
\frac{\de \gt}{\gt}&=& \frac{\de g_2}{g_2} +\frac{\de\rt}{\rt}
.
\eeqar
These relations are  obtained by the first derivatives of the lowest-order relations, 
and by absorbing additional effects originating from tadpole  renormalization (\cf\citere{Denner:1995xt}) 
into a redefinition of the   counterterm $\de \rH \to \de \rH^\eff$. These tadpole effects are discussed in  \refse{se:scalarsectorren}.

In the following sections all counterterms are determined in LA. 
If one adopts the set of parameters \refeq{onshellparamset}, then the
parameter-renormalization  corrections to a specific Born matrix element  are 
obtained from
\beq
\de^\pre \M = \left. \left\{\frac{\partial\M_0}{\partial e}\de e 
+ \frac{\partial\M_0}{\partial\cw}\de\cw 
+ \frac{\partial\M_0}{\partial\rH}\de\rH^{\eff} 
+ \frac{\partial\M_0}{\partial\rt}\de\rt \right\} 
\, \right|_{\mu^2=s},
\eeq
or by the replacements $e\to e+\de e$, $\cw\to\cw+\de\cw$,
$\sw\to\sw+\de\sw$, $\rH\to\rH+\de\rH^{\eff}$,  and $\rt\to\rt+\de\rt$ in $\M_0(e,\cw,\rH,\rt)$.
Equivalently, one can use  the parameters \refeq{symmparamset}
and the counterterms \refeq{gaugecouplings1CT}--\refeq{topyukcouplingsCT}. 
In any case, the lowest-order matrix elements need to be known in the high-energy limit only.
For processes with longitudinal gauge bosons, 
the renormalization of the parameters has to be applied to 
the matrix elements resulting from the GBET.


\section{Renormalization of gauge interactions}
In the on-shell scheme  \cite{Denner:1993kt},  
the gauge interactions with matter are determined by the parameters
$e, \MW,\MZ$.
The electric charge is
defined as the photon--electron coupling constant in the limit of zero momentum transfer.
The  renormalization condition for the corresponding vertex function reads
\beq\label{eRC}
\left.\bar{u}(p)\Gamma_\mu^{Ae\bar{e}}(0,p,-p)u(p)\right|_{p^2=m_\Pe^2}= 
e \bar{u}(p)\gamma_\mu u(p).
\eeq
The resulting counterterm $\de e$ 
is related  to the gauge-boson FRC's \refeq{gbdeltaZmat}
by Ward identities \cite{Denner:1993kt} and  reads 
\beqar \label{chargerenorm}
\frac{\de e}{e}&=& -\frac{1}{2}\left[\de Z_{AA}+\frac{\sw}{\cw}\de
  Z_{ZA}\right]
\nl
&\LA&\frac{1}{2}\left[-\frac{\alpha}{4\pi}
\bew_{AA}\log{\frac{\mu^2}{\MW^2}}+\Dealpha
\right],
\eeqar
where  $\bew_{AA}=-11/3$ is the  one-loop coefficient of the electroweak beta function (see \refapp{app:betafunction}) 
corresponding to the running electromagnetic coupling,
and $\Dealpha$ is given in \refeq{runningal} 
and represents the running of the electromagnetic coupling constant
from zero momentum transfer to the electroweak scale.

The physical masses of the weak gauge-bosons are fixed by the renormalization conditions \refeq{gbmassRC}
and their counterterms 
are given in \refeq{massCT} in LA. 
The resulting counterterm for the weak mixing angle \refeq{mixingangle} is
\beqar \label{weinbergrenorm}
\frac{\delta \cw^2}{\cw^2}&=&\frac{\de\MW^2}{\MW^2} -\frac{\de\MZ^2}{\MZ^2}\nl
&\LA&\frac{\alpha}{4\pi}\left\{\frac{\sw}{\cw}\bew_{AZ}\log{\frac{\mu^2}{\MW^2}}
+\frac{5}{6\cw^2}\log{\frac{\MH^2}{\MW^2}}
-
\frac{9+6\sw^2-32\sw^4}{18\sw^2\cw^2}\log{\frac{\Mt^2}{\MW^2}}
\right\}.\nln
\eeqar
where 
we have used the relation 
\beq
\bew_{AZ}=\frac{\cw}{\sw}(\bew_{ZZ}-\bew_{WW})=-\frac{19+22\sw^2}{6\sw\cw}
\eeq
between coefficients  of the electroweak beta function, which follows from \refeq{betarelations}.
Note that the large  Yukawa contributions $({\Mt^2}/{\MW^2})\log({\mu^2}/{\Mt^2})$ in the mass counterterms \refeq{massCT}
cancel in \refeq{weinbergrenorm}.

For the counterterms \refeq{gaugecouplings1CT} and \refeq{gaugecouplings2CT}
to the $\Uone$ and $\SUtwo$ gauge couplings  we obtain
\beqar \label{gCTs}
\frac{\de g_1}{g_1} 
&\LA&
 \frac{1}{2}\left\{ \frac{\alpha}{4\pi}\left[
-\besw_{B}\log{\frac{\mu^2}{\MW^2}}
-\frac{5}{6\cw^2}\log{\frac{\MH^2}{\MW^2}}
\right.\right.\nl&&\left.\left.\hspace{12mm}{}
+\frac{9+6\sw^2-32\sw^4}{18\sw^2\cw^2}\log{\frac{\Mt^2}{\MW^2}}
\right]+\Dealpha \right\}
,\nl
\frac{\de g_2}{g_2}
&\LA&
 \frac{1}{2}\left\{ \frac{\alpha}{4\pi}\left[
-\besw_{W}\log{\frac{\mu^2}{\MW^2}}
+\frac{5}{6\sw^2}\log{\frac{\MH^2}{\MW^2}}
\right.\right.\nl&&\left.\left.\hspace{12mm}{}
-\frac{9+6\sw^2-32\sw^4}{18\sw^4}\log{\frac{\Mt^2}{\MW^2}}
\right]+ \Dealpha \right\},\nln
\eeqar
where we have used the relations \refeq{betarelations}, and 
$\besw_{B}=-41/(6\cw^2)$ and $\besw_{W}=19/(6\sw^2)$ are the one-loop coefficients of the electroweak beta function 
corresponding to the  $\Uone$ and $\SUtwo$ running couplings, respectively.

\section{Renormalization in the scalar sector}\label{se:scalarsectorren}
We consider now the scalar sector and discuss the renormalization of the dimensionless scalar self-interaction $\gH$ that 
enters the  scalar potential 
\beq
V(\Phi)= -\mu^2|\Phi+\vev|^2+\frac{\gH}{4}|\Phi+\vev|^4,\qquad
\Phi+\vev =\left(\begin{array}{c} 
\phi^+\\ \frac{1}{\sqrt{2}}(v+H+\ri\chi)
\end{array}\right).
\eeq
In order to  relate  $\gH$
to the  masses of the physical Higgs boson and  the weak gauge bosons,
we need to define and to  renormalize  all parameters in the scalar potential, including the dimension-full parameters $\mu$ and $v$.
Note that the vev $v$  and its counterterm $\de v$ are related to the parameters in the gauge sector 
and the corresponding counterterms by 
\beq \label{vevdef}
\MW=\frac{g_2}{2}v,\qquad
\frac{\de \MW}{\MW}=\frac{\de g_2}{g_2}+\frac{\de v}{v}.
\eeq
In order to specify physical renormalization conditions for the 
two remaining parameters in the Higgs potential we
consider those terms  that are linear and bilinear in the Higgs field $H$ 
\beq
V(\Phi)= -t H +\frac{1}{2} \MH^2 H^2 +\dots \,.
\eeq
The tadpole $t$ and the Higgs mass $\MH$ are related to $(\mu,v,\gH)$ by 
\beqar \label{tadpolet}
t&=& v\left(\mu^2-\frac{\gH}{4}v^2\right)
,\\
\MH^2&=&-\mu^2+\frac{3\gH}{4}v^2
=2\mu^2-3\frac{t}{v}.
\label{Hmass}
\eeqar
In the on-shell scheme, one adopts  $t$ and $\MH$ as independent parameters
and one renormalizes them  as follows:
\begin{itemize}
\item The Higgs mass \refeq{Hmass} is defined as physical mass  through 
the renormalization condition \refeq{Hmassrencond}
and the resulting counterterm is given in \refeq{higgsmassCT}.
Evaluating the  Higgs self-energy diagrams \refeq{Hselfenergydiag} in LA we obtain\footnote{In this section, the scale of dimensional regularization is denoted by $\mu_\rD$.}
\beqar
\frac{\de \MH^2}{\MH^2} &\LA&\frac{\alpha}{4\pi}\left\{
\biggl[\frac{9}{2\sw^2}\frac{\MW^2}{\MH^2}\left(1+\frac{1}{2\cw^4}\right)
-\frac{3}{2}\cew_\Phi
+ \frac{15}{8\sw^2}\frac{\MH^2}{\MW^2}\biggr]
\log{\frac{\mu_\rD^2}{\MH^2}}
\right.\nl &&{}+\left.
\frac{\NCt}{2\sw^2}\frac{\Mt^2}{\MW^2}\left(1-6\frac{\Mt^2}{\MH^2}\right)
\log{\frac{\mu_\rD^2}{\Mt^2}}
-\frac{\NCt}{2\sw^2}\frac{\Mt^2}{\MW^2}\left(1-4\frac{\Mt^2}{\MH^2}\right)
\log{\frac{M_{\Htop}^2}{\Mt^2}}
\right\}
,\nln
\eeqar
where $M_{\Htop}=\max{(\MH,\Mt)}$  and $\cew_\Phi=(1+2\cw^2)/(4\cw^2\sw^2)$ [see \refeq{casimirew}].
\item
For the renormalized tadpole \refeq{tadpolet} one requires $t=0$
such  that, in lowest order,  $v$ corresponds to the minimum of the potential 
and to the vev of the Higgs field.
Since this condition is not protected by any symmetry, 
the vev of the Higgs field is shifted  by radiative corrections. 
In order to compensate this shift 
one introduces a tadpole counterterm $\de t$ and one requires 
\beq\label{tadrencond}
\Gamma^{H}(0)=\Gamma_0^{H}(0)+\de t = 0,
\eeq
for the 1PI Higgs 1-point function.
This guarantees  that  $v$ corresponds to the minimum of the effective potential.
Evaluating  the tadpole diagrams
\beqar\label{tadpolediag}
\ri \Gamma_0^{H}&=& \quad
\quad
\vcenter{\hbox{\begin{picture}(50,60)(-10,-10)
\Text(-5,23)[bc]{\scriptsize $H$} 
\Text(32,40)[b]{\scriptsize $V^a$} 
\Vertex(15,20){2}
\DashLine(-10,20)(15,20){2} 
\PhotonArc(30,20)(15,0,360){1}{10} 
\end{picture}}}
\quad +\quad
\vcenter{\hbox{\begin{picture}(50,60)(-10,-10)
\Text(-5,23)[bc]{\scriptsize $H$} 
\Text(32,40)[b]{\scriptsize$u^a$} 
\Vertex(15,20){2}
\DashLine(-10,20)(15,20){2} 
\DashArrowArc(30,20)(15,-180,180){1} 
\end{picture}}}
\quad +\quad
\vcenter{\hbox{\begin{picture}(50,60)(-10,-10)
\Text(-5,23)[bc]{\scriptsize $H$} 
\Text(32,40)[b]{\scriptsize $\Phi_k$} 
\Vertex(15,20){2}
\DashLine(-10,20)(15,20){2} 
\DashCArc(30,20)(15,0,360){4} 
\end{picture}}}
\quad +\quad
\vcenter{\hbox{\begin{picture}(50,60)(-10,-10)
\Text(-5,23)[bc]{\scriptsize $H$} 
\Text(32,40)[b]{\scriptsize $f_{j,\si}$} 
\Vertex(15,20){2}
\DashLine(-10,20)(15,20){2} 
\ArrowArc(30,20)(15,-180,180) 
\end{picture}}}
\nln
\eeqar

\hspace{-1cm}
in LA we obtain 
\beqar
\de t = -\Gamma_0^{H}(0)&\LA&\frac{\alpha}{4\pi}
\frac{2\sw\MW\MH^2}{e}\left\{\biggl[-\frac{3}{2\sw^2}\frac{\MW^2}{\MH^2}\left(1+\frac{1}{2\cw^4}\right)
-\frac{1}{2} \cew_\Phi
\biggr]\log{\frac{\mu_\rD^2}{\MW^2}}
\right.\nl &&\left.{}
-\frac{3}{8\sw^2}\frac{\MH^2}{\MW^2}\log{\frac{\mu_\rD^2}{\MH^2}}
+\NCt\frac{\Mt^4}{\sw^2\MW^2\MH^2}\log{\frac{\mu_\rD^2}{\Mt^2}}\right\}.
\eeqar
\end{itemize}
The  scalar self-coupling $\gH$ is now determined by $v,t,\MH$ through \refeq{tadpolet},
\refeq{Hmass} and reads 
\beq\label{eq:scalarselfcoupling}
\gH=\frac{2}{v^2}\left(\MH^2+ \frac{t}{v}\right).
\eeq
The corresponding counterterm is given by the derivative of \refeq{eq:scalarselfcoupling} at $t=0$ and reads 
\beq\label{scalarcouplrenCT}
\frac{\de\gH}{\gH}= - \frac{\de v^2}{v^2}+\frac{\de\MH^2}{\MH^2}+ \frac{\de t}{v \MH^2}
= 2\frac{\de g_2}{g_2}- \frac{\de \MW^2}{\MW^2}+\frac{\de\MH^2}{\MH^2}+ \frac{e\de t}{2\sw\MW \MH^2}.
\eeq
We stress that besides the naive renormalization of the lowest-order relation \refeq{tophiggscouplings} 
additional tadpole contributions appear in \refeq{scalarcouplrenCT}.
In LA we have
\beqar
\frac{\de\gH}{\gH} 
&\LA&
\frac{\alpha}{4\pi}\left\{
\biggl[\frac{3}{\sw^2}\frac{\MW^2}{\MH^2}\left(1+\frac{1}{2\cw^4}\right)
-6\cew_\Phi\biggr]\log{\frac{{\mu_\rD}^2}{\MW^2}}
+ \frac{3}{2\sw^2}\frac{\MH^2}{\MW^2} \log{\frac{{\mu_\rD}^2}{\MH^2}}
\right.\nl&&\left.\hspace{5mm} {}
+\frac{\NCt}{\sw^2}\frac{\Mt^2}{\MW^2}\left(1-2\frac{\Mt^2}{\MH^2}\right)\log{\frac{{\mu_\rD}^2}{\Mt^2}}
\right.\nl&&\left.\hspace{5mm} {}
+
\biggl[-\frac{9}{2\sw^2}\frac{\MW^2}{\MH^2}\left(1+\frac{1}{2\cw^4}\right)
+\frac{3}{2}\cew_\Phi
+\frac{5}{3\sw^2}\biggr]\log{\frac{\MH^2}{\MW^2}}
\right.\nl&&\left.\hspace{5mm} {} 
-\frac{9-12\sw^2-32\sw^4}{18\sw^4} \log{\frac{\Mt^2}{\MW^2}}
- \frac{\NCt}{2\sw^2}\frac{\Mt^2}{\MW^2}\left(1-4\frac{\Mt^2}{\MH^2}\right)\log{\frac{M_{\Htop}^2}{\Mt^2}}
\right\}+ \Dealpha.\nln
\eeqar
In the parametrization \refeq{onshellparamset}, the effect of the tadpole renormalization
can be absorbed into the counterterm for the  mass ratio $\rH$ as in 
\refeq{higgscouplingsCT}. This is then given by
\beq
\frac{\de\rH^{\eff}}{\rH}=
-\frac{\de \MW^2}{\MW^2}+\frac{\de\MH^2}{\MH^2}+ \frac{e\de t}{2\sw\MW \MH^2},
\eeq
and in  LA 
\beqar
\frac{\de\rH^{\eff}}{\rH} 
&\LA& \frac{\alpha}{4\pi}\left\{
\biggl[\bew_{WW}+
\frac{3}{\sw^2}\frac{\MW^2}{\MH^2}\left(1+\frac{1}{2\cw^4}\right)
-6\cew_\Phi\biggr]\log{\frac{\mu_\rD^2}{\MW^2}}
+ \frac{3}{2\sw^2}\frac{\MH^2}{\MW^2} \log{\frac{\mu_\rD^2}{\MH^2}}
\right.\nl&&\left.\hspace{5mm} {}
+\frac{\NCt}{\sw^2}\frac{\Mt^2}{\MW^2}\left(1-2\frac{\Mt^2}{\MH^2}\right)\log{\frac{\mu_\rD^2}{\Mt^2}}
\right.\nl&&\left.\hspace{5mm} {}
+
\biggl[-\frac{9}{2\sw^2}\frac{\MW^2}{\MH^2}\left(1+\frac{1}{2\cw^4}\right)
+\frac{3}{2}\cew_\Phi
+\frac{5}{6\sw^2}\biggr]\log{\frac{\MH^2}{\MW^2}}
\right.\nl&&\left.\hspace{5mm} {} 
+\frac{\NCt}{3\sw^2} \log{\frac{\Mt^2}{\MW^2}}
- \frac{\NCt}{2\sw^2}\frac{\Mt^2}{\MW^2}\left(1-4\frac{\Mt^2}{\MH^2}\right)\log{\frac{M_{\Htop}^2}{\Mt^2}}
\right\}.
\eeqar

\section{Renormalization of the top-quark Yukawa coupling}
The renormalization of the top-quark Yukawa coupling is determined by the top-quark mass.
In the on-shell scheme, $m_t$ is defined as the physical top-quark mass through the 
renormalization condition \refeq{fermassRC}. The corresponding counterterm is given by
\refeq{fermmassCT}, and  evaluating the top self-energy \refeq{ferselfenergydiag}
in LA we obtain
\newcommand{\Qt}{Q_{\Pt}}
\beq
\frac{\de \Mt}{\Mt} \LA
\frac{\alpha}{4\pi}\left\{
\frac{1}{2}(\cew_{\Pt^\rL}-\cew_{\Pt^\rR})
\log{\frac{\mu^2}{\Mt^2}}
+\frac{3}{8\sw^2}\frac{\Mt^2}{\MW^2}
\log{\frac{\mu^2}{M_{\Htop}^2}}
\right\},
\eeq
where $M_{\Htop}=\max{(\MH,\Mt)}$  and the Casimir operator $\cew$ is defined in \refeq{casimirew}.  
Combining this with the counterterm for the \PW--boson mass \refeq{massCT}
and using  the relation
\beq
\cew_\Phi=\cew_{\Pt^\rL}+\frac{1}{2}\cew_{\Pt^\rR},
\eeq
we obtain the counterterm to the mass ratio $\rt$ in \refeq{Htopratio} 
\beqar
\frac{\de\rt}{\rt} &=& \frac{\de \Mt}{\Mt}- \frac{1}{2}\frac{\de
  \MW^2}{\MW^2}\nl
&\LA&
\frac{\alpha}{4\pi}\left\{
\frac{1}{2}\bew_{WW}\log{\frac{\mu^2}{\MW^2}}
+\biggl[\frac{3+2\NCt}{8\sw^2}\frac{\Mt^2}{\MW^2}
-\frac{3}{2}(\cew_{\Pt^\rL}+\cew_{\Pt^\rR})\biggr]
\log{\frac{\mu^2}{\Mt^2}}
\right.\nl&&\left.{}
+\left(\frac{1}{2\sw^2}-2\cew_\Phi\right)\log{\frac{\Mt^2}{\MW^2}}
-\frac{3}{8\sw^2}\frac{\Mt^2}{\MW^2}\log{\frac{M_{\Htop}^2}{\Mt^2}}
+\frac{5}{12\sw^2}\log{\frac{\MH^2}{\MW^2}}
\right\}.
\eeqar
Finally, for the counterterm  \refeq{topyukcouplingsCT}  to the top-quark Yukawa coupling
we have
\beqar
\frac{\de\gt}{\gt} 
&\LA&\frac{\alpha}{4\pi}
\left\{\biggl[
-\frac{3}{2}(\cew_{\Pt^\rL}+\cew_{\Pt^\rR})
+\frac{3+2\NCt}{8\sw^2}\frac{\Mt^2}{\MW^2}
\biggr]\log{\frac{\mu^2}{\Mt^2}}
\right.\nl&&\left.{}
-\left(\frac{9-12\sw^2-32\sw^4}{36\sw^4}+2\cew_\Phi\right)\log{\frac{\Mt^2}{\MW^2}}
-\frac{3}{8\sw^2}\frac{\Mt^2}{\MW^2}\log{\frac{M_{\Htop}^2}{\Mt^2}}
+\frac{5}{6\sw^2}\log{\frac{\MH^2}{\MW^2}}
\right\}
\nl&&
{}+\frac{1}{2}\Dealpha 
.
\eeqar

\chapter{Applications to simple processes}
\label{ch:applicat}
In this chapter, our generic results for the electroweak 
logarithmic corrections  are applied to the processes
$\Pep\Pem\to \mathrm{f}\mathrm{\bar{f}}$,
$\Pep\Pem\to \PWp\PWm$, $\Pep\Pem\to \PZ\PZ,\PZ\gamma,\gamma\gamma$ and
$\bar{\mathrm{d}}\mathrm{u}\to \PWp\PZ,\PWp\gamma$.
We give explicit analytical and numerical expressions for the relative corrections 
\beq \label{relco}
\de_{\varphi_{i_1}\varphi_{i_2}\to\varphi_{i_3}\varphi_{i_4}}(p_1,p_2,p_3,p_4)
=\frac{\de \M^{\varphi_{i_1}\varphi_{i_2}\bar{\varphi}_{i_3}\bar{\varphi}_{i_4} }(p_1,p_2,-p_3,-p_4)}{\M_0^{\varphi_{i_1}\varphi_{i_2}\bar{\varphi}_{i_3}\bar{\varphi}_{i_4} }(p_1,p_2,-p_3,-p_4)}
\eeq  
to the Born matrix elements. Recall that according to our convention \refeq{process},
the predictions for $2\to 2$ processes have to be extracted
using crossing symmetry from our formulas for $4\to 0$ processes. 
Note that the corrections to the cross sections are twice as large as \refeq{relco}.

The complete logarithmic corrections are first presented in analytic form, 
expressing the coefficients of the various logarithms\footnote{In this chapter the logarithms $\log{(\MH/\MW)}$ and $\log{(\Mt/\MW)}$ have been omitted.} in terms of the gauge couplings and of other  group-theoretical quantities that are defined in \refapp{app:representations}.
The  coefficients of the {\em genuine
electroweak}\footnote{These are obtained by omitting the 
pure electromagnetic contributions that result from the gap between the
electromagnetic and the weak scale as well as the contributions $\log{(\MH/\MW)}$ and $\log{(\Mt/\MW)}$. Accordingly they include the
symmetric-electroweak contributions and the 
$I^Z$  terms originating from $\PZ$-boson loops in \refeq{deSC}.} (ew) logarithms are then  evaluated numerically using the parameters
\begin{displaymath}
\MW=80.35 \GeV,\qquad \MZ=91.1867 \GeV,\qquad  \Mt=175 \GeV,
\end{displaymath}
\beq
\alpha=\frac{1}{137.036},\qquad \sw^2=1-\frac{\MW^2}{\MZ^2}\approx0.22356.
\eeq
For the DL and SL contributions we use the shorthands \refeq{dslogs},
and in order to keep track of the origin of the various  
SL terms 
we introduce different
subscripts
\beq
\lsl=\lYuk=\lpr=\lZ=\ls=\frac{\alpha}{4\pi}\log{\frac{s}{\MW^2}}, 
\eeq
where $\lsl$, $\lYuk$, $\lpr$, and $\lZ$, denote  collinear\footnote{By collinear corrections we mean the corrections \refeq{subllogfact2}
which involve the collinear mass singularities from truncated loop diagrams as well as the mass singularities from field-renormalization.}, Yukawa, PR contributions, 
and the $\PZ$-boson contributions from \refeq{deSC},
respectively.

\section{Four-fermion neutral-current processes}
The Sudakov DL corrections \refeq{deSC} and the collinear or soft SL
corrections \refeq{deccfer} depend only on the quantum numbers of the
external legs, and can be applied to 4-fermion processes in a
universal way. However, we are interested also in the $\SS$ corrections
\refeq{4fsubdl} and $\pre$ corrections, 
which depend on the specific properties of the process. A
general description of these corrections requires a decomposition of
the Born matrix element into neutral-current (NC) and charged-current
(CC) contributions.  In order to simplify the discussion we restrict
ourselves to pure NC transitions.  To simplify notation, we consider
processes involving a lepton--antilepton and a quark--antiquark pair.
However, our analysis applies to the more general case of two
fermion--antifermion pairs of different isospin doublets.
The 4 external states and their momenta 
are chosen to be incoming, so that the process reads 
\beq \label{4fprocess} 
\bar{l}^\kappa_\si l^\kappa_\si q^{\la}_{\rho} \bar{q}^{\la}_{\rho}\rightarrow 0, 
\eeq
where $\kappa,\la=\rR,\rL$ are the chiralities and $\si,\rho=\pm$  the isospin indices. All formulas for the $4\rightarrow 0$ process \refeq{4fprocess}
are expressed in terms of the  eigenvalues $I^\NB_{l^\kappa_\si}$ and  $I^\NB_{q^{\la}_{\rho}}$ of the fermions (see \refapp{app:representations}). 

\newcommand{\NC}{R}
\newcommand{\NCew}{\NC_{\Pe^-_\kappa\PW^-_\la}}
\newcommand{\NCep}{\NC_{\Pe^-_\kappa\phi^-}}
\newcommand{\deNClq}{\Delta_{l^\kappa_\si q^\la_\rho}}
\newcommand{\deNCew}{\Delta_{\Pe^-_\kappa\PW^-_\rT}}
\newcommand{\deNCep}{\Delta_{\Pe^-_\kappa\phi^-}}
\newcommand{\NClq}{\NC_{l^\kappa_\si q^\la_\rho}}
\newcommand{\NCll}{\NC_{l^\kappa_\si l^\kappa_\si}}
\newcommand{\NCqq}{\NC_{q^\la_\rho q^\la_\rho}}
\newcommand{\NClmq}{\NC_{l^\kappa_{-\si} q^\la_\rho}}
\newcommand{\NClqm}{\NC_{l^\kappa_\si q^\la_{-\rho}}}

\begin{list}{\bf \arabic{listnumber}.\hspace{1mm}}{\usecounter{listnumber}
\setlength{\leftmargin}{0mm} \setlength{\labelsep}{0mm}
\setlength{\itemindent}{4mm}
}

\item{\bf Born matrix elements}\nopagebreak

In the high-energy limit, the Born amplitude is given by
\beq \label{born4fnc} 
\M_{0}^{\bar{l}^\kappa_\si l^\kappa_\si q^{\la}_{\rho} \bar{q}^{\la}_{\rho}}
=e^2\NClq \frac{A_{12}}{r_{12}},
\eeq
where
\beq \label{NCcorr}
\NC_{\phi_i\phi_k}:=\sum_{\NB=A,Z}I^\NB_{\phi_i}I^\NB_{\phi_k}=\frac{1}{4\cw^2} Y_{\phi_i} Y_{\phi_k}+\frac{1}{\sw^2} T^3_{\phi_i}T^3_{\phi_k},
\eeq
and terms of order $\MZ^2/r_{12}$, originating from the difference between the photon and the Z-boson mass, are neglected. 
Note that \refeq{NCcorr} and the following formulas have an
important chirality
dependence, owing to the different values of the group-theoretical
operators in the representations for right-handed and left-handed
fermions.

\item{\bf Leading soft-collinear corrections}\nopagebreak

The Sudakov soft--collinear corrections give according to \refeq{deSC}
the leading contribution
\beq
\de^{\SC}_{\bar{l}^\kappa_\si l^\kappa_\si q^{\la}_{\rho} \bar{q}^{\la}_{\rho}}
=-\sum_{f^\mu_\tau=l^\kappa_\si,q^{\la}_\rho} \left[\cew_{f^\mu_\tau}\Ls-2(I^Z_{f^\mu_\tau})^2 \log{\frac{\MZ^2}{\MW^2}}\,\lZ
+Q_{f_\tau}^2\Lemftau \right].
\eeq

\item{\bf Subleading soft-collinear corrections}\nopagebreak

The angular-dependent $\SS$ corrections are obtained from
\refeq{4fsubdl}.  The contribution of the neutral gauge bosons $\NB=A,Z$
is diagonal in the $\SUtwo$ indices, and factorizes into
the Born matrix element \refeq{born4fnc} times the relative correction
\beqar \label{4fncss} 
\sum_{\NB=A,Z}\de^{\NB,\SS}_{\bar{l}^\kappa_\si l^\kappa_\si q^{\la}_{\rho} \bar{q}^{\la}_{\rho}}
&=&-2\ls\left\{(\NCll+\NCqq)\log{\frac{|r_{12}|}{s}}
+2\NClq\log{\frac{|r_{13}|}{|r_{14}|}}
\right\}\nl
&&{}-
\frac{\alpha}{2\pi}\log{\frac{\MW^2}{\la^2}}
\left[(Q_{l_\si}^2+Q_{q_\rho}^2)\log{\frac{|r_{12}|}{s}}+2Q_{l_\si}Q_{q_\rho}\log{\frac{|r_{13}|}{|r_{14}|}}\right],
\eeqar
where $I^\NB_{\bar f \bar f} = -I^\NB_{ff}= -I^\NB_{f}$ has been used and 
terms involving $\log{(\MW/\MZ)}\log{(s/\MW^2)}$ have been omitted. The contribution 
of the charged gauge bosons to \refeq{4fsubdl} gives  
\beqar \label{4fncss2} 
\sum_{V^a=W^\pm}\de^{V^a,\SS} \M^{\bar{l}^\kappa_\si l^\kappa_\si q^{\la}_{\rho} \bar{q}^{\la}_{\rho}}&=&-\frac{1}{\sw^2}\ls\left\{\left(
\de_{\kappa\rL}\M_0^{\bar{l}^\kappa_{-\si} l^\kappa_{-\si} q^{\la}_{\rho} \bar{q}^{\la}_{\rho}}
+\de_{\la\rL}\M_0^{\bar{l}^\kappa_\si l^\kappa_{\si} q^{\la}_{-\rho} \bar{q}^{\la}_{-\rho}}
\right)\log{\frac{|r_{12}|}{s}}\right.\nl
&&\left.{}+\de_{\kappa\rL}\de_{\la\rL}\left[\de_{\si\rho}\left(
\M_0^{\bar{l}^\kappa_{-\si} l^\kappa_\si q^{\la}_{-\rho} \bar{q}^{\la}_{\rho}}
+\M_0^{\bar{l}^\kappa_\si l^\kappa_{-\si} q^{\la}_{\rho} \bar{q}^{\la}_{-\rho}}
\right)\log{\frac{|r_{13}|}{s}}
\right.\right.\nl &&\left.\left.{}
-\de_{-\si\rho}\left(
\M_0^{\bar{l}^\kappa_{-\si} l^\kappa_\si q^{\la}_{\rho} \bar{q}^{\la}_{-\rho}}
+\M_0^{\bar{l}^\kappa_\si l^\kappa_{-\si} q^{\la}_{-\rho} \bar{q}^{\la}_{\rho}}
\right)\log{\frac{|r_{14}|}{s}}
\right]\right\},
\eeqar
where the non-diagonal couplings \refeq{ferpmcoup} have been used\footnote{Note that the non-diagonal effects \
related to the quark-mixing matrix have not been considered in this section.}. 
On the rhs, the $\SUtwo$-trans\-formed Born matrix elements
involving the isospin partners $l^\kappa_{-\si}$, $q^\la_{-\rho}$,
have to be evaluated explicitly. The NC matrix elements (first line) are
obtained from \refeq{born4fnc}, and for the CC amplitudes
we find up to mass-suppressed terms using the non-diagonal couplings
\refeq{ferpmcoup},
\beq
\M_0^{\bar{l}^\kappa_{\si'} l^\kappa_{-\si'} q^{\la}_{\rho'} \bar{q}^{\la}_{-\rho'}}=\frac{e^2}{2\sw^2}\frac{A_{12}}{r_{12}}.
\eeq
Then, dividing \refeq{4fncss2} by the Born matrix element, we obtain
the relative correction
\beqar  \label{4fncss3} 
\sum_{V^a=W^\pm}\de^{V^a,\SS}_{\bar{l}^\kappa_\si l^\kappa_\si q^{\la}_{\rho} \bar{q}^{\la}_{\rho}} 
= -\frac{1}{\sw^2\NClq}\,\ls&&\left\{\left(\de_{\kappa\rL}\NClmq+ \de_{\la\rL}\NClqm\right)\log{\frac{|r_{12}|}{s}}\right.\\
&&\left.{}+\frac{\de_{\kappa\rL}\de_{\la\rL}}{\sw^2}\left[\de_{\si\rho}\log{\frac{|r_{13}|}{s}}-\de_{-\si\rho}\log{\frac{|r_{14}|}{s}}\right]\right\}.\nn
\eeqar

The angular-dependent corrections for $2\rightarrow 2$ processes, like
those depicted in \reffi{ffborn}, are directly obtained from
\refeq{4fncss} and \refeq{4fncss3} by substituting the invariants
$r_{kl}$ 
by 
the corresponding Mandelstam variables $s,t,u$.
For the $s$-channel processes
$\bar{l}^\kappa_{\si} l^\kappa_\si \rightarrow \bar{q}^{\la}_{\rho}
q^{\la}_{\rho}$, we have to substitute $r_{12}=s,r_{13}=t,r_{14}=u$,
and the $\SS$ corrections simplify to
\beqar \label{4fsSS} 
\de^\SS_{\bar{l}^\kappa_\si l^\kappa_\si \rightarrow \bar{q}^{\la}_{\rho} q^{\la}_{\rho}}&=&- \ls\left[4\NClq\ltu+\frac{\de_{\kappa\rL}\de_{\la\rL}}{\sw^4\NClq}\left(\de_{\si\rho}\lts-\de_{-\si\rho}\lus\right) \right]\nl
&&{}- 
\frac{\alpha}{\pi}Q_{l_\si}Q_{q_\rho}\log{\frac{\MW^2}{\la^2}}
\ltu.
\eeqar
If one subtracts the photonic contributions from \refeq{4fsSS} one
finds agreement with eq.~(50) of \citere{Kuhn:2000}.  For the $t$-channel
processes $ \bar{q}^{\la}_{\rho} l^\kappa_\si \rightarrow
\bar{q}^{\la}_{\rho}l^\kappa_{\si}$, the substitution in \refeq{4fncss} and \refeq{4fncss3} reads
$r_{12}=t,r_{13}=s,r_{14}=u$, whereas for $ q^{\la}_{\rho}
l^\kappa_\si \rightarrow q^{\la}_{\rho}l^\kappa_{\si}$ one has to
choose $r_{12}=t,r_{13}=u,r_{14}=s$.

\begin{figure}
\begin{center}
\begin{picture}(300,80)
\put(0,0){
\begin{picture}(120,80)
\ArrowLine(20,30)(0,60)
\ArrowLine(0,0)(20,30)
\Vertex(20,30){2}
\Photon(20,30)(80,30){2}{6}
\Vertex(80,30){2}
\ArrowLine(80,30)(100,0)
\ArrowLine(100,60)(80,30)
\Text(-5,0)[r]{$l^\kappa_\si$}
\Text(-5,60)[r]{$\bar l^\kappa_\si$}
\Text(105,0)[l]{$q^\la_\rho$}
\Text(105,60)[l]{$\bar q^\la_\rho$}
\Text(50,35)[b]{$A,Z$}
\end{picture}}
\put(180,0){
\begin{picture}(120,80)
\ArrowLine(40,45)(10,60)
\Photon(40,15)(40,45){2}{3}
\ArrowLine(10,0)(40,15)
\Vertex(40,15){2}
\Vertex(40,45){2}
\ArrowLine(40,15)(70,0)
\ArrowLine(70,60)(40,45)
\Text(5,0)[r]{$l^\kappa_\si$}
\Text(5,60)[r]{$\bar q^\la_\rho$}
\Text(45,30)[l]{$A,Z$}
\Text(75,0)[l]{$l^\kappa_\si$}
\Text(75,60)[l]{$\bar q^\la_\rho$}
\end{picture}}
\end{picture}
\end{center}
\caption{Lowest-order diagrams for
  $ \bar l^\kappa_\si l^\kappa_\si \to  \bar q^\la_\rho q^\la_\rho$ and $ \bar q^\la_\rho l^\kappa_\si \to  \bar q^\la_\rho l^\kappa_\si$}
\label{ffborn}
\end{figure}

\item{\bf SL corrections associated to  external particles}\nopagebreak

The  SL corrections \refeq{deccfer} give  
\beqar
\de^{\cc}_{\bar{l}^\kappa_\si l^\kappa_\si q^{\la}_{\rho} \bar{q}^{\la}_{\rho}}&=&\sum_{f^\mu_\tau=l^\kappa_\si,q^\la_\rho} \left[3 \cew_{f^\mu_\tau}\lsl -\frac{\Mt^2}{4\sw^2\MW^2}
\left((1+\delta_{\mu R})\de_{f_\tau\Pt}
+\delta_{\mu L}\de_{f_\tau\Pb}
\right)\lYuk \right.\nl
&& \hspace{1.2cm}\left. {}+
\frac{\alpha}{4\pi}
Q_{f_\tau}^2\left(\log{\frac{\MW^2}{m^2_{f_{\tau}}}}
    +2\log{\frac{\MW^2}{\la^2}}\right)
\right],
\eeqar
where the Yukawa corrections contribute only in the case of external top ($f_\tau=\Pt$) or bottom ($f_\tau=\Pb$) quarks and depend on their chirality  $\mu$.

\item{\bf Logarithms from parameter renormalization}\nopagebreak

The PR logarithms for NC processes are obtained from the
renormalization of the electric charge and the weak mixing angle in the
Born amplitude \refeq{born4fnc}. Using \refeq{weinbergrenorm} and
\refeq{chargerenorm} this gives the relative correction
\beq
\de^\pre_{\bar{l}^\kappa_\si l^\kappa_\si q^{\la}_{\rho} \bar{q}^{\la}_{\rho}}=\left[\frac{\sw}{\cw}\bew_{AZ}\deNClq-\bew_{AA}\right]\lpr+\Dealpha
,
\eeq
where
\beq \label{NCren}
\Delta_{\phi_i\phi_k}:=\frac{-\frac{1}{4\cw^2}Y_{\phi_i}
  Y_{\phi_k}+\frac{\cw^2}{\sw^4}
  T^3_{\phi_i}T^3_{\phi_k}}{\NC_{\phi_i\phi_k}}
\eeq
gives a chirality-dependent contribution owing to mixing-angle
renormalization of \refeq{NCcorr}, and $\bew_{AA}$ represents the
universal contribution of electric charge renormalization.  

\item{\bf Numerical evaluation}\nopagebreak

In order to give an impression of the size 
of the  genuine electroweak part of
the corrections, we consider the relative corrections $\de^{\kappa_\Pe
  \kappa_f,\ew}_{\Pep \Pem \rightarrow \bar{f}f}$ to NC processes
$\Pep\Pem\rightarrow \bar{f}f$ with chiralities
$\kappa_\Pe,\kappa_f=\rR$ or $\rL$, and give the numerical
coefficients of the electroweak  logarithms for the cases $f=\mu,\Pt,\Pb$. 
For muon-pair production we have
\newcommand{\eemumuRR}{\de^{\rR\rR,\ew}_{\Pep \Pem \rightarrow \mu^+\mu^-}}
\newcommand{\eemumuRL}{\de^{\rR\rL,\ew}_{\Pep \Pem \rightarrow \mu^+\mu^-}}
\newcommand{\eemumuLR}{\de^{\rL\rR,\ew}_{\Pep \Pem \rightarrow \mu^+\mu^-}}
\newcommand{\eemumuLL}{\de^{\rL\rL,\ew}_{\Pep \Pem \rightarrow \mu^+\mu^-}}
\beqar
\eemumuRR &=&   -2.58\,\Ls-5.15\left(\ltu\right)\ls+0.29\,\lZ+7.73\,\lsl+8.80\,\lpr   ,\nl
\eemumuRL &=&   -4.96\,\Ls-2.58\left(\ltu\right)\ls+0.37\,\lZ+14.9\,\lsl+8.80\,\lpr   ,\nl
\eemumuLL &=&   -7.35\,\Ls-\left(5.76\ltu+13.9\lts\right)\ls+0.45\,\lZ\nl
&&{}+22.1\,\lsl-9.03\,\lpr,
\eeqar
and $\eemumuLR =\eemumuRL$. For top-quark-pair production we find
\newcommand{\eettRR}{\de^{\rR\rR,\ew}_{\Pep \Pem \rightarrow  \bar{\Pt}\Pt}}
\newcommand{\eettRL}{\de^{\rR\rL,\ew}_{\Pep \Pem \rightarrow  \bar{\Pt}\Pt}}
\newcommand{\eettLR}{\de^{\rL\rR,\ew}_{\Pep \Pem \rightarrow  \bar{\Pt}\Pt}}
\newcommand{\eettLL}{\de^{\rL\rL,\ew}_{\Pep \Pem \rightarrow  \bar{\Pt}\Pt}}
\beqar \label{toppair}
\eettRR &=&   -1.86\,\Ls+3.43\left(\ltu\right)\ls+0.21\,\lZ+5.58\,\lsl-10.6\,\lYuk+8.80\,\lpr   ,\nl
\eettRL &=&   -4.68\,\Ls+0.86\left(\ltu\right)\ls+0.50\,\lZ+14.0\,\lsl-5.30\,\lYuk+8.80\,\lpr   ,\nl
\eettLR &=&   -4.25\,\Ls+1.72\left(\ltu\right)\ls+0.29\,\lZ+12.7\,\lsl-10.6\,\lYuk+8.80\,\lpr   ,\nl
\eettLL &=&   -7.07\,\Ls+\left(4.90\ltu-16.3\lus\right)\ls+0.58\,\lZ \nl
&&{}+21.2\,\lsl-5.30\,\lYuk-12.2\,\lpr,
\eeqar
and for bottom-quark-pair production we obtain
\newcommand{\eebbRR}{\de^{\rR\rR,\ew}_{\Pep \Pem \rightarrow \bar{\Pb}\Pb}}
\newcommand{\eebbRL}{\de^{\rR\rL,\ew}_{\Pep \Pem \rightarrow \bar{\Pb}\Pb}}
\newcommand{\eebbLR}{\de^{\rL\rR,\ew}_{\Pep \Pem \rightarrow \bar{\Pb}\Pb}}
\newcommand{\eebbLL}{\de^{\rL\rL,\ew}_{\Pep \Pem \rightarrow \bar{\Pb}\Pb}}
\beqar \label{bpair}
\eebbRR &=&   -1.43\,\Ls-1.72\left(\ltu\right)\ls+0.16\,\lZ+4.29\,\lsl+8.80\,\lpr   ,\nl
\eebbRL &=&   -4.68\,\Ls+0.86\left(\ltu\right)\ls+0.67\,\lZ+14.0\,\lsl-5.30\,\lYuk+8.80\,\lpr   ,\nl
\eebbLR &=&   -3.82\,\Ls-0.86\left(\ltu\right)\ls+0.24\,\lZ+11.5\,\lsl+8.80\,\lpr   ,\nl
\eebbLL &=&   -7.07\,\Ls-\left(4.04\ltu+19.8\lts\right)\ls+0.75\,\lZ \nl
&&{}+21.2\,\lsl-5.30\,\lYuk-16.6\,\lpr.
\eeqar 
The Mandelstam variables are defined
as usual, \ie
$s=(p_\Pep+p_\Pem)^2$, $t=(p_\Pep-p_{\bar{f}})^2$ and $u=(p_\Pep-p_{f})^2$.
Note that the corrections to light quark-pair production $f=\Pu,\Pc\,
(\Pd,\Ps)$ are obtained from the results for heavy quarks $f=\Pt\,
(\Pb)$ by omitting the Yukawa contributions. Independently of the
process and of the 
chirality,
the DL and SL terms appear in the
combination $(-\Ls+3\lsl)$, so that the negative DL contribution
becomes dominating only above $400 \GeV$, and at $\sqrt{s}=1 \TeV$ the
cancellation between SL and DL corrections is still important. The
$\SUtwo$ interaction, which is stronger than the $\Uone$ interaction, 
generates large corrections for left-handed fermions.  Also the PR
logarithms show a strong chirality dependence: the RR and RL
transitions receive positive corrections from the running of the
abelian $\Uone$ coupling, whereas the LL transition is dominated by
the non-abelian $\SUtwo$ interaction and receives negative PR
corrections.
\end{list}

\section{Production of \PW-boson pairs in \Pep\Pem  annihilation}
\label{eeWW}
We consider the polarized scattering process\footnote{The momenta and
  fields of the initial states are incoming, and those of the
  final states are outgoing.}  
$\Pe^+_\kappa\Pe^-_\kappa \rightarrow \PW^+_{\la_+} \PW^-_{\la_-}$, where
$\kappa=\rR,\rL$ is the
electron chirality, and $\la_\pm=0,\pm$ represent the gauge-boson
helicities. In the high-energy limit only the following helicity
combinations are non-suppressed 
\cite{eeWWhe,Denn1}: the purely longitudinal final state
$(\la_+,\la_-)=(0,0)$, which we denote by $(\la_+,\la_-)=(\rL,\rL)$,
and the purely transverse and opposite final state  $(\la_+,\la_-)=(\pm,\mp)$,
which we denote by $(\la_+,\la_-)=(\rT,\rT)$. All these final states,
can be written as $(\la_+,\la_-)=(\la,-\la)$.
The Mandelstam
variables are $s=(p_\Pep+p_\Pem)^2$, $t=(p_\Pep-p_\PWp)^2\sim
-s(1-\cos{\theta})/2$, and $u=(p_\Pep-p_\PWm)^2\sim
-s(1+\cos{\theta})/2$, where $\theta$ is the angle between $\Pep$ and
$\PWp$.  

\begin{list}{\bf \arabic{listnumber}.\hspace{1mm}}{\usecounter{listnumber}
\setlength{\leftmargin}{0mm} \setlength{\labelsep}{0mm}
\setlength{\itemindent}{4mm}
}

\item{\bf Born matrix elements}\nopagebreak

The Born amplitude gets contributions of the $s$- and
$t$-channel diagrams in \reffi{WWborn} and reads
\beqar \label{borneeww} 
\M_{0}^{ \Pe^+_\kappa\Pe^-_\kappa \rightarrow \PW^+_\rL\PW^-_\rL}&=&
e^2\NCep \frac{A_s}{s},\nl 
\M_{0}^{\Pe^+_\rL\Pe^-_\rL \rightarrow
  \PW^+_\rT\PW^-_\rT}&=&\frac{e^2}{2\sw^2}\frac{A_t}{t} 
\eeqar
up to terms of order $\MW^2/s$, where $\NC$ is defined in \refeq{NCcorr}.
The amplitude involving
longitudinal gauge bosons $\PW_\rL$ is expressed by the amplitude
involving would-be Goldstone bosons $\phi^\pm$ and is dominated by the
$s$-channel exchange of neutral gauge bosons. The
amplitude for transverse gauge-boson production is dominated by the
$t$-channel contribution, which involves only the $\SUtwo$ interaction.
Therefore, it is non-vanishing only for left-handed electrons in the
initial state.
\begin{figure}
\begin{center}
\begin{picture}(300,80)
\put(0,0){
\begin{picture}(120,80)
\ArrowLine(20,30)(0,60)
\ArrowLine(0,0)(20,30)
\Vertex(20,30){2}
\Photon(20,30)(80,30){2}{6}
\Vertex(80,30){2}
\DashLine(100,0)(80,30){2}
\DashLine(80,30)(100,60){2}
\Text(-5,0)[r]{$\Pem$}
\Text(-5,60)[r]{$\Pep$}
\Text(105,0)[l]{$\phi^-$}
\Text(105,60)[l]{$\phi^+$}
\Text(50,35)[b]{$A,Z$}
\end{picture}}
\put(180,0){
\begin{picture}(120,80)
\ArrowLine(40,45)(10,60)
\ArrowLine(40,15)(40,45)
\ArrowLine(10,0)(40,15)
\Vertex(40,15){2}
\Vertex(40,45){2}
\Photon(70,0)(40,15){2}{3}
\Photon(40,45)(70,60){-2}{3}
\Text(5,0)[r]{$\Pem$}
\Text(5,60)[r]{$\Pep$}
\Text(47,30)[l]{$\nu_\Pe$}
\Text(75,0)[l]{$\PW^-_\rT$}
\Text(75,60)[l]{$\PW^+_\rT$}
\end{picture}}
\end{picture}
\end{center}
\caption{Dominant  lowest-order diagrams for
  $\Pe^+\Pe^-\to\phi^+\phi^-$ and $\Pe^+\Pe^-\to \PW^+_\rT \PW^-_\rT$}
\label{WWborn}
\end{figure}

\item{\bf Leading soft-collinear corrections}\nopagebreak

The leading DL corrections \refeq{deSC} yield 
\beq\label{SCWW}
\de^{\SC}_{\Pe^+_\kappa\Pe^-_\kappa \rightarrow \PW^+_\la\PW^-_{-\la}}=-\sum_{\varphi=\Pe^-_\kappa,\PW^-_\la} \left[\cew_\varphi\Ls-2(I^Z_\varphi)^2 \log{\frac{\MZ^2}{\MW^2}}\,\lZ
+Q_\varphi^2\Lemphi \right].
\eeq
Here and in the following formulas,
for longitudinally polarized gauge bosons $\PW^\pm_\rL$ the quantum
numbers of the would-be Goldstone bosons $\phi^\pm$ have to be used.
\item{\bf Subleading soft-collinear corrections}\nopagebreak

The $\SS$ corrections are obtained by applying \refeq{4fsubdl} to the
crossing symmetric process $\Pe^+_\kappa\Pe^-_\kappa
\PW^-_\la\PW^+_{-\la}\rightarrow 0$.  The contribution of the neutral
gauge bosons $\NB=A,Z$ gives
\beqar 
\sum_{\NB=A,Z}\de^{\NB,\SS}_{\Pe^+_\kappa\Pe^-_\kappa \rightarrow \PW^+_\la\PW^-_{-\la}}
=-4\NCew\ls\ltu -\frac{\alpha}{\pi}Q_{\Pem}Q_{\PW^-}\log{\frac{\MW^2}{\la^2}}
\ltu,
\eeqar
and corresponds to the result \refeq{4fncss} for 4-fermion $s$-channel NC processes. 
The contribution of soft $\PW^\pm$ 
bosons to \refeq{4fsubdl} yields
\beqar \label{eewwpmssc}
\sum_{V^a=W^\pm}\de^{V^a,\SS} \M^{\Pe^+_\kappa\Pe^-_\kappa\phi^-\phi^+}
&=&-\frac{2\ls\de_{\kappa\rL}}{\sqrt{2}\sw}\sum_{S=H,\chi}\left[
I^+_S \M_0^{\bar{\nu}_\kappa\Pe^-_\kappa S \phi^+}
-I^-_S\M_0^{\Pe^+_\kappa\nu_\kappa\phi^-S}
\right]
\log{\frac{|t|}{s}},\nl
\sum_{V^a=W^\pm}\de^{V^a,\SS} \M^{\Pe^+_\rL\Pe^-_\rL\PW_\rT^-\PW_\rT^+}
&=&-\frac{2\ls}{\sqrt{2}\sw}\sum_{\NB=A,Z}\left[
I^+_\NB\M_0^{\bar{\nu}_\rL\Pe^-_\rL \NB_\rT \PW_\rT^+}
-I^-_\NB\M_0^{\Pe^+_\rL\nu_\rL\PW_\rT^-\NB_\rT}
\right]
\log{\frac{|t|}{s}},\nl
\eeqar
where, depending on the polarization of the final states, one has to
use the non-diagonal $\PW^\pm$ couplings to would-be Goldstone bosons
($I^\pm_S$) defined in \refeq{scapmcoup} or the $\PW^\pm$ couplings
to gauge bosons ($I^\pm_\NB$) defined in \refeq{gaupmcoup}.  The
$\SUtwo$-transformed Born matrix elements on the rhs  of
\refeq{eewwpmssc} have to be evaluated explicitly. For would-be Goldstone
bosons, we have $s$-channel CC amplitudes
\beq
\M_0^{\bar{\nu}_\kappa\Pe^-_\kappa S \phi^+}=-e^2I^-_S\frac{\de_{\kappa\rL}}{\sqrt{2}\sw} \frac{A_s}{s},\qquad
\M_0^{\Pe^+_\kappa\nu_\kappa\phi^-S}=e^2I^+_S\frac{\de_{\kappa\rL} }{\sqrt{2}\sw} \frac{A_s}{s},
\eeq
similar to the NC Born amplitude in \refeq{borneeww}, whereas for
transverse gauge bosons we have
\beq \label{borntransf}
\M_0^{\bar{\nu}_\kappa\Pe^-_\kappa \NB_\rT \PW_\rT^+}=
\M_0^{\Pe^+_\kappa\nu_\kappa\PW_\rT^-\NB_\rT}=
e^2\frac{\de_{\kappa\rL}}{\sqrt{2}\sw} \left(I^\NB_{\nu_\kappa}\frac{A_t}{t}+I^\NB_{e^-_\kappa}\frac{A_u}{u}\right),
\eeq
where  $A_t=A_u$ 
up to mass-suppressed contributions. In contrast to
\refeq{borneeww}, the transformed amplitude \refeq{borntransf}
receives contributions from both $t$ and $u$ channels.  Expressing
\refeq{eewwpmssc} as relative corrections to the Born matrix elements
we obtain
\beqar
\sum_{V^a=W^\pm}\de^{V^a,\SS}_{\Pe^+_\kappa\Pe^-_\kappa\rightarrow \PW_\rL^+\PW_\rL^-}
&=&-\ls\frac{\de_{\kappa\rL}}{\sw^4 \NC_{\Pe^-_\rL\phi^-}}\log{\frac{|t|}{s}},\nl
\sum_{V^a=W^\pm}\de^{V^a,\SS}_{\Pe^+_\rL\Pe^-_\rL\rightarrow \PW_\rT^+\PW_\rT^-}
&=&-\frac{2}{\sw^2}\left(1-\frac{t}{u}\right)\ls\log{\frac{|t|}{s}}.
\eeqar 

\item{\bf SL corrections associated to external particles}\nopagebreak

The SL corrections can be read off from \refeq{deccfer},
\refeq{deccWT}, and \refeq{longeq:coll},
\beqar
\de^\cc_{\Pe^+_\kappa\Pe^-_\kappa \rightarrow \PW^+_\rL\PW^-_\rL} &=& \left[3 \cew_{\Pe_\kappa}+ 4\cew_\Phi\right]\lsl-\frac{3}{2\sw^2}\frac{m^2_t}{\MW^2}\lYuk +
\frac{\alpha}{4\pi}
\sum_{\varphi=\Pe,\PW}
\left(\log{\frac{\MW^2}{m^2_{\varphi}}}
+2\log{\frac{\MW^2}{\la^2}}\right),\nl
\de^\cc_{\Pe^+_\rL\Pe^-_\rL \rightarrow \PW^+_\rT\PW^-_\rT} &=& \left[3 \cew_{\Pe_\rL}+\bew_{W}\right]\lsl+
\frac{\alpha}{4\pi}
\sum_{\varphi=\Pe,\PW}
\left(\log{\frac{\MW^2}{m^2_{\varphi}}}
+2\log{\frac{\MW^2}{\la^2}}\right).
\eeqar
Despite of their different origin, the $\lsl$ contributions for
longitudinal and transverse gauge bosons have similar numerical values
$4\cew_\Phi=14.707$ and $\bew_{W}=14.165$. 
The strong W-polarization dependence of $\de^\cc$ is due
to the large Yukawa contributions occurring only for longitudinal
gauge bosons.

\item{\bf Logarithms from parameter renormalization}\nopagebreak

The PR logarithms are obtained from the renormalization of
\refeq{borneeww} and read according to
\refeq{chargerenorm}
--\refeq{gCTs}
\beqar\label{RGWW}
\de^\pre_{\Pe^+_\kappa\Pe^-_\kappa \rightarrow \PW^+_\rL\PW^-_\rL} &=&\left[\frac{\sw}{\cw}\bew_{AZ}\deNCep -\bew_{AA}\right]\lpr+\Dealpha
,\nl
\de^\pre_{\Pe^+_\rL\Pe^-_\rL \rightarrow \PW^+_\rT\PW^-_\rT} &=&-\bew_{W}\lpr+\Dealpha
,
\eeqar
where $\Delta$ is defined in \refeq{NCren}. 
Note that for transverse polarizations, the 
symmetric-electroweak 
parts
of the PR corrections 
($-\bew_W\lpr$) 
and the collinear SL corrections originating from external gauge bosons 
($\bew_W\lsl$) 
cancel. 
As illustrated in \refapp{app:transvRG}, this kind of cancellation
takes place for all processes with production of arbitrary many charged or
neutral transverse gauge bosons in fermion--antifermion annihilation.  

The results \refeq{SCWW}--\refeq{RGWW} can be compared with those of
\citere{eeWWhe}. After subtracting the real soft-photonic corrections
from the results of \citere{eeWWhe} we find complete agreement for the
logarithmic corrections.

\item{\bf Numerical evaluation}\nopagebreak

The coefficients for the various 
electroweak logarithmic contributions to the relative
corrections 
$\de^{\kappa\la}_{\Pe^+\Pe^-\rightarrow \PW^+\PW^-}$ 
read
\newcommand{\eeWWLL}{\de^{\rL\rL,\ew}_{\Pe^+\Pe^-\rightarrow \PW^+\PW^-}}
\newcommand{\eeWWRL}{\de^{\rR\rL,\ew}_{\Pe^+\Pe^-\rightarrow \PW^+\PW^-}}
\newcommand{\eeWWLT}{\de^{\rL\rT,\ew}_{\Pe^+\Pe^-\rightarrow \PW^+\PW^-}}
\beqar
\eeWWLL &=&   -7.35\,\Ls-\left(5.76\ltu+13.9\lts\right)\ls+0.45\,\lZ \nl 
&&{}+25.7\,\lsl-31.8\,\lYuk-9.03\,\lpr   ,\nl
\eeWWRL &=&   -4.96\,\Ls-2.58\left(\ltu\right)\ls+0.37\,\lZ \nl &&
{}+18.6\,\lsl-31.8\,\lYuk+8.80\,\lpr   ,\nl
\eeWWLT &=&   -12.6\,\Ls-8.95\left[\ltu+\left(1-\frac{t}{u}\right)\lts\right]\ls+1.98\,\lZ \nl &&{}+25.2\,\lsl-14.2\,\lpr.
\eeqar 
Recall that the 
pure electromagnetic contributions have been omitted.
These correction factors are shown in  \reffis{plotWWan} and
\ref{plotWWen} as a function of the scattering angle and the energy, respectively.
If the electrons are left-handed, large negative DL and PR
corrections originate from the $\SUtwo$ interaction. Instead, for
right-handed electrons the DL corrections are smaller, and the PR
contribution is positive. 
For transverse \PW~bosons, there are no Yukawa contributions and the
other contributions are in general larger than for longitudinal
\PW~bosons.  Nevertheless, for energies around $1\TeV$, the
corrections are similar.
Finally, note that the angular-dependent contributions are
very important for the LL and LT corrections: at $\sqrt{s}\approx 1\TeV$
they vary from $+15\%$ to $-5\%$ for scattering angles
$30^\circ<\theta<150^\circ$, whereas the angular-dependent part of the
RL corrections remains between $\pm 2\%$.

\begin{figure}
\centerline{
\setlength{\unitlength}{1cm}
\begin{picture}(10,8.3)
\put(0,0){\includegraphics{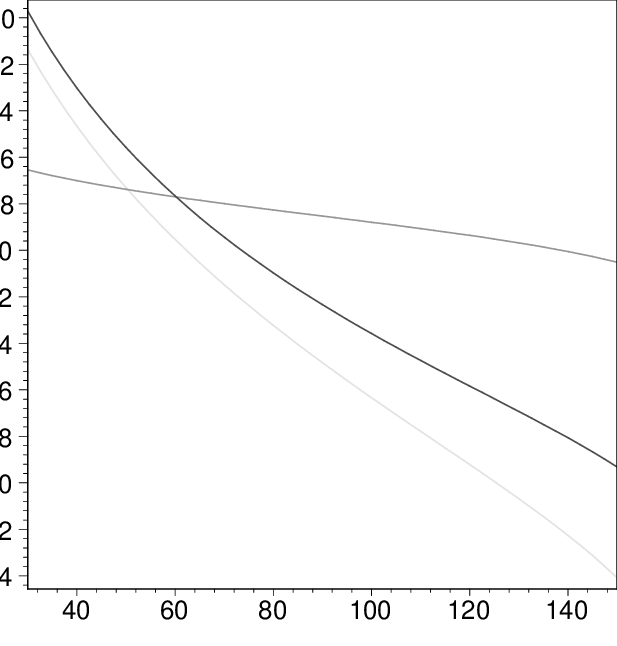}}
\put(2.5,0){\makebox(6,0.5)[b]{$\theta\,[^\circ]$}}
\put(-2.5,4){\makebox(1.5,1)[r]{$\delta^\ew\,[\%]$}}
\put(10,4.5){\makebox(1.5,1)[r]{$\rR\rL$}}
\put(10,2.0){\makebox(1.5,1)[r]{$\rL\rL$}}
\put(10,0.5){\makebox(1.5,1)[r]{$\rL\rT$}}
\end{picture}}
\caption[WWang]{Dependence of the electroweak correction factor
  $\de^{\ew}_{\Pe_\kappa^+\Pe_\kappa^-\rightarrow \PW_\la^+\PW_{-\la}^-}$ on
  the scattering angle $\theta$ at $\sqrt{s}=1\TeV$ for polarizations
  $\kappa\la=\rR\rL$, $\rL\rL$, and $\rL\rT$} 
\label{plotWWan}
\end{figure}%
\begin{figure}
  \centerline{ \setlength{\unitlength}{1cm}
\begin{picture}(10,8.3)
\put(0,0){\includegraphics{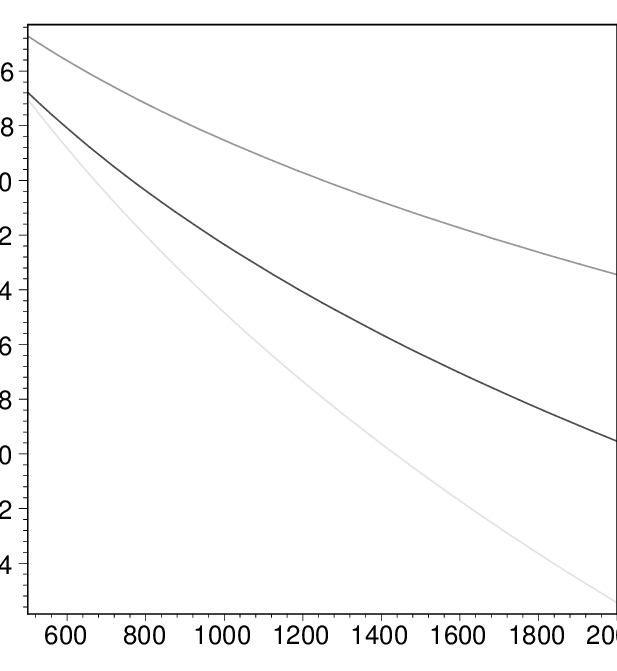}}
\put(2.5,0){\makebox(6,0.5)[b]{$\sqrt{s}\,[\GeV]$}}
\put(-2.5,4){\makebox(1.5,1)[r]{$\delta^\ew\,[\%]$}}
\put(10,4.5){\makebox(1.5,1)[r]{$\rR\rL$}}
\put(10,2.6){\makebox(1.5,1)[r]{$\rL\rL$}}
\put(10,0.8){\makebox(1.5,1)[r]{$\rL\rT$}}
\end{picture}}
\caption[WWang]{Dependence of the electroweak correction factor
  $\de^{\ew}_{\Pe_\kappa^+\Pe_\kappa^-\rightarrow \PW_\la^+\PW_{-\la}^-}$ on
  the energy  $\sqrt{s}$ at $\theta=90^\circ$ for polarizations
  $\kappa\la=\rR\rL$, $\rL\rL$, and $\rL\rT$} 
\label{plotWWen}
\end{figure}
\end{list}

\section{Production of neutral gauge-boson pairs in \Pep\Pem an\-ni\-hila\-tion}
We consider the polarized scattering process $\Pe^+_\kappa\Pe^-_\kappa
\rightarrow \NB_\rT^1 \NB_\rT^2$ with incoming electrons of chirality $\kappa=\rR,\rL$
and outgoing gauge bosons $\NB^k=\PA,\PZ$. The amplitude is
non-suppressed only for transverse and opposite gauge-boson
polarizations $(\la_1,\la_2)=(\pm,\mp)$ \cite{Denn2}. 
\begin{list}{\bf \arabic{listnumber}.\hspace{1mm}}{\usecounter{listnumber}
\setlength{\leftmargin}{0mm} \setlength{\labelsep}{0mm}
\setlength{\itemindent}{4mm}
}

\item{\bf Born matrix elements}\nopagebreak

In lowest order
the $t$- and $u$-channel diagrams (\reffi{ZZborn}) yield
\beq \label{borneevv} 
\M_{0}^{ \Pe^+_\kappa\Pe^-_\kappa \rightarrow \NB_\rT^1\NB_\rT^2}= e^2 I^{\NB^1}_{\Pe^-_\kappa} I^{\NB^2}_{\Pe^-_\kappa}\left[ \frac{A_t}{t}+\frac{A_u}{u}\right]
\eeq
up to terms of order $\MW^2/s$, where the Mandelstam variables are
defined as in \refse{eeWW}. 
In the ultra-relativistic limit  the
amplitude is symmetric with respect to exchange of the gauge bosons,
and if we restrict ourselves to the combinations of helicities that are 
not suppressed we have
\beq \label{eevvchannel}
A_t=A_u,
\eeq 
up to mass-suppressed contributions.
\begin{figure}
\begin{center}
\begin{picture}(300,80)
\put(0,0){
\begin{picture}(120,80)
\ArrowLine(40,45)(10,60)
\ArrowLine(40,15)(40,45)
\ArrowLine(10,0)(40,15)
\Vertex(40,15){2}
\Vertex(40,45){2}
\Photon(70,0)(40,15){2}{3}
\Photon(40,45)(70,60){-2}{3}
\Text(5,0)[r]{$\Pem$}
\Text(5,60)[r]{$\Pep$}
\Text(75,0)[l]{$\NB^2$}
\Text(75,60)[l]{$\NB^1$}
\end{picture}}
\put(180,0){
\begin{picture}(120,80)
\ArrowLine(40,45)(10,60)
\ArrowLine(40,15)(40,45)
\ArrowLine(10,0)(40,15)
\Vertex(40,15){2}
\Vertex(40,45){2}
\Photon(70,0)(40,45){2}{5}
\Photon(40,15)(70,60){-2}{5}
\Text(5,0)[r]{$\Pem$}
\Text(5,60)[r]{$\Pep$}
\Text(75,0)[l]{$\NB^2$}
\Text(75,60)[l]{$\NB^1$}
\end{picture}}
\end{picture}
\end{center}
\caption{Lowest-order diagrams for $\Pe^+\Pe^-\to \NB^1\NB^2$}
\label{ZZborn}
\end{figure}

\item{\bf Leading soft-collinear corrections}\nopagebreak

The leading DL corrections read [cf. \refeq{deSC}]
\beqar
\lefteqn{\de^{\SC}\M^{\Pe^+_\kappa\Pe^-_\kappa \rightarrow
    \NB_\rT^1\NB_\rT^2}=}\quad\nl
&&{}-\left[\cew_{\Pe_\kappa}\Ls-2(I^Z_{\Pe_\kappa})^2
    \log{\frac{\MZ^2}{\MW^2}}\,\lZ+\Leme
  \right]\M_{0}^{\Pe^+_\kappa\Pe^-_\kappa \rightarrow \NB_\rT^1\NB_\rT^2}\nl
  &&{}-\frac{1}{2}\left[\cew_{\NB^{'}\NB^1}\M_{0}^{\Pe^+_\kappa\Pe^-_\kappa
      \rightarrow
      \NB_\rT^{'}\NB_\rT^2}+\cew_{\NB^{'}\NB^2}\M_{0}^{\Pe^+_\kappa\Pe^-_\kappa
      \rightarrow \NB_\rT^1\NB_\rT^{'}} \right] \Ls
\eeqar
with a  non-diagonal contribution associated with the external neutral gauge bosons. Using
\beq \label{cewIid}
\cew_{\NB'\NB}I^{\NB'}=\frac{2}{\sw^2} U_{\NB W^3}(\thw) \tilde{I}^{W^3}=\frac{2}{\sw^2} U_{\NB W^3}(\thw)\frac{T^3}{\sw},
\eeq
where $U_{\NB \sNB}(\thw)$ is the Weinberg rotation defined in 
\refeq{defweinrot},
we can derive a correction relative to the Born matrix element,
\beqar \label{vvssc}
\de^{\SC}_{\Pe^+_\kappa\Pe^-_\kappa \rightarrow \NB_\rT^1\NB_\rT^2}&=&-\left[\cew_{\Pe_\kappa}\Ls-2(I^Z_{\Pe_\kappa})^2 \log{\frac{\MZ^2}{\MW^2}}\,\lZ+\Leme \right]\nl
&&{}-\frac{T^3_{\Pe^-_\kappa}}{\sw^3}\sum_{k=1,2} \frac{U_{\NB^k W^3}(\thw)}{I^{\NB^k}_{\Pe^-_\kappa}} \Ls.
\eeqar
Note that only the $\SUtwo$ component of the neutral gauge bosons is
self-interacting and can exchange soft gauge bosons. For this reason, only  left-handed electrons ($T^3\neq 0$) yield a contribution to \refeq{cewIid} and to the corresponding term in \refeq{vvssc}.

\item{\bf Subleading soft-collinear corrections}\nopagebreak

Angular-dependent DL corrections  \refeq{4fsubdl}
arise only from the exchange of soft $\PW^\pm$ bosons between initial and final states, and using the non-diagonal couplings  \refeq{gaupmcoup} we obtain
\beqar \label{eevvpmssc}
\de^{\SS} \M^{\Pe^+_\kappa\Pe^-_\kappa \NB^1_\rT \NB^2_\rT}
&=&\frac{2\ls\de_{\kappa\rL}}{\sqrt{2}\sw}\left\{\left[
I^+_{\NB^1}\M_0^{\bar{\nu}_\kappa\Pe^-_\kappa \PW_\rT^+\NB^2_\rT}
-I^-_{\NB^2}\M_0^{\Pe^+_\kappa\nu_\kappa \NB^1_\rT\PW_\rT^-}
\right]\log{\frac{|t|}{s}}\right.\nl
&&\left.+\left[
I^+_{\NB^2}\M_0^{\bar{\nu}_\kappa\Pe^-_\kappa \NB^1_\rT\PW_\rT^+}
-I^-_{\NB^1}\M_0^{\Pe^+_\kappa\nu_\kappa \PW_\rT^-\NB^2_\rT}
\right]
\log{\frac{|u|}{s}}\right\}.
\eeqar
The $\SUtwo$-transformed Born matrix elements on the rhs 
are given by
\beq
\M_0^{\bar{\nu}_\kappa\Pe^-_\kappa \PW_\rT^+ \NB_\rT}=
\M_0^{\Pe^+_\kappa\nu_\kappa \NB_\rT \PW_\rT^-}=
e^2\frac{\de_{\kappa\rL}}{\sqrt{2}\sw}
\left(I^\NB_{\Pe^-_\kappa}\frac{A_t}{t}+I^\NB_{\nu_\kappa}\frac{A_u}{u}\right)
\eeq
and by \refeq{borntransf} with $A_t=A_u$. Expressing the correction
\refeq{eevvpmssc} relative to the Born matrix element \refeq{borneevv},
we obtain
\beq 
\de^{\SS}_{\Pe^+_\kappa\Pe^-_\kappa\rightarrow \NB^1_\rT \NB^2_\rT}=
\frac{\de_{\kappa\rL}}{\sw^2}\ls \sum_{k=1}^2 \sum_{r=t,u} \frac{I^{\NB^k}_{\PWm}}{I^{\NB^k}_{\Pe^-_\rL}}\left(\frac{r'}{s}+\frac{r}{s}\frac{I^{\NB^{k'}}_{\nu_\rL}}{I^{\NB^{k'}}_{\Pe^-_\rL}}\right)\log{\frac{|r|}{s}},
\eeq
where $r'=(t,u)$ for  $r=(u,t)$, and  $k'=(1,2)$ for  $k=(2,1)$. 

\item{\bf SL corrections associated to  external particles}\nopagebreak

Using \refeq{deccfer} and \refeq{deccWT} we obtain for the SL
corrections relative to the Born matrix element 
\beq
\de^\cc_{\Pe^+_\kappa\Pe^-_\kappa \rightarrow \NB_\rT^1\NB_\rT^2}= 3\cew_{\Pe_\kappa}\lsl
+\frac{\alpha}{4\pi}
\left(\log{\frac{\MW^2}{m^2_{\Pe}}}
    +2\log{\frac{\MW^2}{\la^2}}\right)
+\de^\cc_{\NB_\rT^1}+\de^\cc_{\NB_\rT^2}
\eeq
with
\beqar\label{N1N2deC}
\de^\cc_A &:=&\de^\cc_{AA}=\frac{1}{2}\bew_{AA}\lsl -\frac{1}{2}\Dealpha 
,\nl
\de^\cc_Z &:=&\de^\cc_{ZZ}+\de^\cc_{AZ}\frac{\M_0^{\Pe^+_\kappa\Pe^-_\kappa \rightarrow A_\rT \NB_\rT}}{\M_0^{\Pe^+_\kappa\Pe^-_\kappa \rightarrow Z_\rT \NB_\rT}} 
=\left[\frac{1}{2}\bew_{ZZ}+\bew_{AZ}\frac{I^A_{\Pe^-_\kappa}}{I^Z_{\Pe^-_\kappa}}\right]\lsl.
\eeqar

\item{\bf Logarithms from parameter renormalization}\nopagebreak

The PR logarithms result from the renormalization of \refeq{borneevv}.
As shown in \refapp{app:transvRG}, they are opposite to the 
collinear SL corrections \refeq{N1N2deC}
up to pure electromagnetic logarithms. 
Relative to the Born matrix element they read
\beq
\de^\pre_{\Pe^+_\kappa\Pe^-_\kappa \rightarrow \NB_\rT^1\NB_\rT^2}= \de^\pre_{\NB_\rT^1}+\de^\pre_{\NB_\rT^2}
\eeq
with (see \refeq{PRpluscc})
\beq
\de^\pre_A :=-\de^\cc_A,\qquad
\de^\pre_Z :=-\de^\cc_Z +\frac{1}{2}\Dealpha.
\eeq

\item{\bf Numerical evaluation}\nopagebreak

For right-handed electrons, $\kappa=\rR$, the various 
electroweak
logarithmic contributions to the relative corrections
$\de^{\kappa\rT}_{\Pe^+\Pe^-\rightarrow \NB^1\NB^2}$ give
\newcommand{\eeAAL}{\de^{\rL\rT,\ew}_{\Pe^+\Pe^-\rightarrow \PA\PA}}
\newcommand{\eeAAR}{\de^{\rR\rT,\ew}_{\Pe^+\Pe^-\rightarrow \PA\PA}}
\newcommand{\eeAZL}{\de^{\rL\rT,\ew}_{\Pe^+\Pe^-\rightarrow \PA\PZ}}
\newcommand{\eeAZR}{\de^{\rR\rT,\ew}_{\Pe^+\Pe^-\rightarrow \PA\PZ}}
\newcommand{\eeZAL}{\de^{\rL\rT,\ew}_{\Pe^+\Pe^-\rightarrow \PZ\PA}}
\newcommand{\eeZAR}{\de^{\rR\rT,\ew}_{\Pe^+\Pe^-\rightarrow \PZ\PA}}
\newcommand{\eeZZL}{\de^{\rL\rT,\ew}_{\Pe^+\Pe^-\rightarrow \PZ\PZ}}
\newcommand{\eeZZR}{\de^{\rR\rT,\ew}_{\Pe^+\Pe^-\rightarrow \PZ\PZ}}
\newcommand{\UTTU}{F_1(t)}
\newcommand{\TTUU}{F_2(t)}
\beqar
\eeAAR &=&   -1.29\,\Ls+0.15\,\lZ+0.20\,\lsl+3.67\,\lpr   ,\nl
\eeAZR &=&   -1.29\,\Ls+0.15\,\lZ-11.3\,\lsl+15.1\,\lpr   ,\nl
\eeZZR &=&   -1.29\,\Ls+0.15\,\lZ-22.8\,\lsl+26.6\,\lpr    .
\eeqar
Note that there is no angular dependence. The PR contributions are
numerically compensated by the SL and DL Sudakov contributions, and at
$\sqrt{s}=1 \TeV$ the electroweak logarithmic corrections are less than
1$\%$. For left-handed electrons, we find
\beqar
\eeAAL &=&-8.15\,\Ls+8.95\UTTU\ls+0.22\,\lZ+7.36\,\lsl+3.67\,\lpr   ,\\
\eeAZL &=& -12.2\,\Ls+(17.0\UTTU-8.09\TTUU)\ls+0.22\,\lZ+28.1\,\lsl-17.1\,\lpr ,\nl
\eeZZL &=&-16.2\,\Ls+(25.1\UTTU-45.4\TTUU)\ls+0.22\,\lZ+48.9\,\lsl-37.9\,\lpr
\nn
\eeqar
with the $(t,u)$-symmetric angular-dependent functions
\beq
\UTTU:=\frac{u}{s}\lts+\frac{t}{s}\lus ,\qquad
\TTUU:= \frac{t}{s}\lts+\frac{u}{s}\lus.
\eeq
For left-handed electrons all contributions are larger than for
right-handed electrons owing to the $\SUtwo$ interaction. The
non-abelian effects are particularly strong for Z-boson-pair production
(see Figs. \ref{plotAZan}, \ref{plotAZen}), where the total
corrections are almost $-25\%$ for $\sqrt{s}=1 \TeV$ and
$\theta=90^\circ$. The angular-dependent contribution is
forward--backward symmetric, and for $\PZ\PZ$ production it
varies from $+15\%$ to $-5\%$ for scattering angles
$30^\circ<\theta<90^\circ$.

\begin{figure}
\centerline{
\setlength{\unitlength}{1cm}
\begin{picture}(10,8.3)
\put(0,0){\includegraphics{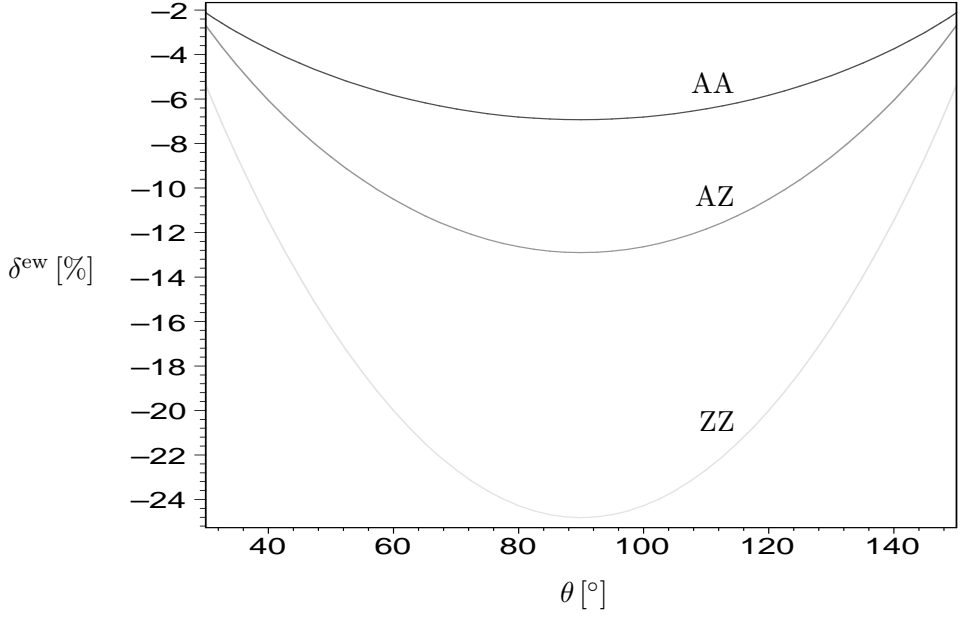}}
\put(2.5,0){\makebox(6,0.5)[b]{$\theta\,[^\circ]$}}
\put(-2.5,4){\makebox(1.5,1)[r]{$\delta^\ew\,[\%]$}}
\put(6,6.5){\makebox(1.5,1)[r]{$\PA\PA$}}
\put(6,5){\makebox(1.5,1)[r]{$\PA\PZ$}}
\put(6,2){\makebox(1.5,1)[r]{$\PZ\PZ$}}
\end{picture}}
\caption[AZLang]{Angular dependence of the electroweak corrections to
  $\Pe_\rL^+\Pe_\rL^-\rightarrow \PA\PA,\PA\PZ,\PZ\PZ$ at $\sqrt{s}=1\TeV$}
\label{plotAZan}
\end{figure}%
\begin{figure}
  \centerline{ \setlength{\unitlength}{1cm}
\begin{picture}(10,8.3)
\put(0,0){\includegraphics{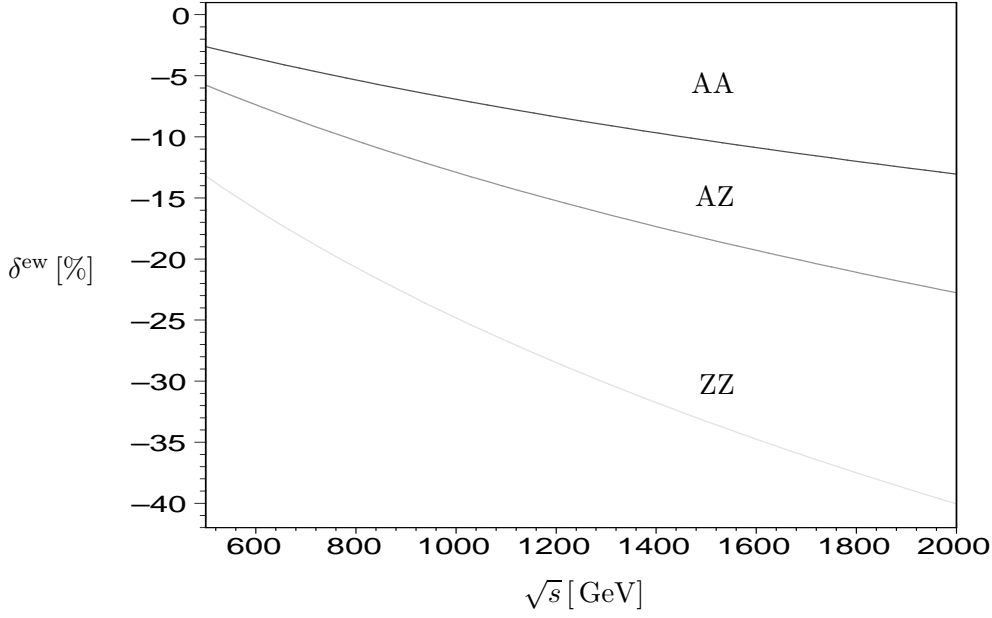}}
\put(2.5,0){\makebox(6,0.5)[b]{$\sqrt{s}\,[\GeV]$}}
\put(-2.5,4){\makebox(1.5,1)[r]{$\delta^\ew\,[\%]$}}
\put(6,6.5){\makebox(1.5,1)[r]{$\PA\PA$}}
\put(6,5){\makebox(1.5,1)[r]{$\PA\PZ$}}
\put(6,2.5){\makebox(1.5,1)[r]{$\PZ\PZ$}}
\end{picture}}
\caption[AZLen]{Energy dependence of the electroweak corrections to
  $\Pe_\rL^+\Pe_\rL^-\rightarrow \PA\PA,\PA\PZ,\PZ\PZ$ at $\theta=90^\circ$}
\label{plotAZen}
\end{figure}
\end{list}

\section{Production of $\mathrm{W}^+\mathrm{A}$ and $\mathrm{W}^+\mathrm{Z}$ in $\bar{\mathrm{d}}\mathrm{u}$ an\-ni\-hila\-tion}
In this section, we consider the logarithmic corrections to the partonic subprocesses 
\newcommand{\duwz}{\bar{\Pd}_\rL\Pu_\rL \rightarrow \PW^+_{\la_\PW} \PZ_{\la_\PZ}}
\newcommand{\duwn}{\bar{\Pd}_\rL\Pu_\rL \rightarrow \PW^+_{\la_\PW} N_{\la_N}}
\beq\label{duprocess}
\duwn,\qquad N=\PA,\PZ,
\eeq
of the proton--proton scattering processes  $\Pp\Pp\to \PWp\PA$ and $\Pp\Pp\to \PWp\PZ$. 
The following results have been used in  \citere{Accomando:2001fn} to obtain the logarithmic corrections to the complete  hadronic processes $\Pp\Pp\to \PWpm\gamma\to l\nu_l\gamma$ and $\Pp\Pp\to\PWpm\PZ \to l\nu_l \bar{l}'l'$.

In the partonic processes \refeq{duprocess},
the charged gauge boson \PWp~  in the final state can be created only from left-handed anti-down and up quarks $\bar{\Pd}_\rL$, $\Pu_\rL$ in the initial state through the $\SUtwo$ interaction.
In the high-energy limit, as in the previous examples,  only the following combinations of the  gauge-boson helicities
 $\la_{\PW,N}=0,\pm$ are non-suppressed: 
  the purely transverse and opposite final state  $(\la_\PW,\la_N)=(\pm,\mp)$, which we denote by $(\la_\PW,\la_N)=(\rT,\rT)$, and, in the case of $\PWp\PZ$ production,  the purely longitudinal final state
$(\la_\PW,\la_\PZ)=(0,0)$, which we denote by $(\la_\PW,\la_\PZ)=(\rL,\rL)$.
All these final states,
can be written as $(\la_\PW,\la_N)=(\la,-\la)$.
The Mandelstam
variables read $s=(p_{\bar{\Pd}}+p_\Pu)^2$, $t=(p_{\bar{\Pd}}-p_\PWp)^2\sim
-s(1-\cos{\theta})/2$, and $u=(p_{\bar{\Pd}}-p_N)^2\sim
-s(1+\cos{\theta})/2$, where the momenta of the initial and final states are incoming and outgoing, respectively, and $\theta$ is the angle\footnote{Note that the angle  used in \citere{Accomando:2001fn} is not $\theta$ but  $\hat{\theta}=180^\circ -\theta$, \ie the angle between $\vec{p}_{\bar{\Pd}}$ and $\vec{p}_N$.} between $\vec{p}_{\bar{\Pd}}$ and $\vec{p}_\PWp$.

\begin{list}{\bf \arabic{listnumber}.\hspace{1mm}}{\usecounter{listnumber}
\setlength{\leftmargin}{0mm} \setlength{\labelsep}{0mm}
\setlength{\itemindent}{4mm}
}

\item{\bf Born matrix elements}\nopagebreak

The Born amplitudes get contributions of the $s-$,  $t-$ and  $u-$channel diagrams in \reffi{duborn}. In the high-energy limit, \ie up to corrections of order $\MW^2/s$,  the Born amplitudes\footnote{According to our convention \refeq{process}, we 
give the  amplitudes for the crossing symmetric process $\bar{\Pd}_\rL\Pu_\rL \PW^-_{\la_\PW} N_{\la_N}\to 0$, which are equivalent to the amplitudes for the process  \refeq{duprocess}.} 
for the non-suppressed helicities read 
\newcommand{\Hdu}{H}
\newcommand{\sN}{\tilde{N}}
\beqar\label{bornWN} 
\M_0^{\bar{\Pd}_\rL\Pu_\rL\PW^-_\rL\PZ_\rL}&=&
-\ri\M_0^{\bar{\Pd}_\rL\Pu_\rL\phi^-\chi}=
-\frac{\ri e^2}{\sqrt{2}\sw}I^+_\chi \frac{A_s}{s}=
-\frac{e^2}{2\sqrt{2}\sw^2}\frac{A_s}{s},\nl
\M_0^{\bar{\Pd}_\rL\Pu_\rL \PW_\rT^- N_\rT }&=&
\frac{e^2}{\sqrt{2}\sw} \Hdu^N, \qquad 
\Hdu^N=\left(I^N_{\Pu_\rL}\frac{A_t}{t}+I^N_{\Pd_\rL}\frac{A_u}{u}\right),
\eeqar
\begin{figure}
\begin{center}
\begin{picture}(300,80)
\put(-30,0){
\begin{picture}(120,80)
\ArrowLine(20,30)(0,60)
\ArrowLine(0,0)(20,30)
\Vertex(20,30){2}
\Photon(20,30)(80,30){2}{6}
\Vertex(80,30){2}
\DashLine(100,0)(80,30){2}
\DashLine(80,30)(100,60){2}
\Text(-5,0)[r]{$\Pu_\rL$}
\Text(-5,60)[r]{$\bar{\Pd}_\rL$}
\Text(105,0)[l]{$\chi$}
\Text(105,60)[l]{$\phi^+$}
\Text(50,35)[b]{$\PWp$}
\end{picture}}
\put(120,0){
\begin{picture}(120,80)
\ArrowLine(40,45)(10,60)
\ArrowLine(40,15)(40,45)
\ArrowLine(10,0)(40,15)
\Vertex(40,15){2}
\Vertex(40,45){2}
\Photon(70,0)(40,15){2}{3}
\Photon(40,45)(70,60){-2}{3}
\Text(5,0)[r]{$\Pu_\rL$}
\Text(5,60)[r]{$\bar{\Pd}_\rL$}
\Text(37,30)[r]{$\Pu_\rL$}
\Text(75,0)[l]{$N_\rT$}
\Text(75,60)[l]{$\PW^+_\rT$}
\end{picture}}
\put(240,0){
\begin{picture}(120,80)
\ArrowLine(40,45)(10,60)
\ArrowLine(40,15)(40,45)
\ArrowLine(10,0)(40,15)
\Vertex(40,15){2}
\Vertex(40,45){2}
\Photon(70,0)(40,45){2}{5}
\Photon(70,60)(40,15){-2}{5}
\Text(5,0)[r]{$\Pu_\rL$}
\Text(5,60)[r]{$\bar{\Pd}_\rL$}
\Text(37,30)[r]{$\Pd_\rL$}
\Text(75,0)[l]{$N_\rT$}
\Text(75,60)[l]{$\PW^+_\rT$}
\end{picture}}
\end{picture}
\end{center}
\caption{Dominant  lowest-order diagrams for
  $\bar{\Pd}_\rL\Pu_\rL\to\phi^+\chi$ and $\bar{\Pd}_\rL\Pu_\rL\to\PW^+_\rT N_\rT$}
\label{duborn}
\end{figure}
The Born amplitude involving longitudinal gauge bosons $\PW^\pm_\rL,\PZ_\rL$ is expressed by the amplitude involving would-be Goldstone bosons $\phi^\pm,\chi$ and is dominated by the $s$-channel exchange of \PW~ bosons.
The production of transverse  gauge bosons is dominated by the $t$- and $u$-channel contributions and, as noted in \refeq{eevvchannel}, the amplitudes $A_t$ and $A_u$ are equal for non-suppressed helicities and in the high-energy limit. Therefore, using the fact that left-handed up and down quarks have opposite weak isospin and equal weak hypercharge,  the transverse amplitude $\Hdu^N$ can be easily expressed in terms of the symmetric and antisymmetric functions
\beq
F_{\pm}=A_t\left(\frac{1}{t}\pm\frac{1}{u}\right)
\eeq
as 
\beq
\Hdu^N=U_{NW^3}(\thw)\tilde{I}_{\Pu_\rL}^{W^3}F_-+U_{NB}(\thw)\tilde{I}^{B}_{\Pu_\rL}F_+,
\eeq
where we have used \refeq{rotcouplings}. The quantities  $\tilde{I}_{\Pu_\rL}^{W^3}$ and $\tilde{I}_{\Pu_\rL}^{B}$  are the eigenvalues of the $\SUtwo$ and $\Uone$ gauge couplings  \refeq{symmgenerat}, and $U_{N\sN}(\thw)$ are the components of the Weinberg rotation \refeq{defweinrot}.
The explicit expressions for $N=A,Z$ read
\beqar\label{sbornWXtra} 
\Hdu^A&=&
-\frac{F_-+Y_{q_\rL}F_+}{2},\qquad
\Hdu^Z=
\frac{\cw^2F_--\sw^2Y_{q_\rL}F_+ }{2\sw\cw}.
\eeqar

\item{\bf Leading soft-collinear corrections}\nopagebreak

The leading (angular-independent) Sudakov soft-collinear corrections are obtained from \refeq{deSC}. For longitudinally polarized final states, the resulting relative correction to the Born matrix element reads 
\beqar\label{SCWZcorr}
\de^{\SC}_{\bar{\Pd}_\rL\Pu_\rL \rightarrow \PW^+_{\rL} \PZ_{\rL}}&=&
-\frac{1}{2}\sum_{\varphi=\bar{\Pd}_\rL,\Pu_\rL,\phi^-,\chi} \left[\cew_\varphi\Ls-2(I^Z_\varphi)^2 \log{\frac{\MZ^2}{\MW^2}}\,\lZ
+Q_\varphi^2\Lemphi\right],\nl
&=& -\left[\cew_{q_\rL}+\cew_\Phi \right]\Ls
+\sum_{\varphi=\bar{\Pd}_\rL,\Pu_\rL,\phi^-,\chi} 
(I^Z_\varphi)^2 \log{\frac{\MZ^2}{\MW^2}}\,\lZ
\nl&&
-\frac{1}{2}\sum_{\varphi=\bar{\Pd}_\rL,\Pu_\rL,\phi^-}
Q_\varphi^2\Lemphi,
\eeqar
whereas for transversely polarized final states we have the relative correction
\beqar\label{duSCtra}
\de^{\SC}_{\bar{\Pd}_\rL\Pu_\rL \rightarrow \PW^+_{\rT} N_{\rT}}&=&
-\frac{1}{2}\sum_{\varphi=\bar{\Pd}_\rL,\Pu_\rL,\PW^-} \left[\cew_\varphi\Ls-2(I^Z_\varphi)^2 \log{\frac{\MZ^2}{\MW^2}}\,\lZ
+Q_\varphi^2\Lemphi\right]\nl
&&
-\frac{1}{2}\Ls\sum_{N'=A,Z}\cew_{N'N}
\frac{\M_0^{\bar{\Pd}_\rL\Pu_\rL \rightarrow \PW^+_{\rT} {N'}_{\rT}}}{\M_0^{\bar{\Pd}_\rL\Pu_\rL \rightarrow \PW^+_{\rT} N_{\rT}}}.
\eeqar
Since the Casimir operator \refeq{physadjointcasimir} is non-diagonal in the neutral gauge sector, the contribution associated with the neutral gauge boson $N$ (last line) involves also
the transformed Born matrix element with a neutral gauge boson $N'\neq N$.
This contribution can be simplified  using\footnote{As one can see from the rhs of \refeq{simplif}, 
the soft-collinear corrections associated with the final-state neutral gauge bosons originate only from their 
$W^3$ component.}
\beq\label{simplif}
\sum_{N'=A,Z}\cew_{N'N}\M_0^{\bar{\Pd}_\rL\Pu_\rL \rightarrow \PW^+_{\rT} {N'}_{\rT}}= 
\cew_\PW U_{NW^3}(\thw)\frac{e^2}{\sqrt{2}\sw}\tilde{I}^{W^3}_{\Pu_\rL}F_-,
\eeq
so that \refeq{duSCtra} can be written as 
\newcommand{\factG}{G}
\beqar
\de^{\SC}_{\bar{\Pd}_\rL\Pu_\rL \rightarrow \PW^+_{\rT} N_{\rT}}&=&
-\left[\cew_{q_\rL}+\frac{1}{2}\cew_W\left(1+ \factG^N_- \right)\right]\Ls
\nl&&+\sum_{\varphi=\bar{\Pd}_\rL,\Pu_\rL,\PW^-}\left[(I^Z_\varphi)^2 \log{\frac{\MZ^2}{\MW^2}}\,\lZ
-\frac{1}{2} Q_\varphi^2\Lemphi\right].
\eeqar
The factor $\factG^N_-$ is defined by 
\beq\label{Gfact1}
\factG_\pm^N=U_{NW^3}(\thw)\frac{\tilde{I}^{W^3}_{\Pu_\rL}}{\Hdu^N}F_\pm,
\eeq
and for $N=A,Z$ this reads
\beq\label{Gfact2}
\factG_\pm^A=\frac{F_\pm}{F_-+Y_{q_\rL}F_+},\qquad
\factG_\pm^Z=\frac{\cw^2F_\pm}{\cw^2F_--\sw^2 Y_{q_\rL}F_+}.
\eeq

\item{\bf Subleading corrections from soft-collinear photon  and $\PZ$--boson exchange }\nopagebreak

The angular-dependent Sudakov ($\SS$) corrections are obtained by applying the complete formula \refeq{4fsubdl} to the crossing symmetric process $\bar{\Pd}_\rL\Pu_\rL \PW^-_{\la_W} N_{\la_N}\rightarrow 0$. We first consider the contribution originating from the exchange of soft-collinear  neutral gauge bosons $V^a=A,Z$ between initial and final states. This yields
\beqar 
\sum_{V^a=A,Z}\de^{V^a,\SS}_{\bar{\Pd}_\rL\Pu_\rL \rightarrow \PW^+_{\la_\PW} N_{\la_N}}
&=&
\sum_{V^a=A,Z}2\left[\ls +
\frac{\alpha}{4\pi}\log{\frac{\MW^2}{M^2_{V^a}}}
\right]I^{V^a}_{\PW^-_{\la_\PW}}\left(I^{V^a}_{\bar{d}_\rL}\lts+I^{V^a}_{u_\rL}\lus\right)
\nl&&\hspace{1cm}{}
+2\ls I^Z_{N_{\la_N}}\left(I^Z_{\bar{d}_\rL}\lus+I^Z_{u_\rL}\lts\right)=
\nl&=&
2\ls\left\{\left[\frac{T^3_{\PW^-_{\la_\PW}}T^3_{u_\rL}}{\sw^2} +I^Z_{N_{\la_N}}\frac{\cw}{\sw}T^3_{u_\rL}\right]\left(\lts+\lus\right)
\right.\nl&&\hspace{1cm}\left.{}
- \left[\frac{Y_{\PW^-_{\la_\PW}}Y_{u_\rL}}{4\cw^2}+I^Z_{N_{\la_N}}\frac{\sw}{2\cw}Y_{q_\rL}\right]\left(\lts-\lus\right)\right\}\nl
&&{}-
\frac{\alpha}{2\pi}
Q_{\PW^-}\left[Q_\Pd \lts - Q_\Pu\lus \right]
\log{\frac{\MW^2}{\la^2}}.
\eeqar
In this formula, for longitudinal final states $\PW^+_{\rL} \PZ_{\rL}$ the quantum numbers of the corresponding would-be Goldstone bosons $\phi^+,\chi$ have to be used, in particular  $T^3_{\phi^-} =-\frac{1}{2}$,  $Y_{\phi^-} =-1$. For the transverse final states, one has to use the quantum numbers 
$T^3_{W^-}=-1$, $Y_{W^-}=0$.
The factor $I^Z_{N_{\la_N}}$ describes the coupling of the soft \PZ-boson with the final-state gauge boson $N_{\la_N}$, and is given by
\beq\label{du:Zchicoupling0}
I^Z_{N_{\la}}:=
\sum_{N'_{\la}}I^Z_{N'_{\la}N_{\la}}\frac{\M_0^{\bar{\Pd}_\rL\Pu_\rL\PW^-_{\la_\PW} N'_{\la} }}{\M_0^{\bar{\Pd}_\rL\Pu_\rL\PW^-_{\la_\PW} N_{\la}} },
\eeq
with
\beq\label{du:Zchicoupling}
I^Z_{Z}=I^Z_{A}=0,\qquad I^Z_\chi=
I^Z_{H\chi}\frac{I^+_{H}}{I^+_{\chi}} =-\frac{1}{2\sw\cw}.
\eeq
For transverse final states $I^Z_{N}$ vanishes since there is no $\PZ\PZ\PA$ or $\PZ\PZ\PZ$ gauge coupling. For longitudinal final states it is non-vanishing owing to the $I^Z_{H\chi}$ coupling, that gives rise to mixing between $\chi$ and $\PH$. The effect of this mixing is included into the definition \refeq{du:Zchicoupling} of $I^Z_\chi$.   
The resulting $\SS$ corrections from soft neutral gauge bosons read
\beqar \label{duSSCsoftNGB}
\sum_{V^a=A,Z}\de^{V^a,\SS}_{\bar{\Pd}_\rL\Pu_\rL \rightarrow \PW^+_{\rL} \PZ_{\rL}}
&=&
-\frac{\ls}{\sw^2}\left[\left(\lts+\lus\right)-\frac{\sw^2}{\cw^2}Y_{q_\rL}
\left(\lts-\lus\right)
\right]\nl
&&-
\frac{\alpha}{2\pi}
Q_{\PW^-}\left[Q_\Pd \lts - Q_\Pu\lus \right]
\log{\frac{\MW^2}{\la^2}},\nl
\sum_{V^a=A,Z}\de^{V^a,\SS}_{\bar{\Pd}_\rL\Pu_\rL \rightarrow \PW^+_{\rT} N_{\rT}}
&=&-\frac{\ls}{\sw^2}\left(\lts+\lus\right)\nl
&&-
\frac{\alpha}{2\pi}
Q_{\PW^-}\left[Q_\Pd \lts - Q_\Pu\lus \right]
\log{\frac{\MW^2}{\la^2}}.
\eeqar

\item{\bf Subleading corrections from soft-collinear $\PW$--boson exchange}\nopagebreak

We come now to the part of the $\SS$ corrections \refeq{4fsubdl} that originates from exchange of soft-collinear charged gauge bosons $V^a=\PW^\pm$ between initial and final states. This part reads
\beqar \label{duwzWssc}
\lefteqn{\sum_{V^a=W^\pm}\de^{V^a,\SS} \M^{\bar{\Pd}_\rL\Pu_\rL \PW^-_{\la_\PW} N_{\la_N}}=}\qquad&&\nl
&=&-\frac{2\ls}{\sqrt{2}\sw}\left\{
\left[\sum_{N'_{\la_\PW}}
I^+_{N'_{\la_\PW}} \M_0^{\bar{\Pu}_\rL\Pu_\rL {N'}_{\la_\PW} N_{\la_N}}
+I^+_{N_{\la_N}}\M_0^{\bar{\Pd}_\rL\Pd_\rL \PW^-_{\la_\PW} \PW^+_{\la_N}}
\right]\log{\frac{|t|}{s}}\right.\nl&&\left.
\hspace{1.3cm}{}-\left[\sum_{N'_{\la_\PW}}
I^+_{N'_{\la_\PW}} \M_0^{\bar{\Pd}_\rL\Pd_\rL {N'}_{\la_\PW} N_{\la_N}}
+I^+_{N_{\la_N}}\M_0^{\bar{\Pu}_\rL\Pu_\rL \PW^-_{\la_\PW} \PW^+_{\la_N}}
\right]\log{\frac{|u|}{s}}\right\}.\nln
\eeqar
In the case of longitudinal polarizations  $\la_\PW=\la_N=\rL$,  longitudinal gauge bosons $\PW^\pm_\rL,\PZ_\rL$ have to be substituted by the would-be Goldstone bosons $\phi^\pm,\chi$, the sums on the rhs run over the neutral scalar fields $N'_\rL=H,\chi$, and  the corresponding  non-diagonal $\PW^\pm$ couplings \refeq{scapmcoupB}  have to be used. The transformed Born matrix elements for longitudinal final states are given by 
\beqar\label{rotbornWZlong} 
\M_0^{\bar{q}_\rL q_\rL H\chi}&=&e^2I^Z_{q_\rL}I^Z_{\chi H}\frac{A_s}{s},
\qquad
\M_0^{\bar{q}_\rL q_\rL \chi\chi}=0,
\nl
\M_0^{\bar{q}_\rL q_\rL \phi^-\phi^+}&=&e^2\sum_{V^a=A,Z}I^{V^a}_{q_\rL}I^{V^a}_{\phi^-}\frac{A_s}{s}=-e^2 \left(\frac{T^3_{q_\rL}}{2\sw^2}+\frac{Y_{q_\rL}}{4\cw^2}\right) \frac{A_s}{s}.
\eeqar
Inserting these in \refeq{duwzWssc}
and dividing by the Born matrix element for longitudinal gauge bosons we obtain the relative correction 
\beq\label{dulongSSCsoftCGB}
\sum_{V^a=W^\pm}\de^{V^a,\SS}_{\bar{\Pd}_\rL\Pu_\rL \rightarrow \PW^+_{\rL} \PZ_{\rL}}
=-\frac{\ls}{\sw^2}\left\{ \left(\lts+\lus\right)
-\frac{\sw^2}{\cw^2}Y_{q_\rL}\left(\lts-\lus\right)
\right\}.
\eeq
For transverse gauge bosons $\la_\PW=\la_N=\rT$, the sums on the rhs run over the neutral gauge bosons $N'_\rT=A,Z$, and the non-diagonal $\PW^\pm$ couplings  \refeq{gaupmcoup2} have to be used.
The transformed Born matrix elements for transverse final states read
\beqar
\M_0^{\bar{q}_\rL q_\rL {N'}_\rT N_\rT }&=&
e^2 I^{N'}_{q_\rL}I^N_{q_\rL}F_+,\nl
\M_0^{\bar{d}_\rL d_\rL \PW^-_\rT \PW^+_\rT}&=&
\frac{e^2}{2\sw^2}\frac{A_t}{t}=\frac{e^2}{\sw^2}\frac{F_++F_-}{4}
,\nl
\M_0^{\bar{u}_\rL u_\rL \PW^-_\rT \PW^+_\rT}&=&
\frac{e^2}{2\sw^2}\frac{A_u}{u}=\frac{e^2}{\sw^2}\frac{F_+-F_-}{4}.
\eeqar
Inserting these in \refeq{duwzWssc}, and using
\beqar
\sum_{N'}I^+_{N'}\M_0^{\bar{q}_\rL q_\rL \rightarrow {N'}_{\rT} N_{\rT}}
&=&\frac{e^2}{\sw}I^N_{q_\rL}\tilde{I}^{W^3}_{q_\rL}F_+
=\frac{e^2}{\sw^2}
T^3_{\Pu_\rL}\left(U_{NW^3}(\thw)\tilde{I}^{W^3}_{q_\rL}+U_{NB}(\thw)\tilde{I}^{B}_{q_\rL}\right)\nln
\eeqar
and $
I^+_N=2U_{NW^3}(\thw)\tilde{I}^{W^3}_{\Pu_\rL},
$
we obtain
\beqar
\lefteqn{\sum_{V^a=W^\pm}\de^{V^a,\SS}\M^{\bar{\Pd}_\rL\Pu_\rL \rightarrow \PW^+_{\rT} N_{\rT}}=}\quad&&\nl
&=&\frac{2\ls}{\sqrt{2}\sw}\frac{e^2}{\sw^2}\left\{
-\left[T^3_{\Pu_\rL}(U_{NW^3}(\thw)\tilde{I}^{W^3}_{\Pu_\rL}+U_{NB}(\thw)\tilde{I}^{B}_{\Pu_\rL})F_+ +I^+_N\frac{F_++F_-}{4}\right]\lts
\right.\nl&&\left.\hspace{1.3cm}\qquad{}
+\left[T^3_{\Pd_\rL}(U_{NW^3}(\thw)\tilde{I}^{W^3}_{\Pd_\rL}+U_{NB}(\thw)\tilde{I}^{B}_{\Pd_\rL})F_+ +I^+_N\frac{F_+-F_-}{4}\right]\lus
\right\}
\nl&=&
-\frac{2\ls}{\sqrt{2}\sw}\frac{e^2}{\sw^2}\left\{
\frac{1}{2}\Hdu^N\left(\lts+\lus \right)
+U_{NW^3}(\thw)\tilde{I}^{W^3}_{\Pu_\rL} F_+\left(\lts-\lus \right)
\right\}.\nln
\eeqar
Then, dividing by the Born matrix element for transverse gauge bosons we obtain the relative correction
\beq\label{dutraSSCsoftCGB}
\sum_{V^a=W^\pm}\de^{V^a,\SS}_{\bar{\Pd}_\rL\Pu_\rL \rightarrow \PW^+_{\rT} N_{\rT}}
=-\frac{\ls}{\sw^2}\left\{ \left(\lts+\lus\right)
+2\factG^N_+\left(\lts-\lus\right)
\right\},
\eeq
where $G^N_+$,  is given in \refeq{Gfact1}, \refeq{Gfact2}.

\item{\bf Complete subleading  soft-collinear corrections}\nopagebreak

Combining the contributions of soft-collinear  neutral gauge bosons \refeq{duSSCsoftNGB} and of soft-collinear charged gauge bosons \refeq{dulongSSCsoftCGB}, \refeq{dutraSSCsoftCGB} we obtain  the total $\SS$ relative corrections,
\beqar
\de^{\SS}_{\bar{\Pd}_\rL\Pu_\rL \rightarrow \PW^+_{\rL} \PZ_{\rL}}
&=&-\frac{2}{\sw^2}\ls\left\{ \left(\lus+\lts\right)
-\frac{\sw^2}{\cw^2}Y_{q_\rL}\ltu
\right\}\nl
&&-
\frac{\alpha}{2\pi}
Q_\PWm\left[Q_\Pd \lts - Q_\Pu\lus \right]
\log{\frac{\MW^2}{\la^2}},\nl
\de^{\SS}_{\bar{\Pd}_\rL\Pu_\rL \rightarrow \PW^+_{\rT} N_{\rT}}
&=&
-\frac{2}{\sw^2}\ls\left\{
\left(\lts+\lus\right)+ \factG^N_+ \ltu
\right\}\nl
&&-
\frac{\alpha}{2\pi}
Q_\PWm\left[Q_\Pd \lts - Q_\Pu\lus \right]
\log{\frac{\MW^2}{\la^2}}.
\eeqar

\item{\bf SL contributions associated to the  external particles}\nopagebreak

For longitudinal final states, according to  \refeq{deccfer} and \refeq{longeq:coll}, the SL corrections associated to the external particles  give 
\beqar
\de^\cc_{\bar{\Pd}_\rL\Pu_\rL \rightarrow\PW_\rL^+ \PZ_\rL }&=& \left[3 \cew_{q_\rL}+ 4\cew_\Phi\right]\lsl-\frac{3}{2\sw^2}\frac{m^2_t}{\MW^2}\lYuk
+
\sum_{\varphi=\bar{\Pd}_\rL,\Pu_\rL,\PW^-}Q_\varphi^2
\,\lemphi,
\eeqar
with
\beq
\lemphi=\frac{\alpha}{4\pi}
\left(\frac{1}{2}\log{\frac{\MW^2}{m^2_{\varphi}}}
+\log{\frac{\MW^2}{\la^2}}\right).
\eeq
For transverse final states, according to  \refeq{deccfer} and \refeq{deccWT} we have the relative corrections
\beqar\label{Ccorrtrafinstat}
\de^\cc_{\bar{\Pd}_\rL\Pu_\rL \rightarrow\PW_\rT^+ \PZ_\rT }&=& 
\left\{ 3 \cew_{q_\rL}
+\frac{1}{2}\left[ \bew_{WW}+
\sum_{N'=A,Z}\left(\bew_{N'N}+ \antikro_{N'N}\bew_{AZ}\right)
\frac{\M_0^{\bar{\Pd}_\rL\Pu_\rL \to\PW_\rT^+ N'_\rT }}{\M_0^{\bar{\Pd}_\rL\Pu_\rL \to\PW_\rT^+ N_\rT }}
\right]
\right\}\lsl 
\nl&&+
\sum_{\varphi=\bar{\Pd}_\rL,\Pu_\rL,\PW^-}Q_\varphi^2
\,\lemphi,
\eeqar
so that
\beqar
\de^\cc_{\bar{\Pd}_\rL\Pu_\rL \to \PW_\rT^+ A_\rT }
&=& 3 \cew_{q_\rL}\lsl
   +\frac{1}{2}\left[ \bew_{WW}+\bew_{AA}\right]\lsl
+
\sum_{\varphi=\bar{\Pd}_\rL,\Pu_\rL,\PW^-}Q_\varphi^2
\,\lemphi
,\\
\de^\cc_{\bar{\Pd}_\rL\Pu_\rL \to \PW_\rT^+ Z_\rT }
&=& 3 \cew_{q_\rL}\lsl
   +\frac{1}{2}\left[ \bew_{WW}+\bew_{ZZ}+ 2 \bew_{AZ}
\frac{H^{ A}}{H^{Z}}\right]\lsl
+
\sum_{\varphi=\bar{\Pd}_\rL,\Pu_\rL,\PW^-}Q_\varphi^2
\,\lemphi
.\nn
\eeqar

\item{\bf Logarithmic corrections from parameter renormalization}\nopagebreak

The PR corrections are obtained from the renormalization of the $\Uone$ and $\SUtwo$ couplings [see \refeq{gCTs}] in the Born amplitude 
\refeq{bornWN}.
For longitudinally polarized final states they read
\beqar
\de^\pre_{\bar{\Pd}_\rL\Pu_\rL \rightarrow\PW_\rL^+ \PZ_\rL }
 &=&-\bew_{W}\lpr+\Delta\alpha(\MW^2). 
\eeqar
For transverse polarizations of the final states, as shown in \refapp{app:transvRG},
the PR corrections can be associated to the external gauge-boson lines, and according to  
\refeq{deRGgb}--\refeq{WRMAT2}
they are given by
\beqar\label{RCcorrtrafinstat}
\de^\pre_{\bar{\Pd}_\rL\Pu_\rL \to \PW_\rT^+ N_\rT }
&=& \de^\pre_{WW}+
\sum_{N'=A,Z}\de^\pre_{N'N}\frac{\M_0^{\bar{\Pd}_\rL\Pu_\rL \to\PW_\rT^+ N'_\rT }}{\M_0^{\bar{\Pd}_\rL\Pu_\rL \to\PW_\rT^+ N_\rT }}
\\&=&
-\frac{1}{2}\left[ \bew_{WW}+
\sum_{N'=A,Z}\left(\bew_{N'N}+ \antikro_{N'N}\bew_{AZ}\right)
\frac{H^{ N'}}{H^{N}}\right]\lpr
+\Delta\alpha(\MW^2),\nn
\eeqar
so that 
\beqar
\de^\pre_{\bar{\Pd}_\rL\Pu_\rL \to \PW_\rT^+ A_\rT }
&=& 
-\frac{1}{2}\left[ \bew_{WW}+\bew_{AA}\right]\lpr
+\Delta\alpha(\MW^2),\nl
\de^\pre_{\bar{\Pd}_\rL\Pu_\rL \to \PW_\rT^+ Z_\rT }
&=& 
-\frac{1}{2}\left[ \bew_{WW}+\bew_{ZZ}+ 2 \bew_{AZ}
\frac{H^{ A}}{H^{Z}}\right]\lpr
+\Delta\alpha(\MW^2).
\eeqar
Adding the SL corrections \refeq{Ccorrtrafinstat} and \refeq{RCcorrtrafinstat}
for transverse final states, we obtain\footnote{In the following formula $\de_{N\PZ}$ represents the Kronecker symbol.}
\beqar
\lefteqn{\de^\cc_{\bar{\Pd}_\rL\Pu_\rL \to\PW_\rT^+ N_\rT }+
\de^\pre_{\bar{\Pd}_\rL\Pu_\rL \to \PW_\rT^+ N_\rT }
=}\quad&&\nl&=& 
3 \cew_{q_\rL}\lsl +\sum_{\varphi=\bar{\Pd}_\rL,\Pu_\rL,\PW^-}Q_\varphi^2\,\lemphi+\frac{1}{2}(1+\de_{N\PZ})\Delta\alpha(\MW^2),
\eeqar
where only energy-dependent logarithmic corrections \refeq{deccfer} that are associated to the initial states contributes
whereas, as shown in \refeq{PRpluscc}, the energy-dependent logarithms associated to  the final-state gauge bosons are cancelled by the contributions of parameter renormalization.

\item{\bf Numerical evaluation}\nopagebreak

The coefficients for the various 
electroweak logarithmic contributions to the relative
corrections read
\newcommand{\duWAT}{\de^{\ew}_{\bar{\mathrm{d}}_\rL\mathrm{u}_\rL\to \PW^+_\rT\PA_\rT}}
\newcommand{\duWZT}{\de^{\ew}_{\bar{\mathrm{d}}_\rL\mathrm{u}_\rL\to \PW^+_\rT\PZ_\rT}}
\newcommand{\duWZL}{\de^{\ew}_{\bar{\mathrm{d}}_\rL\mathrm{u}_\rL\to \PW^+_\rL\PZ_\rL}}
\newcommand{\GAm}{G_-^A}
\newcommand{\GAp}{G_+^A}
\newcommand{\GZm}{G_-^Z}
\newcommand{\GZp}{G_+^Z}
\newcommand{\ltsplus}{\left(\log{\frac{|t|}{s}}+\log{\frac{|u|}{s}}\right)}
\newcommand{\rHAZ}{\frac{H^A}{H^Z}}
\beqar
\duWAT &=& \left[-7.86 - 4.47 \,\GAm\right] \Ls
         - 8.95 \left[ \lts +\lus +\GAp \ltu\right] \ls
\nl&&{}
         + 1.32 \,\lZ + 15.42 \,\lsl - 5.25 \,\lpr
,\nl
\duWZT &=& \left[-7.86 - 4.47 \,\GZm\right] \Ls
         - 8.95 \left[\lts +\lus + \GZp \ltu\right] \ls
\nl&&{}
         + 1.32 \,\lZ + \left[21.77 - 9.57 \,\rHAZ\right] \,\lsl
         + \left[-11.60 + 9.57 \,\rHAZ\right] \,\lpr, 
\nl
\duWZL &=&  -7.07 \,\Ls
         + \left[-8.95 \,\ltsplus + 0.86 \,\ltu\right] \,\ls
\nl&&{}
         + 0.92 \,\lZ + 24.88 \,\lsl - 31.83 \,\lYuk
         - 14.16 \,\lpr.
\eeqar
The angular dependence of the corrections is plotted in \reffi{plotduWNan} for scattering  angles $30^\circ<\theta<75^\circ$. 
In this plot, the central angular region, where the LO matrix element has zeros, has been omitted. 
The energy dependence of the corrections at $\theta=60^\circ$ is represented in \reffi{plotduWNen}. 
\end{list}

For a detailed discussion of the behaviour of these corrections and their impact on the 
hadronic processes 
$\Pp\Pp\to \PWpm\gamma\to l\nu_l\gamma$ and $\Pp\Pp\to\PWpm\PZ \to l\nu_l \bar{l}'l'$
we refer to \citere{Accomando:2001fn}.

\begin{figure}
\centerline{
\setlength{\unitlength}{1cm}
\begin{picture}(10,8.3)
\put(0,0){\includegraphics{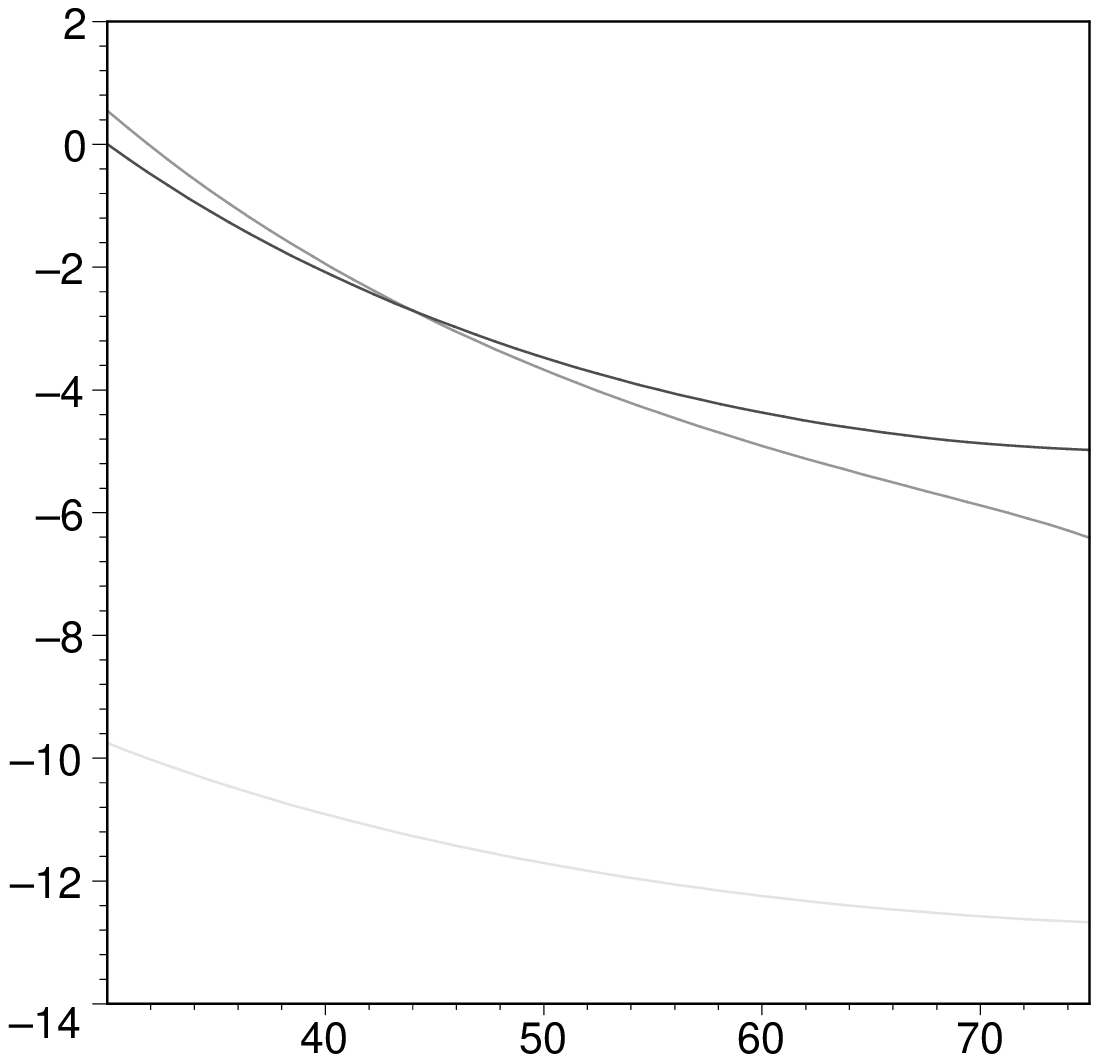}}
\put(2.5,0){\makebox(6,0.5)[b]{$\theta\,[^\circ]$}}
\put(-2.5,4){\makebox(1.5,1)[r]{$\delta^\ew\,[\%]$}}
\put(10.3,4.7){\makebox(1.5,1)[r]{$\PW^+_\rT\PA_\rT$}}
\put(10.3,3.7){\makebox(1.5,1)[r]{$\PW^+_\rT\PZ_\rT$}}
\put(10.3,1.2){\makebox(1.5,1)[r]{$\PW^+_\rL\PZ_\rL$}}
\end{picture}}
\caption[WWang]{Dependence of the electroweak correction factors
$\de^{\ew}_{\bar{\Pd}_\rL\Pu_\rL\rightarrow \PW_\rT^+\PA_{\rT}}$,
$\de^{\ew}_{\bar{\Pd}_\rL\Pu_\rL\rightarrow \PW_\rT^+\PZ_{\rT}}$,
and $\de^{\ew}_{\bar{\Pd}_\rL\Pu_\rL\rightarrow \PW_\rL^+\PZ_{\rL}}$, 
on the scattering angle $\theta$ at $\sqrt{s}=1\TeV$.} 
\label{plotduWNan}
\end{figure}%

\begin{figure}
\centerline{
\setlength{\unitlength}{1cm}
\begin{picture}(10,8.3)
\put(0,0){\includegraphics{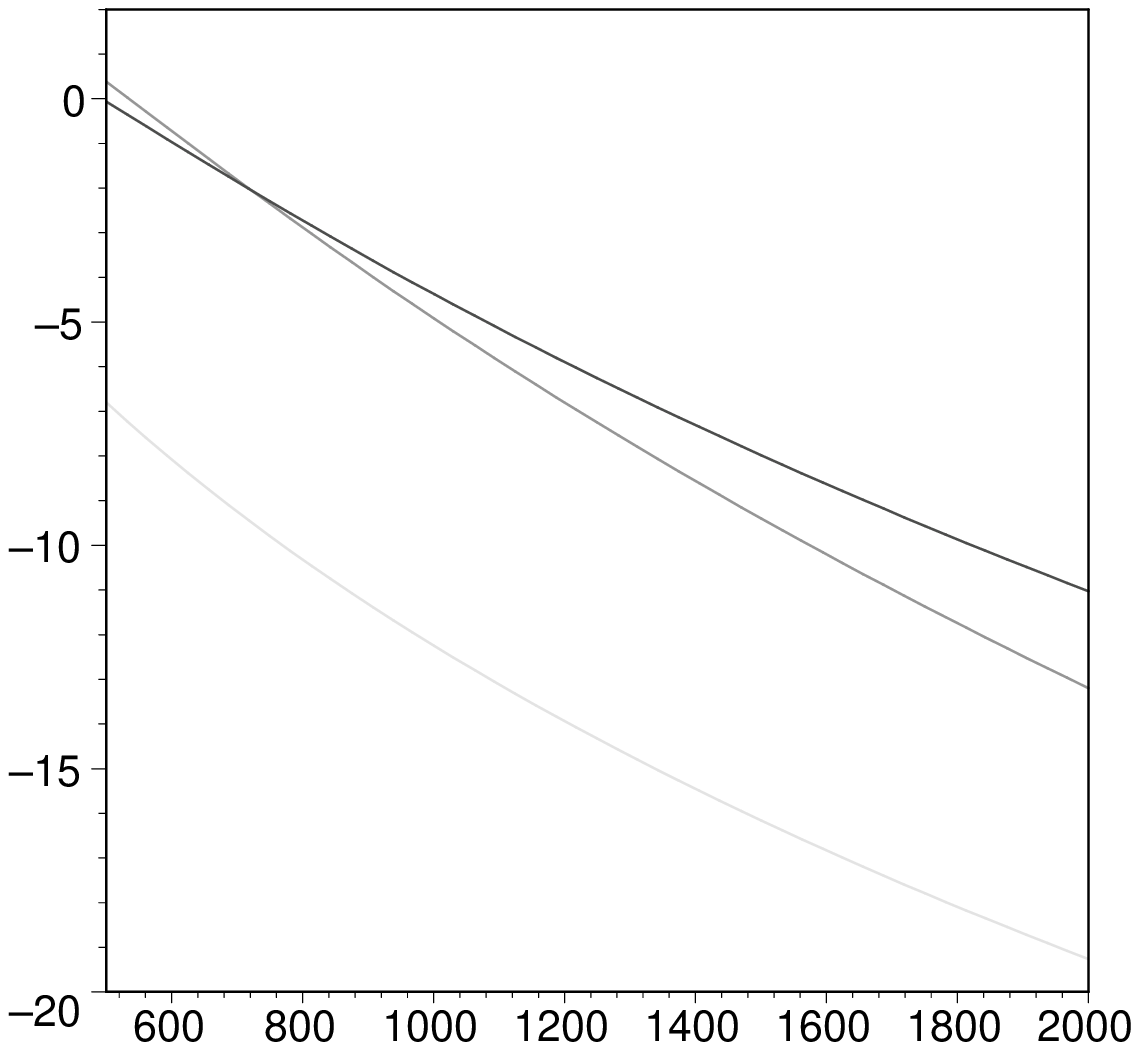}}
\put(2.5,0){\makebox(6,0.5)[b]{$\sqrt{s}\,[\GeV]$}}
\put(-2.5,4){\makebox(1.5,1)[r]{$\delta^\ew\,[\%]$}}
\put(10.3,3.7){\makebox(1.5,1)[r]{$\PW^+_\rT\PA_\rT$}}
\put(10.3,2.7){\makebox(1.5,1)[r]{$\PW^+_\rT\PZ_\rT$}}
\put(10.3,1.0){\makebox(1.5,1)[r]{$\PW^+_\rL\PZ_\rL$}}
\end{picture}}
\caption[WWang]{Dependence of the electroweak correction factors
$\de^{\ew}_{\bar{\Pd}_\rL\Pu_\rL\rightarrow \PW_\rT^+\PA_{\rT}}$,
$\de^{\ew}_{\bar{\Pd}_\rL\Pu_\rL\rightarrow \PW_\rT^+\PZ_{\rT}}$,
and $\de^{\ew}_{\bar{\Pd}_\rL\Pu_\rL\rightarrow \PW_\rL^+\PZ_{\rL}}$, 
on the centre-of-mass energy $\sqrt{s}$ at  $\theta=60^\circ$.} 
\label{plotduWNen}
\end{figure}%

\addcontentsline{toc}{chapter}{Appendix}
\begin{appendix}
\chapter{Conventions for Green functions}
\label{app:GFs} 
Our conventions for Green functions are based on
\citere{DennBohmJos}. In configuration space we use the equivalent
notations 
\beq
\langle 0|T\varphi_{i_1}(x_1)\dots \varphi_{i_n}(x_n)|0\rangle
=
\langle\varphi_{i_1}(x_1)\dots \varphi_{i_n}(x_n)\rangle
=
G^{\varphi_{i_1}\dots\varphi_{i_n}}(x_1,\dots ,x_n)
.
\eeq
The fields $V_\mu^a, \Phi_k, \Psi_{j,\si}$ appearing in the Green functions 
are associated with outgoing particles or incoming antiparticles, because the
corresponding field operators 
create antiparticles and annihilate particles.
The opposite  holds for the charge-conjugate fields $\bar{V}_\mu^a, \Phi^+_k, \bar{\Psi}_{j,\si}$.

Fourier transformation is defined with incoming momenta, and  the
momentum-con\-ser\-va\-ti\-on $\de$-function is factorized as  
\beqar\label{fourier}
\lefteqn{
(2\pi)^4\de^{(4)}\left(\sum_{k=1}^n p_{k}\right)
G^{\varphi_{i_1}\dots\varphi_{i_n}}(p_1,\dots ,p_n)}\quad
\nl&&=
\int\left(\prod_{k=1}^n 
{\mathrm{d}^4x_k}\right)
\exp{\left(-\ri\sum_{j=1}^n x_jp_j\right)}
G^{\varphi_{i_1}\dots\varphi_{i_n}}(x_1,\dots ,x_n).
\eeqar
The vertex functions are given by the functional derivatives of the corresponding generating functional $\Gamma$,
\beq\label{vertexfunct}
\Gamma^{\varphi_{i_1}\dots \varphi_{i_n}}(x_1,\dots,x_n)=\frac{\de^n \Gamma\{\varphi\}}{\de\varphi_{i_1}(x_1) \dots \de \varphi_{i_n}(x_n)},
\eeq
and their Fourier transforms are defined as in \refeq{fourier} for Green functions.
For 2-point Green functions and vertex functions we use the shorthand notations  \beq\label{propagators}
G^{\varphi_{i}\varphi^+_{j}}(p)=G^{\varphi_{i}\varphi^+_{j}}(p,-p),\qquad
\Gamma^{\varphi_{i}\varphi^+_{j}}(p)=\Gamma^{\varphi_{i}\varphi^+_{j}}(p,-p).
\eeq
For {\em anticommuting fields} the sign is inverted when the  field arguments are exchanged, \ie
\beq
G^{\varphi_{i}\varphi^+_{j}}(p)=
-G^{\varphi^+_{j}\varphi_{i}}(-p),\qquad
\Gamma^{\varphi_{i}\varphi^+_{j}}(p)=
-\Gamma^{\varphi^+_{j}\varphi_{i}}(-p),
\eeq
for fermions  $\varphi=\Psi$ or ghost fields $\varphi=u$.

The 2-point vertex functions correspond to the inverse propagators. More precisely
\beq
\sum_{\varphi_k}\Gamma^{\varphi^+_{i}\varphi_{k}}(p)
G^{\varphi_{k}\varphi^+_{j}}(p)=
\sum_{\varphi_k}G^{\varphi^+_{j}\varphi_{k}}(-p)
\Gamma^{\varphi_{k}\varphi^+_{i}}(-p)
=\pm\ri \de_{\varphi_i\varphi_j},
\eeq
with $+$ for bosons and $-$ for fermions and ghosts.
In the case of  fermions with Dirac indices $\alpha,\beta,\gamma$  one has to substitute $\varphi^+_i=\bar{\Psi}_\alpha$, $\varphi^+_j=\bar{\Psi}_\beta$, $\varphi_k=\Psi_\gamma$.

For the truncation of an external leg $\varphi_{i_k}$ in momentum
space we adopt the convention  
\mda
\beq
G^{\varphi_{i_1}\dots\, \varphi_{i_k} \dots\,\varphi_{i_n}}(p_1,\dots,p_k,\dots,p_n)=
\sum_{\varphi^+_{i_k}}G^{\varphi_{i_k}\varphi^+_{i_k}}(p_k)
G^{\varphi_{i_1}\dots \,\underline{\varphi}^+_{i_k} \dots\,\varphi_{i_n}}(p_1,\dots,p_k,\dots,p_n),
\eeq
where the field argument corresponding to the
truncated leg is underlined.  
The fields $V_\mu^a, \Phi_k, \Psi_{j,\si}$  in 
truncated Green functions  and vertex functions are associated 
with  incoming particles or outgoing antiparticles, and the opposite holds for the 
charge-conjugate fields.


\chapter{Representations of the gauge group}
\label{app:representations} 
In this appendix, we define the generators of the gauge group and the other group-theoretical quantities 
that are used in our generic formulas. 
Explicit matrix representations are given in the symmetric basis corresponding to $\Uone$ and $\SUtwo$ gauge fields, as well as in the physical basis corresponding to mass-eigenstate gauge fields.

\section{$\SUtwo\times \Uone$ generators}
The basic group-theoretical matrices are the generators $I^{V^a}$ of global transformations
\beq\label{globtra}
\de \varphi_i= 
\ri e  \sum_{V^a}\sum_{\varphi_{i'}}
I^{V^a}_{\varphi_i\varphi_{i'}}\de\theta^{V^a} \varphi_{i'}.
\eeq
According to the notation introduced in \refse{se:not},
the fields $\varphi_i,\varphi_{i'}$ are understood as the components of a multiplet $\varphi$ which may represent chiral fermions,
gauge-bosons or scalar bosons. 
The representation of the matrices $I^{V^a}_{\varphi_i\varphi_{i'}}$ depends on the multiplet $\varphi$ and is in general not irreducible.
The sums in  \refeq{globtra} run over all gauge fields $V^a$ of the $\ewgroup$ group and all components $\varphi_{i'}$ of the multiplet $\varphi$.

The transformation of the complex-conjugate fields is fixed by the complex conjugation of \refeq{globtra}, thus
\beq\label{complexconjug}
\left(I^{V^a}_{\varphi_j\varphi_i}\right)^*=-I^{\bar{V}^a}_{\varphi^+_j\varphi^+_i},
\eeq
where $\bar{V}^a$ and $\varphi^+_j$ represent the charge-conjugate of  $V^a$ and $\varphi_j$, respectively.
Since the representations are unitary, in a real basis ($\bar{V}^a=V^a$) the generators are self-adjoint, and in general  
\beq\label{selfadjoint}
I^{\bar{V}^a}=\left(I^{V^a}\right)^+,\qquad \mbox{\ie}\quad I^{\bar{V}^a}_{\varphi_i\varphi_j}=\left(I^{V^a}_{\varphi_j\varphi_i}\right)^*.
\eeq
Combining \refeq{selfadjoint} with \refeq{complexconjug} yields the relation 
\beq\label{transprel}
I^{V^a}_{\varphi_i\varphi_j}
=-I^{V^a}_{\varphi^+_j\varphi^+_i},
\eeq
between transposed components.
The generators in the adjoint representation are related to the structure constants [see \refeq{symmadjgen}, \refeq{totalantitens}  and \refeq{phystotalantitens}]
through the commutation relations
\beq\label{adjgendef}
[I^{V^a},I^{V^b}]_{\varphi_i\varphi_j}=\sum_{V^c}I^{V^c}_{\varphi_i\varphi_j}I^{V^a}_{V^cV^b}
=\frac{\ri}{\sw}\sum_{V^c}\varepsilon^{V^a V^b V^c}I^{\bar{V}^c}_{\varphi_i\varphi_j}.
\eeq
For the eigenvalues of diagonal generators (or other diagonal matrices) we use the notation
\beq
I_{\varphi_i\varphi_{i'}}=\de_{\varphi_i\varphi_{i'}}I_{\varphi_i}.
\eeq  

\section{Symmetric and physical basis for gauge fields}
Group-theoretical quantities carrying  gauge-boson indices can be expressed in
the symmetric  or in the physical basis.
The symmetric basis is formed by the $\Uone$ and $\SUtwo$ gauge bosons,
which are an $\SUtwo$ singlet and  triplet, respectively. These are combined into the 4-vector 
\beq\label{GBsymmbasis}
\sV=(B,W^3,W^1,W^2)^\rT,  
\eeq
and all quantities in this basis are denoted by a tilde. 
The physical basis is given by the charge and mass
eigenstates 
\beq
V=(A,Z,W^+,W^-)^\rT,
\eeq
and is related to the symmetric basis by the unitary transformation 
\beq\label{defweinrot}
V =U(\thw)\sV,\qquad  U(\thw)=\left(\begin{array}{c@{\;}c@{\;}c@{\;}c}
\cw & -\sw & 0 & 0 \\ 
\sw & \cw & 0 & 0 \\
0 & 0 & \frac{1}{\sqrt{2}} & \frac{-\ri}{\sqrt{2}} \\
0 & 0 & \frac{1}{\sqrt{2}} & \frac{\ri}{\sqrt{2}}
\end{array}\right),
\eeq 
or
\beq
A=\cw B-\sw W^3,\qquad
Z=\sw B+\cw W^3,\qquad
W^\pm=\frac{W^1\mp\ri W^2}{\sqrt{2}}. 
\eeq
For the Weinberg rotation in the  neutral sector we use the shorthands  $\cw=\cos{\thw}$ and $\sw=\sin{\thw}$. In the on shell renormalization scheme the Weinberg angle is fixed by \refeq{mixingangle}.

\subsection*{Generators of the gauge group}
In the symmetric basis, the generators of the gauge group are given by
\beq\label{symmgenerat}
\tilde{I}^B=-\frac{1}{\cw}\frac{Y}{2},\qquad \tilde{I}^{W^a}=\frac{1}{\sw}T^a,\qquad a=1,2,3,
\eeq
where $Y$ is the weak hypercharge and $T^a$ are the components of the weak isospin.  
These are related to the electric charge by $Q=T^3+Y/2$.
As a convention, the generators are treated  as co-vectors
\beq
\tilde{I}^{\sV}=(\tilde{I}^B,\tilde{I}^{W^3},\tilde{I}^{W^1},\tilde{I}^{W^2})
\eeq
in the symmetric basis, so that they transform to the physical basis
\beq
I^V=(I^A,I^Z,I^{W^+},I^{W^-}),
\eeq
as 
\beq\label{covectortransf}
I^V =\tilde{I}^{\sV} U^+(\thw),\qquad  
U^+(\thw)=U^{-1}(\thw)=
\left(\begin{array}{c@{\;}c@{\;}c@{\;}c}
\cw 
& \sw & 0 & 0 \\ 
-\sw & \cw & 0 & 0 \\
0 & 0 & \frac{1}{\sqrt{2}} & \frac{1}{\sqrt{2}} \\
0 & 0 & \frac{\ri}{\sqrt{2}} & \frac{-\ri}{\sqrt{2}}
\end{array}\right),
\eeq
or equivalently
\beq
I^A=-Q,\qquad I^Z=\frac{T^3-\sw^2 Q}{\sw\cw},\qquad
I^\pm=\frac{1}{\sw}T^\pm=\frac{1}{\sw}\frac{T^1\pm\ri T^2}{\sqrt{2}}.
\eeq

Note that, owing to\footnote{Here the adjoint operation acts as a complex conjugation as well as a transposition of the indices $V^a$ and the representation indices of the matrices $\tilde{I}^{\sV^a}$.
This identity follows from $(\tilde{I}^{\sV^a})^+=\tilde{I}^{\sV^a}$.} $(\tilde{I}^{\sV})^+=(\tilde{I}^{\sV})^\rT$, the $\bar{V}^a$ component of the co-vector $I^V$ can be understood as the $V^a$ component of the vector $I^{\bar{V}}$ defined by 
\beq\label{rotcouplings}
I^{\bar{V}}=\left(I^{V}\right)^+=\left(\tilde{I}^{\sV} U^+(\thw) \right)^+
= U(\thw)\left(\tilde{I}^{\sV}\right)^\rT,\quad\mbox{with}\quad 
I^{\bar{V}^a} =
\sum_{\sV^b} U_{V^a\sV^b}(\thw)\tilde{I}^{\sV^b}.
\eeq

\subsection*{Matrices with gauge-boson indices and their transformation}
As a general convention, all  matrices  that carry two gauge-boson (lower) indices, \ie the $4\times 4$  matrices of the type $M_{V^aV^b}$, are transformed as usual matrices.
By this we mean that the
the first and second indices transform as a vector and a co-vector, respectively, \ie
\beq\label{adjmattransf}
M_{V^aV^b}= \left[U(\thw)\tilde{M}U^+(\thw)\right]_{V^aV^b}.
\eeq 

In the symmetric basis, most of the invariant  $4\times 4$ matrices can be decomposed into the $\Uone$ and $\SUtwo$ Kronecker matrices
\beq\label{sadjKron}
\sdeone_{\sV^a\sV^b}:=\de_{\sV^a B}\de_{\sV^b B},\qquad
\sdetwo_{\sV^a\sV^b}:=\de_{\sV^a\sV^b}-\sdeone_{\sV^a\sV^b},
\eeq
where $\de_{\sV^a\sV^b}$ represents the usual Kronecker symbol.
In the physical basis, \refeq{sadjKron} translate into
\beq\label{adjKron}
\deone=\left(\begin{array}{c@{\;}c@{\;}c@{\quad}c}
\cw^2 & \sw\cw & 0 & 0 \\ 
\sw\cw & \sw^2 & 0 & 0 \\
0 & 0 & 0 & 0 \\
0 & 0 & 0 & 0
\end{array}\right),\qquad
\detwo=\left(\begin{array}{c@{\;}c@{\;}c@{\quad}c}
\sw^2 & -\sw\cw & 0 & 0 \\ 
-\sw\cw & \cw^2 & 0 & 0 \\
0 & 0 & 1 & 0 \\
0 & 0 & 0 & 1
\end{array}\right).
\eeq
We also define the antisymmetric matrix  $\antikro$, whose components are all vanishing except for $\antikro_{AZ}=-\antikro_{ZA}=1$, \ie
\beq\label{adjantiKron}
\antikro:=\left(\begin{array}{c@{\;}c@{\;}c@{\quad}c}
0 & 1 \,& 0 & 0 \\ 
-1 & 0 & 0 & 0 \\
0 & 0 & 0 & 0 \\
0 & 0 & 0 & 0
\end{array}\right).
\eeq
This matrix  is invariant with respect to Weinberg rotations, \ie
\beq
U(\thw)\antikro U^+(\thw)=\antikro.
\eeq

\subsection*{Totally antisymmetric tensor and adjoint representation}
In the symmetric basis, we define the totally antisymmetric tensor as usual by
\beq\label{totalantitens}
\tilde{\varepsilon}^{\sV^a \sV^b\sV^c}=
\left\{\begin{array}{c@{\quad}l} 
(-1)^p & \mbox{if}\quad \sV^a\sV^b\sV^c= \pi(W^1W^2W^3), \\  
0 & \mbox{otherwise,} 
\end{array}\right.
\eeq
where $(-1)^p$ represents the sign of the permutation $\pi$.
In order to preserve a manifestly totally antisymmetric form,
as a convention for the transformation behaviour of \refeq{totalantitens}
we treat all indices  as covariant [in the sense of \refeq{covectortransf}].
As a result, in  the physical  basis we have
\beq\label{phystotalantitens}
\varepsilon^{V^a V^bV^c}=
-\ri \times \left\{\begin{array}{c@{\quad}l} 
 (-1)^p U_{NW^3}(\thw) & \mbox{if}\quad V^aV^bV^c= \pi(N W^+W^-), \\  
0 & \mbox{otherwise,} 
\end{array}\right.
\eeq
where $(-1)^p$ represents the sign of the permutation $\pi$,  $U_{A W^3}(\thw)=-\sw$ and $U_{Z W^3}(\thw)=\cw$. 

Using the above conventions, we can write the well-known relations for the self-contraction of the totally antisymmetric tensor in the invariant form
\beqar
\sum_{V^c} \varepsilon^{V^a V^b V^c}\varepsilon^{V^{a'} V^{b'}\bar{V}^c}
&=&\detwo_{\bar{V}^{a}V^{a'}}\detwo_{\bar{V}^{b}V^{b'}}-\detwo_{\bar{V}^{a}V^{b'}}\detwo_{\bar{V}^{b}V^{a'}}\nl
\sum_{V^b,V^c} \varepsilon^{V^a V^b V^c}\varepsilon^{V^{a'} \bar{V}^{b}\bar{V}^c}
&=&2\detwo_{\bar{V}^{a}V^{a'}},
\eeqar
where  the $\SUtwo$ Kronecker matrices on the rhs are given by \refeq{sadjKron}
or \refeq{adjKron} depending on the basis.

As well-known, the generators of the gauge group
in the adjoint representation $I^{V^a}_{V^cV^b}$ [see \refeq{symmadjgen}] 
are proportional to the totally antisymmetric tensor. 
Their transformation behaviour is as follows:
the upper index must transform as a co-vector \refeq{covectortransf}
whereas the two lower indices must transform as in \refeq{adjmattransf}. 
Therefore we have 
\beq
I^{V^a}_{V^cV^b} \propto \varepsilon^{V^a V^b\bar{V}^c},
\eeq
and we note that care has to be taken in handling the first lower index, which has to be charge conjugated.
In particular, we see that  in the physical basis the generators  are not manifestly totally antisymmetric.

\section{Casimir operator}
The  electroweak Casimir operator is defined as a sum over the squared $\Uone$ and $\SUtwo$ generators
\begin{equation}\label{CasimirEW} 
\cew_{\varphi_i\varphi_{i'}}:=\sum_{V^a=A,Z,W^\pm} \left(I^{V^a}I^{\bar{V}^a}\right)_{\varphi_i\varphi_{i'}}=\frac{1}{\cw^2}\left(\frac{Y}{2}\right)^2_{\varphi_i\varphi_{i'}}+\frac{1}{\sw^2}\ctwo_{\varphi_i\varphi_{i'}},
\end{equation}
where
\beq
\ctwo=\sum_{a=1}^3(T^a)^2
\eeq
is the $\SUtwo$ Casimir operator.
For {\em irreducible representations} with hypercharge $Y_\varphi$ and isospin $T_\varphi$,  the $\SUtwo$ Casimir operator is given by $T_\varphi(T_\varphi+1)$ times the identity matrix. Therefore {\em for fermions and scalars} the electroweak Casimir operator  is {\em diagonal} and reads
\beq\label{casimirew}
\cew_{\varphi_i\varphi_{i'}}=\de_{\varphi_i \varphi_{i'}}
\left[\frac{Y_\varphi^2}{4\cw^2}+
\frac{T_\varphi(T_\varphi+1)}{\sw^2}\right].
\eeq
The {\em adjoint representation} (gauge bosons) {\em is not irreducible}, since the gauge group is semi-simple. In this representation, with $Y_V=0$ and $T_V=1$,  the electroweak Casimir operator is {\em non-diagonal} in the neutral components of the physical basis and reads
\beq\label{physadjointcasimir}
\cew_{V^aV^b}=\frac{2}{\sw^2}\detwo_{V^aV^b},
\eeq
where $\detwo$ is defined in \refeq{adjKron}.
In order to isolate the part of $\cew$ associated with the
charged gauge bosons we also introduce
\begin{equation}
(I^W)^2:=\sum_{V^a=W^\pm} I^{V^a} I^{\bar{V}^a}=\frac{\ctwo-(T^3)^2}{\sw^2}.
\end{equation}

\section{Explicit values for $Y$, $Q$,  $T^3$, $\ctwo$, $(I^A)^2$,
  $(I^Z)^2$, $(I^W)^2$, $C^\ew$, and $I^\pm$} 
Here we list the eigenvalues (or components) of the operators $Y$,
$Q$, $T^3$, $\ctwo$, $(I^A)^2$, $(I^Z)^2$, $(I^W)^2$, $C^\ew$, and
$I^\pm$. In our general results, for incoming
particles or outgoing antiparticles the values of the particles have
to be used, whereas for incoming antiparticles or outgoing particles one has to use the values of the antiparticles.

\subsubsection*{Fermions}
The fermionic doublets $f^\kappa=(f^\kappa_+, f^\kappa_-)^\rT$
transform according to the fundamental or trivial representations of $\SUtwo$, depending on the chirality $\kappa=\rL,\rR$. Except for $I^\pm$, the
above operators are diagonal. For lepton doublets,
$L^\kappa= (\nu^\kappa, l^\kappa)^\rT$, their eigenvalues are






\beq \label{Llept}
\renewcommand{\arraystretch}{1.5}
\begin{array}{c@{\quad}|@{\quad}c@{\quad}c@{\quad}c@{\quad}c@{\quad}c@{\quad}c@{\quad}c@{\quad}c@{\quad}} 
& {Y}/{2}& Q & T^3 &\ctwo  &(I^A)^2&(I^Z)^2&(I^W)^2 & C^\ew\\ 
\hline
 \nu^{\rL}(\phi^*_0),\bar{\nu}^{\rL}(\phi_0) &\mp \frac{1}{2}  &0 & \pm \frac{1}{2} & \frac{3}{4}&  0  & \frac{1}{4\sw^2\cw^2} & \frac{1}{2\sw^2}  & \frac{1+2\cw^2}{4\sw^2\cw^2}  \\

 l^{\rL}(\phi^-),\bar{l}^{\rL}(\phi^+) &\mp \frac{1}{2} &\mp 1 & \mp \frac{1}{2} & \frac{3}{4}   & 1  &  \frac{(\cw^2-\sw^2)^2}{4\sw^2\cw^2} & \frac{1}{2\sw^2} & \frac{1+2\cw^2}{4\sw^2\cw^2}  \\

 l^{\rR},\bar{l}^{\rR}  & \mp 1  & \mp 1 & 0 & 0  & 1  & \frac{\sw^2}{\cw^2} & 0  & \frac{1}{\cw^2} \\
\end{array}
\eeq
where in the first two lines the correspondence \refeq{scaleptcorr} is indicated.  For quark doublets,
$Q^\kappa=(u^\kappa,d^\kappa)^\rT$, the eigenvalues read 
\beq \label{Lquark}
\renewcommand{\arraystretch}{1.5}
\begin{array}{c@{\quad}|@{\quad}c@{\quad}c@{\quad}c@{\quad}c@{\quad}c@{\quad}c@{\quad}c@{\quad}c@{\quad}} 
& {Y}/{2}& Q & T^3 &\ctwo  &(I^A)^2&(I^Z)^2&(I^W)^2 & C^\ew\\ 
\hline
 u^{\rL},\bar{u}^{\rL} &\pm \frac{1}{6} & \pm \frac{2}{3} & \pm \frac{1}{2} & \frac{3}{4}  &  \frac{4}{9}  &  \frac{(3\cw^2-\sw^2)^2}{36\sw^2\cw^2} & \frac{1}{2\sw^2}   & \frac{\sw^2+27\cw^2}{36\cw^2\sw^2} \\

 d^{\rL},\bar{d}^{\rL} &\pm \frac{1}{6} & \mp \frac{1}{3} & \mp \frac{1}{2} & \frac{3}{4}  &  \frac{1}{9}  &  \frac{(3\cw^2+\sw^2)^2}{36\sw^2\cw^2} & \frac{1}{2\sw^2}  & \frac{\sw^2+27\cw^2}{36\cw^2\sw^2} \\

 u^{\rR},\bar{u}^{\rR}  & \pm \frac{2}{3}  & \pm \frac{2}{3} & 0 & 0  & \frac{4}{9}  &  \frac{4}{9}\frac{\sw^2}{\cw^2} & 0 & \frac{4}{9\cw^2}  \\

 d^{\rR},\bar{d}^{\rR}  & \mp \frac{1}{3}  & \mp \frac{1}{3} & 0 & 0  &  \frac{1}{9}  & \frac{1}{9}\frac{\sw^2}{\cw^2} & 0 & \frac{1}{9\cw^2}  \\
\end{array}
\eeq
For left-handed fermions, $I^\pm$ are the usual raising and lowering operators, the non-vanishing components of which read
\beq \label{ferpmcoup}
I^{\si}_{f^\rL_{\si'}f^\rL_{-\si'}}=-I^{\si}_{\bar{f}^\rL_{-\si'}\bar{f}^\rL_{\si'}}=\frac{\de_{\si\si'}}{\sqrt{2}\sw},
\eeq
whereas for right-handed fermions $I^\pm=0$.

\subsubsection*{Scalar fields}
The scalar doublet,
$\Phi= (\phi^+,\phi_0)^\rT$, $\Phi^*= (\phi^-,\phi_0^*)^\rT$,
transforms according to the fundamental representation, and its
quantum numbers correspond to those of left-handed leptons
\refeq{Llept} with 
\beq\label{scaleptcorr}
\phi^+ \leftrightarrow \bar{l}^{\,\rL}, \qquad \phi_0 \leftrightarrow \bar{\nu}^\rL, \qquad
\phi^- \leftrightarrow {l}^\rL, \qquad \phi_0^* \leftrightarrow {\nu}^\rL.
\eeq
In the physical basis, the $\phi_0$ component\footnote{We note that  $\Phi$ denotes the dynamical part of the Higgs doublet whereas the vev is denoted by $\vev$.} is parametrized by the neutral mass-eigenstate fields $H$ and $\chi$, 
\beq \label{Higgschi}
\phi_0=\frac{1}{\sqrt{2}} (H + \ri\chi).
\eeq
In this basis, $S=(H,\chi)$, the operators $Q,\ctwo,(I^A)^2,(I^Z)^2$,
and $\cew$ remain unchanged, while $T^3$ and $Y$ become non-diagonal
in the neutral components
\begin{equation}
T^3_{SS'}=-\left(\frac{Y}{2}\right)_{SS'}=
-\frac{1}{2}\left(\begin{array}{c@{\quad}c}0 & \ri \\  -\ri & 0 \end{array}\right)
,
\end{equation}
so that
\beq \label{ZHcoup}
I^Z_{H\chi}=-I^Z_{\chi H}=\frac{-\ri}{2\sw\cw}.
\eeq
The $I^\pm$ couplings read
\beq \label{scapmcoup}
I^{\si}_{S\phi^{-\si'}}=-I^{\si}_{\phi^{\si'}S}=\de_{\si\si'}I^{\si}_S,
\eeq
with
\beq \label{scapmcoupB}
I^\si_{H}:=-\frac{\si}{2\sw},\qquad
I^\si_{\chi}:=-\frac{\ri}{2\sw}.
\eeq
\subsubsection*{Gauge fields}
In the adjoint representation, \ie for gauge bosons, the generators are fixed by the structure constants of the gauge group through  \refeq{adjgendef} and read
\beq\label{symmadjgen}
I^{V^a}_{V^cV^b}=\frac{\ri}{\sw}\varepsilon^{V^a V^b \bar{V}^c},
\eeq 
where $\varepsilon$ is the totally antisymmetric tensor given by \refeq{totalantitens}  and \refeq{phystotalantitens} in the symmetric and physical basis, respectively.
In particular, in the physical basis we have
\beq\label{physadjgen}
I^{V^a}_{\bar{V}^cV^b}=
\left\{\begin{array}{c@{\quad}l} 
(-1)^{p+1} & \mbox{if}\quad V^aV^bV^c= \pi(AW^+W^-), \\  
(-1)^{p}\frac{\cw}{\sw} & \mbox{if}\quad V^aV^bV^c= \pi(ZW^+W^-), \\  
0 & \mbox{otherwise,} 
\end{array}\right.
\eeq
where $(-1)^p$ represents the sign of the permutation $\pi$
and care  must be taken for the first lower index $\bar{V}^c$, which is charge conjugated.

The  eigenvalues of the  gauge fields in the symmetric basis read
\begin{equation}\label{gaugeeigenvalues}
\renewcommand{\arraystretch}{1.5}
\begin{array}{c@{\quad}|@{\quad}c@{\quad}c@{\quad}c@{\quad}c@{\quad}c@{\quad}c@{\quad}c@{\quad}c@{\quad}} 
& {Y}/{2}&Q & T^3 &\ctwo &(I^A)^2&(I^Z)^2&(I^W)^2 & \cew  \\
\hline 

 W^\pm  & 0  & \pm 1  & \pm 1 & 2   & 1  & \frac{\cw^2}{\sw^2} &\frac{1}{\sw^2}  & \frac{2}{\sw^2}   \\

 W^3  &  0  &  0  & 0  & 2  & 0  & 0 & \frac{2}{\sw^2}  & \frac{2}{\sw^2}  \\

 B  &  0  & 0  & 0 & 0  & 0  & 0 & 0  & 0  \\
\end{array}
\end{equation}
In the physical basis, the operators  in \refeq{gaugeeigenvalues} that have vanishing eigenvalues
in the neutral sector remain unchanged, \ie 
\begin{equation}\label{gaugeeigenvalues2}
\renewcommand{\arraystretch}{1.5}
\begin{array}{c@{\quad}|@{\quad}c@{\quad}c@{\quad}c@{\quad}c@{\quad}c@{\quad}c@{\quad}c@{\quad}c@{\quad}} 
& {Y}/{2}&Q & T^3  &(I^A)^2&(I^Z)^2  \\
\hline 

 W^\pm  & 0  & \pm 1  & \pm 1   & 1  & \frac{\cw^2}{\sw^2}   \\

 Z  &  0  &  0  & 0    & 0  & 0  \\

 A  &  0  & 0  & 0 & 0  & 0   \\
\end{array}
\end{equation}
whereas  the remaining operators  become non-diagonal in the neutral sector 
and read
\beq
\cew=\frac{\ctwo}{\sw^2}=(I^W)^2=\frac{2}{\sw^2}\detwo, 
\eeq
with the Kronecker matrix $\detwo$ defined in \refeq{adjKron}.

Finally, for the non-vanishing physical components of the $I^\pm$ couplings we introduce the notation
\beq \label{gaupmcoup}
I^{\si}_{N W^{-\si'}}=-I^{\si}_{W^{\si'}N}=\de_{\si\si'}I^\si_N,
\eeq
where $I^\si_N=\si U_{NW^3}(\thw)/\sw$, with
\beq\label{gaupmcoup2}
I^\si_A=-\si,\qquad
I^\si_Z=\si\frac{\cw}{\sw}.
\eeq

\section{Dynkin operator}
The group-theoretical object appearing in gauge-boson self-energy diagrams with internal particles $\varphi_i,\varphi_{i'}$ is the Dynkin operator
\begin{equation}\label{Dew}
\dew_{V^aV^b}(\varphi):=\Tr_\varphi\left\{I^{\bar{V}^a}I^{V^b}\right\}=\sum_{\varphi_i,\varphi_{i'}}I^{\bar{V}^a}_{\varphi_i\varphi_{i'}}I^{V^b}_{\varphi_{i'}\varphi_{i}},
\end{equation}
which  depends on the representation of the multiplet $\varphi$.
In the symmetric basis $\dew$ is diagonal and proportional to the Kronecker matrices \refeq{sadjKron}. 
In the physical basis it can be decomposed into the $\Uone$ and $\SUtwo$ parts \refeq{adjKron} as
\beq
\dew_{V^aV^b}(\varphi)=\dew_B(\varphi)\deone_{V^aV^b}+\dew_W(\varphi)\detwo_{V^aV^b}.
\eeq
For the scalar doublet and the left-handed fermionic doublets, $\varphi =\Phi,f^\rL$, 
\beq
\dew_{B}(\varphi)=\frac{Y_{\varphi}^2}{4\cw^2} \Tr_\varphi\{1\}
,\qquad \dew_{W}(\varphi)=\frac{1}{4\sw^2}\Tr_\varphi\{1\}
,
\eeq
with 
\beq
\Tr_{f^\rL}\{1\}=2,\qquad 
\Tr_\Phi\{1\}=4,
\eeq
\ie the left-handed doublet is treated as two complex Dirac fields, whereas the scalar doublet is treated 
as four real scalar fields\footnote{Note that in \citere{Denner:2001jv} also the scalar doublet  has been treated as two 
complex fields. There, we had $\Tr_\Phi\{1\}=2$, so that the Dynkin operator $\dew(\Phi)$ was
half as large as here
but the factor in front of it in \refeq{betafunction}
was twice as large.}.
For the right-handed fermionic singlets, $\varphi =f^\rR$, the eigenvalues read
\beq
\dew_{B}(f^\rR)=\frac{Y_{f^\rR_+}^2+Y_{f^\rR_-}^2}{4\cw^2},\qquad
\dew_{W}(f^\rR)=0,
\eeq
and include the sum over the $f^\rR_+$ (up) and $f^\rR_-$ (down) fermions. 
The explicit values of the  components of the Dynkin operator for the leptonic doublets 
and for the scalar doublet are
\begin{equation} \label{llDynkin} 
\renewcommand{\arraystretch}{1.5}
\begin{array}{c@{\quad}|@{\quad}c@{\quad}c@{\quad}c@{\quad}c@{\quad}} 
&   \dew_{AA} & \dew_{AZ} & \dew_{ZZ} & \dew_{W}\\ 
\hline
 \Phi& 2 & \frac{\sw^2-\cw^2}{\sw\cw}  & \frac{\sw^4+\cw^4}{\sw^2\cw^2} & \frac{1}{\sw^2} \\
 L^{\rL}& 1 & \frac{\sw^2-\cw^2}{2\sw\cw}  & \frac{\sw^4+\cw^4}{2\sw^2\cw^2} & \frac{1}{2\sw^2} \\
 L^{\rR}& 1 & \frac{\sw}{\cw}  & \frac{\sw^2}{\cw^2} & 0 \\
 L^{\rL}+L^{\rR}& 2 & \frac{3\sw^2-\cw^2}{2\sw\cw}  & \frac{3\sw^4+\cw^4}{2\sw^2\cw^2} & \frac{1}{2\sw^2} \\
\end{array}
\end{equation}
and for the quark doublets
\begin{equation}
\renewcommand{\arraystretch}{1.5}
\begin{array}{c@{\quad}|@{\quad}c@{\quad}c@{\quad}c@{\quad}c@{\quad}} 
&   \dew_{AA} & \dew_{AZ} & \dew_{ZZ} & \dew_{W}\\ 
\hline
 Q^{\rL}& \frac{5}{9} & \frac{\sw^2-9\cw^2}{18\sw\cw}  & \frac{\sw^4+9\cw^4}{18\sw^2\cw^2} & \frac{1}{2\sw^2} \\
 Q^{\rR}& \frac{5}{9} &\frac{5}{9} \frac{\sw}{\cw}  &\frac{5}{9} \frac{\sw^2}{\cw^2} & 0 \\
 Q^{\rL}+Q^{\rR}& \frac{10}{9} & \frac{11\sw^2-9\cw^2}{18\sw\cw}  & \frac{11\sw^4+9\cw^4}{18\sw^2\cw^2} & \frac{1}{2\sw^2} \\
\end{array}
\end{equation}
In the adjoint representation the Dynkin operator corresponds to the electroweak Casimir operator \refeq{physadjointcasimir},
\beq 
\dew_{V^aV^b}(V)=\cew_{V^aV^b},
\eeq
with eigenvalues
\beq
\dew_{B}(V)=0,\qquad \dew_{W}(V)=\frac{2}{\sw^2}.
\eeq

\section{$\beta$-function coefficients}\label{app:betafunction}
In gauge-boson self-energies and mixing-energies, the sum of
gauge-boson, scalar, and fermionic loops give the following
combination of Dynkin operators
\beq \label{betafunction}
\bew_{V^aV^b}:=\frac{11}{3}\dew_{V^aV^b}(V)-\frac{1}{6}\dew_{V^aV^b}(\Phi)-\frac{2}{3}\sum_{f=Q,L}\sum_{j=1,2,3} \NCf \sum_{\la=\rR,\rL}\dew_{V^aV^b}(f_j^\lambda),
\eeq
which is proportional to the one-loop coefficients of the
$\beta$-function. This can be decomposed into the $\Uone$ and $\SUtwo$ invariant parts 
\beq \label{vvbetafunctionx}
\bew_{V^aV^b}=\bew_B\deone_{V^aV^b}+\bew_W\detwo_{V^aV^b},
\eeq
with the Kronecker matrices \refeq{adjKron}. The eigenvalues are given by
\beq
\bew_{B}= -\frac{41}{6\cw^2},\qquad 
\bew_{W}=\frac{19}{6\sw^2},
\eeq
and describe the running of the hypercharge and weak-isospin coupling constants [\cf\refeq{gCTs}], respectively. 
In the physical basis \refeq{adjKron}, the single components read
\beqar \label{betarelations}
&&\bew_{AA}=\cw^2\bew_{B}+\sw^2\bew_{W}=-\frac{11}{3},\qquad
\bew_{AZ}=\cw\sw(\bew_{B}-\bew_{W})=-\frac{19+22\sw^2}{6\sw\cw},\nl
&&\bew_{ZZ}=\sw^2\bew_{B}+\cw^2\bew_{W}=\frac{19-38\sw^2-22\sw^4}{6\sw^2\cw^2},\qquad
\bew_{WW}=\bew_{W}=\frac{19}{6\sw^2}.
\eeqar
The $AA$ component determines the running of the electric
charge, and the $AZ$ component is associated with the running of the
weak mixing angle [\cf\refeq{chargerenorm} and \refeq{weinbergrenorm}].

\chapter{Electroweak Lagrangian and Feynman rules}\label{Feynrules}
In this appendix, we describe the electroweak Lagrangian \cite{DennBohmJos} and the corresponding Feynman rules using our 
conventions\footnote{The parametrization and the conventions we adopt are equivalent to those of \citere{DennBohmJos}. 
However, we use the generic group-theoretical quantities introduced in \refapp{app:representations}.}.
This formulation is invariant with respect to unitary mixing transformations in the gauge sector. 
Explicit  expressions in the physical or in the symmetric basis can be obtained using the corresponding representations given in \refapp{app:representations}.

In \refses{firstsector}--\ref{lastsector} we present the various parts of the electroweak Lagrangian
\beq\label{Lagrangian}
\L_\ew=\L_{\mathrm{gauge}}+\L_{\mathrm{scalar}}+\L_{\mathrm{Yukawa}}+\L_{\mathrm{ferm.}}+\L_{\mathrm{fix}}+\L_{\mathrm{ghost}},
\eeq
and  list the corresponding vertices.
Our notation for 3-point vertex functions is
\beqar
\ri\Gamma^{\varphi_{i_1}\varphi_{i_2} \varphi_{i_3}}(p_1,p_2,p_3)=
\vcenter{\hbox{
\begin{picture}(110,90)(-50,-45)
\Text(-50,5)[lb]{$\varphi_{i_1}(p_1)$}
\Text(35,30)[b]{$\varphi_{i_2}(p_2)$}
\Text(35,-30)[t]{$\varphi_{i_3}(p_3)$}
\Vertex(0,0){2}
\Line(0,0)(35,25)
\Line(0,0)(35,-25)
\Line(0,0)(-45,0)
\end{picture}}},
\eeqar
where all momenta are incoming. A similar notation holds for $n$-point functions.
The conventions and the Feynman rules for propagators are summarized in \refse{se:propagators}. 

The covariant derivative, which generates the gauge interactions, reads 
\beq\label{covder}
\left(\D_\mu\right)_{\varphi_i\varphi_j}=\de_{\varphi_i\varphi_j}\partial_\mu-\ri e \sum_{V^a=A,Z,W^\pm}I^{V^a}_{\varphi_i\varphi_j}V^a_\mu,
\eeq
in our notation.

\section{Gauge sector}\label{firstsector}
The gauge sector of the Lagrangian,
\beq\label{gaugeLagrangian}
\L_{\mathrm{gauge}}=-\frac{1}{4}\sum_{V^a}\F^{V^a,\mu\nu}\F^{\bar{V}^a}_{\mu\nu},
\eeq
with the field strength tensor
\beq
\F_{\mu\nu}^{V^a}=\frac{\ri}{e}\left[\D_\mu,\D_\nu\right]^{V^a}=\partial_\mu V^a_\nu-\partial_\nu V^a_\mu-\ri e\sum_{V^b,V^c}I^{V^b}_{V^aV^c}V^b_\mu V^c_\nu,
\eeq
gives rise to the triple and to the quartic gauge-boson vertices
\beqar
\begin{array}{l}
\vcenter{\hbox{
\begin{picture}(110,90)(-50,-45)
\Text(-50,5)[lb]{$V^{a_1}_{\mu_1}(k_1)$}
\Text(35,30)[b]{$V^{a_2}_{\mu_2}(k_2)$}
\Text(35,-30)[t]{$V^{a_3}_{\mu_3}(k_3)$}
\Vertex(0,0){2}
\Photon(0,0)(35,25){2}{3.5}
\Photon(0,0)(35,-25){2}{3.5}
\Photon(0,0)(-45,0){2}{3.5}
\end{picture}}}
\end{array}
&&
\begin{array}{l}
\\
=
{\displaystyle \frac{e}{\sw}}
\varepsilon^{V^{a_1}V^{a_2}V^{a_3}}
\Bigl[
g_{\mu_1\mu_2}(p_1-p_2)_{\mu_3}
+g_{\mu_2\mu_3}(p_2-p_3)_{\mu_1}
\\
\hphantom{=\frac{e}{\sw}\varepsilon^{V^{a_1}V^{a_2}V^{a_3}}\Bigl[}
+g_{\mu_3\mu_1}(p_3-p_1)_{\mu_2}\Bigr],
\end{array}
\nl\\
\begin{array}{l}
\vcenter{\hbox{
\begin{picture}(110,90)(-50,-45)
\Text(-35,30)[cb]{$V^{a_1}_{\mu_1}$}
\Text(-35,-30)[ct]{$V^{a_3}_{\mu_3}$}
\Text(35,30)[cb]{$V^{a_2}_{\mu_2}$}
\Text(35,-30)[ct]{$V^{a_4}_{\mu_4}$}
\Vertex(0,0){2}
\Photon(0,0)(35,25){2}{3.5}
\Photon(0,0)(35,-25){2}{3.5}
\Photon(0,0)(-35,25){2}{3.5}
\Photon(0,0)(-35,-25){2}{3.5}
\end{picture}}}
\end{array}
&&
\begin{array}{l}
\\
=-\ri{\displaystyle \frac{e^2}{\sw^2}}\Bigl[ 
\detwo_{\bar{V}^{a_1}V^{a_2}}\detwo_{\bar{V}^{a_3}V^{a_4}}
(2g_{\mu_1\mu_2}g_{\mu_3\mu_4}-g_{\mu_1\mu_3}g_{\mu_2\mu_4}
\\
\hphantom{=-\ri\frac{e^2}{\sw^2}\Bigl[}
-g_{\mu_1\mu_4}g_{\mu_2\mu_3})
+(2\leftrightarrow 3)+(2\leftrightarrow 4)\Bigr],
\end{array}
\nonumber\\
\eeqar
where $\varepsilon$ is the totally antisymmetric tensor given by \refeq{totalantitens}, \refeq{phystotalantitens}
and $\detwo$ is the $\SUtwo$ Kronecker matrix defined in \refeq{adjKron}.
The gauge-boson propagators depend on the gauge-fixing Lagrangian and are given in \refeq{gaugeprop1} and \refeq{gaugeprop2}.

\section{Scalar sector}
The  scalar sector of the Lagrangian consists of 
\beqar\label{scalarLagrangian}
\L_{\mathrm{scalar}}&=&\left[\D^\mu (\Phi+\vev) \right]^+ \D_\mu (\Phi+\vev) +\mu^2|\Phi+\vev|^2-\frac{\gH}{4}|\Phi+\vev|^4
\nl&=&
\frac{1}{2}\sum_{\Phi_i,\Phi_j,\Phi_k}(\D^+_\mu)_{\Phi^+_i\Phi_k}(\D^\mu)_{\Phi_k\Phi_j}(\Phi+\vev)_i(\Phi+\vev)_j \nl
&&{}+\frac{\mu^2}{2}\sum_{\Phi_i}(\Phi+\vev)^+_i(\Phi+\vev)_i 
-\frac{\gH}{16}\left(\sum_{\Phi_i}(\Phi+\vev)^+_i(\Phi+\vev)_i   
\right)^2,
\eeqar
where the sums run over the four dynamical components $\Phi_i=H,\chi,\phi^\pm$ of the doublet. The corresponding vev is denoted by 
\beq\label{vev}
\vev_i=v\,\de_{H\Phi_i}
\eeq
and corresponds to the minimum of the potential. In lowest order, the vev and the parameters $\mu$, $\gH$ of the potential are related to the physical parameters $\sw$, $\MW$ and $\MH$ by 
\beq
\mu=\frac{\MH}{\sqrt{2}},\qquad v=\frac{2\sw\MW}{e},\qquad \gH=\frac{e^2\MH^2}{2\sw^2\MW^2}.
\eeq
The kinetic-energy term of \refeq{scalarLagrangian}
gives rise to the vertices
\beqar
\begin{array}{l}
\vcenter{\hbox{
\begin{picture}(110,100)(-50,-50)
\Text(-45,5)[lb]{$V^a_{\mu}$}
\Text(35,30)[cb]{$\Phi_{i_1} (p_1)$}
\Text(35,-30)[ct]{$\Phi_{i_2}(p_2)$}
\Vertex(0,0){2}
\DashLine(0,0)(35,25){5}
\DashLine(0,0)(35,-25){5}
\Photon(0,0)(-45,0){2}{3}
\end{picture}}}
\end{array}
&&
\begin{array}{l}
=\ri e I^{V^a}_{\Phi^+_{i_1}\Phi_{i_2}}(p_2-p_1)_\mu,
\end{array}
\\
\begin{array}{l}
\vcenter{\hbox{
\begin{picture}(110,100)(-50,-50)
\Text(-35,30)[cb]{$\Phi_{i_1}$}
\Text(-35,-30)[ct]{$\Phi_{i_2}$}
\Text(35,30)[cb]{$V^{a_1}_{\mu_1}$}
\Text(35,-30)[ct]{$V^{a_2}_{\mu_2}$}
\Vertex(0,0){2}
\Photon(0,0)(35,25){2}{3}
\Photon(0,0)(35,-25){-2}{3}
\DashLine(0,0)(-35,25){5}
\DashLine(0,0)(-35,-25){5}
\end{picture}}}
\end{array}
&&
\begin{array}{l}
=\ri e^2 g_{\mu_1\mu_2}\left\{I^{V^{a_1}},I^{V^{a_2}}\right\}_{\Phi^+_{i_1}\Phi_{i_2}},
\end{array}
\\
\begin{array}{l}
\vcenter{\hbox{
\begin{picture}(110,100)(-50,-50)
\Text(-45,5)[lb]{$\Phi_i$}
\Text(35,30)[cb]{$V^{a_1}_{\mu_1}$}
\Text(35,-30)[ct]{$V^{a_2}_{\mu_2}$}
\Vertex(0,0){2}
\Photon(0,0)(35,25){2}{3}
\Photon(0,0)(35,-25){2}{3}
\DashLine(0,0)(-45,0){5}
\end{picture}}}
\end{array}
&&
\begin{array}{l}
=\ri e^2 v g_{\mu_1\mu_2}\left\{I^{V^{a_1}},I^{V^{a_2}}\right\}_{H\Phi_i},
\end{array}
\eeqar
where the curly brackets denote anti-commutators. 
{}From the Higgs potential we obtain the vertices
\beqar
\begin{array}{l}
\vcenter{\hbox{
\begin{picture}(110,100)(-50,-50)
\Text(-35,30)[cb]{$\Phi_{i_1}$}
\Text(-35,-30)[ct]{$\Phi_{i_3}$}
\Text(35,30)[cb]{$\Phi_{i_2}$}
\Text(35,-30)[ct]{$\Phi_{i_4}$}
\Vertex(0,0){2}
\DashLine(0,0)(35,25){5}
\DashLine(0,0)(35,-25){5}
\DashLine(0,0)(-35,25){5}
\DashLine(0,0)(-35,-25){5}
\end{picture}}}
\end{array}
&&
\begin{array}{l}
\\
=-\ri{\displaystyle\frac{\gH}{2}}
\Bigl(\de_{\Phi^+_{i_1}\Phi_{i_2}}\de_{\Phi^+_{i_3}\Phi_{i_4}}+\de_{\Phi^+_{i_1}\Phi_{i_4}}
\de_{\Phi^+_{i_3}\Phi_{i_2}}
\\ \hphantom{=-\ri\frac{\gH}{2}\Bigl(}
+\de_{\Phi^+_{i_1}\Phi_{i_3}}\de_{\Phi^+_{i_2}\Phi_{i_4}}\Bigr),
\end{array}
\\
\begin{array}{l}
\vcenter{\hbox{
\begin{picture}(110,100)(-50,-50)
\Text(-45,5)[lb]{$\Phi_{i_1}$}
\Text(35,30)[cb]{$\Phi_{i_2}$}
\Text(35,-30)[ct]{$\Phi_{i_3}$} 
\Vertex(0,0){2}
\DashLine(0,0)(35,25){5}
\DashLine(0,0)(35,-25){5}
\DashLine(0,0)(-45,0){5}
\end{picture}}}
\end{array}
&&
\begin{array}{l}
\\
=-\ri{\displaystyle \frac{\gH v}{2}}\Bigr(\de_{H\Phi_{i_1}}\de_{\Phi^+_{i_2}\Phi_{i_3}}+\de_{H\Phi_{i_2}}\de_{\Phi^+_{i_3}\Phi_{i_1}}
\\ \hphantom{=-\ri\frac{\gH v}{2}\Bigr(}
+\de_{H\Phi_{i_3}}\de_{\Phi^+_{i_1}\Phi_{i_2}}\Bigl),
\end{array}
\eeqar
where $\de_{\Phi_i\Phi_j}$ is the usual Kronecker symbol. 
The propagators for Higgs and would-be Goldstone bosons are given in \refeq{scaprop}.

\section{Yukawa  sector}
In the Yukawa sector, the Lagrangian reads
\beq\label{YukLagrangian}
\L_{\mathrm{Yukawa}}=-\sum_{f=Q,L}\sum_{i,j=1}^3\left(G^{f_-}_{ij} \bar{f}^\rL_i (\Phi+\vev) f^\rR_{j,-} +G^{f_+}_{ij} \bar{f}^\rL_i (\Phi+\vev)^c f^\rR_{j,+} +\mathrm{h.c.}\right),
\eeq
where $\Phi=(\phi^+,\phi_0)^\rT$, $\Phi^c=(\phi^*_0,-\phi^-)$, multiply the left-handed doublets $f^\rL_i$ with flavour $i=1,2,3$, and $G^{f_\pm}$ are the Yukawa matrices for $f_+$ (up) and $f_-$ (down) fermions. These matrices can be diagonalized by the unitary transformations
\beqar
f^\kappa_{i,\si}&\rightarrow& \sum_{i'=1}^3U^{f^\kappa_\si}_{ii'}f^\kappa_{i',\si}\,,\qquad f=Q,L,\quad \kappa=\rR,\rL,\quad \si=\pm,\nl
G^{f_\si}_{ij}&\rightarrow& \de_{ij}\la_{f_{i,\si}},
\eeqar  
and the resulting eigenvalues are related to the fermion masses by 
\beq
\la_{f_{j,\si}}=\frac{e\, m_{f_{j,\si}}}{\sqrt{2}\sw\MW},
\eeq
in lowest order. The mixing transformation enters the interaction through the 
CKM matrix
\beq
\ckm_{ij}=\sum_{k=1}^3U^{Q^\rL_+}_{ik}\left(U^{Q^\rL_-}\right)^+_{kj}. 
\eeq
In order to describe the corresponding effects in a generic way we introduce the matrix\footnote{In the calculations we often use the shorthand $U^{V^a}_{jj'}=U^{f^\kappa,V^a}_{jj'}$, where the dependence on $f^\kappa$ is implicitly understood.}
\beq\label{Umatrix}
U^{f^\kappa,V^a}_{jj'}:=\left\{\begin{array}{c@{\quad}l} 
\ckm_{jj'} & \mbox{if}\quad V^a=W^+ \quad\mbox{and}\quad f^\kappa=Q^\rL, \\
\ckm^+_{jj'} & \mbox{if}\quad V^a=W^- \quad\mbox{and}\quad f^\kappa=Q^\rL, \\
 \de_{jj'} & \mbox{otherwise.} 
\end{array}\right.
\eeq
The Yukawa vertices resulting from \refeq{YukLagrangian}
can be written as\footnote{The following vertices correspond to $\Gamma^{\Phi_i\bar{\Psi}_{j,\si}\Psi_{j',\si'}}=-\Gamma^{\Phi_i\Psi_{j',\si'}\bar{\Psi}_{j,\si}}$.} 
\beqar
\begin{array}{l}
\vcenter{\hbox{
\begin{picture}(110,100)(-50,-50)
\Text(-45,5)[lb]{$S$}
\Text(35,30)[cb]{$\bar{\Psi}_{j,\si}$}
\Text(35,-30)[ct]{$\Psi_{j',\si'}$}
\Vertex(0,0){2}
\ArrowLine(0,0)(35,25)
\ArrowLine(35,-25)(0,0)
\DashLine(0,0)(-45,0){5}
\end{picture}}}
\end{array}
&&
\begin{array}{l}
=\de_{\si\si'}\de_{jj'}{\displaystyle \frac{\la_{f_{j,\si}}}{\sqrt{2}}}\times 
\left\{\begin{array}{c@{\quad}l} 
\ri & \mbox{if}\quad S=H, \\  
\si \gamma^5 & \mbox{if}\quad S=\chi,
\end{array}\right.
\end{array}
\\
\begin{array}{l}
\vcenter{\hbox{
\begin{picture}(110,100)(-50,-50)
\Text(-45,5)[lb]{$\phi^\pm$}
\Text(35,30)[cb]{$\bar{\Psi}_{j,\si}$}
\Text(35,-30)[ct]{$\Psi_{j',\si'}$}
\Vertex(0,0){2}
\ArrowLine(0,0)(35,25)
\ArrowLine(35,-25)(0,0)
\DashLine(0,0)(-45,0){5}
\end{picture}}}
\end{array}
&&
\begin{array}{l}
=\mp\ri
\de_{\si\pm}\de_{\si'\mp}U^{W^\pm}_{jj'}(\omega_- \la_{f_{j,\pm}}- \omega_+ \la_{f_{j',\mp}}),
\end{array}
\eeqar
with the chiral projectors
\beq
\omega_\pm=\frac{1}{2}(1\pm\gamma^5).
\eeq

\section{Fermionic  sector}
The  fermionic gauge interactions originate from 
\beq\label{fergaugeLagrangian}
\L_{\mathrm{ferm.}}=\ri \sum_{f=Q,L}\sum_{j,j'=1}^3\sum_{\kappa=\rR,\rL}\sum_{\si,\si'=\pm}
\bar{f}^\kappa_{j,\si} \left[\gamma_\mu(\D^\mu)_{f^\kappa_{j,\si} f^\kappa_{j',\si'}} \right]f^\kappa_{j',\si'}\, ,
\eeq
where the generators in the covariant derivative \refeq{covder} for $\varphi_i=f^\kappa_{j,\si}$ have to be understood as  generalized generators\footnote{In the calculations, we often use the shorthand notation $I^{V^a}_{\si{\si'}}=I^{V^a}_{f^\kappa_\si f^\kappa_{\si'}}$,  where the dependence on $f^\kappa$ is implicitly understood.}
\beq\label{fermgenerators}
I^{V^a}_{f^\kappa_{j,\si}f^\kappa_{j',\si'}}:=I^{V^a}_{f^\kappa_\si f^\kappa_{\si'}}U^{f^\kappa,V^a}_{jj'},
\eeq
involving the flavour-mixing matrix \refeq{Umatrix}.
The resulting fermionic gauge vertices read\footnote{The following vertices correspond to $\Gamma_\mu^{V^a\bar{\Psi}_{j,\si}\Psi_{j',\si'}}=-\Gamma_\mu^{V^a\Psi_{j',\si'}\bar{\Psi}_{j,\si}}$.}
\beqar
\begin{array}{l}
\vcenter{\hbox{
\begin{picture}(110,100)(-50,-50)
\Text(-45,5)[lb]{$V^a_{\mu}$}
\Text(35,30)[cb]{$\bar{\Psi}_{j,\si}$}
\Text(35,-30)[ct]{$\Psi_{j',\si'}$}
\Vertex(0,0){2}
\ArrowLine(0,0)(35,25)
\ArrowLine(35,-25)(0,0)
\Photon(0,0)(-45,0){2}{3}
\end{picture}}}
\end{array}
\begin{array}{l}
=-\ri e \gamma_\mu \left(\omega_+ I^{V^a}_{f^\rR_{j,\si}f^\rR_{j',\si'}}
+\omega_- \, I^{V^a}_{f^\rL_{j,\si}f^\rL_{j',\si'}}\right).
\end{array}
\eeqar
The fermionic propagators originate from $\L_{\mathrm{ferm.}}+\L_{\mathrm{Yukawa}}$ and are given by \refeq{fermprop}.

\section{Gauge-fixing  Lagrangian}
The gauge-fixing Lagrangian is given by
\beq\label{gfixlagrangian}
\mathcal{L}_{\mathrm{fix}}=
-\sum_{V^a=A,Z,W^\pm}\frac{1}{2\xi_a}C^{{V}^a}C^{\bar{V}^a},
\eeq
and depends on the gauge parameters  $\xi_A$, $\xi_Z$, and
$\xi_+=\xi_-$.
A 't~Hooft gauge fixing is given by
\beqar  \label{Gfix}
C^{\bar{V}^a}\{V,\Phi,x\}&=&
\partial^\mu \bar{V}^a_\mu-
\ri e v \xi_{{a}}\sum_{\Phi_{i}=H,\chi,\phi^\pm}I^{V^a}_{H\Phi_i}\Phi_i
=\partial^\mu \bar{V}^a_\mu-
\ri^{Q_{\bar{V}^a}}\xi_{{a}} 
M_{V^a}\Phi^+_a.
\eeqar
Here, the Higgs gauge couplings $I^{V^a}_{H\Phi_i}$ 
relate the gauge fields $V^a=Z,W^\pm$ to the associated
would-be Goldstone boson  fields $\Phi_a=\chi,\phi^\pm$ through
\beq\label{Higgsgaugecoup}
\ri e v I^{V^a}_{H\Phi_j^+}= 
-\ri e v I^{V^a}_{\Phi_j H}= \ri^{Q_{\bar{V}^a}} \de_{\Phi_j\Phi_a} M_{V^a}
,\qquad  |I^{V^a}_{H\Phi_j}|^2 = \de_{\Phi_j\Phi_a} \frac{M^2_{V^a}}{4\sw^2\MW^2}.
\eeq
In the 't~Hooft gauge 
the contributions of the would-be Goldstone bosons to the gauge-fixing terms cancel the lowest-order  mixing between gauge bosons and would-be Goldstone bosons, such that the lowest-order propagators are diagonal. 
These are given by \refeq{gaugeprop1},\refeq{gaugeprop2} for gauge bosons and  by \refeq{scaprop} for Higgs bosons and would-be Goldstone bosons.

\section{Ghost Lagrangian}\label{lastsector}
The  ghost Lagrangian corresponding to the gauge-fixing \refeq{gfixlagrangian}, \refeq{Gfix} is given by 
\beqar\label{ghostLagrangian}
\L_{\mathrm{ghost}}&=&
-\sum_{V^a,V^b}\int\rd^4 y\, \bar{u}^a(x)
\frac{\de C^{V^a}(x)}{\de\theta^{V^b}(y)}u^b(y)\\
&=&\sum_{V^a,V^b}\left[\left(\partial_\mu \bar{u}^a\right)\left(\D^\mu\right)_{V^aV^b} u^b
-\xi_a e^2 v \sum_{\Phi_i}\left(I^{\bar{V}^a}I^{V^b}\right)_{H\Phi_i} \Phi_i\bar{u}^a u^b
\right],\nn
\eeqar
where the ghost and antighost fields are denoted by\footnote{The ghosts and antighosts transform  
as vectors  and covectors, respectively, under mixing transformations  \refeq{defweinrot}. 
In particular, $u^\pm=(u^1\mp\ri u^2)/\sqrt{2}$ and $\bar{u}^\pm=(\bar{u}^1\pm\ri\bar{u}^2)/\sqrt{2}$.} $u^a=u^{V^a}$ and $\bar{u}^a=\bar{u}^{V^a}$.
The Lagrangian \refeq{ghostLagrangian} generates the vertices
\beqar
\vcenter{\hbox{
\begin{picture}(110,100)(-50,-50)
\Text(-45,3)[lb]{$V^a_{\mu}$}
\Text(35,30)[cb]{$\bar{u}^b(p_b)$}
\Text(35,-30)[ct]{$u^c(p_c)$}
\Vertex(0,0){2}
\DashArrowLine(0,0)(35,25){1}
\DashArrowLine(35,-25)(0,0){1}
\Photon(0,0)(-45,0){2}{3}
\end{picture}}}
&&
={\displaystyle \frac{e}{\sw}}\,
\varepsilon^{V^{a}\bar{V}^{b}V^{c}}
p_{b\mu}
,
\\
\vcenter{\hbox{
\begin{picture}(110,100)(-50,-50)
\Text(-45,5)[lb]{$\Phi_i$}
\Text(35,30)[cb]{$\bar{u}^a(p_a)$}
\Text(35,-30)[ct]{$u^b(p_b)$}
\Vertex(0,0){2}
\DashArrowLine(0,0)(35,25){1}
\DashArrowLine(35,-25)(0,0){1}
\DashLine(0,0)(-45,0){5}
\end{picture}}}
&&=\ri \xi_a e^2 v  \left(I^{\bar{V}^a}I^{V^b}\right)_{H\Phi_i}.
\eeqar
The propagators for ghost fields are given by \refeq{ghostprop}.

\section{Feynman rules for propagators}\label{se:propagators}
For the propagators of commuting fields we use the notation
\beq
G^{\varphi_i \varphi^+_j}(p)=
\vcenter{\hbox{
\begin{picture}(100,40)(-40,-20)
\Line(35,0)(-35,0)
\Vertex(35,0){2}
\Vertex(-35,0){2}
\Text(-34,5)[cb]{$\varphi_{i}(p)$}
\Text(34,5)[cb]{$\varphi^+_{j}(-p)$}
\end{picture}}}.
\eeq 
The gauge-boson propagators read 
\beq\label{gaugeprop1}
\vcenter{\hbox{
\begin{picture}(100,40)(-40,-20)
\Photon(35,0)(-35,0){2}{4}
\Vertex(35,0){2}
\Vertex(-35,0){2}
\Text(-34,5)[cb]{$V^a_{\mu}(p)$}
\Text(34,5)[cb]{$\bar{V}^b_{\nu}(-p)$}
\end{picture}}}
= \left(g_{\mu\nu}-\frac{p_\mu p_\nu}{p^2}\right)G^{V^a\bar{V}^b}_{\rT}(p^2)
+\frac{p_\mu p_\nu}{p^2}G^{V^a\bar{V}^b}_{\rL}(p^2),
\eeq
with
\beq\label{gaugeprop2}
G^{V^a\bar{V}^b}_{\rT}(p^2)= 
\frac{-\ri\de_{V^aV^b}}{p^2-M_{V^a}^2}
,\qquad
G^{V^a\bar{V}^b}_{\rL}(p^2)= 
\frac{-\ri\xi_a\de_{V^aV^b}}{p^2-\xi_a M_{V^a}^2}.
\eeq
The propagators for Higgs bosons and would-be Goldstone bosons $\Phi_{a,b}=H,\chi,\phi^\pm$ are given by
\beqar\label{scaprop}
\vcenter{\hbox{
\begin{picture}(100,40)(-40,-20)
\DashLine(35,0)(-35,0){5}
\Vertex(35,0){2}
\Vertex(-35,0){2}
\Text(-34,5)[cb]{$\Phi_a(p)$}
\Text(34,5)[cb]{$\Phi^+_b(-p)$}
\end{picture}}}
&=&
\frac{\ri\de_{\Phi_a\Phi_b}}{p^2-M^2_{\Phi_a}},
\eeqar
with
\beq
M^2_{\Phi_a}=
\left\{\begin{array}{c@{\quad}l} 
\MH^2 & \mbox{if}\quad \Phi_a=H, \\  
\xi_{a} M_{V^a}^2 & \mbox{if}\quad \Phi_a=\chi,\phi^\pm,
\end{array}\right.
\eeq
where  $V^a=Z,W^\pm$ are  the weak gauge bosons associated to the would-be Goldstone bosons $\Phi_a=\chi,\phi^\pm$. 
For the propagators of anticommuting fields we adopt the notation
\beq
G^{\varphi_i \varphi^+_j}(-p)
=\pm G^{\varphi^+_j\varphi_i}(p)=
\vcenter{\hbox{
\begin{picture}(100,40)(-40,-20)
\ArrowLine(35,0)(-35,0)
\Vertex(35,0){2}
\Vertex(-35,0){2}
\Text(-34,5)[cb]{$\varphi_{i}(-p)$}
\Text(34,5)[cb]{$\varphi^+_{j}(p)$}
\end{picture}}}.
\eeq 
The fermionic propagators are given by
\beq\label{fermprop}
\vcenter{\hbox{
\begin{picture}(100,40)(-40,-20)
\ArrowLine(35,0)(-35,0)
\Vertex(35,0){2}
\Vertex(-35,0){2}
\Text(-34,5)[cb]{$\Psi_\alpha(-p)$}
\Text(34,5)[cb]{$\bar{\Psi}_\beta(p)$}
\end{picture}}}
=\frac{\ri(\ps+m)_{\alpha\beta}}{p^2-m^2},
\eeq 
where $\alpha,\beta$ are the Dirac indices, and the remaining indices are implicitly understood.
Finally, the ghost-field propagators read
\beq \label{ghostprop}
\vcenter{\hbox{
\begin{picture}(100,40)(-40,-20)
\DashArrowLine(35,0)(-35,0){1}
\Vertex(35,0){2}
\Vertex(-35,0){2}
\Text(-34,5)[cb]{$u^a(-p)$}
\Text(34,5)[cb]{$\bar{u}^b(p)$}
\end{picture}}}
=\frac{\ri\de_{V^aV^b}}{p^2-\xi_a M_{V^a}^2}.
\eeq

\chapter{BRS transformations}\label{BRStra}
In this appendix we summarize the explicit form of the gauge transformations and BRS transformations in the electroweak Standard Model. 
As in \refapp{Feynrules} we follow the conventions of
\citere{DennBohmJos} but we use the generic notation introduced in \refapp{app:representations}.

\section{Gauge transformations}
The classical Lagrangian of the electroweak Standard Model,
\beq\label{clLagrangian}
\L_\mathrm{cl.}=\L_{\mathrm{gauge}}+\L_{\mathrm{scalar}}+\L_{\mathrm{Yukawa}}+\L_{\mathrm{ferm.}},
\eeq
is invariant with respect to
gauge transformations of the physical fields and would-be Goldstone bosons 
which can generically be written as 
\beq \label{physgaugetra}
\de \varphi_i(x)= \sum_{V^a=A,Z,W^\pm}\left[\linbrs^{V^a}_{\varphi_{i}}\de\theta^{V^a}(x)
+\ri e  \sum_{\varphi_{i'}}
I^{V^a}_{\varphi_i\varphi_{i'}}\de\theta^{V^a}(x) \varphi_{i'}(x)\right].
\eeq
The matrices $I^{V^a}_{\varphi_i\varphi_{i'}}$ are the $\ewgroup$ generators 
in the
representation of the fields $\varphi_{i}$ (see \refapp{app:representations}), 
and the linear operator
$\linbrs^{V^a}_{\varphi_{i}}$ represents the
transformation of free fields.
\subsubsection*{Scalar bosons}
For scalar bosons, $\varphi_i=\Phi_{i}+\vev_{i}$, only the dynamical part $\Phi_{i}=H,\chi,\phi^\pm$ transforms
\beq
\de \vev_{i}=0,\qquad
\de \Phi_{i}(x)=
\ri e  \sum_{V^a,\Phi_{i'}}
I^{V^a}_{\Phi_i\Phi_{i'}}\de\theta^{V^a}(x) \left[\Phi_{i'}(x)+\vev_{i'}\right],
\eeq
so that the operator $\linbrs^{V^a}_{\Phi_{i}}$ in \refeq{physgaugetra} is
determined by the contribution of the vev \refeq{vev},
and reads  
\beq
\linbrs^{V^a}_{\Phi_{i}}\de\theta^{V^a}(x)=
\ri ev I^{V^a}_{\Phi_iH}\de\theta^{V^a}(x)=
-\ri^{Q_{\bar{V}^a}}  \de_{\Phi_i\Phi_a}M_{V^a} \de\theta^{V^a}(x),
\eeq
where  $\Phi_a=\chi,\phi^\pm$ are the would-be Goldstone bosons associated to the  gauge fields $V^a=Z,W^\pm$.

\subsubsection*{Gauge bosons}
For gauge bosons, $\varphi_i=V^b=A,Z,W^\pm$, we have
\beq
\linbrs^{V^a}_{V^b_\mu}\de\theta^{V^a}(x)=\de_{V^aV^b}\partial_\mu \de\theta^{V^a}(x),
\eeq
which in momentum space leads to 
\beq
\linbrs^{V^a}_{V^b_\mu}\de\theta^{V^a}(p) = \ri p_\mu\de_{V^aV^b}\de\theta^{V^a}(p) .
\eeq

\subsubsection*{Fermions}
For fermions, $\varphi_i=\Psi^\kappa_{j,\si}$,
\beq
\linbrs^{V^a}_{\Psi^\kappa_{j,\si}}\de\theta^{V^a}(x)=0,
\eeq
and the gauge transformation of the mass-eigenstate fermions 
is determined by the generalized 
generators \refeq{fermgenerators}, which involve the flavour-mixing matrix \refeq{Umatrix}.

\section{BRS transformations}
The  gauge-fixing terms \refeq{gfixlagrangian}
and the corresponding ghost terms \refeq{ghostLagrangian}
break the  gauge invariance of the classical electroweak Lagrangian. However, 
the complete electroweak Lagrangian is invariant 
with respect to BRS transformations \cite{BRS}
of the ghost  and physical fields.

The BRS transformation of the physical fields 
corresponds to a local gauge transformation \refeq{physgaugetra} with
gauge-transformation parameters
$\de\theta^{V^a}(x)=\de\la u^a(x)$ determined by the ghost
fields $u^a(x)$ and the infinitesimal Grassman parameter $\de\la$. To
be precise, the BRS variation $\brs \varphi_i(x)$ is defined as
left derivative\footnote{The product rule for a Grassman left derivative is
$
\brs (\varphi_i\varphi_j) =
(\brs \varphi_i) \varphi_j
+(-1)^{N_i}
\varphi_i \brs \varphi_j ,
$
where $N_i$ is given by the ghost plus the fermion number of
the field $\varphi_i$.} with respect to the Grassman
parameter $\de \la$, \ie 
$\de\varphi_i(x)=\de\la \,\brs \varphi_i(x)$, and reads
\beq \label{physbrstra}
\brs \varphi_i(x)= \sum_{V^a=A,Z,W^\pm}\left[\linbrs^{V^a}_{\varphi_{i}}u^a(x)
+\ri e  \sum_{\varphi_{i'}}
I^{V^a}_{\varphi_i\varphi_{i'}}u^a(x) \varphi_{i'}(x)\right].
\eeq
The BRS variation for charge-conjugate fields is obtained from the
adjoint of \refeq{physbrstra} as 
\beq \label{ccphysbrstra}
\brs \varphi^+_i(x)= \sum_{V^a=A,Z,W^\pm}\left[\linbrs^{V^a}_{\varphi^+_{i}}u^a(x)
-\ri e \sum_{\varphi_{i'}}
u^a(x) \varphi^+_{i'}(x)I^{V^a}_{\varphi_{i'}\varphi_i}\right],
\eeq
where we have used \refeq{selfadjoint}.

The BRS variation of the ghost fields is given by
\beq  \label{ghostbrstra}
\brs u^b(x)=\frac{\ri e}{\rm 2}  \sum_{V^a,V^c=A,Z,W^\pm}
I^{V^a}_{V^bV^c}u^a(x)u^c(x),
\eeq
and the BRS variation of the antighost fields is 
determined by the gauge-fixing functionals \refeq{Gfix} and reads
\beqar  \label{antighostbrstra}
\brs \bar{u}^a(x)&=&-\frac{\rm 1}{\xi_{{a}}}C^{\bar{V}^a}\{V,\Phi,x\}
=
-\frac{\rm 1}{\xi_{{a}}}\partial^\mu \bar{V}^a_\mu
+\ri e v \sum_{\Phi_{i}=H,\chi,\phi^\pm}I^{V^a}_{H\Phi_i}\Phi_i
\nl&=& 
-\frac{\rm 1}{\xi_{{a}}}\partial^\mu \bar{V}^a_\mu
+\ri^{Q_{\bar{V}^a}}
M_{V^a}\Phi^+_a.
\eeqar

\chapter{Production of transverse gauge bosons in fermion--antifermion annihilation}
\label{app:transvRG}
\newcommand{\matg}{G}
\newcommand{\smatg}{\tilde{\matg}}
For transverse W-pair production, we have observed in \refse{eeWW}
that the symmetric
electroweak parts of the PR contributions \refeq{PRbornmel}
and of the mass-singular SL corrections  \refeq{completeCCcorr} which are 
associated to the external transverse gauge bosons cancel exactly.
Here we illustrate how this cancellation takes place 
for all processes of the type
\beq\label{GBprodproc}
f^\kappa_{j,\si} \bar{f}^\kappa_{j',\si'}\rightarrow 
V^{a_1}_{\rT}\cdots V^{a_n}_{\rT},
\eeq
where an arbitrary number $n$ of neutral or charged transverse gauge bosons 
$V^{a_k}_\rT=\PA_\rT,\PZ_\rT,\PW^\pm_\rT$ are produced in fermion--antifermion annihilation.
The mass-singular  SL corrections for the process \refeq{GBprodproc}  which are associated to the external particles
read  
\beqar\label{gbprodcccorr}
\de^{\cc} \M^
{f^\kappa_{j,\si} \bar{f}^\kappa_{j',\si'} V^{a_1}_\rT \ldots\, V^{a_n}_\rT}
&=& \left(\de^\cc_{f^\kappa_{j,\si} f^\kappa_{j,\si}}
+\de^\cc_{f^\kappa_{j',\si'} f^\kappa_{j',\si'}}\right)
\M_0^{f^\kappa_{j,\si} \bar{f}^\kappa_{j',\si'} V^{a_1}_\rT \ldots\, V^{a_n}_\rT}
\nl&&{}+
\sum_{k=1}^n \delta^\cc_{V^{a'_k} V^{a_k}} 
\M_0^{f^\kappa_{j,\si} \bar{f}^\kappa_{j',\si'} 
V^{a_1}_\rT \ldots\, V^{a'_k}_\rT \ldots\, V^{a_n}_\rT},
\eeqar
where the correction matrix  associated to the external transverse gauge bosons 
is given in \refeq{deccWT}. Recall that, owing to  the non-diagonal $\delta^\cc_{AZ}$ component 
\refeq{CCAZmixing}, this matrix  gives rise to mixing between 
matrix elements involving external \PZ~bosons and photons.

For the processes of type \refeq{GBprodproc}, it turns out that in the high-energy
limit also the contribution of 
coupling-constant renormalization can be written as a sum over the
external gauge bosons. This can be easily shown, 
relating the physical gauge bosons in \refeq{GBprodproc} 
to their gauge-group eigenstate components \refeq{GBsymmbasis}.
In the high-energy limit, if one neglects all mass terms in the propagators,
the Born matrix element for the process \refeq{GBprodproc} can be written 
as\footnote{The gauge-boson indices of matrix elements transform 
as co-vectors \refeq{covectortransf}.}
\beq\label{GBprodrot}
\M_0^{f\bar{f} V^{a_1}_\rT \ldots\, V^{a_n}_\rT}=
\sum_{\sV^{a'_1},\dots,\sV^{a'_n}}
\widetilde{\M}_0^{f \bar{f} \sV^{a'_1}_\rT \ldots\, \sV^{a'_n}_\rT}
\prod_{k=1}^n U^{-1}_{\sV^{a'_k}V^{a_k}}(\thw),
\eeq
where the indices of the fermions have been suppressed.
The Born matrix elements on the rhs correspond to the processes
\beq\label{sGBprodproc}
f^\kappa_{j,\si} \bar{f}^\kappa_{j,\si'}\rightarrow 
\sV^{a'_1}_{\rT}\cdots\, \sV^{a'_n}_{\rT},
\eeq
where gauge-group eigenstates $\tilde{V}^{a'_k}=W^i$ or $B$ are produced.
The dependence of these processes on the coupling constants can be easily related to 
the external gauge-boson lines.
The Born matrix elements for \refeq{sGBprodproc}
are proportional to a factor $g_1$ for each $\Uone$ gauge boson
and a factor $g_2$ for  each $\SUtwo$ gauge boson. 
Therefore, using the diagonal matrix
\beq\label{couplmatg}
\smatg_{\sV^{a'}\sV^{a}}
= g_{1}\sdeone_{\sV^{a'}\sV^{a}}
+g_{2}\sdetwo_{\sV^{a'}\sV^{a}}, 
\eeq
with the $\Uone$ and $\SUtwo$ Kronecker matrices defined in \refeq{sadjKron}, 
we can write
\beq \label{symmbornampl} 
\widetilde{\M}_0^
{f\bar{f} \sV^{a_1}_\rT \ldots\, \sV^{a_n}_\rT}
=\sum_{\sV^{a'_1},\dots,\sV^{a'_n}}
\tilde{A}_0^
{f \bar{f} \sV^{a'_1}_\rT \ldots\, \sV^{a'_n}_\rT}
\prod_{k=1}^n \smatg_{\sV^{a'_k}\sV^{a_k}},
\eeq 
and \refeq{GBprodrot} becomes 
\beq\label{GBprodrot2}
\M_0^{f \bar{f}V^{a_1}_\rT \ldots\, V^{a_n}_\rT}
=\sum_{\sV^{a'_1},\dots,\sV^{a'_n}}
\tilde{A}_0^{f \bar{f} \sV^{a'_1}_\rT \ldots\, \sV^{a'_n}_\rT}
\prod_{k=1}^n \left[\smatg \U^{-1}(\thw)\right]_{\sV^{a'_k}V^{a_k}},
\eeq
where the amplitudes $\tilde{A}_0$ are independent of the coupling constants.
The renormalization of the coupling constants (and of the mixing angle) 
gives 
\beqar
\lefteqn{\hspace{-5mm}\de \left(\prod_{k=1}^n \left[\smatg \U^{-1}(\thw)\right]_{\sV^{a''_k}V^{a_k}}\right)
=}\quad&&\nl&=&
\sum_{l=1}^n\prod_{k\neq l} \left[\smatg \U^{-1}(\thw)\right]_{\sV^{a''_k}V^{a_k}}
\left[\de \smatg \U^{-1}(\thw)+
\smatg \de\U^{-1}(\thw)
\right]_{\sV^{a''_l}V^{a_l}}
\nl&=&
\sum_{V^{a'_1},\dots,V^{a'_n}}
\prod_{k=1}^n \left[\smatg \U^{-1}(\thw)\right]_{\sV^{a''_k}V^{a'_k}}
\times\nl&&
\sum_{l=1}^n
\left(\prod_{j\neq l} \de_{V^{a'_j}V^{a_j}}\right)
\left[U(\thw)\frac{\de\smatg}{\smatg} \U^{-1}(\thw)
+U(\thw)\de\U^{-1}(\thw)
\right]_{V^{a'_l}V^{a_l}}.
\eeqar
Therefore, the correction to \refeq{GBprodrot2} originating from 
the parameter renormalization
can be written as a sum over the external gauge bosons
\beqar \label{deRGgb}
\de^{\pre}
\M^{f \bar{f} V^{a_1}_\rT \ldots\, V^{a_n}_\rT}
&=& \sum_{k=1}^n \sum_{V^{a'_k}}
\delta^\pre_{V^{a'_k} V^{a_k}} 
\M_0^{f\bar{f} V^{a_1}_\rT \ldots\, V^{a'_k}_\rT \ldots\, V^{a_n}_\rT},
\eeqar
with 
\beq
\delta^\pre_{V^{a'} V^{a}} =
\left.\left\{\left[U(\thw)\frac{\de\smatg}{\smatg} \U^{-1}(\thw)
\right]_{V^{a'} V^{a}} 
+\left[U(\thw) \de U^{-1}(\thw)
\right]_{V^{a'} V^{a}} 
\right\}\right|_{\mu^2=s}.
\eeq
The first term originates from the renormalization of \refeq{couplmatg} and
is rotated as in  \refeq{adjmattransf}. With the rotated $\Uone$ and $\SUtwo$ 
Kronecker matrices \refeq{adjKron}, and according to \refeq{gCTs} it yields 
\beqar
\left[U(\thw)\frac{\de\smatg}{\smatg} \U^{-1}(\thw)\right]_{V^{a'} V^{a}} 
&=&\frac{\de g_1}{g_{1}}\,\deone_{V^{a'} V^{a}} 
+\frac{\de g_2}{g_{2}}\,\detwo_{V^{a'} V^{a}} 
\nl
&\LA&-\frac{1}{2}\frac{\alpha}{4\pi}\bew_{V^{a'}V^a}\log{\frac{\mu^2}{\MW^2}}+\frac{1}{2}\De\alpha (\MW^2) \de_{V^{a'}V^a},
\eeqar
where the matrix $\bew$, corresponding to the one-loop coefficients of the beta function, 
is defined in \refapp{app:betafunction}.
The second term is related to the renormalization of the Weinberg angle
\refeq{weinbergrenorm}, and in LA it yields 
\beq \label{WRMAT2}
\left[U(\thw) \de U^{-1}(\thw)\right]_{V^{a'} V^{a}}
=-\frac{\cw}{2\sw}\frac{\delta \cw^2}{\cw^2}\antikro_{V^{a'} V^{a}} 
\LA-\frac{1}{2}\frac{\alpha}{4\pi}\bew_{AZ}\antikro_{V^{a'} V^{a}}\log{\frac{\mu^2}{\MW^2}}, 
\eeq
where $\antikro$ is the antisymmetric matrix defined in \refeq{adjantiKron}.

Adding the contributions of parameter renormalization \refeq{deRGgb}
and the  mass-singular SL corrections  \refeq{deccWT} associated with  the 
external transverse gauge bosons, we obtain
\beq\label{PRpluscc}
\delta^\pre_{V^{a'} V^{a}} 
+ \delta^\cc_{V^{a'} V^{a}} =
\frac{1}{2}
\left(\de_{V^{a'} V^{a}}- \de_{V^{a}A} \de_{V^{a'}A} \right)\De\alpha (\MW^2)
+
\de_{V^{a'} V^{a}}Q^2_{V^{a}}\frac{\alpha}{4\pi}\log{\frac{\MW^2}{\la^2}}.
\eeq
In this sum all 
symmetric-electroweak logarithms, \ie all $\log{(s/\MW^2)}$ terms which grow with energy cancel.
Therefore, apart for the contributions in \refeq{gbprodcccorr}
that are associated with the fermionic initial states,   
only large logarithms of pure electromagnetic origin contribute to the
the complete SL corrections.
Note that this cancellation between PR and collinear logarithms 
is a consequence of Ward identities, 
like the identity between the electric charge and the photonic FRC in QED. 

\chapter{Leading logarithms from 2-point functions and their derivatives}
\label{app:2pointlogapp}
In this appendix, we present explicit results for the logarithmic approximation of 2-point functions and their derivatives. These have been used in \refch{FRCSllogs} and \refch{Ch:PRllogs} for the evaluation of the logarithmic contributions to field renormalization constants (FRC's),  and parameter-renormalization counterterms, respectively.

\section{Definitions}
For $n$-point functions we use the same notation as in \citere{Denner:1993kt}. In this appendix we restrict ourselves to the scalar 1-point function $A_0$, and the scalar and tensor 2-point functions $B_0,B_{\mu}$, and $B_{\mu\nu}$, which are defined by
\beqar
\frac{\ri}{(4\pi)^2} A_0(m_0)&:=&\mu^{4-D}\int \ddq \frac{1}{(q^2-m_0^2+\ri\varepsilon)},\\
\frac{\ri}{(4\pi)^2}B_{\{0,\mu,\mu\nu\}}(p,m_0,m_1)&:=&\mu^{4-D}\int \ddq\frac{\{1,q_\mu,q_\mu q_\nu\}}{(q^2-m_0^2+\ri\varepsilon) [(q+p)^2-m_1^2+\ri\varepsilon]}.\nn
\eeqar
These integrals are evaluated in $D=4-2\varepsilon $ dimensions, and $\mu$ is the mass scale introduced by the procedure of dimensional regularization. 
For completeness we also give the definition of the scalar 3-point function
\beqar\label{Cfunction}
\lefteqn{\frac{\ri}{(4\pi)^2}C_{0}(p_1,p_2,m_0,m_1,m_2):=} \\
&&\mu^{4-D}\int \ddq\frac{1}{(q^2-m_0^2+\ri\varepsilon) [(q+p_1)^2-m_1^2+\ri\varepsilon][(q+p_2)^2-m_2^2+\ri\varepsilon]},\nn
\eeqar
which is used in \refse{se:eikapp}.
The tensor 2-point integrals have the following Lorentz--invariant decompositions:
\beqar
B_\mu(p,m_0,m_1)&:=&p_\mu B_1(p^2,m_0,m_1),\nn\\
B_{\mu\nu}(p,m_0,m_1)&:=&g_{\mu\nu} B_{00}(p^2,m_0,m_1)+p_\mu p_\nu B_{11}(p^2,m_0,m_1).
\eeqar
Derivatives are denoted by
\beq
B'(p^2,m_0,m_1):=\frac{\partial}{\partial p^2}B(p^2,m_0,m_1).
\eeq

\section{Logarithmic approximation}
Generic massive self-energy diagrams
\beqar\label{selfenergydiagram}
\vcenter{\hbox{\begin{picture}(85,70)(0,-15)
\Text(30,-2)[t]{$m_0$} 
\Text(30,40)[b]{$m_1$} 
\Text(4,25)[b]{$p$} 
\Line(-10,20)(15,20) 
\Line(45,20)(70,20) 
\GCirc(30,20){15}{1}
\end{picture}}}
\eeqar
depend on four mass scales: the external mass $p^2$, the internal masses $m^2_0$ and $m^2_1$ 
and the scale $\mu^2$ of dimensional regularization.
In the following, we restrict ourselves to those diagrams
which contribute to the electroweak FRC's and to the coupling-constant counterterms. 
The corresponding scalar and tensor 2-point functions and their derivatives
are evaluated in the limit 
\beq
\mu^2\sim s \gg p^2,m_0^2,m_1^2,
\eeq 
in logarithmic approximation (LA). In this approximation, only logarithms involving  large  ratios of the scales $\mu^2, p^2,m_0^2,m_1^2$ are considered. The ultraviolet $1/\varepsilon$ poles as well as constant and mass-suppressed contributions are neglected. Also the imaginary part of the integrals is neglected, since we restrict ourselves to one-loop approximation.

As basic input for the logarithmic approximation we have used 
\beq\label{A0integral}
A_0(m_0)\LA m^2_0\log{\frac{\mu^2}{m_0^2}},
\eeq
for the 1-point function and the explicit expressions \cite{Denner:1993kt} 
\beqar
\label{B0integral}
B_0(p,m_0,m_1)&\LA&
\log{\frac{\mu^2}{m_0^2}} +\frac{1}{2}\left(1-\frac{m_0^2-m_1^2}{p^2}\right)\log{\frac{m_0^2}{m_1^2}} +\frac{1}{2}\frac{m_0m_1}{p^2}\left(r-\frac{1}{r}\right)\log{r^2},\nl
p^2 B'_0(p,m_0,m_1)&\LA&\frac{1}{2} \left\{ \frac{m_1^2-m_0^2}{p^2}\log{\frac{m_1^2}{m_0^2}}-\left[ \frac{m_0m_1}{p^2}\left(r-\frac{1}{r}\right)+\frac{r^2+1}{r^2-1}\right]\log{r^2}\right\},\nl
\eeqar
for the scalar 2-point function and its derivative, where $r$ is determined by 
\beq
r+\frac{1}{r}=\frac{m_0^2+m_1^2-p^2-\ri\varepsilon}{m_0m_1}.
\eeq
The  components $B_1,B_{00},B_{11}$ of tensor 2-point functions and the derivatives  $B'_1,B'_{00}$ can be obtained from \refeq{A0integral} and \refeq{B0integral} using simple reduction formulas \cite{Denner:1993kt}.

The formulas \refeq{B0integral} and the resulting formulas for the tensor components can be further simplified if some ratios of the masses  $p^2,m_0^2,m_1^2$  are very small (or very large). 
In this case, the result \refeq{B0integral} has to be expanded up to the needed power in the small mass ratios, and care has to  be taken if inverse powers of the small expansion parameters occur, especially in the reduction formulas.
 
Different expansions have  to be performed, depending on the hierarchy of the 
masses $p^2,m_0^2,m_1^2$. For this reason, we have to distinguish the following four cases
\beqar\label{hierarchies}
&(a)&\, m_i^2\ll p^2 \,\,\mbox{and}\,\, p^2-m_{1-i}^2\ll p^2 \quad\mbox{for}\,\, i=0 \,\,\mbox{or}\,\, i=1,\nl
&(b)&\, \mbox{not}\,\, (a)\quad\mbox{and}\quad m_i^2\not\gg p^2\quad\mbox{for}\,\, i=0,1,  \nl
&(c)&\, m_0^2=m_1^2\gg p^2,\nl
&(d)&\, m_i^2\gg p^2 \not\ll m_{1-i}^2\quad \mbox{for}\,\, i=0 \,\,\mbox{or}\,\, i=1,
\eeqar
which include the possible hierarchies  occurring   in electroweak self-energy diagrams.
Note that, in practice, the case ($a$)  occurs  only for diagrams involving a virtual photon with infinitesimal mass $m_i=\la$. Therefore,  we can restrict ourselves to the special case
\beq\label{photoniccase}
(a')\quad m_{i}^2=\la^2\ll p^2=m_{1-i}^2
\quad\mbox{for}\,\, i=0 \,\,\mbox{or}\,\, i=1.
\eeq

\section{Results}
It turns out that each 2-point function considered in this section has the same logarithmic approximation in all four cases \refeq{hierarchies}, apart from the derivatives $B'_0$ and  $B'_1$, which give rise to additional infrared logarithms in the case $(a')$.

In the explicit results presented below, the scale of the logarithms is determined by
\beq\label{scales}
M^2:=\max{(p^2,m_0^2,m_1^2)},\qquad
m^2:=\max{(m_0^2,m_1^2)},
\eeq
and the logarithmic approximation for $B_{11}$ can be read off from the results  for  $B_{00}$ and
\beq
g^{\mu\nu}B_{\mu\nu}(p,m_0,m_1)= D B_{00}(p^2,m_0,m_1)+p^2 B_{11}(p^2,m_0,m_1).
\eeq

\subsection*{Results for 2-point functions}
The  logarithmic approximations in this section are valid in all four cases \refeq{hierarchies}.
\beqar
B_0(p,m_0,m_1)&\LA&\log{\frac{\mu^2}{M^2}},\nl
B_1(p^2,m_0,m_1)&\LA&-\frac{1}{2}\log{\frac{\mu^2}{M^2}},\nn\\
\frac{1}{p^2}B_{00}(p^2,m_0,m_1)&\LA&\frac{3m_0^2+3m_1^2-p^2}{12p^2}\log{\frac{\mu^2}{M^2}},\nn\\
\frac{1}{p^2}g^{\mu\nu}B_{\mu\nu}(p^2,m_0,m_1)&\LA&\frac{m_0^2+m_1^2}{p^2}\log{\frac{\mu^2}{M^2}}.
\eeqar
For $p^2=0$ we have 
\beqar
g^{\mu\nu}B_{\mu\nu}(0,m_0,m_1)&=&D B_{00}(0,m_0,m_1)\LA(m_0^2+m_1^2)\log{\frac{\mu^2}{m^2}},\nn\\
B_0(0,m_0,m_1)&\LA&\log{\frac{\mu^2}{m^2}},\nn\\
B_1(0,m_0,m_1)&\LA&-\frac{1}{2}\log{\frac{\mu^2}{m^2}},
\eeqar
and since  $(m_0^2+m_1^2)/p^2 \log{(M^2/m^2)}$ is always suppressed, it follows that 
\beqar
B_0(p,m_0,m_1)-B_0(0,m_0,m_1)&\LA&-\log{\frac{M^2}{m^2}},\nn\\
B_1(p^2,m_0,m_1)-B_1(0,m_0,m_1)&\LA&\frac{1}{2}\log{\frac{M^2}{m^2}},\nn\\
\frac{1}{p^2}\left[B_{00}(p^2,m_0,m_1)-B_{00}(0,m_0,m_1)\right]&\LA&-\frac{1}{12}\log{\frac{\mu^2}{M^2}},\nn\\
\frac{1}{p^2}\left[g^{\mu\nu}B_{\mu\nu}(p^2,m_0,m_1)-g^{\mu\nu}B_{\mu\nu}(0,m_0,m_1) \right]&\LA&0.
\eeqar
Note that all results in this section are explicitly symmetric with respect to exchange of the internal masses $m_0$ and $m_1$.
\subsection*{Results for derivatives of 2-point functions}

In the cases $(b),(c),(d)$ in \refeq{hierarchies} the derivatives of $B_0$ and $B_1$ are given by 
\beq
p^2 B'_0(p,m_0,m_1)=p^2 B'_1(p^2,m_0,m_1)\LA 0,
\eeq
in logarithmic approximation. 
In the case $(a')$ in \refeq{photoniccase}, \ie for photonic diagrams, we have the additional infrared logarithms
\beqar
p^2 B'_0(p,m_0,m_1)&\LA&\frac{1}{2}\log{\frac{m_{1-i}^2}{m_{i}^2}}=\frac{1}{2}\log{\frac{p^2}{\la^2}},\nn\\
p^2 B'_1(p^2,m_0,m_1)+\frac{1}{2}p^2 B'_0(p,m_0,m_1)&\LA&-\frac{1}{4}\log{\frac{m_0^2}{m_{1}^2}}.
\eeqar
Note that the first expression is symmetric  with respect to exchange of the internal masses, whereas the second one  is antisymmetric. 
Finally,
\beq
B'_{00}(p^2,m_0,m_1)\LA-\frac{1}{12}\log{\frac{\mu^2}{M^2}}
\eeq
in all four cases  \refeq{hierarchies}.

\end{appendix}


\newcommand{\vj}[4]{{\sl #1~}{\bf #2~}\ifnum#3<100 (19#3) \else (#3) \fi #4}
 \newcommand{\ej}[3]{{\bf #1~}\ifnum#2<100 (19#2) \else (#2) \fi #3}
 \newcommand{\vjs}[2]{{\sl #1~}{\bf #2}}

 \newcommand{\am}[3]{\vj{Ann.~Math.}{#1}{#2}{#3}}
 \newcommand{\ap}[3]{\vj{Ann.~Phys.}{#1}{#2}{#3}}
 \newcommand{\app}[3]{\vj{Acta~Phys.~Pol.}{#1}{#2}{#3}}
 \newcommand{\cmp}[3]{\vj{Commun. Math. Phys.}{#1}{#2}{#3}}
 \newcommand{\cnpp}[3]{\vj{Comments Nucl. Part. Phys.}{#1}{#2}{#3}}
 \newcommand{\cpc}[3]{\vj{Comp. Phys. Commun.}{#1}{#2}{#3}}
 \newcommand{\epj}[3]{\vj{Eur. Phys. J.}{#1}{#2}{#3}}
 \newcommand{\fp}[3]{\vj{Fortschr. Phys.}{#1}{#2}{#3}}
 \newcommand{\hpa}[3]{\vj{Helv. Phys.~Acta}{#1}{#2}{#3}}
 \newcommand{\ijmp}[3]{\vj{Int. J. Mod. Phys.}{#1}{#2}{#3}}
 \newcommand{\jetp}[3]{\vj{JETP}{#1}{#2}{#3}}
 \newcommand{\jetpl}[3]{\vj{JETP Lett.}{#1}{#2}{#3}}
 \newcommand{\jmp}[3]{\vj{J.~Math. Phys.}{#1}{#2}{#3}}
 \newcommand{\jp}[3]{\vj{J.~Phys.}{#1}{#2}{#3}}
 \newcommand{\lnc}[3]{\vj{Lett. Nuovo Cimento}{#1}{#2}{#3}}
 \newcommand{\mpl}[3]{\vj{Mod. Phys. Lett.}{#1}{#2}{#3}}
 \newcommand{\nc}[3]{\vj{Nuovo Cimento}{#1}{#2}{#3}}
 \newcommand{\nim}[3]{\vj{Nucl. Instr. Meth.}{#1}{#2}{#3}}
 \newcommand{\np}[3]{\vj{Nucl. Phys.}{#1}{#2}{#3}}
 \newcommand{\npbps}[3]{\vj{Nucl. Phys. B (Proc. Suppl.)}{#1}{#2}{#3}}
 \newcommand{\pl}[3]{\vj{Phys. Lett.}{#1}{#2}{#3}}
 \newcommand{\prp}[3]{\vj{Phys. Rep.}{#1}{#2}{#3}}
 \newcommand{\pr}[3]{\vj{Phys.~Rev.}{#1}{#2}{#3}}
 \newcommand{\prl}[3]{\vj{Phys. Rev. Lett.}{#1}{#2}{#3}}                       
 \newcommand{\ptp}[3]{\vj{Prog. Theor. Phys.}{#1}{#2}{#3}}                     
 \newcommand{\rpp}[3]{\vj{Rep. Prog. Phys.}{#1}{#2}{#3}}                       
 \newcommand{\rmp}[3]{\vj{Rev. Mod. Phys.}{#1}{#2}{#3}}                        
 \newcommand{\rnc}[3]{\vj{Revista del Nuovo Cim.}{#1}{#2}{#3}}                 
 \newcommand{\sjnp}[3]{\vj{Sov. J. Nucl. Phys.}{#1}{#2}{#3}}                   
 \newcommand{\sptp}[3]{\vj{Suppl. Prog. Theor. Phys.}{#1}{#2}{#3}}             
 \newcommand{\zp}[3]{\vj{Z. Phys.}{#1}{#2}{#3}}                                
 \renewcommand{\and}{and~}

\newpage
\addcontentsline{toc}{chapter}{Acknowledgements}          
\centerline{\huge{
\textsc{Acknowledgements}}}
\vspace{2.5cm}
\begin{itemize}
\item I am very grateful to Dr.~Ansgar Denner
who proposed the interesting topic of my thesis and 
guided me during my work.
Thanks to his helpfulness I could profit a lot from his great experience.
His professional and human support was fundamental.

\item I am also grateful to Prof.~Daniel Wyler for his  support 
and for the stimulating atmosphere at the Institute of Theoretical Physics of the University of Z\"urich.

\item I would like to thank my colleagues of the Paul Scherrer Institut, 
in particular Elena Accomando and Michael Melles for their interest in my work.

\item I am thankful to Wim Beenakker and Anja Werthenbach for valuable discussions.

\item I am indebted to my parents
for their strong and continuous support during my studies.

\item I am very  grateful to Serena Fiscalini, 
who shared with me the difficulties and the nice moments
during the years of the studies and of the doctoral thesis.
I am particularly thankful for her help 
in the period when this work was written.

\end{itemize}

\end{document}